\documentclass[a4, nocrop, rm]{editorialupv}

\usepackage{rotating}
\usepackage{multirow}
\usepackage[export]{adjustbox}

\usepackage{lmodern}

\usepackage{setspace}
\usepackage{blindtext}
\usepackage{quotchap}

\usepackage{indentfirst} 

\usepackage{subcaption}
\usepackage{graphicx}
\usepackage{amsmath,mathtools}
\usepackage{amssymb}
\usepackage{epstopdf}
\usepackage{siunitx}
\usepackage{color}
\usepackage{float}
\usepackage{empheq}
\usepackage[ngerman,english]{babel}
\usepackage{fancyhdr}
\usepackage{multirow}
\usepackage{feynmp}
\usepackage{makecell}
\usepackage[toc,page]{appendix}

\newcommand{\cenuns}{CE$\nu$NS}
\newcommand{\ganess}{GanESS}

\newcommand{\fe}{$^{55}$Fe} 
\newcommand{\kr}{$^{83m}$Kr}

\newcommand{\mum}{$\mu$m}
\newcommand{\mus}{$\mu$s}
\newcommand{\gtwo}{$g_2$}

\newcommand{\etdCopyrightText}{
    \thispagestyle{empty} 
	\null\vfill
	\centerline{Copyright \copyright\ \number\year\ by \myName}
	\centerline{All Rights Reserved}
	\vskip 15pt\relax
}

\newcommand{\etdFrontMatter}[1]{
	\section*{#1}}

\newcommand{\etdPreTOCMatter}[1]{
	\chapter*{}
	\thispagestyle{empty}
	\begin{center}
		#1
	\end{center}
}

\newcommand{\dedication}[1]{\newcommand{\etdDedicationText}{#1}}

\newcommand{\makededication}{\etdPreTOCMatter{\etdDedicationText}}

\newcommand{\abstract}{\etdFrontMatter{Abstract}}

\newcommand{\makebibliography}{
	\cleardoublepage
	\phantomsection
    \renewcommand\bibname{References}
	\addcontentsline{toc}{chapter}{References}
	\begin{singlespace}
		\bibliography{PhD/refs.bib}
		\bibliographystyle{ieeetr} 
	\end{singlespace}
}

\newif\iffull\fulltrue

\graphicspath{{./figs/}}

\newcommand{\myName}{Leire Larizgoitia Arcocha}
\newcommand{\mySupervisor}{Francesc Monrabal Capilla}

\title{
	Towards high-pressure noble gaseous detectors for coherent elastic neutrino-nucleus scattering
    }

\author{ 
    A dissertation submitted in total fulfilment\\
    of the requirements for the degree of\\
    {Doctor of Philosophy}\\[0.5cm]
    by\\[0.5cm]
    \textbf{\myName}\\[1cm]
    under the supervision of\\
    \textbf{\mySupervisor}
}

\begin{document}

\frontmatter

\setstretch{1.05}
\iffull
	\maketitle	
\fi


\etdCopyrightText

\dedication{A mi aitete, sabiduría hecha abrazo.} 
\makededication

\clearpage

\setstretch{1.3}


\abstract{Coherent elastic neutrino-nucleus scattering (\cenuns) is a dominant low-energy neutrino interaction that remains experimentally challenging due to its weak-scale cross section and the small nuclear recoil energies involved. This thesis explores the scientific motivations and technical feasibility of \cenuns\ detection, emphasizing the use of diverse neutrino sources—reactor, spallation, and solar—and multiple target materials. The European Spallation Source (ESS) is identified as a particularly promising site, with simulations indicating an optimal signal-to-background ratio achievable with minimal shielding at a location $\sim$24 meters from the tungsten target. Alternative facilities, such as JPARC-MFL, are also evaluated given construction delays at ESS.

A significant contribution of this work is the development of a compact, low-cost, 4$\pi$ neutron scatter camera with integrated optical imaging, capable of characterizing neutron backgrounds and localizing sources using advanced discrimination algorithms and neural networks. 

Additionally, the thesis presents the design and early testing of GanESS, a novel high-pressure noble gas time projection chamber with electroluminescence amplification, optimized for \cenuns\ detection using argon, xenon, or krypton. The  Gaseous Prototype (GaP) demonstrates a promising energy threshold of 0.42$\pm$0.04 $\rm{keV}_{\rm{ee}}$ at 8.62 bar of argon and stable high-pressure performance.

Detailed simulations using Garfield++ and COMSOL provide insights into electroluminescence behavior and threshold estimation, although some non-linear detector responses at low E/p remain unresolved. 

Overall, this work establishes a robust foundation for \cenuns\ studies at spallation sources and advances detection technologies with broader implications for neutrino physics and rare-event searches.}

\bigbreak

\bigbreak
\noindent \textbf{Acknowledgments} \\ 
I was supported by the predoctoral training program non-doctoral research
personnel of the Department of Education of the Basque Government. I warmly thank the Fulbright Program too, as the neutron scatter camera project developed at the University of Chicago was conducted with the support of a US-Spain Fulbright grant.

\cleardoublepage


\ifEPUB
\else
	\cleardoublepage
	\phantomsection
	\tableofcontents
    \listoffigures
    \listoftables
\fi

\mainmatter


\renewcommand{\arraystretch}{1.5}

\begin{savequote}[65mm]
``Hitza, bihotzaren giltza"
\qauthor{-- Old proverb in Basque language.}
\end{savequote}

\chapter{Introduction}
\label{sec:introduction}

Coherent elastic neutrino-nucleus scattering (\cenuns) is a neutral-current process in the Standard Model predicted in 1974, in which a neutrino scatters off an entire nucleus via the exchange of a Z boson. The coherence condition is fulfilled when the momentum transfer is small enough that the wavelength associated with the exchanged boson exceeds the size of the nucleus. Under such condition, all nucleons contribute in phase, enhancing the cross-section approximately with the square of the neutron number, $\sigma (E_\nu) \propto N^2$. This renders \cenuns\ a powerful probe for studying weak interactions and nuclear structures, as well as a sensitive tool in searches for new physics, such as non-standard neutrino interactions, the neutrino magnetic moment, or sterile neutrinos.

Despite its relatively large cross-section, the experimental observation of \cenuns\ is challenging due to the extremely low recoil energies (a few keV) imparted to the nucleus. Only in recent years has detector technology advanced sufficiently to enable its detection. This has sparked growing interest in exploiting \cenuns\ as a platform for both fundamental physics and practical applications, including reactor monitoring and astrophysical neutrino detection, as the recent measurements in the XENONnT and PandaX-4T detectors.

Among the various neutrino sources available for \cenuns\ studies, spallation neutron sources are particularly well-suited. Facilities like the Spallation Neutron Source (SNS) in the United States and the European Spallation Source (ESS) in Sweden produce intense, pulsed beams of neutrinos from the decay-at-rest (DAR) of pions and muons generated by high-energy protons colliding with a heavy metal target.

The well-defined timing structure of these pulsed beams provides a powerful handle for background rejection. Prompt signals associated with pion decay can be temporally separated from delayed muon-decay neutrinos and random backgrounds. Additionally, the energy spectrum of the emitted neutrinos, peaking at tens of MeV, together with the large cross section of the channel, is ideal for \cenuns\ observation in smaller mass target nuclei compared to those oriented to the charge-current (CC) searches.

The European Spallation Source (ESS), currently under construction in Lund, Sweden, is poised to become the world’s most powerful neutron source. With its design of a 2 GeV proton beam and a high duty factor, it provides an unprecedented flux of low-energy neutrinos. This makes ESS a promising candidate for a long-term \cenuns\ experimental program as the GanESS project that is introduced in this thesis in chapter~\ref{sec:ganess}. 

However, to realize the full potential of \cenuns\ detection at ESS requires careful evaluation of the surrounding experimental environment. One of the main challenges is the presence of neutron-induced backgrounds. Spallation processes inherently generate a large number of neutrons, which can mimic the nuclear recoils expected from \cenuns, thus complicating the signal extraction.

To address this, the characterization of the neutron background at ESS is essential. A dedicated instrument, the Neutron Scatter Camera (NSC), has been developed for this purpose. The neutron camera will enable precise mapping of the fast neutron flux and angular distributions around potential detector sites. This information is crucial not only for background mitigation strategies but also for the optimal design and placement of future \cenuns\ detectors. Chapter~\ref{sec:ncamera} discusses the construction and the first validation of the neutron camera built, which enables the localization of a neutron source placed in every direction of a 4$\pi$ vision.

Among the various technologies under consideration for \cenuns\ detection, this thesis proposes for the first time the use of a high-pressure noble gaseous time projection chamber (TPC), the GanESS detector, built upon the technology pioneered by the NEXT experiment. The NEXT (Neutrino Experiment with a Xenon TPC) program has demonstrated the feasibility of achieving excellent energy resolution and background discrimination through electroluminescence (EL) amplification in xenon gas. The adaptation of this technology for \cenuns\ detection offers several advantages. The use of several gaseous targets within the same detector allows for a comprehensive study of \cenuns\ with a minimal increase in cost. Additionally, the EL signal amplification allows for low-noise charge readout, which is necessary as the process requires low energy threshold and high energy resolution.

To test and validate the proposed detection concept, a prototype detector named GaP (Gaseous Prototype) has been constructed at the Donostia International Physics Center (DIPC). Introduced in chapter~\ref{sec:gap}, GaP serves as a scaled-down version of the envisioned full detector and it is designed to evaluate the operational principles of a high-pressure TPC under conditions relevant for \cenuns\ detection. 

This thesis presents the first experimental results obtained with GaP, focusing in particular on the electroluminescence process, which is central to signal amplification in this technology. The measurements discussed in chapter~\ref{sec:gapresultschapter} are complemented by detailed simulations using the customized Garfield++ toolkit (chapter~\ref{sec:garfield}), which models the transport of electrons and behavior of photons in the gas under varying electric fields and pressures.

The overall structure of the thesis is organized as follows: Chapter~\ref{sec:cenuns} reviews the theoretical foundations of \cenuns\ and summarizes the current experimental status. Chapter~\ref{sec:cevnsess} discusses the advantages of spallation sources and especially evaluates the viability of observing \cenuns\ at the ESS. To do so, chapter~\ref{sec:ncamera} presents the Neutron Scatter Camera developed which will be used to characterize the neutron environment at ESS. Chapter~\ref{sec:ganess} introduces a novel detector approach: High-Pressure Noble Gas TPC, which stands by the name GanESS. A detector prototype (GaP) to validate the technology for studying \cenuns\ is detailed in chapter~\ref{sec:gap}. Chapter~\ref{sec:gapresultschapter} details its first experimental measurements, especially related to electroluminescence amplification and their interpretation. Chapter~\ref{sec:garfield} presents Garfield simulations and compares them with experimental data. Finally, chapter~\ref{sec:conclusion} summarizes the main results and outlines prospects for future work.				
\renewcommand{\arraystretch}{1.5}

\begin{savequote}[65mm]
``Lehen hala, orain hola, gero ez jakin nola"
\qauthor{-- Old proverb in Basque language.}
\end{savequote}

\chapter{Coherent elastic neutrino-nucleus scattering}
\label{sec:cenuns}

First described in 1974 \cite{freedman1974coherent}, Daniel Z. Freedman introduced the coherent elastic neutrino nucleus scattering (CE$\nu$NS) as a new Standard Model (SM) process. This scattering mechanism is a neutral-current process mediated by a \textit{Z}-boson and arises in the low energy regime when the momentum transfer is small enough that the neutrino scatters off the nucleus as a whole. 

\begin{figure}
\centering
\includegraphics[width=0.4\textwidth]{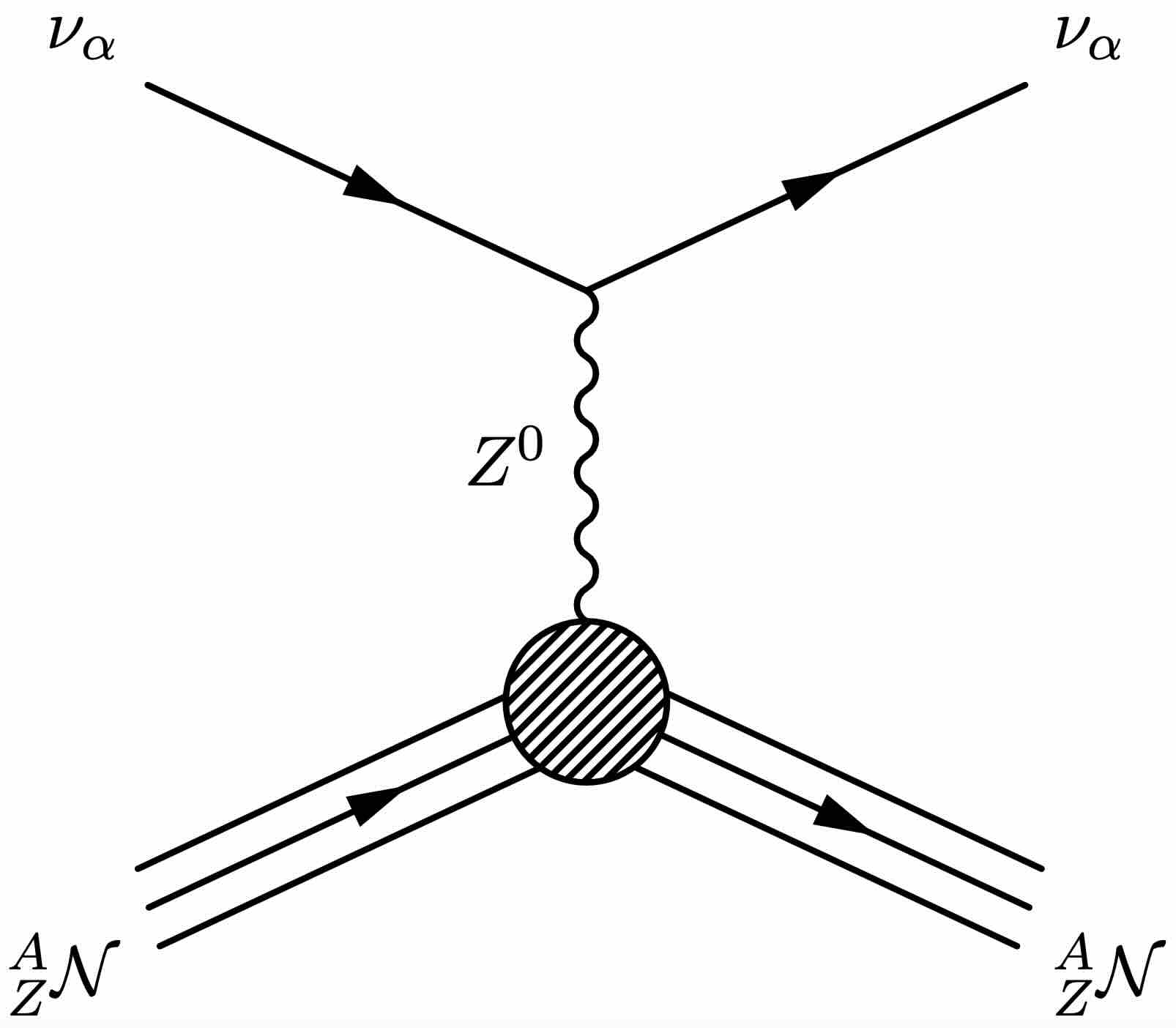}
\caption{Diagram of coherent elastic neutrino-nucleus scattering (CE$\nu$NS) in the SM. $\nu_\alpha$ here denotes any neutrino or anti-neutrino with flavor  $\alpha=e,\,\mu,\,\text{or}\,\tau$, and $^A_Z\mathcal{N}$ a nucleus with $A$ nucleons. $Z^0$ is the neutral vector boson mediator in neutral-current interactions in the SM.}
\label{fig:feyman}
\end{figure}

CE$\nu$NS interactions in the Standard Model can be represented as in equation~\ref{eq:cevns} and illustrated as in figure~\ref{fig:feyman}:

\begin{equation}
    \nu_\alpha + ^A_Z\mathcal{N} \rightarrow \nu_\alpha + ^A_Z\mathcal{N},
    \label{eq:cevns}
\end{equation} where $\nu_\alpha$ is any neutrino or anti-neutrino with flavor $\alpha=e,\,\mu,\,\text{or}\,\tau$ and $^A_Z\mathcal{N}$ exemplifies a nucleus with $A$ nucleons and $Z$ protons. This is a neutral-current process in the SM mediated by a $Z^0$ boson, hence, no electric charge is exchanged. 

The interaction between a neutrino and a nucleus depends on the nuclear radius $R$ and the three-momentum transfer scale $|\vec{q}|$, which is related to the de Broglie wavelength of the $Z^0$ neutral vector boson mediator: $\lambda_{Z^0}= h/|\vec{q}|$ ($h$ is the Planck's constant). At high energies, when $\lambda_{Z^0} \ll 2R$, the neutrino may fragment the nucleus ejecting a single nucleon via an inelastic incoherent scattering. When $\lambda_{Z^0} \lesssim 2R$, the neutrino interacts incoherently with individual nucleons, resulting in an elastic incoherent scattering on single nucleons. The specific nature of \cenuns\ happens in the regime where $\lambda_{Z^0} \gtrsim 2R$, enabling an interaction with the nucleus as a whole. This coherency condition results in nucleons recoiling in phase with one another. 

As the only observable of this process is the recoiling nucleus, one can relate $|\vec{q}|$ with the measurable kinetic energy ($T$) of the recoiling nucleus. In a typical two-body elastic scattering, the velocity transferred to the nucleus at rest can be kinematically derived and therefore, the momentum transfer and the recoil energy. If $E_{\nu}$ is the energy of the incident neutrino, the minimum neutrino energy to recoil the nucleus with energy $T$ is $E_\nu^{\min}(T) = \frac{1}{2} (T + \sqrt{T^2 + 2T m_N})$, where $m_N$ is the mass of the nucleus. In the same way, for a given neutrino energy, assuming $m_N \gg m_\nu$, a back-scatter is the limiting case, where $|\vec{q}|$ is maximized and the kinetic recoil energy too (equation~\ref{eq:Tmax}):

\begin{equation}
    T_{\text{max}} =  \frac{2 E_{\nu}^2}{2 E_{\nu} + m_N}.
\label{eq:Tmax}
\end{equation}

Since the squared momentum transfer of the process can be written as $|\vec{q}|^2 = 2 E_\nu^2 m_N T / (E_\nu^2 - E_\nu T)$, the coherence condition is satisfied only in the low energy regime. \cenuns\ will be produced by low-energy neutrinos and the recoil energies will be very small too. From the experimental perspective, if heavy nuclei are used as the detector targets, such as xenon (A=132) or CsI (A=133, 127), neutrino scatters will lose coherence at $E_\nu \gtrsim 30$ MeV, which implies $T \lesssim 15$ keV. Consequently, the detection energy thresholds ($T_{th}$) achievable by different technologies, along with the nuclei-dependent coherence limit, restrict the type of detectors sensitive to this process. This also constrains the suitable neutrino sources. There are various experiments sensitive to natural low-energy neutrino sources, however, to get precision measurements, laboratory sources (nuclear reactors and spallation sources) are more adequate as they provide a controlled and intense low-energy neutrino flux. Section~\ref{sec:observation} provides an extensive overview of the observation of \cenuns\ up to date.

\section{Coherent elastic neutrino-nucleus scattering cross section}

Insensitive to neutrino flavor conversion, the SM CE$\nu$NS differential cross section on a spin-0 nucleus with $Z$ protons and $N$ neutrons at tree-level (radiative corrections set to zero) is \cite{formfactor}:

\begin{equation}
    \frac{d\sigma(T,E_\nu)}{dT} = \frac{G_F^2 m_N}{2\pi} F^2(|\vec{q}|^2) \frac{Q_W^2(Z,N)}{4} \left[2 - \frac{2T}{E_\nu} + \left(\frac{T}{E_\nu}\right)^2 - \frac{m_N T}{E_\nu^2}\right] ,
    \label{eq:crosssection}
\end{equation} where $G_F$ is the Fermi coupling constant and $F(|\vec{q}|^2)$ is the form factor of the nucleus evaluated at the squared momentum transfer of the process $|\vec{q}|$. $Q_W(Z,N) = N + (4\sin^2\!\vartheta_{W} - 1)Z$ is the weak hypercharge of a nucleus and $\vartheta_{W}$ is defined as the weak mixing angle, with $\sin^2\!\vartheta_{W} \simeq 0.2386$ at small momentum transfer \cite{weakmixing}. From this expression, the proton contribution is suppressed ($4\sin^2\!\vartheta_{W} -1 \approx 0$), and hence, the \cenuns\ cross-section is essentially proportional to $N^2$.

The form factor is the Fourier transform of the distribution of neutrons and protons within the nucleus. It captures the coherence loss in momentum transfer due to the size of the nucleus. Assuming a spherical nucleus, it is possible to write the form factor as \cite{formfactor}:

\begin{equation}
    F(|\vec{q}|) = \frac{1}{Q_W} \left[ F_n(|\vec{q}|) + (4\sin^2\!\vartheta_{W} - 1)F_p(|\vec{q}|) \right],
\end{equation} where $F_n(|\vec{q}|)$
and $F_p(|\vec{q}|)$ are the neutron and proton form factors, respectively. Since the proton contribution is heavily suppressed at low momentum transfer, the neutron form factor dominates the determination of CE$\nu$NS cross section \cite{formfactor}, so the neutron form factors can be explored.

Different form factor models based on phenomenological approximations of the nuclear charge density can be used. The Helm \cite{formfactorhelm1956inelastic}, Klein-Nystrand \cite{formfactorklein1999} or Skyrme \cite{formfactorBartel1982-SkM} parametrizations are the most extended models. In the context of the CE$\nu$NS cross sections, for the energy ranges handled, the influence that each form factor model has is similar because it is normalized to fully coherent zero-momentum transfers: $F^2(|\vec{q}|=0) = 1$. 
The parametrization used to illustrate this chapter on \cenuns\ interactions is the Klein-Nystrand distribution. This is obtained from the convolution of a Yukawa potential of range $a = 0.7$ fm over a Woods-Saxon nuclear density profile of radius $R_A = A^{1/3}\cdot 1.2$ fm \cite{formfactorklein1999}. The analytical expression of the form factor is described in equation~\ref{eq:klein}:

\begin{equation}
    F_{\text{KN}} (|\vec{q}|) = \frac{4\pi \rho_0}{A|\vec{q}|^3} \cdot \left[  \sin{(|\vec{q}|R_A)} - (|\vec{q}|R_A) \cos{(|\vec{q}|R_A)}\right]  \cdot \frac{1}{1+a^2|\vec{q}|^2}
    \label{eq:klein}
\end{equation} 

with the normalization density given by

\begin{align}
    \rho_0 = \frac{A}{\frac{4}{3}\pi R_A^3}.
\end{align}

\begin{figure}[htb!]
    \begin{tabular}{@{}cc@{}}
    \includegraphics{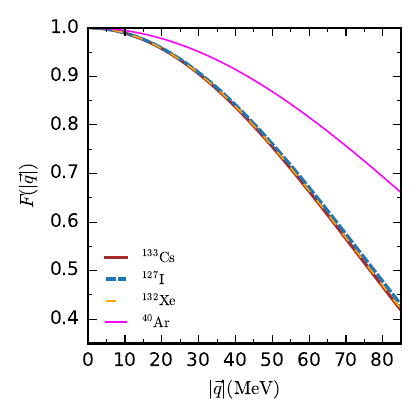}  &
    \hspace{0.1in}
    \includegraphics{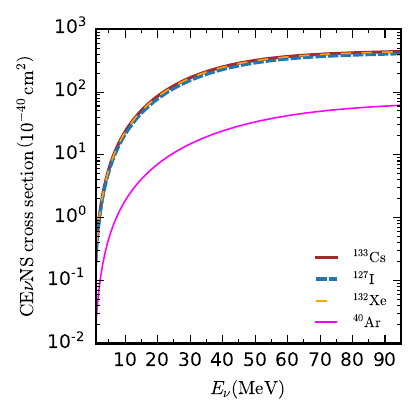}
    \end{tabular}
\caption{Representation for $^{133}\text{Cs}$, $^{127}\text{I}$, $^{132}\text{Xe}$ and $^{40}\text{Ar}$ isotopes. Left: The nuclear form factor according to the Klein-Nystrand parametrization. Right: The \cenuns\ cross section for as a function of the incoming neutrino energy. Note the overlap between Cs, I and Xe as they have similar atomic numbers, and the smaller cross-section for Ar as it is a lighter nucleus.}
\label{fig:formfactor_cross-sec}
\end{figure}

To present the distributions of the form factor in figure~\ref{fig:formfactor_cross-sec} left, 
the target elements of the first two \cenuns\ observations were evaluated, CsI and argon (Ar), as well as xenon (Xe). The similar atomic numbers in Cs, I and Xe suggest a comparable form factor, which therefore hints at an analogous response for the cross section. Thus, including this element in the study is interesting. For illustration only the most abundant isotope per element is adopted in figure~\ref{fig:formfactor_cross-sec} left: $^{133}\text{Cs}$, $^{127}\text{I}$, $^{40}\text{Ar}$ and $^{132}\text{Xe}$. Nevertheless, for \cenuns\ target materials like argon or xenon, the weighted contribution of every present isotope should be accounted for in the total cross section calculation. Note that in the same way, for composite nuclei, such as CsI, the total nuclear recoil rate from \cenuns\ is the sum of events expected from the weighted interaction with each individual element.  

Despite this form factor model being valid for the low energy regime, at higher energies the distribution goes to zero ($F^2(|\vec{q}|) \rightarrow 0$), indicating the loss of coherence. The neutrino will start to interact with individual nucleons, so at these energies the CE$\nu$NS channel does not occur.

The total elastic cross section for a given element as a function of the neutrino energy, $\sigma (E_\nu)$, is then found by integrating equation~\ref{eq:crosssection} from the detector recoil energy threshold ($T_{th}$) to the maximum recoil energy expressed in equation~\ref{eq:Tmax}, as described in equation~\ref{eq:sigma} and shown figure~\ref{fig:formfactor_cross-sec} right. 

\begin{equation}
    \sigma (E_\nu) =\int_{T_{th}}^{T_{\max}}  \frac{d\sigma(T,E_\nu)}{dT} dT.
    \label{eq:sigma}
\end{equation}

Paying attention to the dependence of the cross section with respect to the neutrino energy in the \cenuns\ process is key to understand the difficulty of observing this process. On the one hand, for higher energies we expect higher recoil energies and easier detection capabilities, however, the suppression of the form factor at higher $|\vec{q}|$, which translates as large recoil energies induced by higher neutrino energies, results in a negligible contribution to the total cross-section. 
On the other hand, we get a similar situation with the selection of the isotope for our experimental measurement. Light isotopes will produce larger recoil energies, thus easier to observe, but with a cross section much smaller than for heavier nuclei, as it depends quadratically with the square of the number of neutrons: $\sigma (E_\nu) \propto N^2$. 

\begin{figure}[tbp]
    \centering
    \includegraphics{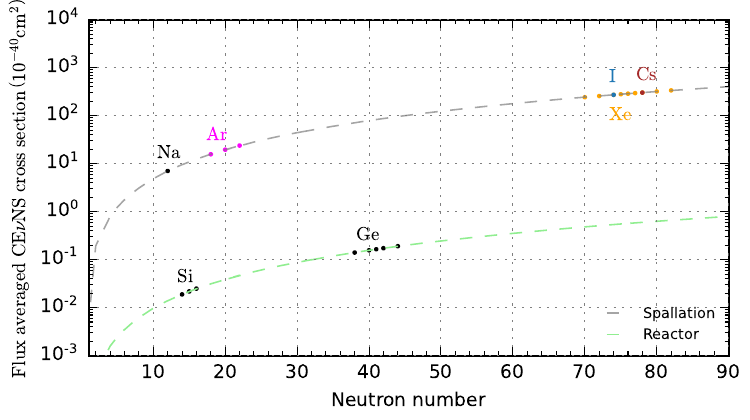} 
    \caption{Flux-averaged cross section as a function of the neutron number for both spallation and reactor. Specific isotopes of conventional target materials for \cenuns\ are also marked. Klein-Nystrand parametrization is used for the nuclear form factor and setting this to unity will have a visible steeper deviation as the neutron number increases.}
    \label{fig:N2}
\end{figure}

Precisely, figure~\ref{fig:N2} shows the flux-averaged cross section dependence on the neutron number for both spallation and reactor energy regimes. Specific isotopes of conventional target materials for \cenuns\ are also shown. A deviation from a $N^2$ prediction can be a signature of beyond-the-SM physics.

\begin{figure}[tbp]
    \centering
    \includegraphics[width=0.6\textwidth]{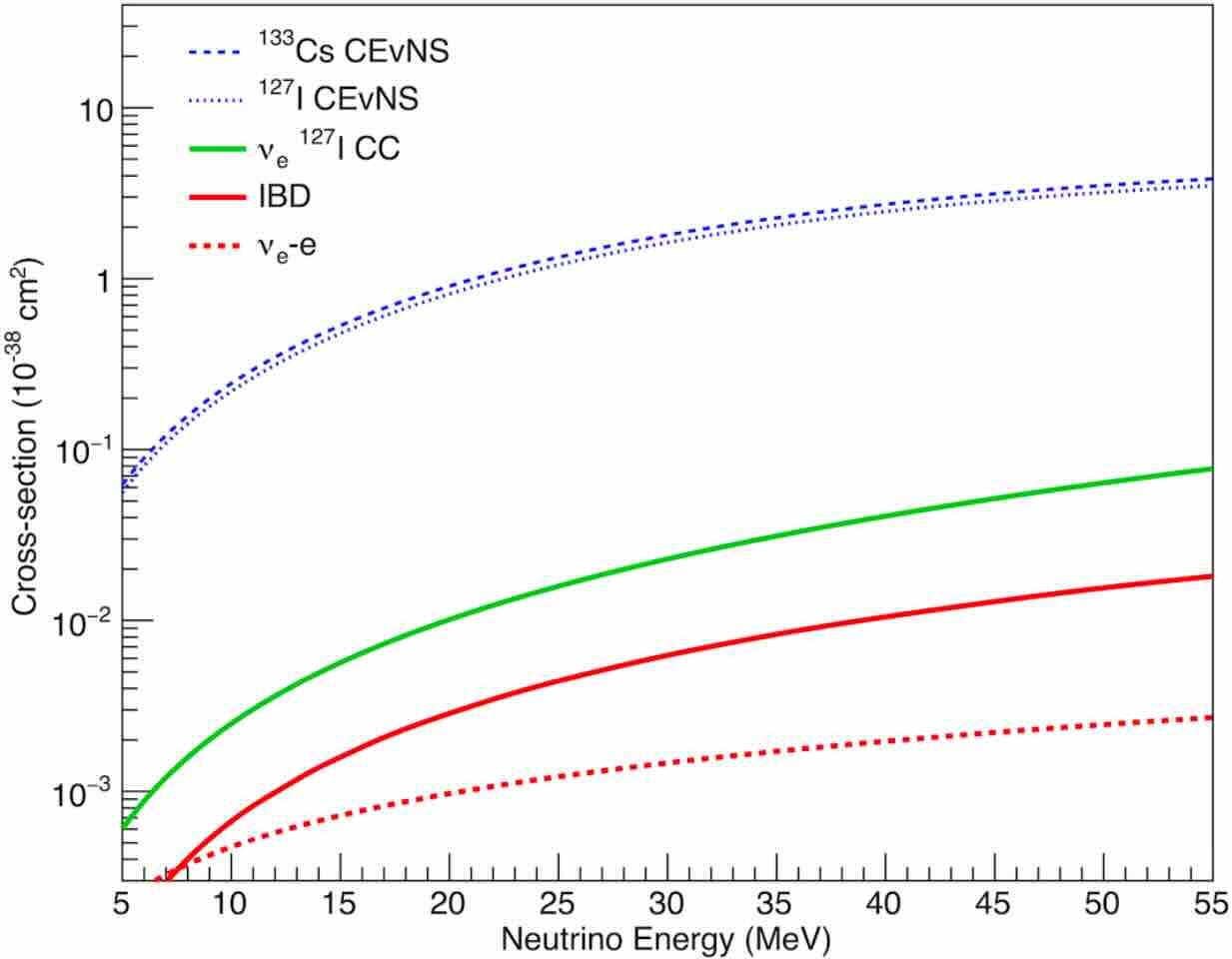} 
    \caption{Illustration of the total cross-section for several neutrino interactions present at spallation and reactor neutrino energy ranges. \cenuns\ cross-section with cesium and iodine (blue), charged-current (CC) interaction with iodine (green), inverse beta decay (red) and neutrino-electron scattering (dotted red) are shown. \cenuns\ dominates over any charged-current interaction for incoming neutrino energies of less than 55 MeV. Plot extracted from \cite{scholz2018first}.}
    \label{fig:neutrino_cross-sec}
\end{figure}

The result of approximately scaling  with the square of the target nucleus’ neutron number is that the scattering cross-section is several orders of magnitude larger than any other neutrino-nucleus coupling as figure~\ref{fig:neutrino_cross-sec} illustrates. Consequently, \cenuns\ is the dominant channel for neutrino interactions at these low energies. 

A good experimental approach should combine then a technology with heavy nuclei and ultra-low threshold, and the appropriate neutrino source.

\section{Coherent elastic neutrino-nucleus scattering signal}
\label{sec:cenunssignal}
In the nuclear recoil energy space, the time-integrated differential number of events for the \cenuns\ signal reads:

\begin{equation}
    \frac{dN}{dT} = \sum_{\nu_{\alpha}} \mathcal{N} \int_{E_\nu^{\min}(T)} dE_\nu f_{\nu_{\alpha}}(E_\nu) \frac{d\sigma(T,E_\nu)}{dT}, 
    \label{eq:dndt}
\end{equation} where the sum is over all present neutrino and antineutrino flavors. $\mathcal{N}$ is a normalization constant that depends on the number of nuclei in the detector, and on the number of protons on target (POT) in spallation sources. The minimum neutrino energy ($E_\nu^{\min}$) required to induce a nuclear recoil depends on the recoil energy itself together with the mass of the target nucleus. $f_{\nu_{\alpha}}(E_\nu)$ is the neutrino flux, dependent on the neutrino source. $d\sigma(T,E_\nu) / dT$ is the coherent elastic neutrino-nucleus scattering cross section defined in equation~\ref{eq:crosssection}.

It is straightforward to determine the total number of \cenuns\ events expected (before accounting for detector characteristics like efficiency, energy resolution or quenching factor) in an interaction by integrating the differential scattering rate (equation~\ref{eq:dndt}) above a given nuclear recoil energy threshold ($T_{th}$), which depends on the detector technology. 

\section{Coherent elastic neutrino-nucleus scattering observation}
\label{sec:observation}

The observation of \cenuns\ must be approached taking into account two aspects. On the one hand, the identification of sufficiently intense neutrino sources at the low energy regime so that the coherence condition is preserved. On the other hand, the use of detection techniques that allow the observation of small nuclear recoils. 

In this section, an overview of \cenuns\ observation is described from the experimental side. The possible neutrino sources that may interact via \cenuns\ are introduced as well as the technological requirements for the detectors. Moreover, a brief narrative of the current experiments that have observed this interaction is included. In section~\ref{sec:potential} the interest in the observation \cenuns\ for its physics potential is brought out. 

As a global picture of the needs and difficulties to observe and study CE$\nu$NS, table~\ref{tab:threshold} summarizes the maximum nuclear recoil energy that reactor and spallation neutrinos may originate in different nuclei detectors: $T_{\max} (E_\nu^{\max})$. Considering that reactors bring neutrinos with energies up to $\sim$10 MeV and spallation sources up to $\sim 53$ MeV, even at laboratory sources which provide a controlled neutrino flux, the detector threshold that should give at least a visible signal for the maximally recoiled nucleus (equation~\ref{eq:Tmax}) is very low. From equation~\ref{eq:Tmax} one can estimate the maximum nuclear recoil energy that will be generated for a given neutrino energy. In fact, for precision measurements, the detectors should provide a wide range of visible recoil energies below that maximum recoil energy considered in each case. Table~\ref{tab:threshold} indicates that limit case for various targets to be sensitive to \cenuns\ nuclear recoil. As argon is the lightest element shown, its maximum nuclear recoil energy will be the largest for a given neutrino energy. Note the large recoil energy difference that spallation sources and reactor neutrinos induce in every case, as the maximum neutrino energy in the latter is more than six times smaller. 
These values are considered to be very low compared to other physics detectors, but as it is described in section~\ref{sec:technologies} there are sophisticated technologies capable of reaching thresholds around $\sim$1 $\text{keV}_{\text{nr}}$. Additionally, as it will be later introduced in the chapter, the visible energy in the detector is lower than that of the nuclear recoil due to some energy loss that it is quantified by the quenching factor (QF). 

\begin{table}[h!]
\centering
\begin{tabular}{|c || c | c|} 
 \hline
 Target & $T^{\text{spallation}}_{\max}(\text{keV}_{\text{nr}})$ & $T^{\text{reactor}}_{\max}(\text{keV}_{\text{nr}})$\\ [0.5ex] 
 \hline\hline
 Argon & 149.58 & 3.45  \\ [0.5ex] 
 \hline
 Xenon & 45.60  & 1.05  \\ [0.5ex] 
 \hline
 Germanium & 82.38 & 1.90  \\  [0.5ex] 
 \hline
 Cesium Iodide   & 46.09  & 1.06  \\  [0.5ex] 
 \hline
\end{tabular}
\caption{Theoretical calculations of the maximum nuclear recoil energy for neutrinos interacting with several detector targets at the maximum neutrino energy from spallation sources and reactors, 52.8 MeV and 8.01 MeV respectively. The technologies used need thresholds lower than these values in order to be sensitive to the \cenuns\ process. Note that for argon, xenon and germanium all the stable isotopes with their corresponding abundance are considered. Furthermore, for xenon the isotope $^{136}\text{Xe}$ and for germanium $^{76}\text{Ge}$ have also been considered. These are isotopes that undergo double beta decay with a very long half-life.}
\label{tab:threshold}
\end{table}

\subsection{Neutrino sources}

The total effective cross section of \cenuns\ increases with the incident neutrino energy, up to a limit imposed by the coherence condition of the process (see figure~\ref{fig:formfactor_cross-sec} right). An adequate neutrino source for \cenuns\ measurements should provide an intense flux of low-energy neutrinos and preferably, with a well-known energy spectrum.

There are natural sources of neutrinos with energies in the suitable range for CE$\nu$NS which include: geoneutrinos ($\nu_e$ originated from $\beta$ decays of elements present in the core of the Earth), core-collapse supernova neutrinos, solar neutrinos (specifically those produced from $^8$B) and the low energy tail of atmospheric neutrinos. However, the flux of these neutrinos is low ($\sim 10^6 \text{cm}^{-2}\text{s}^{-1}$), except for neutrinos from remanent supernovae. This unpredictable source can contribute to higher fluxes of the order of $\sim 10$ MeV, which are emitted after the gravitational core-collapse of a star. Experiments dedicated to direct dark matter (DM) searches, especially, are sensitive to \cenuns\ events induced by these natural neutrino sources \cite{dutta2019darkmatterdetectors}. Indeed, candidates like Weakly Interacting Massive Particles (WIMPs) with masses of $\mathcal{O}(10\,\mathrm{GeV/}c^2)$ induce nuclear recoils in the keV range \cite{cadeddu2023view}, analogous to the \cenuns\ process. Recent data from the PandaX-4T and XENONnT experiments (section~\ref{sec:technologies}) provide the first indications of direct measurement of CE$\nu$NS nuclear recoils from solar neutrinos in DM experiments, as well as the first \cenuns\ observation on a xenon target. Furthermore, these constitute the first observations entering the \textit{neutrino fog} \cite{NeutrinoFog}, an irreducible background in direct dark matter experiments, as potential dark matter signals become indistinguishable from neutrino background signals.
 
Nuclear power reactors also produce neutrinos in the region of interest. Here, $\bar{\nu}_e$ with energies up to $\sim 10$ MeV are released via $\beta$ decay of neutron-rich nuclei. $^{235}$U, $^{238}$U, $^{239}$Pu or $^{241}$Pu are the main fission products. To have an estimate of the flux, the $\bar{\nu}_e$ flux that a nuclear power plant produces with a power of 1 GW is $\sim 2 \cdot 10^{20} \, \bar{\nu}_e \, \text{s}^{-1}$ \cite{giunti2007fundamentals}. This powerful source will provide precision measurements using an adequate detector. Moreover, \cenuns\ channel can be used for reactor monitor operation. Although there are different experiments and proposals for the measurement of \cenuns\ in nuclear reactors, the only experiments that have reported compatible signals so far are the Dresden-II experiment (NCC-1701) and the CONUS experiment as outlined in section~\ref{sec:reactor}. 

Particle accelerators offer complementary opportunities to reactors. In particular neutron spallation sources, which rely on colliding a proton linear beam with a target of high atomic number. This provides a flux of three types of neutrinos ($\nu_\mu, \bar{\nu}_\mu, \nu_e$) from pion-decay-at-rest ($\pi$-DAR) with energies below $\sim 53$ MeV. In 2017, the COHERENT experiment was the first to report a \cenuns\ measurement as described in section~\ref{sec:coherent}. 

In this direction, the European Spallation Source (ESS), currently under construction, represents a new opportunity to exploit \cenuns\ physics as presented in chapter~\ref{sec:cevnsess}.

\subsection{Detector technologies}
\label{sec:technologies}
The Standard Model predicts that the effective \cenuns\ cross section is proportional to $N^2$, being $N$ the number of neutrons for a given atom (see figure~\ref{fig:N2}). Therefore, the use of high neutron number nuclei as the target material maximizes the probability of interaction. Yet, it does not ensure the observation of the interaction. The only observable of the \cenuns\ channel is the recoil of the nucleus. As mentioned earlier in this chapter, for heavy nuclei, nuclear recoils of the order of few-tens-of-keV are induced. However, the energy observed by the detection systems (keVee, electron-equivalent energy) is even lower due to energy dissipation in the form of heat or atomic motion, for example. The quenching factor (QF) quantifies the observed reduction in yield produced by a nuclear recoil compared to an electron recoil, which is energy and nuclei-dependent. Even if the theoretical prediction by Lindhard et al. \cite{lindhard1963integral} is one of the approaches, dedicated QF measurements are required at such low energies to analyze \cenuns\ signals properly for the techniques mentioned in this work, in contrast to bolometer-type detectors that can measure the total nuclear recoil energy. This emphasizes the need of developing technologies with very low detection thresholds, which is a challenging requirement.

To develop Dark Matter experiments, over the last decades several technologies to read low-energy nuclear recoils have been developed: CsI or NaI scintillators; germanium semiconductors; liquid argon (LAr) time projection chambers (TPCs); dual-phase (liquid-gas) detectors or low-background CCDs sensors. So far, these technologies have been brought to the \cenuns\ field, however, this work proposes for the first time the use of a noble gas high-pressure TPC (operable with different gases). Even if gaseous phase detectors had not been considered due to the relatively low density compared to solids (CsI, Ge) or liquids (LAr), a sufficiently intense neutrino flux will ensure high statistics. This project is detailed in chapter~\ref{sec:ganess}.

\subsubsection{The COHERENT experiment}
\label{sec:coherent}

The COHERENT experiment is developed at the Spallation Neutron Source (SNS) in Oak Ridge Laboratory (Tennessee, USA). Briefly describing the operation of the SNS (see chapter~\ref{sec:cevnsess} for details on spallation sources), a $\sim$1 GeV proton short-pulsed beam collides with a mercury (Hg) target spallating neutrons. In the process pions ($\pi^{\pm}$) are produced too and the positive pions decay at rest ($\sim 26$ ns) into $\mu^{+}$ and $\nu_\mu$. These $\mu^{+}$ later decay again ($\sim 2.2 \, \mu$s) with the process  $ \mu^{+} \rightarrow e^{+} + \bar{\nu}_\mu + \nu_e$ generating a delayed flux of neutrinos. For the SNS, the total neutrino flux is about $4.3 \cdot 10^{7} \nu \, \text{cm}^{-2} \, \text{s}^{-1}$ and the neutrino energies are below $\sim 53$ MeV, which enables to develop a \cenuns\ experiment with small detectors in comparison to typical neutrino detectors (multiple of tons mass scales). In 2017, the COHERENT experiment observed for the first time \cenuns, with 6.7$\sigma$ confidence, using a 14.6 kg CsI[Na] scintillator detector \cite{akimov2017observation,akimov2018coherent}. In following years, the collaboration approached a multi-target perspective, claiming in 2020 a positive signal with 3$\sigma$ on a 24 kg liquid argon scintillator detector \cite{akimov2021firstAr,akimov2020coherent}. Recently, in 2024, they also reported the observation of CE$\nu$NS signal in a germanium detector with an effective mass of 10.66 kg and a significance of 3.9$\sigma$ \cite{adamski2024firstGe}.  

\subsubsection{CE$\nu$NS at nuclear reactors}
\label{sec:reactor}

The Dresden-II experiment obtained a measurement compatible with the presence of \cenuns\ events with evidence $>$ 3$\sigma$ \cite{lewis2023particle} exposing a 2.924 kg ultra low-noise p-type point contact (PPC) germanium detector (NCC-1701) to the Dresden-II nuclear reactor (Illinois, USA) \cite{colaresi2021first,colaresi2022measurement}. The flux of $\bar{\nu}_e$ from the reactor, with energies below $\sim$10 MeV, is estimated to be about $4.8 \cdot 10^{13}  \, \bar{\nu}_e \, \text{cm}^{-2} \text{s}^{-1}$ with a $\sim 2 \%$ uncertainty \cite{huber2011determination,mueller2011improved}. 

The CONUS experiment \cite{bonet2021constraints} which took data until 2022 at a nuclear reactor power plant (KBR) in Germany, moved to Switzerland (KKL) to continue as CONUS+ \cite{ackermann2024conus+}. The experimental setup is composed of several high-purity germanium (HPGe) detectors with a total fiducial germanium mass of 3.73 $\pm$ 0.02 kg. In 2025, CONUS announced their first possible \cenuns\ measurement with reactor antineutrinos using a germanium target claiming a statistical significance of 3.7$\sigma$ \cite{conus2025}. 

The observation of \cenuns\ reported (concerning the background-only hypothesis) strongly depends on the quenching factor considered. Both QF models considered at the Dresden-II experiment are large compared to the Lindhard et al. \cite{lindhard1963integral} theoretical prediction below 1.35 $\text{keV}_{\text{nr}}$ \cite{collar2021germanium}. In the case of CONUS, however, the Lindhard model has been assumed. Thus, an incompatibility is shown between the two results. 

\subsubsection{CE$\nu$NS at Dark Matter experiments}

Published at the same time, November 2024, the first measurement of \cenuns\ nuclear recoils from solar $^8$B neutrinos was announced in the PandaX-4T experiment~\cite{PandaXPhysRevLett133191001} and the XENONnT experiment~\cite{aprile2024first}. The main goal of these experiments is the detection of Dark Matter particles. However, other applications include the measurement of the solar $^8$B neutrino flux via \cenuns\ with xenon nuclei. While PandaX-4T is a dual-phase time-projection chamber filled with 3.7 tonnes of liquid xenon (LXe) at the China Jinping Underground Laboratory, XENONnT is a filled with 5.9 tonnes of active mass LXe situated in the Laboratori Nazionali del Gran Sasso (LNGS).

The background-only hypothesis is disfavored at 2.64$\sigma$ and 2.73$\sigma$ significance for PandaX-4T and XENONnT, respectively. 

The result also provides a complementary measurement of $^8$B solar neutrinos flux to dedicated neutrino experiments \cite{aharmim2013combined,abe2011measurement,agostini2020improved,abe2016solar}. PandaX-4T reported a flux of $8.4\pm 3.1 \cdot 10^{6}  \, \nu  \, \text{cm}^{-2} \text{s}^{-1}$ which is also consistent with the standard solar model prediction \cite{vinyoles2017new}. XENONnT indicated that, assuming no new physics, the measured $^8$B solar neutrinos flux is $4.7^{+3.6}_{-2.3} \cdot 10^{6}  \, \nu  \, \text{cm}^{-2} \text{s}^{-1}$, consistent with results from the Sudbury Neutrino Observatory (SNO) \cite{aharmim2013combined}. This results in a neutrino flux-weighted \cenuns\ xenon cross section of $1.1 ^{+0.8}_{-0.5} \cdot 10^{-39} \text{cm}^{2}$, consistent with the SM prediction.

\section{Physics potential}
\label{sec:potential}

\cenuns\ is a neutral current neutrino-matter interaction with a large physics potential. There is an ongoing effort to explore this channel within and beyond the Standard Model (BSM). In quantum field theory, several quantities like the value of $\sin^2\!\vartheta_{W}$ or the neutron distribution radii depend on the energy scale at which they are measured. \cenuns\ is a new channel that will allow to explore wider physics models that are not accessible through traditional neutrino experiments.

\subsection{Electroweak physics}

The determination of the weak mixing angle at small momentum transfer may show deviations from the SM prediction. For the moment, high-precision measurements are performed at the energy scale of 100 GeV in high energy colliders \cite{PDGPhysRevD.98.030001}, and now \cenuns\ offers a new channel to characterize the weak mixing angle at lower energies \cite{weakmixing}. 

\subsection{Nuclear physics}

The compatibility of the cross-section to the square of the number of neutrons in the target nucleus accounts for full coherence. Therefore, in a realistic case where a small amount of incoherency is expected, powerful information can be obtained on the nuclear structure and as a consequence, on the neutron distribution in the nucleus by quantifying a realistic form factor. Indeed, the proton distribution is well known from electromagnetic measurements \cite{angeli2013chargeradii}, while the neutron distribution requires measurements mediated by weak or strong forces. Information on the effective charge radii of neutron distribution in the nucleus can be extracted from the \cenuns\ recoil spectrum \cite{krauss1991low}, which can later be combined with Atomic Parity Violation (APV) experiments \cite{wood1997measurement,guena2005measurement} to improve its determination. The measurements available so far from \cenuns\ by the COHERENT experiment are in agreement with the Nuclear Shell Model (NSM) predictions \cite{hoferichter2020coherent,cadeddu2021new}. However, measurements with higher statistics are required to improve this comparison.

\subsection{New physics}

A neutrino-matter interaction sensitive to all active neutrino flavors is a unique resource to go BSM and explore possible neutral current oscillations or non-standard interactions (NSI). Any observed oscillation would introduce the so-called sterile neutrino \cite{giunti2011sterileneutrino}, providing a feasible answer to the anomalies presented in \cite{aguilar2001evidence} and \cite{aguilar2018significant}. Therefore, evidence of eV-scale sterile neutrino can be explored \cite{Blanco2020sterileneutrino}.

One could also look for contributions to \cenuns\ from BSM interactions where a flavor-changing neutral current is involved. Among the least exotic models, different possible mediators lead to this scenario like a photon ($\gamma$), a new massive vector boson ($Z'$) or a scalar boson ($\Phi$). A photon mediator considers the possibility that neutrinos interact electromagnetically. This allows searches for neutrino properties like the small electric charge (for an overview, see \cite{giunti2015neutrino}) called ``millicharge" or like the neutrino magnetic moment \cite{vogel1989neutrino}. 

\begin{figure}[tbp]
    \centering
    \includegraphics[width=1\textwidth]{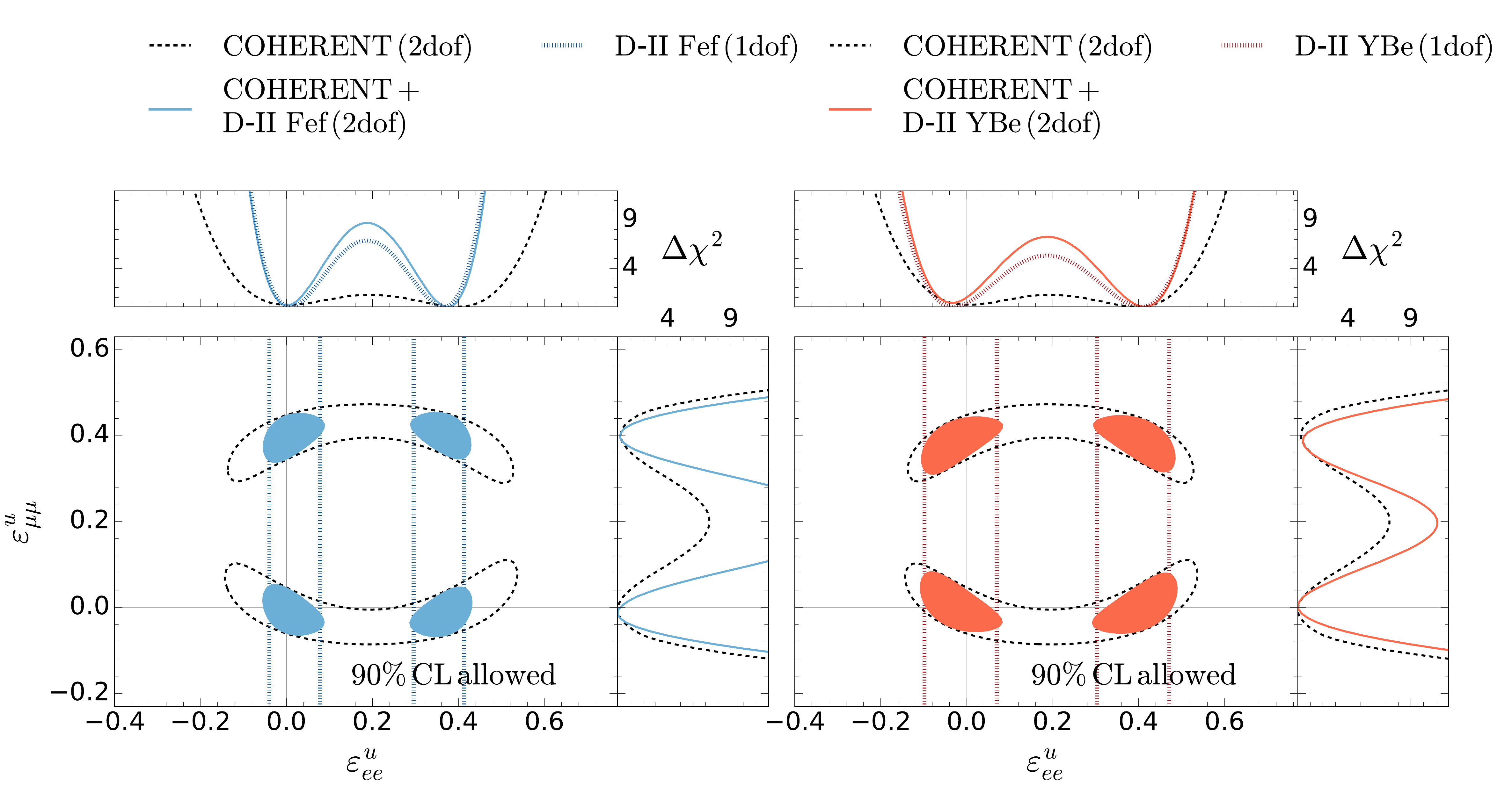} 
    \caption{Extracted from \cite{Coloma2022}. The analysis of COHERENT CsI and Ar data, the Dresden-II reactor data (Fef and YBe quenching factor in the respective panels) and their combination for 90$\%$ CL allowed regions on flavor diagonal NSI with up-quarks (for zero values of all other NSI coefficients). Note that, in the two-dimensional panels, the results are obtained for 2 dof ($\Delta \chi^2 = 4.61$) except for the Dresden-II reactor experiment data which is obtained for 1 dof ($\Delta \chi^2 = 2.71$).}
    \label{fig:NSI}
\end{figure}

To maximize the physics potential of CE$\nu$NS, combining the information from different neutrino sources and nuclei is necessary \cite{baxter2020coherent}. Reactor neutrinos ($\lesssim 10$ MeV) will provide results of CE$\nu$NS interactions, as those of the Dresden-II experiment~\cite{colaresi2022measurement}, ensuring full coherence with a single flavor, $\bar{\nu}_e$, as opposed to spallation sources. This will be complementary to measurements in spallation sources, like the COHERENT experiment at the SNS. In the same way, a multi-target approach will help break degeneracies and set bounds for new physics models. In particular, we published an analysis that combined data from these two experiments: ``Bounds on new physics with data of the Dresden-II reactor experiment and COHERENT" \cite{Coloma2022}. Extracted from there, figure~\ref{fig:NSI} demonstrates that a joint analysis helps constrain models BSM. 

The recent evidence of CE$\nu$NS recoil signal in a dark matter experiment announced by both the XENONnT and PandaX-4T collaboration could be included in the analysis too, as it is already done in various phenomenological studies \cite{de2024bounds} for the determination of the weak mixing angle at low momentum transfer \cite{maity2024first}, new physics in the form of NSI \cite{sierra2024implications,li2024constraints} or light mediators~\cite{de2024bounds}.

\renewcommand{\arraystretch}{1.5}

\begin{savequote}[65mm]
``A ze parea karakola eta barea" 
\qauthor{-- Old proverb in Basque language.}
\end{savequote}

\chapter{Coherent elastic neutrino-nucleus scattering at the European Spallation Source}
\label{sec:cevnsess}
The European Spallation Source (ESS) is a multi-disciplinary research facility located in Lund, Sweden, which is currently approaching its first-stage completion. The ESS is being built to provide one of the world's most intense pulsed neutron beams. The Spallation Neutron Source (SNS) in the Oak Ridge National Laboratory (ORNL), USA, was the first operational spallation facility in 2006. The Japan Proton Accelerator Research Centre (J-PARC) in Tokai, Japan, followed in 2009. In Europe, the development of a world-leading neutron spallation source was approved in 2003. However, it was not until 2009 that Lund was established as the site for the project \cite{garoby2017european}. 

This chapter overviews the feasibility of performing \cenuns\ measurements at the European Spallation Source considering the ramp-up plan and the available areas for neutrino experiments. These are locations with no assigned role but where the neutron background does not blind the \cenuns\ signal.

\section{Neutrinos at the European Spallation Source}
\label{sec:facility}

The essentials of any spallation research facility are an accelerator and a target. The ESS linear accelerator (linac) has a normal conducting region followed by three superconducting cavities and a high-energy beam transport (HEBT) that will inject protons into the spallation target. The ESS target is a helium-cooled rotating wheel with a diameter of 2.5 m and 36 tungsten sectors to ensure high beam current density \cite{garoby2017european}.

The ESS was originally designed to provide a long-pulse with an average beam power of 5 MW in bunches of 2.86 ms long at a pulse repetition rate of 14 Hz, bombarding 2 GeV protons at full operation. This adopted a nominal proton rate of $\sim 1.6 \cdot 10^{16} \, \text{p}/s$. However, as it is described later in section~\ref{sec:future}, ESS has encountered some inconveniences. Now, the ESS intends to operate in the near term providing a proton beam with energies between 800 MeV and 2 GeV at a beam power of 2 MW \cite{ESSreport2023}.

The accelerated protons will impact the target nuclei with such high energy that will interact only with individual nucleons. 
Via elastic collisions the proton energy will be transferred into the nucleons, ensuing an intra-nuclear cascade of nucleon collisions. The course of the nucleon cascade lasts about $10^{-22}$ s, where the nucleus is left in an excited state. It is followed by a $\sim 10^{-16}$ s energy dissipation, mainly via neutron evaporation. Over this fast process, a few tens of high energy neutrons per incoming proton are spallated. These are beam-related neutrons, hereafter, the neutrons spallated directly from beam-target collisions are labeled as \textit{prompt} neutrons.

During the process, not only neutrons are produced but both $\pi^{-}$ and $\pi^{+}$ too. The $\pi^{-}$ are mostly absorbed by the target nuclei before they can decay, while $\pi^{+}$ propagate before they decay-at-rest (DAR). Also known as $\pi$-DAR, the process is described in equation~\ref{eq:pidar}:

\begin{equation}
\begin{split}
    & \pi^{+} \longrightarrow \mu^{+} + \nu_{\mu} \\
    & \mu^{+} \longrightarrow e^{+} + \bar{\nu}_{\mu} + \nu_e,
\end{split}
\label{eq:pidar}
\end{equation} where $\mu^{+}$ decays in-flight and emits two delayed neutrinos. The $\nu_{\mu}$ is considered as the \textit{prompt} neutrino since its emission timescale follows the beam profile. The $\bar{\nu}_{\mu}$ and $\nu_e$ are therefore called \textit{delayed} neutrinos, as the muon decay ($\sim$ 2.2 $\mu$s) is much longer. The emission spectra can be analytically derived. The \textit{prompt} $\pi^{+}$-DAR, for a vanishing neutrino mass, is a two-body decay problem that results in a mono-energetic $\nu_{\mu}$ of 

\begin{equation}
    E_{\nu_{\mu}} = \frac{m_{\pi}^2 - m_{\mu}^2 }{2 m_{\pi}} \simeq 29.8 \mathrm{MeV},
    \label{eq:Enumu}
\end{equation} where $m_{\pi}$ and $m_{\mu}$ refer to the pion and muon masses, respectively. Therefore, the energy distribution that $\nu_{\mu}$ follows is described in equation~\ref{eq:fEnumu}:

\begin{equation}
    f_{\nu_{\mu}} (E_{\nu}) = \delta (E_{\nu} - 29.8 \mathrm{MeV}).
    \label{eq:fEnumu}
\end{equation}

The \textit{delayed} neutrinos from the $\mu^{+}$ decay follow continuous flux distributions described in equations~\ref{eq:fEnuantimu} and \ref{eq:fEnue} for energies $E_{\bar{\nu}_{\mu}, \nu_{e} } < m_{\mu}/2 \simeq 52.8$ MeV. Normalized to one,

\begin{equation}
    f_{\bar{\nu}_{\mu}} (E_{\nu}) = \frac{64}{m_{\mu}} \left[ \left(\frac{E_{\nu}}{m_{\mu}} \right)^2 \left( \frac{3}{4} - \frac{E_{\nu}}{m_{\mu}} \right) \right]
    \label{eq:fEnuantimu}
\end{equation}

\begin{equation}
    f_{\nu_{e}} (E_{\nu}) = \frac{192}{m_{\mu}} \left[ \left(\frac{E_{\nu}}{m_{\mu}} \right)^2 \left( \frac{1}{2} - \frac{E_{\nu}}{m_{\mu}} \right) \right].
    \label{eq:fEnue}  
\end{equation}

\begin{figure}
    \centering
\includegraphics{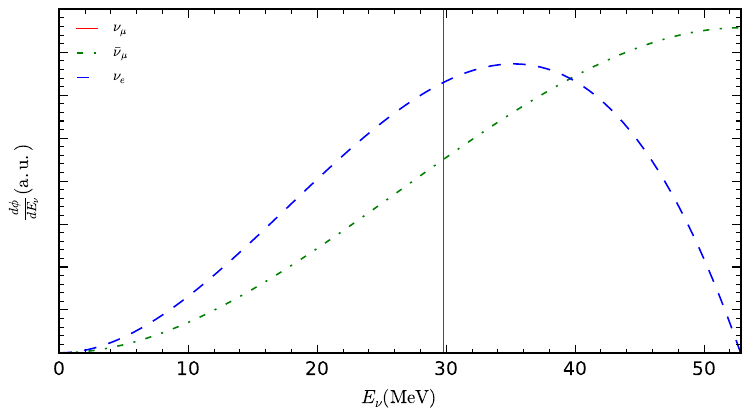} 
    \caption{Normalized neutrino flux spectra expected, in arbitrary units (a.u.) as a function of the neutrino energy from a stopped-pion source, for instance, spallation sources. Each of the components corresponds to a neutrino flavor.}
    \label{fig:ESSflux}
\end{figure}

\section{Expected coherent elastic neutrino-nucleus scattering signal}
\label{sec:expectedsignal}

For the calculations done in this section the original design values were used. They can easily be adjusted to the present conditions scaling the corresponding normalization factor. 

\begin{figure}[htb!]
    \begin{tabular}{@{}cc@{}}
    \hspace{0.1in}
\includegraphics{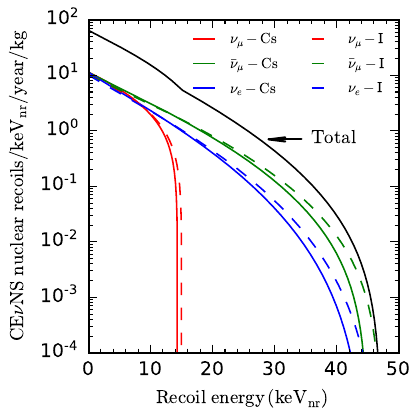}  &
    \hspace{0.1in}
\includegraphics{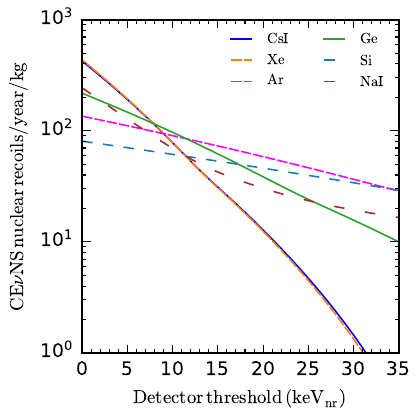}
    \end{tabular}
\caption{Left: Calculated \cenuns\ nuclear recoil rate placing a pure CsI detector 20 m away from the ESS target. The solid lines show the contribution of cesium recoils from each neutrino flavor and the dashed lines for iodine. Right: Expected \cenuns\ events above a nuclear recoil threshold for several detector materials. The detector is 20 m away from the ESS target.}
\label{fig:recCsI-recthreshold}
\end{figure}

Considering the discussion in section~\ref{sec:cenunssignal}, one can calculate the expected \cenuns\ rate at ESS for a given detector medium. The total nuclear recoil rate from \cenuns\ interaction is the sum of the differential recoil spectra from each isotope or element, in the case of composite nuclei, evaluated for every emitted neutrino flavor (equation~\ref{eq:dndt}). The ESS is scheduled to operate 5000 hours per year with a design proton delivery rate of $\sim 1.6 \cdot 10^{16} \, \text{p}/s$. Adopting a yield of 0.3 neutrinos of each flavor per proton as in \cite{baxter2020coherent}, $\sim 8.5 \cdot 10^{22}$ neutrinos of each spallation-flavor per year are expected at the ESS. As it will later be discussed, available ESS sites $\sim 20$ m away from the tungsten target indicate viability for \cenuns\ studies. Considering this as the location reference for a detector, one can calculate the total recoil rate in the detector. As an example, figure~\ref{fig:recCsI-recthreshold} left shows the contribution of each spallation neutrino type interacting with each CsI component coherently and the total nuclear recoil rate of a pure CsI detector. 

The integration of the nuclear recoil rate above an energy threshold estimates the total \cenuns\ events visible in a detector (without any signal acceptance cuts). Figure~\ref{fig:recCsI-recthreshold} right shows the expected \cenuns\ events visible for different detector mediums 20 m from the ESS target as proposed in \cite{baxter2020coherent}. Each of the \cenuns\ targets will provide unique advantages and the use of a variety of targets will contribute to better sensitivity in constraining physics BSM. Concretely, it is interesting to mention the overlap between the expected signal in a xenon and a CsI detector. Their similar number of neutrons per nucleus adopts this behavior, as the \cenuns\ cross section is directly related to the square of the number of neutrons. This ideal pairing is a very interesting tool to exploit neutrino interactions via \cenuns\ and look for physics BSM, as the systematics from each detector technology are completely different but a similar response is expected. 

\section{Coherent elastic neutrino-nucleus scattering backgrounds at the ESS}
\label{sec:bckg}

A major problem in \cenuns\ measurements at spallation sources is the beam-related backgrounds, but there are order types of backgrounds expected too, as the steady-state backgrounds or neutrino-induced neutrons (NIN). 

The latter is predominantly produced by the charge-current interaction of the delayed $\nu_e$ in heavy elements like Pb, Fe, or Cu \cite{lewis2023particle}, particularly $^{208}\text{Pb}(\nu_e,e^{-})^{208}\text{Bi}$ reaction in equation~\ref{eq:pb208}, which is usually used for gamma shielding \cite{collar2015coherent}. The energies of the neutrinos from $\pi$-DAR are above the threshold for the activation of these common shielding elements and result in neutron emissions. It has been shown that the NIN's contribution to \cenuns\ background is negligible \cite{akimov2017observation,scholz2018first}. Nevertheless, it is reducible with the addition of a layer of few tens of centimeters of high-density polyethylene (HDPE) to the detector. 

\begin{equation}
\begin{split}
    \nu_e \thinspace + \thinspace ^{208}\text{Pb} \thinspace \rightarrow \thinspace {}&^{208}\text{Bi}^{*} \thinspace + \thinspace e^{-} \\
    &\downarrow\\
    {}&^{208-y}\text{Bi} \thinspace + \thinspace x \times \gamma \thinspace + \thinspace y \times  n,
\end{split}
\label{eq:pb208}
\end{equation}where $x$ and $y$ are the gamma and neutron multiplicity, respectively  \cite{scholz2018first}.

Steady-state contributions are those induced by cosmic-ray or environmental radioactivity. Since the ESS will provide a pulsed beam, the neutrino emission profile will be pulsed too. Therefore, the \cenuns\ signals can only arise directly after the protons-on-target (POT) trigger (section~\ref{sec:facility}). This background can be continuously monitored and then subtracted, as it is not correlated with the beam time.

Another contribution and the one evaluated in this chapter are beam-related backgrounds. Although spallated neutrons lose energy as they propagate through the monolith surrounding the target, some will escape the shielding and reach certain locations in the facility. In particular, the higher energy neutrons are of special interest, since they can provide nuclear recoils temporally coincident with the neutrino emissions and possible interactions, and therefore, this background cannot be suppressed. Full-scale simulations were performed in MCNP~\cite{MCNP6} (see \cite{lewis2023particle} for details) whereas in this chapter the Geant4~\cite{geant4} (section~\ref{sec:nsim}) simulations developed to evaluate the average neutron flux per proton sent down the beamline are described. Geant4 is a broadly used toolkit to simulate the interactions of particles through matter, especially in the high-energy regime. In recent years there has been an effort to implement higher precision neutron interaction calculations. Therefore, it is interesting to push towards characterizing the neutron background for \cenuns\ experiments employing Geant4.

To determine the feasibility of the ESS for the next-generation \cenuns\ detectors, beam-related neutron backgrounds at available locations have been compared to the expected \cenuns\ signal in section~\ref{sec:shielding}. As in every experiment simulations must be corroborated, so eventually, dedicated neutron background measurements will be performed at the ESS experimental site. To assess this, a full-coverage imaging neutron scatter camera has been developed and it is ready to characterize the neutron spectrum when the ESS becomes operational. Chapter~\ref{sec:ncamera} describes this in detail. 

\subsection{Available locations}

There are several sites at the ESS facility without assigned roles \cite{ESSpersonnel} and as such, possible locations for \cenuns\ experiments. 

\begin{figure}
    \centering
    \includegraphics[width=0.8\textwidth]{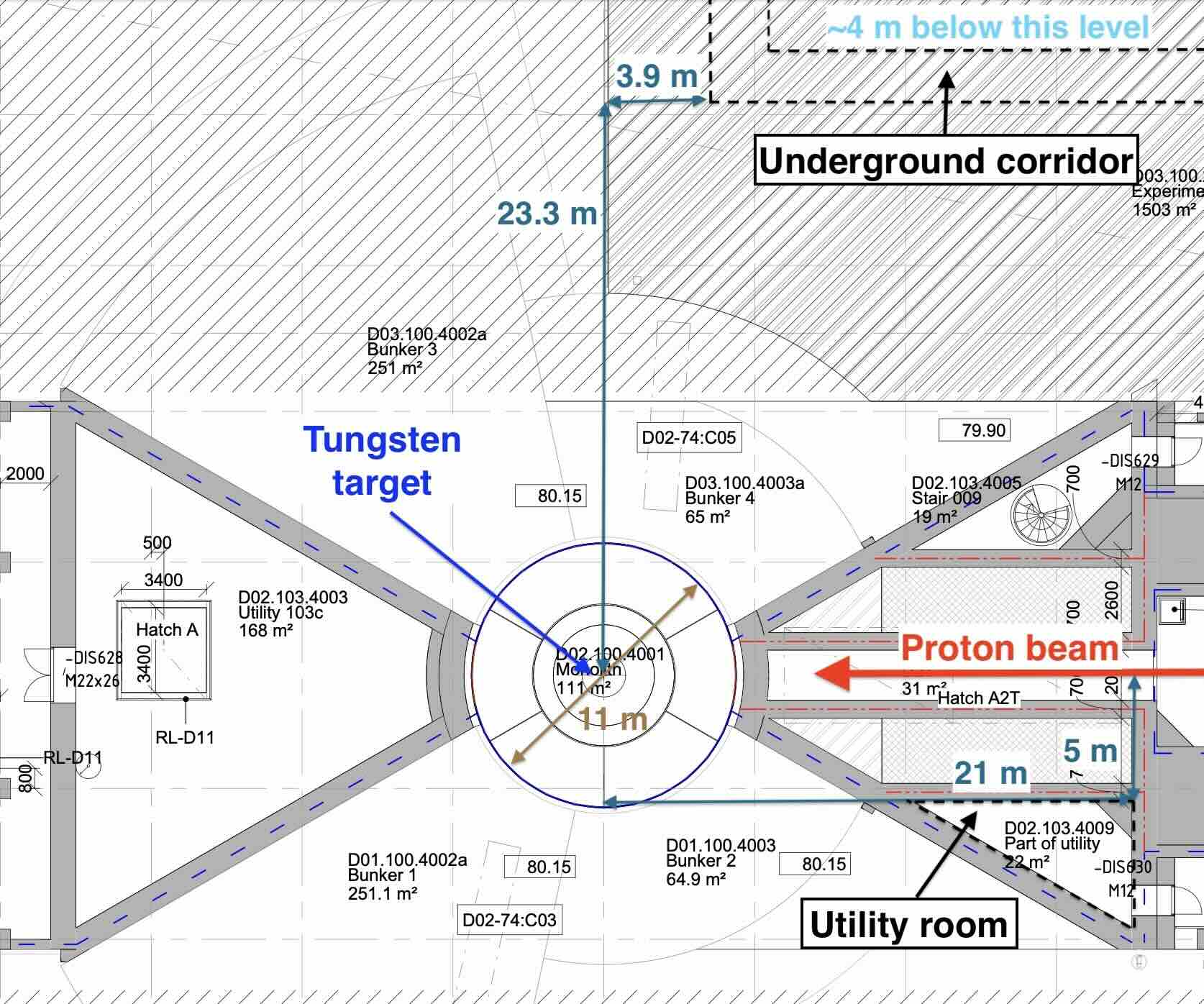} 
    \caption{Modified from the blueprint of the central level at the ESS facility. This target station technical plan was updated in November 2017. Marked in black the utility room and the underground corridor available for neutrino experiments are shown. The proton beam (red) and tungsten target (blue) are indicated too.}
    \label{fig:ESSphoto}
\end{figure}

An indispensable consideration for determining whether these locations are appropriate for neutrino physics is that the \cenuns\ signal stays unmasked by the neutron background associated with the beam and particles propagated along the target building. Simulations can reasonably resolve the beam-related neutron backgrounds at those locations, even if dedicated measurements are planned with a neutron scatter camera. Preliminary studies of available sites at the facility identified two possible locations. Later, an extensive study for the two locations of interest was performed using MCNP6~\cite{MCNP6} to model the process of 2 GeV protons impacting the tungsten target and the transport of the spallated neutrons  \cite{lewis2023particle}. Visible in figure~\ref{fig:ESSphoto}, which shows the blueprint at the central level of the building, the first location is a room with $\sim$21 $\text{m}^2$ utile, located next to the beamline separated by 38$\%$ enriched in magnetite high-density concrete~\cite{MagnaDense}. Because of its shape, this utility room is also labeled as \textit{triangle room}. The second space of interest is an auxiliary corridor in a sub-level $\sim$4 m below the ESS target (see dashed lines in the drawing from figure~\ref{fig:ESSphoto}) and $\sim$24 meters from the target separated by part of the building foundations and the monolith.  

\subsection{Neutron background simulations}
\label{sec:nsim}

ESS personnel V. Santoro, L. Zanini, and Z. Lazic provided architectural information and relevant dimensions on the various sites of the ESS installation together with MCNP parts of the relevant geometry. Putting this together \cite{LewisPrivate}, the analogous simulation in MCNP to model the beam-related backgrounds has been performed in Geant4. This is detailed in the following sections.

The two main sources of spallation identified are the protons impacting the tungsten target and protons scattering off residual gases in the beamline.

\subsubsection{Geometry and materials}

\begin{figure}[ph!]
    \begin{tabular}{@{}cc@{}}
    \includegraphics[width=.5\textwidth]{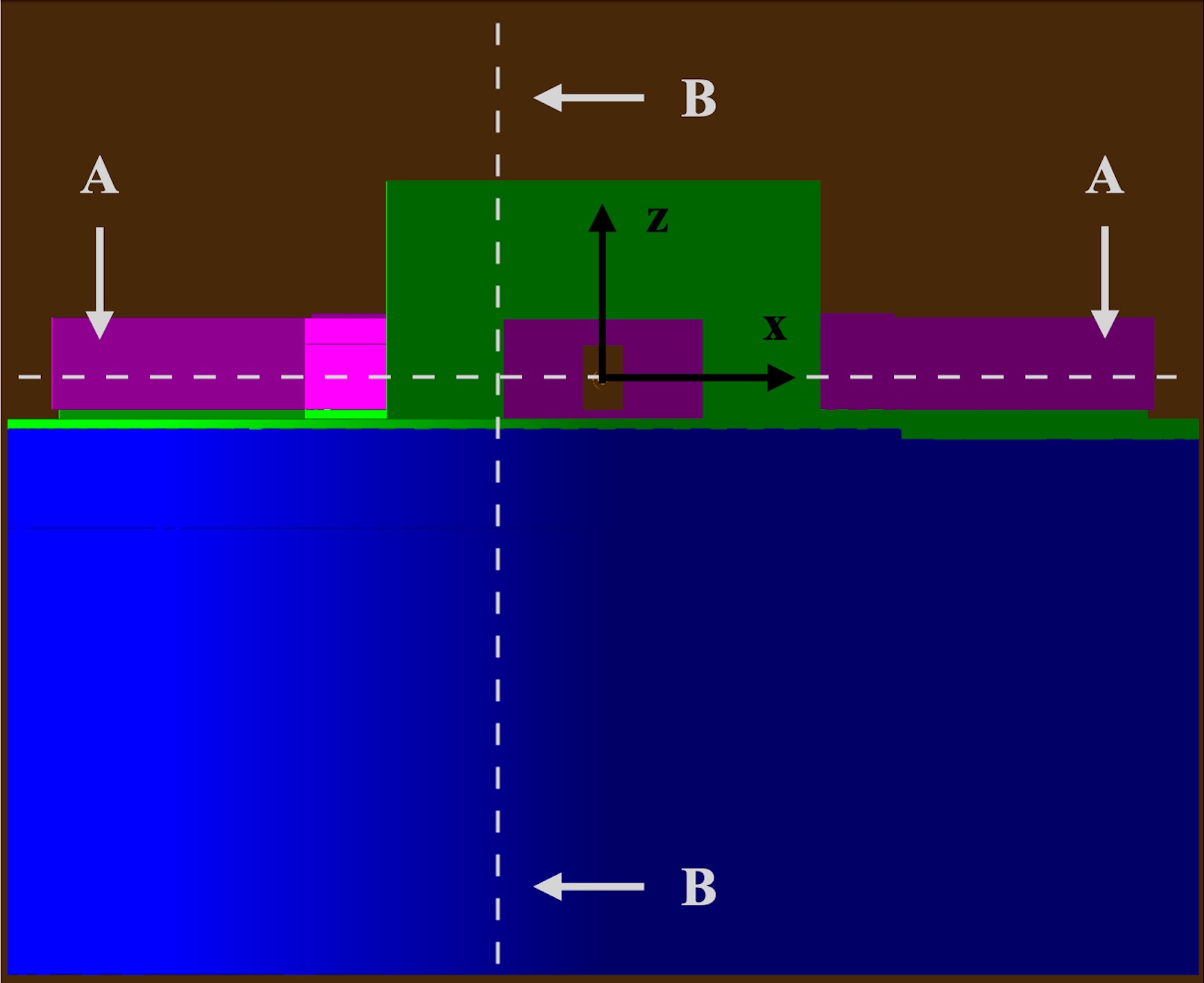} &
    \includegraphics[width=.5\textwidth]{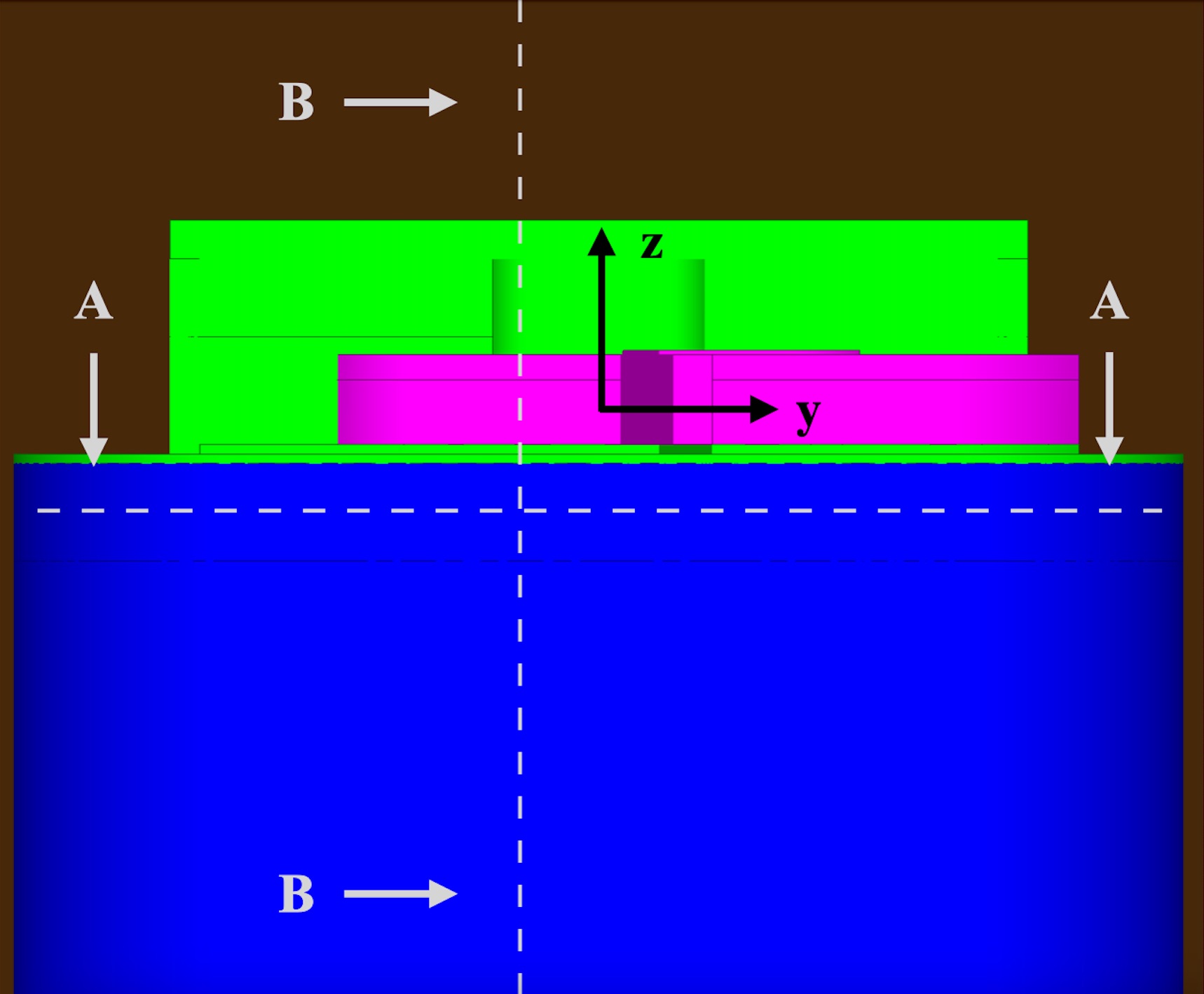}
    \end{tabular}
  \vspace{\floatsep}
    \begin{tabular}{@{}cc@{}}
    \includegraphics[width=.5\textwidth]{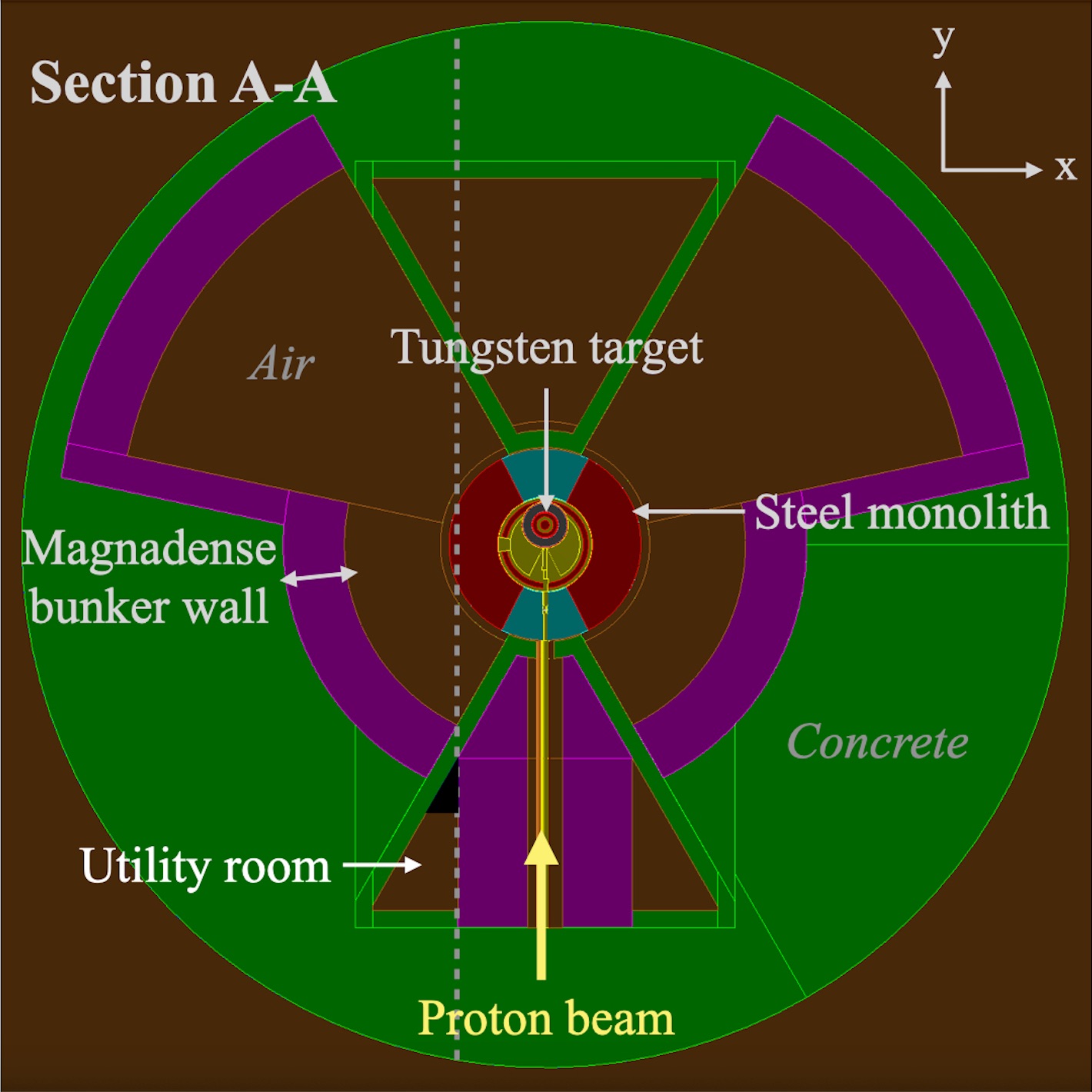} &
    \includegraphics[width=.5\textwidth]{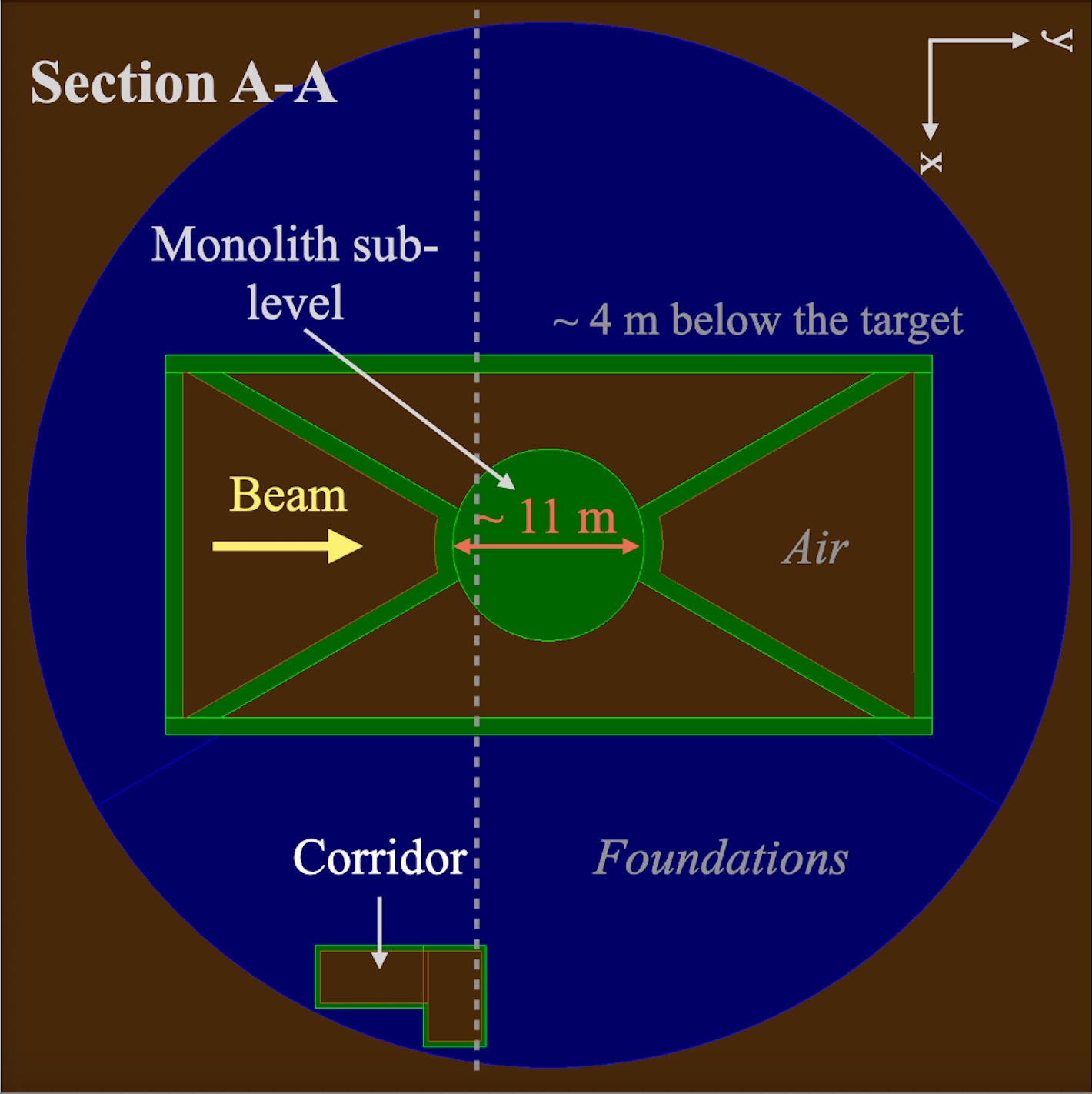}
    \end{tabular}
  \caption{The upper images show the frontal views of the ESS model developed in Geant4. There are two cutting planes indicated on each image that correspond to the sectioning drawings of horizontal cuts (Section A-A) at different heights shown on the corresponding lower panels. Vertical plane cuts are also shown in the top images (Section B-B) and their sectioning drawings are illustrated in figure~\ref{fig:frontview}. Colors express the material composition of the facility following the description in the text. Gray dashed lines in sectioning images mark the intersection plane of the partner sectioning image. As illustrated in the lower-left, the utility or triangle room is next to the proton beam-line separated by MagnaDense shielding. The corner of the utility room colored in black defines an unusable region from ventilation system pipes that are already installed. In the lower-right, the top view after a plane cut at the mid-height of the corridor is indicated.}
  \label{fig:topview}
\end{figure}

\begin{figure}[phbt!]
    \begin{tabular}{@{}c@{}}
    \includegraphics[width=.5\textwidth]{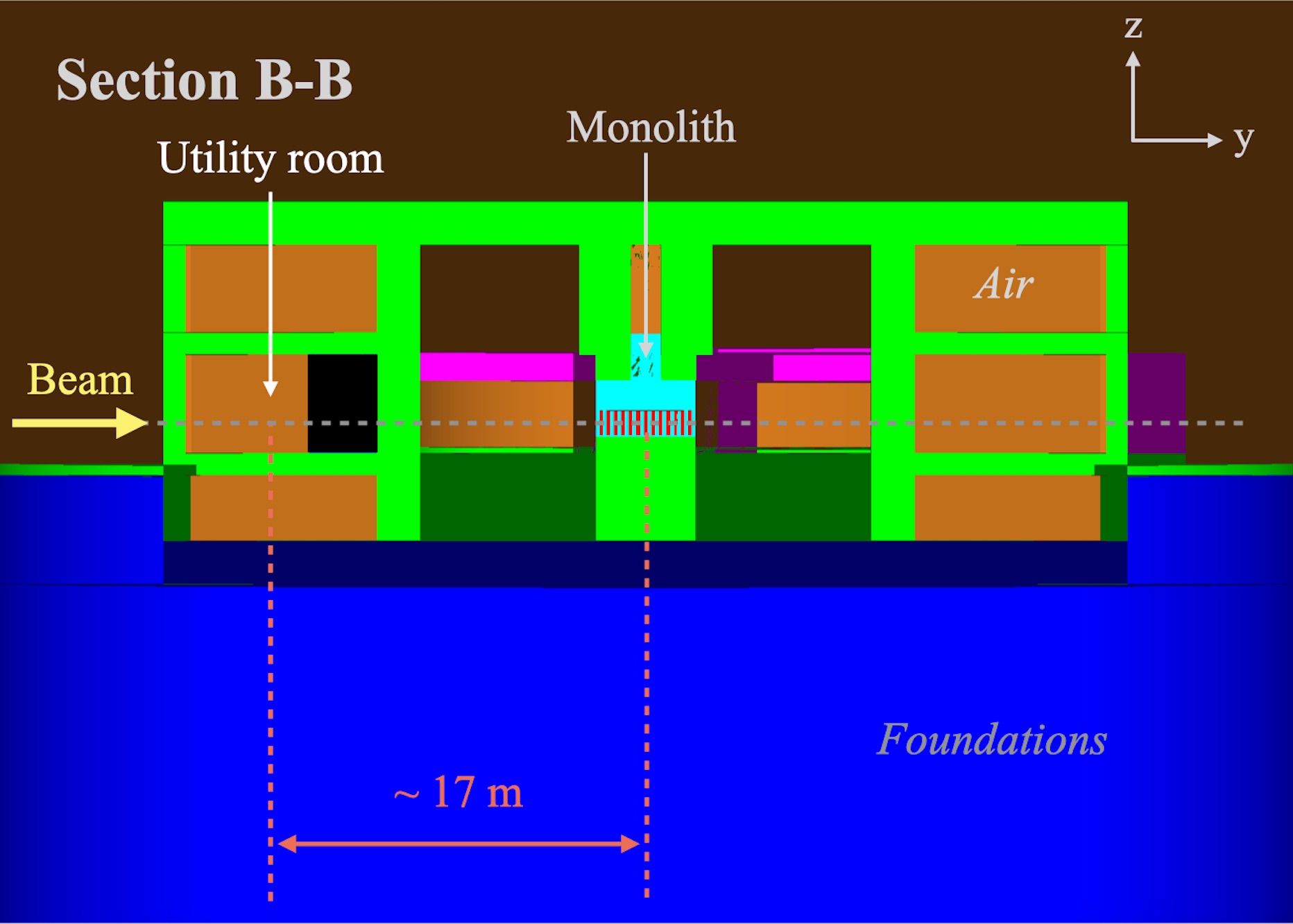} 
    \end{tabular}
  \vspace{\floatsep}
    \begin{tabular}{@{}c@{}}
    \includegraphics[width=.5\textwidth]{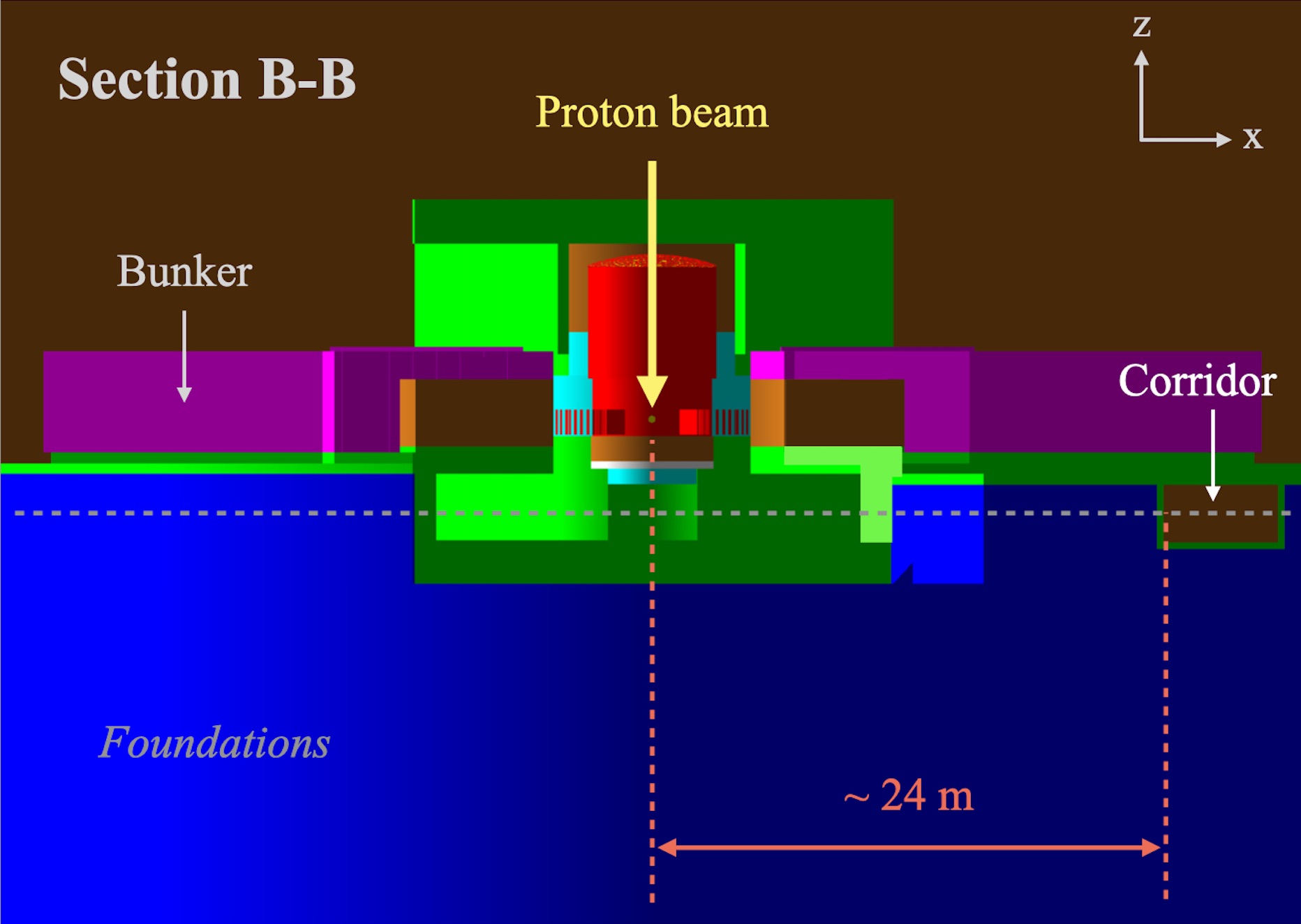}
    \end{tabular}
  \caption{Taking as a reference the vertical sectioning planes illustrated in the top panels of figure~\ref{fig:topview} (Section B-B), the analogous Section A-A  outlines are shown. The perspective in the left image focuses on the location of the utility room and the right image focuses on the location of a corridor that has no assigned role for ESS operations. The material composition is described using the same color palette. Gray dashed lines in sectioning images also mark the intersection plane of the partner sectioning image. For Section B-B emphasizing the room (left), one can see that it is at the same level as the target and $~\sim$17 m away. Other candidates as the top room are also drawn. Section B-B in the right, however, shows the cut on the Y axes for the closest part of the corridor to the target ($\sim$24 m).}
  \label{fig:frontview}
\end{figure}

Figures~\ref{fig:topview} and \ref{fig:frontview} visualize parts of the ESS geometry used in Geant4 to model the neutron flux. Each color illustrates the material of the corresponding volume. Shown in magenta, the bunker wall is composed of concrete 38$\%$ enriched in magnetite high-density concrete which is named as MagnaDense \cite{MagnaDense}. Regular concrete in green, stainless steel in red and a mixture of stainless steel with iron in light blue. Brown and orange indicate air and yellow is used to identify very-low density air. Within the monolith, the tungsten target is shown in gray. Dark blue coincides with the foundations of the facility, which is composed of moraine clay and approximates to 85$\%$ soil and 15$\%$ regular concrete. This concrete is only used to support the pylons \cite{ESSpersonnel}. 

From the solid angle of the tungsten target to the underground corridor, the latter is heavily protected by the foundations in addition to the different layers in the monolith. As a result, the main contribution to the neutron background in the corridor are scattered skyshine particles \cite{lewis2023particle}. This effect may happen because the top of the facility is less shielded. Looking at figures~\ref{fig:topview} and \ref{fig:frontview} the utility room is overall less shielded than the underground corridor. Apart from the skyshine particles that are not stopped in the upper-level room, there are two main contributions of neutrons reaching the room. The first contribution are the neutrons penetrating the steel monolith, which may also go through part of the bunker. The second contribution are the neutrons that propagate along the beamline and escape its shielding. Even if the focus of this work are the two areas previously mentioned, the room above the one already identified as the utility room has recently been known to still be unused, therefore, it could be another candidate to exploit as it is much larger than the utility room and further away from the target.

Therefore, to simulate the neutron background and its propagation over several meters of high-density moderators, a method that breaks into smaller sections each part of the facility is described in the next section, making the Monte Carlo simulation computationally feasible. 

\subsubsection{Weighted window technique}

The main challenge for the neutron flux simulation is to populate the potential neutrino experimental halls in a computational efficient way. The tallies of this study are the neutrons that reach the utility room and the underground corridor, individually. However, due to the materials that constitute the monolith (mainly stainless steel and concrete) and other shielding in the beamline for example, many of the neutrons will be stopped before reaching the areas of interest, which computationally has its limitations. Hence, some methods balance computational efficiency and sufficient statistics. These are often called variance reduction techniques~\cite{van2011easy}, as it is the case of weight windows \cite{booth1984importance,hendricks2000assessment,stenander2015application}. 

In the case of this work, the weight window method relies on separate preparatory simulations of the neutron propagation along the facility accounting for the geometry and materials. The physical geometry is sampled in regions, \textit{cells}, creating a parallel mesh overlaid with the physical geometry modeled. While toolkits such as MCNP rely on the usage of a weight-window generator~\cite{evans1997enhanced}, for Geant4 the ``WeightWindow built-in" classes have been used.  Therefore, an automated weight-window generator has been implemented. An initial analog run is performed to obtain the relative errors and fluxes of each cell in the mesh. This generator statistically estimates the importance of the spatially distributed cells. Individually to generate a maximized neutron flux reaching the utility room and corridor, each region has tailored upper and lower weight bounds. 
 
Note that the weight bounds are numerically inversely equivalent to the cell’s importance. Accordingly, cells along parts of the geometry where a weight modification or sample splitting would benefit the tally (neutrons surviving into the areas of interest) are the regions with higher estimated importance and consequently, regions with lower weight bounds. Particles passing into regions of higher importance (particle weights above the upper bound) are split into more samples, each then uniquely propagating with a reduced weight per generated particle. Each weight is contained within the window, overall conserving the weight of the undivided particle. Particles with weights below the lower bound are no longer tracked based on a fixed probability (probabilistic kill). However, some survive and continue their propagation with an increased weight, correspondingly to conserve the total weight of the cell. Examples of these regions are areas geometrically further away from the room or corridor. This will make the simulation more efficient as it stops unnecessary particle tracks.

Summarizing the usage of weight windows: a specific analysis is performed to cover the splitting importance throughout the geometry and ensure dedicated high-statistics evaluation of the neutron flux in the possible neutrino experimental halls. To tune the weight assignment, iterative weight window estimations were carried out. A simple way to start is to lower the density of the moderating volumes so that sufficient flux from surviving particles is obtained. The weight window generator can then continue to be tuned by iteratively repeating this process, slightly increasing the densities until reaching the full density of the geometry. A cumulative set of weight windows is used in the final neutron propagation simulation.

\subsubsection{Simulated neutron flux and differences between Geant4 and MCNP estimations}

Here the analogous neutron flux simulation carried out in~\cite{lewis2023particle} is described, using Geant4 instead of MCNP. This is a benchmarking test to describe the suitability of Geant4 for spallation energy neutrons. Different Monte Carlo studies have already been performed for the neutron production and propagation through the ESS~\cite{dijulio2016benchmarking,dijulio2020simulating,dijulio2022radiation}.

Simulating with sufficient statistics the neutron generation at the spallation source and its transport along the ESS facility needs computationally large resources. The high-performance computing (HPC) system provided by the Supercomputation Center (SCC) at Donostia International Physics Center (DIPC) was used. The full-scale simulation was performed isolating two main sources of spallation: neutrons produced after accelerated protons hit the tungsten target (\textit{monolith neutrons}) and the ones generated from protons scattering off residual gases in the beamline (\textit{beamline neutrons}). Overall, the average neutron flux per proton injected in the beamline (POT or protons-on-target) has been estimated in the locations of interest: utility room and underground corridor. The geometry and material based weighted window technique, described in the last section, has been tailored to each simulation allowing for the tracking of neutrons through heavy shielding.

In short, discrepancies between from Geant4 (this work) and MCNP~\cite{lewis2023particle} for neutron propagation at the ESS environment have been observed. To untangle this, thorough tests have been carried out.

The first aspect to verify was the consistency of the neutron production per proton penetrating the target. A simplified model of the spallation source was simulated, where a 2 GeV proton beam hit a simple tungsten target. This first test proved the excellent reproduction of the MCNP version in Geant4.

The second test aimed to demonstrate the compatibility of the cross sections for every material in the facility between both toolkits. To do so, parallel simulations were executed for isotropic neutrons with uniformly distributed energy between 0 and 200 MeV propagating along a solid sphere of a given material. The neutrons that made it out of the volume were tracked obtaining an estimated neutron flux. The test was repeated varying the radius of the volume and a very interesting outcome was obtained: while there is good agreement between the spectra for a small volume, differences start to appear for thicker spheres. In particular, stainless steel showed a visible underestimation of the neutron flux in Geant4 compared to MCNP for radii larger than $\sim$5 m, likewise the full-scale simulation. These discrepancies are expected to increase as the radius enlarges, since the neutron propagation distances increase as well. This finding could be an answer to the differences between the analogous simulations developed in MCNP and Geant4, given that the monolith is composed of about 5.5 m radius of stainless steel. 

\begin{figure}
    \centering
    \includegraphics[width=0.8\textwidth]{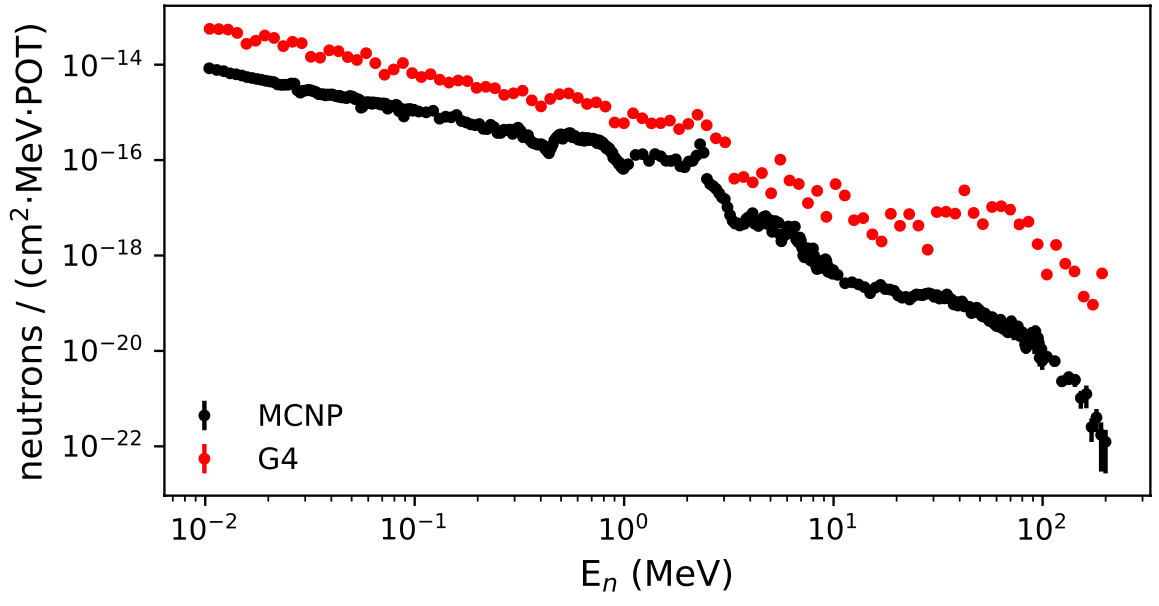} 
    \caption{Simulated neutron flux as a function of energy in the triangle room currently being considered for \cenuns\ experiments at the ESS, only for the neutrons proceeding from the monolith. Error bars are statistical uncertainties. This work's result obtained using Geant4 (red) are shown compared to those in MCNP (black).}
    \label{fig:troomcompare}
\end{figure}

On top of that, it is worth noting the updates that have been carried out by the Geant4 developers group, in particular those related to processes involving neutrons. At this moment in time, we are trying to understand the impact that the Geant4 version itself and the different libraries may have on the results. Work is still underway but with all the simulations and material studies carried out, we have seen that the differences oscillate as the neutrons propagates through the installation, and that the physics lists also have a large impact. In particular, we have seen that the use of the ``Binary Light Ion Cascade" (BIC) physics list instead of the ``Bertini intranuclear cascade" gives an expected flux in the triangle room closer to the MCNP results. To give an overview, in the absence of more statistics, figure~\ref{fig:troomcompare} shows the expected neutron flux in the triangular room from MNCP and Geant4 having used the ``BIC" physics list.

Therefore, one can conclude that although Geant4 is becoming increasingly competitive in simulating spallation energy neutrons that propagate along small geometries, there are still improvements to be made. Therefore, for the next section, the neutron spectrum obtained in MCNP will be used as input. In general, a more robust estimation of background with different Monte Carlo studies are encouraged. Nevertheless, differences between the simulated neutron flux in the rooms enforce the need to measure the neutron background at the facility. For that, a neutron camera has been built and it is introduced in chapter~\ref{sec:ncamera}.

\subsection{Detector shielding strategy}
\label{sec:shielding}

Feasibility studies for performing \cenuns\ measurements at spallation sources not only require an understanding of the neutron flux and propagation along the facility but also the prompt neutron background contribution in a \cenuns\ detector. Neutron interactions may produce comparable signals to the only observable from a \cenuns\ interaction, which is the recoiling of the nucleus. 

Elastic scattering ($n,n$) is one of the main interactions between neutrons and matter, where part of the energy is transferred into the atomic nucleus initiating a nuclear recoil. Neutron inelastic scattering ($n,n'$) could also generate an indistinguishable background in a \cenuns\ detector. Here, the neutron is scattered but the energy is not conserved, leaving the atom in an excited state causing the emission of a gamma, mainly. In addition, the neutron could be captured by a nucleus, which can lead to the emission of gammas, charged particles or lower energy neutrons.

Depending on the detector target material the estimated background varies. In~\cite{lewis2023particle} a detailed study of the performance of a $\sim$32 kg CsI detector with different thicknesses of external polyethylene in the assembly, which is a neutron moderator, was described along both the utility room and the underground corridor. This was accomplished using MCNPX-Polimi~\cite{MCNPPoliMi} simulations that estimated the prompt neutron background considering a previous MCNP simulation that evaluated the contributions from neutrons generated at the target and through the beamline.

Additionally, initial studies on Geant4 were completed in \cite{mariaTFM} considering the same MCNP simulated prompt neutron background provided by \cite{lewis2023particle}, because as seeing in the previous section~\ref{sec:nsim} the Geant4 prediction still differs from this one. This work analyzes the neutron-induced background in gaseous detectors using a Geant4 simulation. A 20 bar gas cylinder of 60 cm in height and 30 cm wide was considered filled with argon, xenon or krypton, precisely. Surrounding the detector with high-density polyethylene (HDPE) of different thicknesses and combining this with an extra shielding of lead (gamma shielding) for secondary $\gamma$'s from neutron-HDPE interactions, the expected recoil rate induced by the simulated neutron background was modeled for the detector placed in the two locations of interest. 

The studies carried out indicate that apart from the moderators and passive shielding of the ESS facility, additional shielding is required in both locations. At first glance, the utility room needs a deeper shielding study, as the \cenuns\ signal/background ratio is too low with the shielding configurations considered. The thicker the layers, the larger the background reduction, though the available space limits the shielding design for each detector. On the contrary, adding a minimal shielding of $\sim$20 cm of HDPE, the underground corridor starts to be suitable for \cenuns\ measurements. Adding an extra 10 cm of Pb to this, enables a broader energy range for \cenuns\ exploration as the secondary gammas (1-10 MeV) from neutron-HDPE interactions are efficiently stopped. The results of these simulations demonstrate that the combination of the additional shielding designed for the detector together with the background reduction in the sub-level corridor due to the ESS design itself, provide an optimal signal-to-background ratio for \cenuns\ measurements $\sim 24$ m away from the target in the underground level. In any case, these neutron background estimations will be complemented with in-situ measurements. Neutron flux characterizations will be carried out using the imaging neutron scatter camera introduced in chapter~\ref{sec:ncamera}.

\section{Future directions of coherent elastic neutrino-nucleus scattering experiment venues}
\label{sec:future}

\begin{figure}
    \centering
    \includegraphics[width=0.8\textwidth]{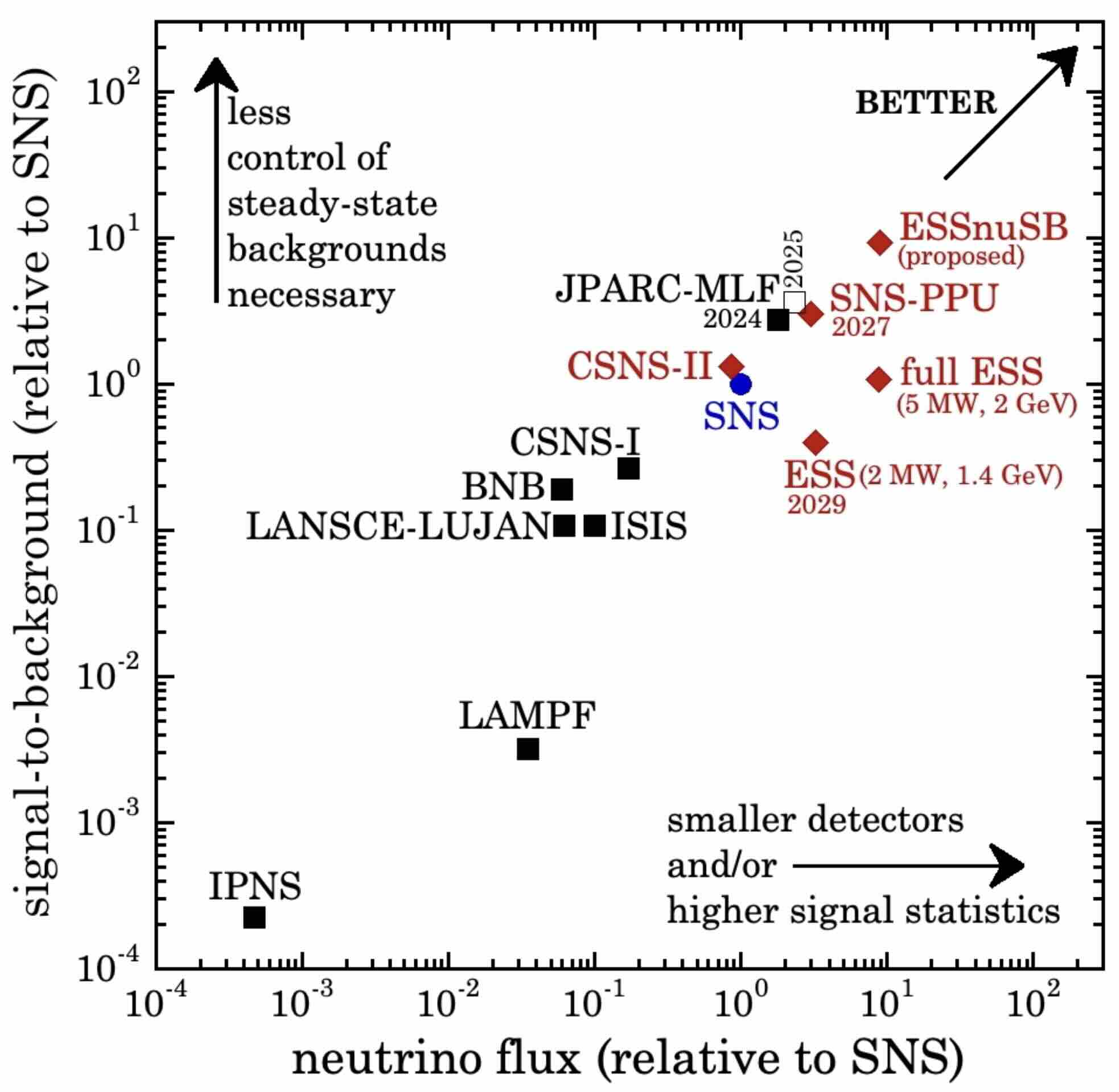} 
    \caption{Representation of past (black square) and future (red diamond) pulsed spallation facilities as \cenuns\ neutrino sources relative to the SNS (blue circle). The estimated improvement on JPARC-MLF during 2025 (white square) is also illustrated~\cite{IvanJPARC}. The proton power update at SNS is captured as well as the short and long-term status of the ESS.}
    \label{fig:neutrinofacilities}
\end{figure}

There is a worldwide effort to develop high-performance pulsed neutron sources. An updated outlook of the current and future pulsed spallation facilities is shown in figure~\ref{fig:neutrinofacilities}. SNS, J-PARC and ESS are competing to take the lead in the upcoming years. Between the three powerful facilities, J-PARC's Materials and Life Science Experimental Facility (MLF) or JPARC-MFL achieved what seems to be the world's highest intensity pulsed neutrons accelerating 3 GeV protons with a beam power of 1 MW for 50 continuous days starting from April 2024 as indicated in their webpage \cite{JPARCweb}. 

To date, SNS was designed to deliver a total beam power of up to 1.4 MW with a typical proton energy of $\sim$1 GeV. Now, there is a proton power upgrade (PPU) undergoing at the SNS (SNS-PPU), which aims to double the power capability from 1.4 MW to 2.8 MW. The project is scheduled to perform 2 MW operations by the end of 2026 \cite{SNSPPU}. 

Despite ESS originally being designed to have an average beam power of 5 MW, due to budget constraints the linac power has been reduced to 2 MW. The target, however, has already been installed for a full 5 MW scope. According to the ESS Annual Report 2023, ESS plans the first beam on target by the end of 2025 (in the year of writing this work) and foresees the first operations with low power accelerator for a few users in 2026 \cite{ESSreport2023}, ramping up to 0.8 MW by the end of 2027 and accelerating protons of $\sim$0.8 GeV. Eventually, after the construction phase, the proton energy will be increased up to 2 GeV as in the initial design.

The neutrino production is affected by the power downgrade as shown in figure~\ref{fig:neutrinofacilities}, as the neutrino yield is expected to increase rapidly with the proton energy at spallation sources. Theoretical calculations for the decay-at-rest (DAR) neutrino production \cite{burman1996neutrino,burman1997neutrino} that can later be compared to experimental data~\cite{burman2003neutrino} are in fair agreement with simulations performed in \cite{baxter2020coherent}, for the nominal proton energies of 2 GeV at ESS and 0.94 GeV at SNS, the facility's scenario at the time of the first \cenuns\ measurement. The simulations indicated that while 0.08 neutrinos of each flavor per incident 0.94 GeV proton were adopted for the first \cenuns\ measurement at the SNS \cite{akimov2017observation}, 0.3 neutrinos of each flavor per proton are estimated for ESS operating with 2 GeV protons~\cite{baxter2020coherent}. The corresponding simulations for J-PARC have also been performed obtaining $\sim$0.45 neutrinos of each flavor per proton~\cite{IvanJPARC}, which is compatible with the measured value of 0.48 electron neutrinos per proton at the target~\cite{JSNS22024first}.

The ESS has a unique capability of providing long pulses of 2.86 milliseconds, while SNS provides 60 Hz of 1 $\mu$s-wide POT spills, and at J-PARC protons are produced with a repetition rate of 25 Hz, where each spill contains two 80 ns wide pulses of protons spaced 540 ns apart. If ESS operated with 2 GeV protons even with the updated design of 2 MW beam power, the neutrino yield expected would be larger. However, looking at the present status and ramp-up plan, ESS schedules its initial operations with a nominal proton energy of $\sim$0.8 GeV \cite{ESSreport2023}. One should remember that a very high-intensity neutrino flux is needed to use small detectors for CE$\nu$NS. So at this stage, the ESS would not be competitive compared to JPARC-MLF or SNS-PPU in neutrino production. Furthermore, the signal-to-background ratio starts to be competitive at 5 MW as indicated in figure~\ref{fig:neutrinofacilities}, which is also essential in \cenuns\ experiments. Accordingly, one could explore other venues to perform \cenuns\ physics shortly, apart from the existing neutrino alley at SNS. This work is been carried out analyzing J-PARC as a potential site for the \cenuns\ detectors that are currently been developed at DIPC within the NuESS collaboration (see chapter~\ref{sec:NuESS})~\cite{IvanJPARC}. 

\renewcommand{\arraystretch}{1.5}

\begin{savequote}[65mm]
``Ezina, ekinez egina"
\qauthor{-- Old proverb in Basque language.}
\end{savequote}

\chapter{Full coverage neutron scatter camera}
\label{sec:ncamera}

The figures in this chapter have been updated for better comprehensiveness and style.

Motivated by the need to characterize neutron transportation at the European Spallation Source (ESS), a neutron scatter camera (NSC) was deployed to simultaneously obtain not only the flux but also the directions of neutrons and superimpose it with a real image.

This chapter overviews the operating concept of conventional neutron scatter cameras and empathizes the need to adequate the design to the neutron source of interest, spallation neutrons in this case. The neutron scatter camera final design, assembly and calibration process are introduced. In order to detect only neutrons coming from the source, a discrimination method based on the deposited signal has been implemented. Hereafter, to validate the device and the reconstruction algorithms, prior measurements have been performed in the laboratory using an accessible source, Californium-252 ($^{252}$Cf), which provides an energy spectrum up to $\sim$10 MeV. Accordingly, the feasibility of performing such measurements was studied in detail using Geant4 simulations. Finally, the spectrum and position reconstruction of the source is shown, providing the first validation of the neutron camera built. Indeed, it enables the localization of the source placed in every direction in a 4$\pi$ perspective. Lastly, this chapter demonstrates how the coupling of a 360$^{\circ}$ optical camera allows for an easier visualization of hidden sources.

\section{Operating concept}
\label{sec:ncameraconcept}

Neutron scatter cameras are a set of detectors that allow for the measurement of fast neutrons and their direction of origin. They operate by detecting the double elastic scatter of an incident neutron while a back-projection method allows for the reconstruction of the neutron trajectory and energy. Similar in operating principle to Compton cameras \cite{xu20044Compton}, the determination of the original particle direction relies on the kinematics of elastic scattering, though neutron scatter cameras use fast neutrons instead of gamma rays. 

These devices use either arrays of scintillators attached to photodetectors spatially isolated or a series of photodetectors within the same scintillator volume \cite{weinfurther2018model}. The most common existing designs for spatially separated scintillators are arrays arranged in multiple planes \cite{mascarenhas2006development} or using a radial distribution \cite{goldsmith2016compact}. For this work, spatially separated scintillators with a plane-based design were used. The detailed discussions on kinematics in \cite{mascarenhas2006development} were followed. For reconstruction, the same assumption that the neutron mainly elastically interacts in the scintillator with the hydrogen atoms, or protons, was made. This assumption is made in every work previously mentioned as the hydrogen cross-section dominates over other nuclides normally found in plastic scintillators in the energy range of $^{252}$Cf for instance, which is a usual laboratory neutron source. Indeed, this is the case of study in this chapter to provide a first validation of the device. As it will be explained later, the data taking was performed using an available $^{252}$Cf source and plastic scintillators. Therefore, the detection mechanism described is based on the approximation that the interactions occur with the hydrogen atoms in the scintillators. However, for the future measurements at spallation sources, the kinematics may be determined by the dominant interaction nuclei at the corresponding energy range, or informed by simulations to better understand the interaction ratio of each target nuclei.

In this setup, an incident neutron scatters off a proton in one of the scintillators. If the scattered neutron consecutively interacts with another scintillator leaving an observable signal, the energy and trajectory of the incident neutron can be reconstructed. Assuming a non-relativistic scenario, the energy of the incident neutron ($E_n$) can be determined by the sum of the proton recoil energy ($E_p$) in the first detector and the scattered neutron energy ($E_{ns}$) as shown in equation \ref{eq:En}:

\begin{equation}
    E_n = E_p + E_{ns}.
    \label{eq:En}
\end{equation}

\begin{figure}
    \centering
    \includegraphics[width=0.8\textwidth]{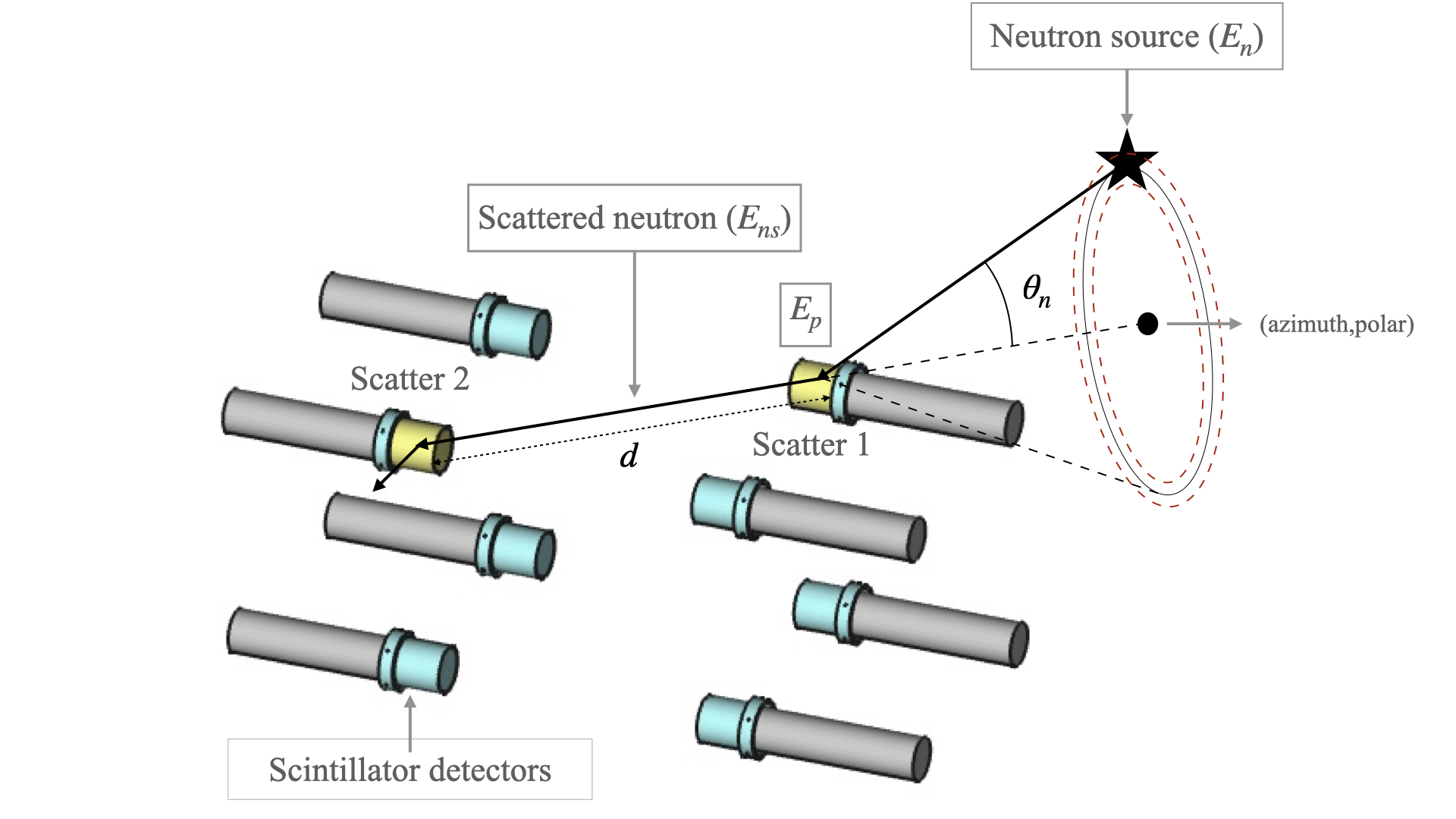}
    \caption{Schematic of a neutron scatter camera with spatially separated scintillator volumes and a back-projected double scatter neutron event. The probability cone after accounting for uncertainties in the cone surface angle $\theta_n$, derived from the reconstructed double scatter neutron event, leads to a widening of $\theta_n$ that can be represented with the red dashed rings.}
    \label{fig:ncamerasigmaangle}
\end{figure}

Since the energy deposition in the second plane is an unknowable fraction of the remaining kinetic energy, the time-of-flight (TOF or $\tau$) between the two volumes is used for a more accurate determination of $E_{ns}$.
Knowing the position of each scatter and the relative distance ($d$), the scattered neutron energy can be described as in equation \ref{eq:Ens} \cite{mascarenhas2006development}:

\begin{equation}
    E_{ns} = \frac{1}{2}m\left( \frac{d}{\tau}\right)^2,
    \label{eq:Ens}
\end{equation}
where $m$ is the mass of the neutron. From kinematic limitations, the particle's initial direction is constrained to a probability cone surface of angle $\theta_n$ back-projected onto a virtual image plane. This is shown in figure \ref{fig:ncamerasigmaangle} and derived from equation \ref{eq:thetan}. For neutrons scattering off protons, this angle cannot be greater than 90$^\circ$ \cite{knoll2010radiation}. The overlap of all the cones encompassing the trajectory of each neutron indicates the most likely location of the neutron source in the form of a hot-spot. 

\begin{equation}
    \theta_{n} = \tan^{-1}\left(\sqrt{\frac{E_{p}}{E_{ns}}}\right).
    \label{eq:thetan}
\end{equation}

\begin{figure}
    \centering
    \hspace{0.3in}
    \includegraphics[width=0.8\textwidth]{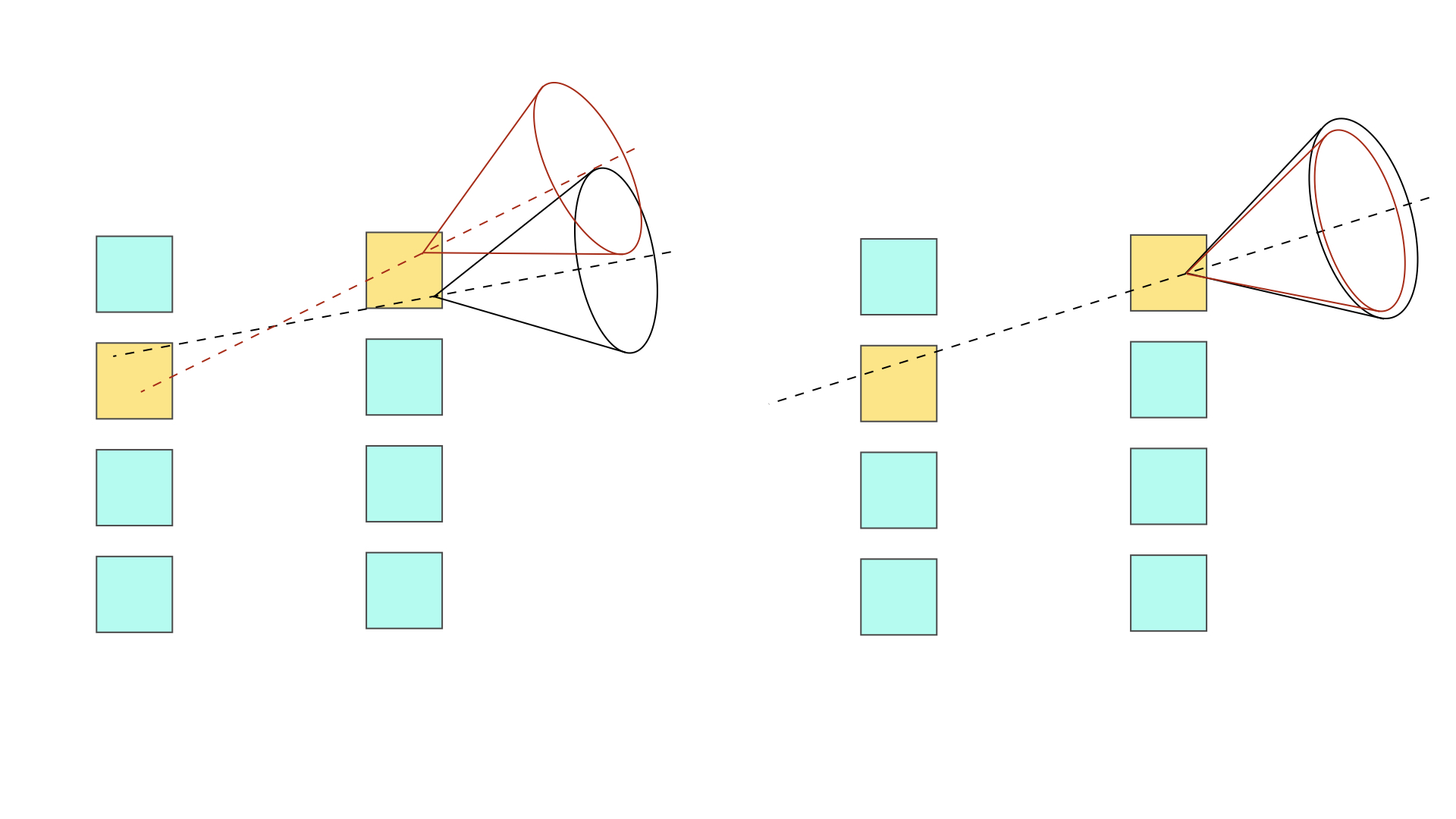} 
    \caption{Schematic example of reconstruction errors. Left: an uncertainty in the back-projected direction is shown due to the finite size of the scintillator. Right: a variation of the scattering angle derived from the uncertainties in the energy measurements.}
    \label{fig:ncameraerror}
\end{figure}

However, the geometry of the set-up and the signal readout introduces some uncertainties to the location of the neutron source. As shown in the left diagram of figure \ref{fig:ncameraerror}, as the scintillators have a finite size and the signal is read by attached photodetectors, the position resolution will introduce an uncertainty in the back-projected direction of the cone axis and in the scattered neutron energy described in equation \ref{eq:Ens}. Combining this with the uncertainty in the initial deposited energy ($E_p$) leads to an uncertainty in the initial scatter angle $\theta_n$ (equation \ref{eq:thetan}) as illustrated in the right panel of figure \ref{fig:ncameraerror}. As a consequence, the probability cone has a thickness and the most likely location of the source will be enlarged as shown in figure~\ref{fig:ncamerasigmaangle} by the red dashed rings.

\section{Camera design}
\label{sec:ncameradesign}

The choice of design determines the performance of the scatter camera. This includes detector cells type (material, size, weight) and distribution. The design has a direct impact on spatial resolution, spectral resolution, sensitivity, and particle discrimination. For instance, since the energy of the scattered neutron ($E_{ns}$) is determined from equation \ref{eq:Ens}, its resolution will be completely dominated by the uncertainties regarding distances between detectors and their sizes, and the time of flight resolution.

The design of a neutron scatter camera is an experiment-specific endeavor, taking into account the motivation of its use, the energy range of the neutron sources, the sites where the measurement will be performed, and the cost. The rest of the chapter will describe a design adequate to map neutrons at spallation sources. Since the goal is determining the neutron background relevant to \cenuns\ detectors at multiple locations along the installation, a compact and mobile design is preferred. 

As mentioned, these devices commonly use scintillators attached to photosensors. For the scintillators, organic ones are usually used instead of inorganic ones, like NaI \cite{al2020neutron}. Even if inorganic scintillators provide a better energy resolution, organic scintillators feature faster timing characteristics and less anisotropy in light transport \cite{schuster2017characterization}. Two kinds of organic scintillators can be found: liquid or plastic. The latter was used in the design because of its relatively low cost, decent timing resolution, easy scalability, and adherence to waste management protocols at the power plants. The device uses arrays of eight plastic scintillators attached to photodetectors spatially distributed as seen in figure~\ref{fig:ncamerasigmaangle}. Concretely, the scintillators used are eight 2" polyvinyl toluene (PVT) scintillators that were, conveniently, already available in the laboratory. Continuing the economic design, standard cylindrical photomultiplier tubes (PMTs), model H6410 from Hamamatsu, were used. These were also operational and well-characterized, so no additional acquisitions were necessary regarding the detector volumes. 

Concerning the geometric configuration of these volumes, some alignments are more efficient than others. Even if the concept of the neutron scatter camera can easily be understood with scintillators placed in two parallel planes forming a front and a backplane (see the illustration in figure~\ref{fig:ncameraerror}), the distribution for the plastic scintillators attached to PMTs is most sensible when forming a cube as it features six surfaces where the detector volumes can easily be mounted. With the appropriate acquisition system, a cube offers full coverage of the space under study providing three-fold symmetry planes to find the consecutive double scatters.  Considering all this, a cube was selected to distribute eight plastic scintillators.

In order to determine the ideal separation between the detector volumes and their alignments, simulations were done considering the expected neutron background flux introduced in the previous chapter for spallation sources. This was studied in detail by S.G. Yoon~\cite{YoonPrivate} using MCNPX-Polimi \cite{MCNPPoliMi} simulations. The response of eight scintillators for a large number of configurations was modeled, varying the distance between them. The work is not published yet as we are working on completing the paper and should be ready soon. 
Spallation neutrons are a continuum with energy ranges up $\sim$1 GeV, but the design was chosen to be most efficient for neutron energies (tens of MeV) that are particularly relevant to the induced background inside \cenuns\ detectors, balancing geometric efficiency (via solid angle) with triggering feasibility, which needs a few ns. Ultimately, a cubic configuration was chosen where the scintillators are center-to-center 40 cm apart. Even if a smaller geometry could have provided larger statistics due to a higher geometric efficiency, the estimated TOF would have been less accurate because of the timing resolution of the system ($\sim$1.5 ns). This distribution balances the large statistics needed and the sensitivity to the energy range of interest. The efficiency of the set-up depends of many factors: geometry, source energy and proximity, and background rejection. 

\begin{figure}
    \centering
    \includegraphics[width=0.8\textwidth]{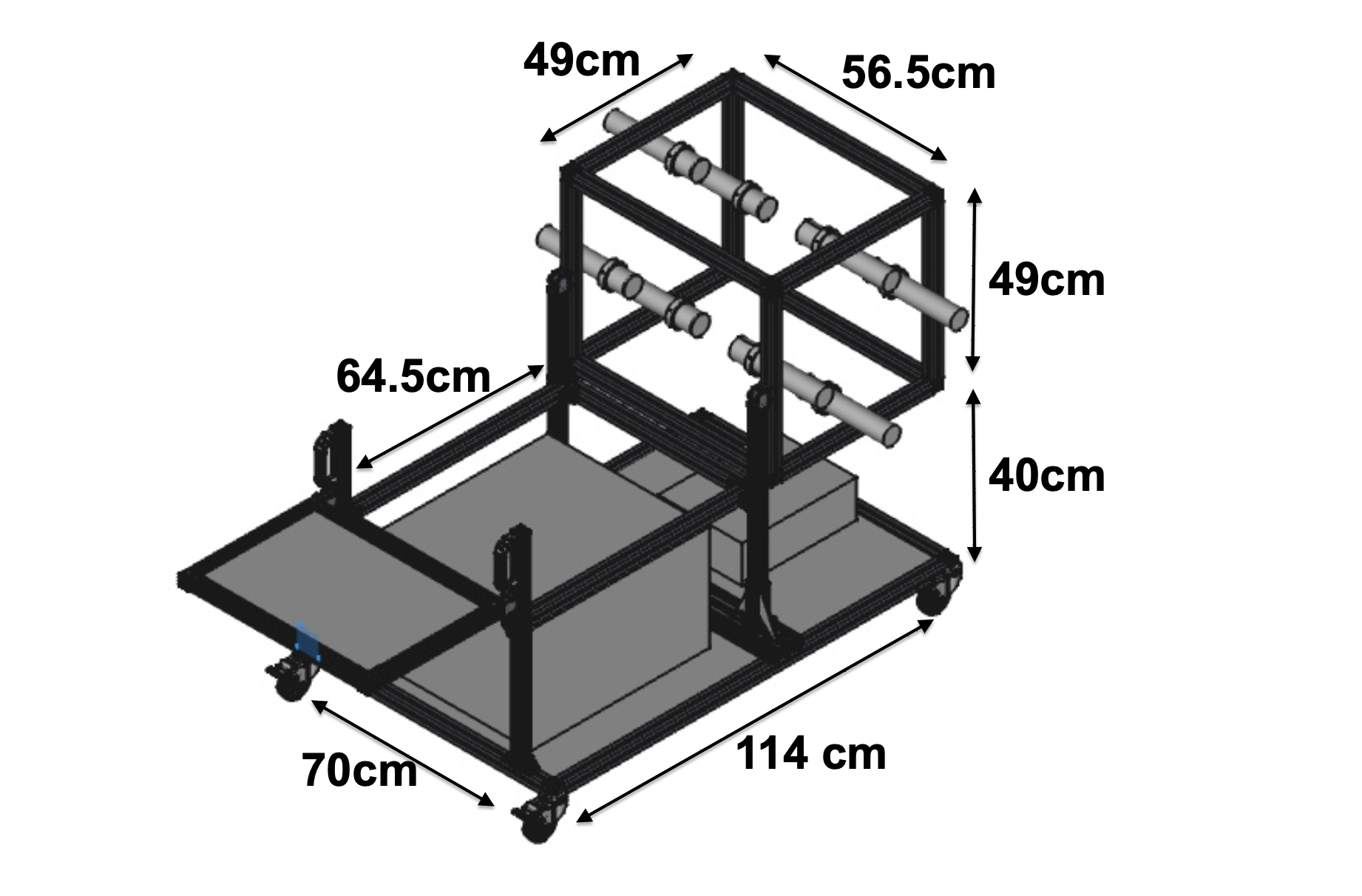}
    \caption{Illustration of the final mechanical design of the neutron scatter camera. The dimensions where chosen to ensure an easy transport along many rooms and corridors at a neutron facility.}
    \label{fig:ncameradesing}
\end{figure}

The CAD drawing in figure \ref{fig:ncameradesing} shows the complete mechanical system of the compact and mobile neutron scatter camera optimized for the characterization of neutron backgrounds at the ESS.

\section{Neutron scatter camera assembly}
\label{sec:ncamerasssembly}

\begin{figure}
    \centering
    \includegraphics[width=0.8\textwidth]{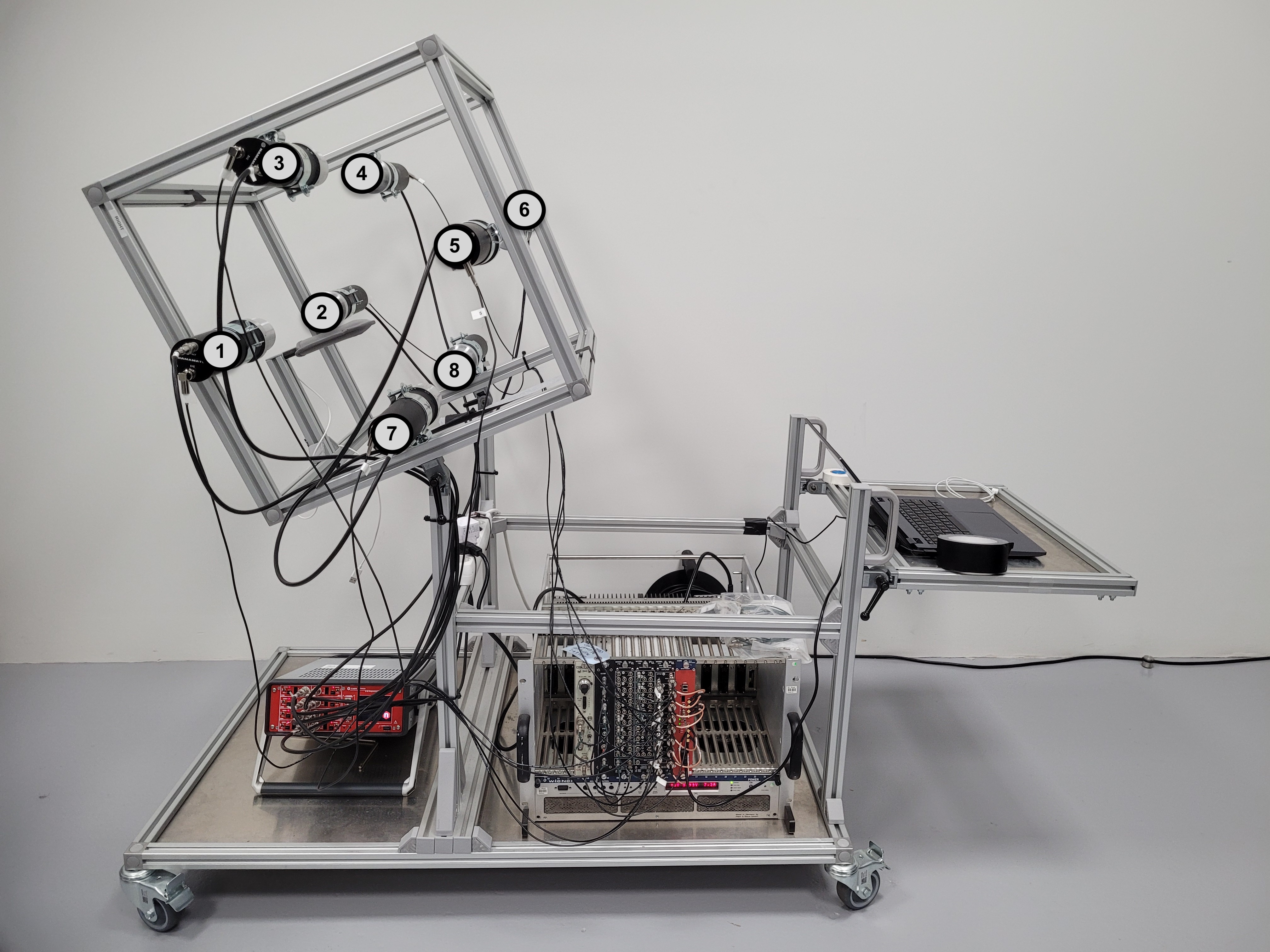} 
    \caption{Photograph of the portable neutron scatter camera for a complete angular reconstruction. All the set-up and electronics needed is mounted in the aluminum chart making it easy to transport. The eight plastic scintillators attached to PMTs are hold in a rotatable aluminum frame with a cubic shape. The numbers assigned to each of the volumes are shown and will also be referenced along the chapter. The laptop for the data acquisition is placed in a table that can be folded to make the device more compact. A 360$^{\circ}$ optical image camera is also installed centered at the cubic detector frame to facilitate the location of sources, as described in section~\ref{sec:resolve}.}
    \label{fig:ncameraimage}
\end{figure}

The detector arrangement has a cubic shape formed by eight plastic scintillators attached to PMTs that are mounted on an aluminum extrusion structure with a rotating axis. This degree of freedom is useful to point at specific locations for a higher efficiency. Moreover, this assembly was placed on a cart to make it portable. An auxiliary, and foldable, table was also added for a laptop to have real-time characterization visible as data accrues. All the electronics needed fit on the cart and only a multi-plug adapter to a single power connection is required. The dimensions of this structure were carefully designed to fit the width of a standard door and to be able to enter many rooms and corridors of the building. A photograph of the neutron scatter camera is shown in figure~\ref{fig:ncameraimage}. 

\subsection{Electronics and Digitizer considerations}
\label{sec:ncameradaq}

The signal output is read by a CAEN-N6730SB multichannel digitizer with a 2 V range, which provides samples every 2 ns and a configurable coincidence window. However, coincidence logic is only available for 4 channel groups, attained by merging channels in groups of 2, which only covers neutron scatterings across 4 of the 6 planes formed by the arrangement. For a 4$\pi$ image reconstruction another external trigger was introduced to cover the final 2 planes. The process works as follows: the output signal of all eight PMTs is amplified by the Phillips Scientific 777 linear-amplifier module. Then, each channel's signal is duplicated so that it can head towards the DAQ directly and go through the internal trigger logic, and at the same time go through the external trigger logic. For this reason, a LabView\textsuperscript{\textcopyright} data acquisition system (DAQ) was developed providing a combined software/hardware trigger readout to be able to observe every combination of channel coincidences.

The internal DAQ trigger logic can only trigger pulse coincidences within a 2 ns window from neutron scatters that happen in scintillators not directly opposite one another. 

\begin{figure}
    \centering
    \includegraphics[width=0.9\textwidth]{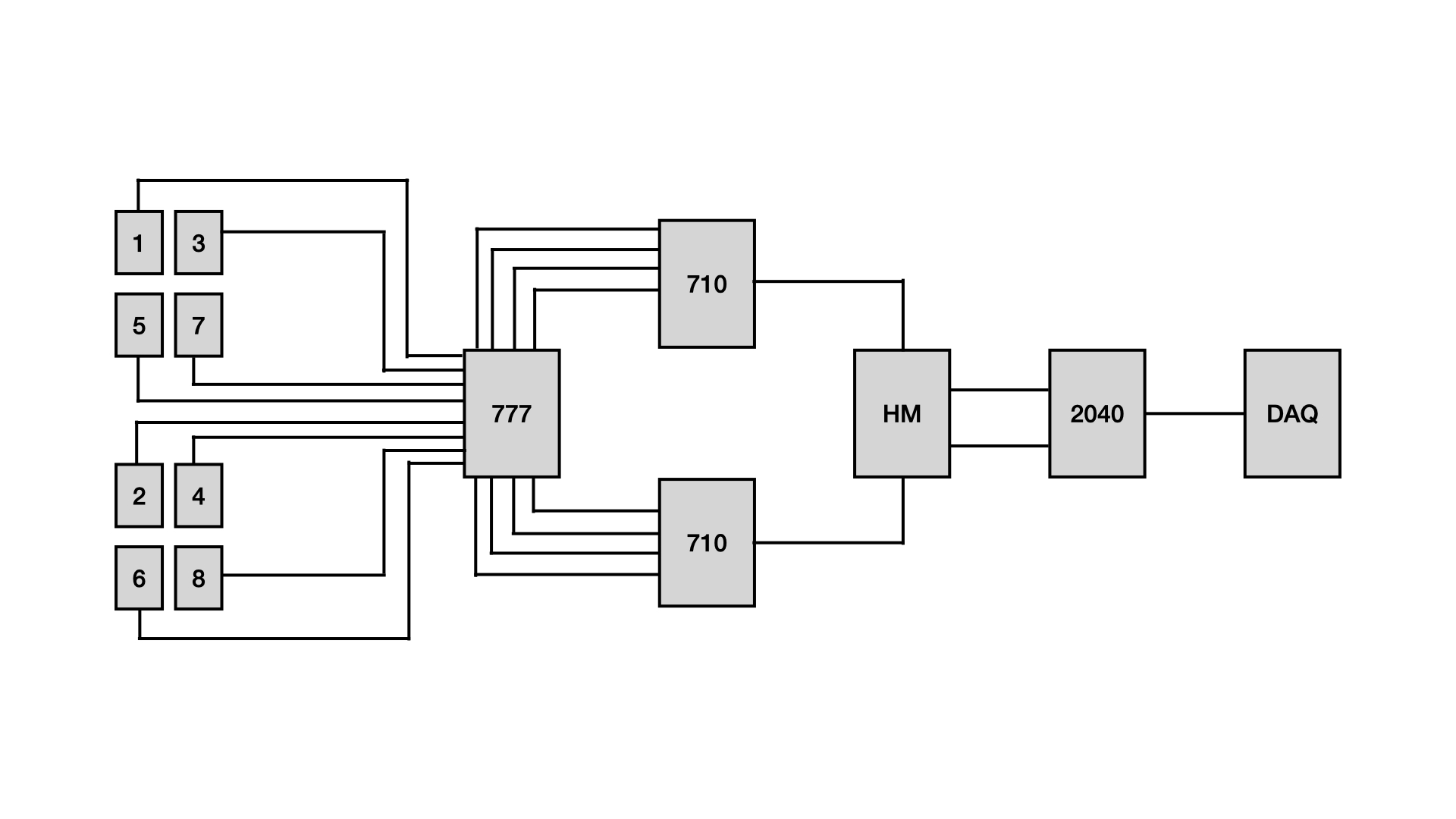}
    \caption{Scheme of the external trigger logic, which performs the coincidence logic for neutron scatters that happen in the detectors facing each other, following the distribution and labeling identified in figure~\ref{fig:ncameraimage}. This coincidence logic formed by a set of NIM modules is later combined with the digitizer trigger logic to ensure that scatters from every pair of scintillator volumes can be covered. Following horizontally, the Phillips Scientific 777 linear-amplifier module amplifies the signal from every PMT. Each Phillips Scientific 710 discriminator groups and sums up the output of four PMTs as indicated, and using a homemade (HM) module, these are converted to readable logic pulses that Canberra 2040 coincidence module analyzes. Finally, the coincidence triggering signal between the two planes is digitized in combination with the internal trigger in the DAQ.}
    \label{fig:ncameratrigger}
\end{figure}

The additional hardware trigger system was linked to cover events that happen in the detectors facing each other. For the external trigger, as described in figure~\ref{fig:ncameratrigger}, a set of NIM modules performs the coincidence logic for these remaining possible neutron paths. After amplifying the signal of every PMT output, the channels are grouped between the two opposite planes that face one another. Separately, each group of channels goes separately through a Phillips Scientific 710 discriminator and the resulting summed output is converted into Canberra 2040 module-readable logic pulses by a homemade (HM) module. The Canberra 2040 is a coincidence analyzer that reads the two inputs and provides a coincidence trigger within a resolving time range of 1.0 microsecond. Finally, the coincidence triggering signal between the two planes will reach the DAQ and be digitized in combination with the internal trigger. Events are triggered by whichever logic signal is first seen, in the case of scatters between scintillator pairs covered by both trigger logic. 

\section{Neutron camera system calibration}
\label{sec:ncameracalib}

After the commissioning of the camera, several types of calibrations are required in order to have an adequate idea of the performance of the system. This section goes through the different calibration procedures. It includes the gain match of PMTs, the energy calibrations, and the time characteristics. 

\subsection{Gain match}

\begin{figure}
    \centering
    \includegraphics[width=1\textwidth]{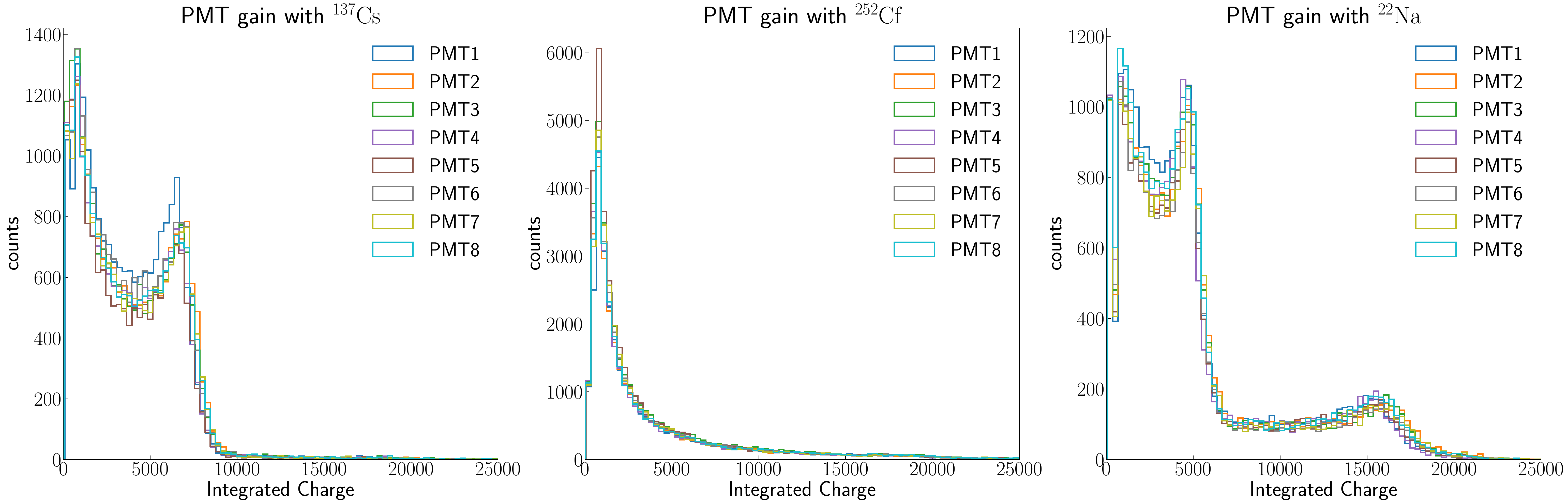}
    \caption{PMT energy spectra shown for $^{22}$Na, $^{137}$Cs and $^{252}$Cf sources for all 8 PMTs. It is visible from all three panels that the corresponding integrated charge, of the same incident energy, were matched, showing that the 8 PMTs were gain matched supplying each one with the voltage indicated in table~\ref{tab:pmtV}.}
    \label{fig:ncamerapmtgainmatch}
\end{figure}

Since the principle of the PMTs is a multiplication process that depends on the applied high voltage (HV), to gain match the PMTs in the neutron camera system the flat-fielding technique was used, which ensures that all of them see the same light. The high voltage of each PMT was adjusted so that the corresponding integrated charge of the same incident energy were matched. To do so, a $^{137}$Cs gamma source was used and corroborated that gain match with $^{22}$Na and $^{252}$Cf. In figure~\ref{fig:ncamerapmtgainmatch} the gain match of the eight PMTs used is illustrated using several sources. Therefore, to have an operational voltage range that keeps signal from higher energy neutrons (tens of MeV) in range and avoiding saturation, the final high voltages were set as in table~\ref{tab:pmtV}.

\begin{table}[h!]
\centering
\begin{tabular}{|c | c |} 
 \hline
 PMT number & Voltage applied (V)\\ [0.5ex] 
 \hline\hline
 1 & 1150  \\ 
 \hline
 2 & 1280  \\
 \hline
 3 & 1235  \\
 \hline
 4 & 1290  \\
 \hline
 5 & 1300  \\  
 \hline
 6 & 1220  \\ 
 \hline
 7 & 1250  \\  
 \hline
 8 & 1197  \\ 
 \hline
\end{tabular}
\caption{High voltage applied to each PMT in the neutron camera system following the labeling of each PMT as in figure~\ref{fig:ncameraimage}.}
\label{tab:pmtV}
\end{table}

\subsection{Energy calibration and energy resolution}

After gain matching all the PMTs, an energy calibration was made to relate the light output signal of each detector cell to visible energy depositions, as an accurate estimation of the energy deposition in the first scatter is needed for operating the device.

The energy calibration was done using $^{22}$Na and $^{137}$Cs gamma ray sources. From $^{22}$Na two photon energies are expected from the atomic photoelectric effect, one at 511 keV from the positron-electron annihilation and the other at 1274.54 keV from the decay itself. $^{137}$Cs decays into $^{137}$Ba, which has a very short half-life, and that immediately decays into the ground state with the emission of a 661.66 keV gamma.

Considering the gamma energies emitted from these calibration sources and that the scintillator is made of hydrogen and carbon, which are low Z elements, the dominant interaction in these detectors is the Compton scattering. 
In addition, the photoelectric effect cross section is suppressed at the corresponding photon energies and if an unlikely characteristic photon is detected it will be overshadowed by the Compton shelve. 

Interacting via Compton scattering, the incident gammas do not undergo a full energy deposition in the scintillators due to their size and density, and thus a Compton continuum is the expected response. For a single Compton scattering event the maximum energy transfer will happen in a complete back-scatter (180 degrees) and it is described in equation~\ref{eq:Ecompton},

\begin{equation}
    E_{c} = \frac{2E_{\gamma}}{m_e c^2 + 2E_{\gamma} },
    \label{eq:Ecompton}
\end{equation}
where $E_{\gamma}$ is the incident photon energy, $m_e$ the mass of the electron, $c$ speed of light. This forms the so-called ``Compton edge" corresponding to the highest energy transfer via a single Compton scatter. For the gamma sources named before, table~\ref{tab:CompotonEdge} shows the photopeak ($E_{\gamma}$) and Compton edge ($E_{c}$).
 
\begin{table}[h!]
\centering
\begin{tabular}{|c || c | c|} 
 \hline
 Gamma ray source & $E_{\gamma}$ (keV) & $E_{c}$ (keV)\\ [0.5ex] 
 \hline\hline
 $^{22}$Na & 511 & 340.67  \\ [0.5ex] 
 \hline
 $^{22}$Na & 1274.54 & 1061.71  \\ [0.5ex] 
 \hline
 $^{137}$Cs & 661.66 & 477.34  \\  [0.5ex] 
 \hline
\end{tabular}
\caption{Theoretical calculations of the Compton edge for $^{22}$Na and $^{137}$Cs sources.}
\label{tab:CompotonEdge}
\end{table}

This sharp upper edge of the well-known Compton distribution is used for energy calibration. However, the sharpness of this feature is modified due to the limited energy resolution of these plastic scintillators. The position of the Compton edge in the continuum is determined by the detector resolution~\cite{hristova1990compton}. Moreover, the size of the detector will also have a strong impact and thus, we have set the position of the Compton edge at the FWHM in the pulse spectra because small detectors were used~\cite{dietze1982gamma}. 

\begin{figure}[!ht]
\centering
  \hspace{-0.1in}
    \includegraphics{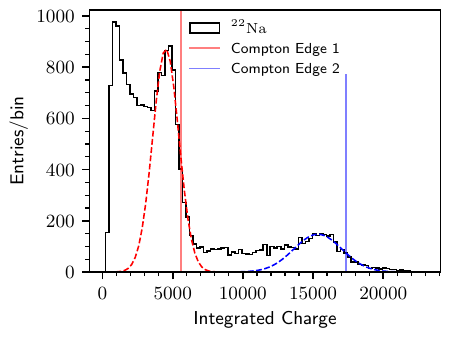}
  \hspace{0.1in}
    \includegraphics{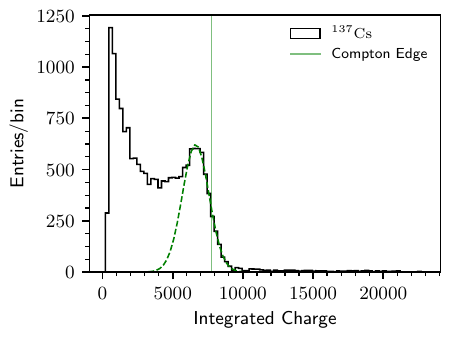}

    \caption{Example of the fit to Compton edge peaks for one of the scintillator volumes, the left panel showing that for the $^{22}$Na source, and the right panel for $^{137}$Cs.}
    \label{fig:ncameracomptonedgefitpmt8}
\end{figure}

\begin{figure}
    \centering
    \includegraphics{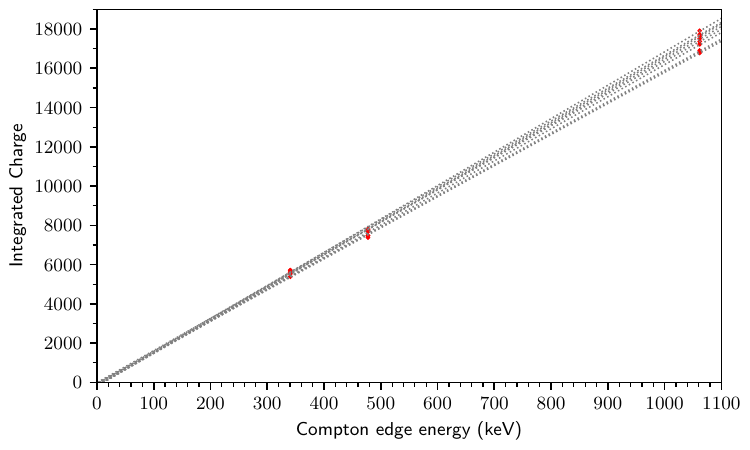}
    \caption{Energy calibration linear curve for all the 8 scintillator volumes repeating the fit-process in figure~\ref{fig:ncameracomptonedgefitpmt8} and using the Compton edge energies indicated in table~\ref{tab:CompotonEdge}.}
    \label{fig:ncameracomptonedge}
\end{figure}

Both plots in figure~\ref{fig:ncameracomptonedgefitpmt8} illustrate the Compton edge finding method for one of the detector volumes, the left showing the response to $^{22}$Na and the right one to the output signal for $^{137}$Cs. A linear fit to the calibration points is done for each of the volumes, which is shown in figure~\ref{fig:ncameracomptonedge}. This provides a unique charge-energy conversion for each detector that is indicated in table~\ref{tab:ecalib}.

\begin{table}[!ht]
\centering
\begin{tabular}{|c || c | c|}
 \hline 
 \thead{PMT number} & \thead{Charge-energy conversion ($\text{keV}_\text{ee}$)} & \thead{R}  \\ 
 \hline\hline
 1 & $E$ = 0.063$Q$ + 3.163& 0.999616\\ [0.5ex] 
 \hline
 2 & $E$ = 0.059$Q$ + 11.211 & 0.999850\\ [0.5ex] 
 \hline
 3 & $E$ = 0.060$Q$ + 11.451& 0.999880\\  [0.5ex] 
 \hline
 4 & $E$ = 0.063$Q$ + 2.796& 0.999978\\  [0.5ex] 
 \hline
 5 & $E$ = 0.061$Q$ + 9.962& 0.999719 \\  [0.5ex] 
 \hline
 6 & $E$ = 0.060$Q$ + 8.703& 0.999863 \\ [0.5ex] 
 \hline
 7 & $E$ = 0.060$Q$ + 5.484& 0.999842 \\  [0.5ex] 
 \hline
 8 & $E$ = 0.061$Q$ + 0.912& 0.999981 \\ [0.5ex] 
 \hline
\end{tabular}
\caption{Energy calibration from a linear fit to the Compton edge calibration points for each scintillator volume.}
\label{tab:ecalib}
\end{table}

Accordingly, the energy resolution of the scintillator volumes can be derived from the spread of the fits to the Compton peaks. As each detector volume has a slightly different energy resolution, a average one will be used (equation~\ref{eq:Eres}), which is in good agreement with dedicated measurements as in \cite{dietze1982gamma,van2016high}. The spectrum features a resolved photopeak with an energy resolution of about 16$\%$ (FWHM) for 662 keV photons from $^{137}$Cs.

\begin{equation}
    \sigma (E) = 1.35 \sqrt{E(\text{keV}_\text{ee})} + 23.80.
    \label{eq:Eres}
\end{equation}

\subsection{PMT time characteristics}
\label{sec:transittime}

The time response of fast response PMTs, which are the ones used, is mainly determined by the electron transit speed along the multiplication process in the dynodes. The type of dynode and the voltage applied to the PMT will each have an impact. An upturn in the supply voltage will increase the electric field and result in a faster electron transit. 

The fluctuations in electron transit time will result in time fluctuations of single-photon output signals, which is called Transit Time Spread (TTS). From the Hamamatsu specifications, the H6410 Photomultiplier tube has a TTS of 1.1 ns \cite{HamamatsuBasics}. 

For time-of-flight measurements, which is necessary to resolve the energy left in the scattered neutron, fluctuations in transit time for coincident PMTs need to be accounted for. For this, a $^{22}$Na source, emitting annihilation gammas in opposite directions, was placed between a pair of PMTs. The coincident signals are read by the PMTs attached to scintillators and their initial signal onsets are obtained. The evaluation of the difference in onset timing will determine the relative performance of the coincident timing. Fluctuations will appear mainly from the supply voltage of each PMT, which also has an impact on TTS. Higher voltages mean faster electron transit times, smaller TTS, and less deviation in signal onsets. Therefore, from table~\ref{tab:pmtV}, the PMT labeled as ``1" has the lowest voltage applied, so the longest transit time is expected. Nevertheless, PMT ``5" has the greatest supply voltage, providing the lowest transit time. An event that triggers these two in the coincidence will have the largest difference in timing. All the photosensor coincidences have been quantified and a TOF correction matrix has been derived. Therefore, the TOF of an event will be described as in equation~\ref{eq:tcorrection}:

\begin{equation}
    \tau = t_1 - t_2 + t_{\rm{correction}(12)} , 
    \label{eq:tcorrection}
\end{equation}
where $t_1$ and $t_2$ are the onsets of the digitized signals in the first and second volumes respectively within the coincidence window. $t_{\rm{correction}(12)}$ is the transit time correction element for that combination PMT volumes.  

\section{Pulse shape discrimination}
\label{sec:ncamerapsd}

The plastic scintillators used in the system are sensitive to the energy dissipation differences between neutrons (n) and gamma rays ($\gamma$), and as such are capable of n/$\gamma$ discrimination to better isolate fast neutrons in a dataset \cite{knoll2010radiation}. Reported in literature in the 1950s \cite{owen1958decay,brooks1959scintillation}, the most utilized technique is Pulse Shape Discrimination (PSD), which benefits from the different decay times depending on the type of radiation. The scintillator features an excitation by neutrons different from the one by gammas: each generates a particle characteristic scintillation pulse. 

\begin{figure}[ht!]
    \centering
\includegraphics{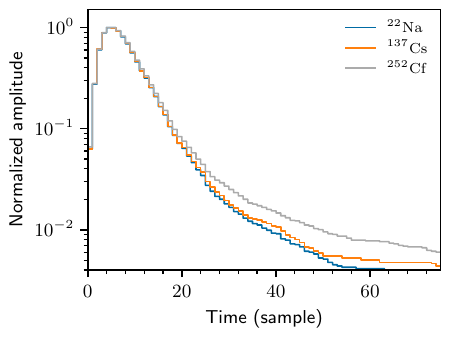} 
    \caption{Illustration of the average waveforms from $^{22}$Na, $^{137}$Cs and $^{252}$Cf. While the first two sources are pure gammas sources, $^{252}$Cf emits both neutrons and gammas. The contribution of the neutrons is present in the average $^{252}$Cf waveform as the neutrons have longer decay times than gammas. This provides the key for discriminating neutron and gammas.}
    \label{fig:ncameraNormAmp}
\end{figure}

The luminescence process in scintillators happens via both fluorescence and phosphorescence. Fluorescence is characterized by its prompt scintillation light while phosphorescence provides slower scintillation emissions and thus, longer decay times \cite{Brooks1979yd}. In particular, neutron interactions will generate more phosphorescence photons than gamma rays, allowing for pulse discrimination due to their longer decay times. Figure~\ref{fig:ncameraNormAmp} shows the average scintillation response shape from $^{22}$Na ($\gamma$), $^{137}$Cs ($\gamma$), and $^{252}$Cf (n-$\gamma$) sources. Although the branching ratio of spontaneous fission in $^{252}$Cf is small, the contribution of longer decay times from neutrons is readily visible by eye in the average $^{252}$Cf waveform. 
 
Among PSD techniques, there are different approaches that one can use to highlight the fraction of light in the slow component from different radiation types \cite{Ranucci1995}. Concretely, the Charge Comparison (CC) method and the Integrated Rise Time (IRT) method are PSD performances with high potential. The Charge Comparison (CC) method \cite{Ranucci1995}, setting a time boundary ($t_{\rm{B}}$), considers the charge ratio between the two-time gaps. The integrated waveform from the beginning of the pulse to $t_{\rm{B}}$ is known as $Q_1$ and the charge after $t_{\rm{B}}$ $Q_2$. Hence, it is sometimes called the $Q_1-Q_2$ method. The Integrated Rise Time (IRT) method \cite{Ranucci1995} quantifies the rise time of the normalized cumulative charge distribution between two bounds ($\Lambda_{\rm{Low}}$ and $\Lambda_{\rm{High}}$). For the work done in this chapter the $Q_1-Q_2$ method has been used, so it is the PSD method described in the next section. 

\subsection{Pulse shape discrimination by $Q_1-Q_2$ method}
\label{sec:ncameraQ1Q2}

\begin{figure}[ph!]
    \centering
    \begin{tabular}{@{}cc@{}}
    \includegraphics{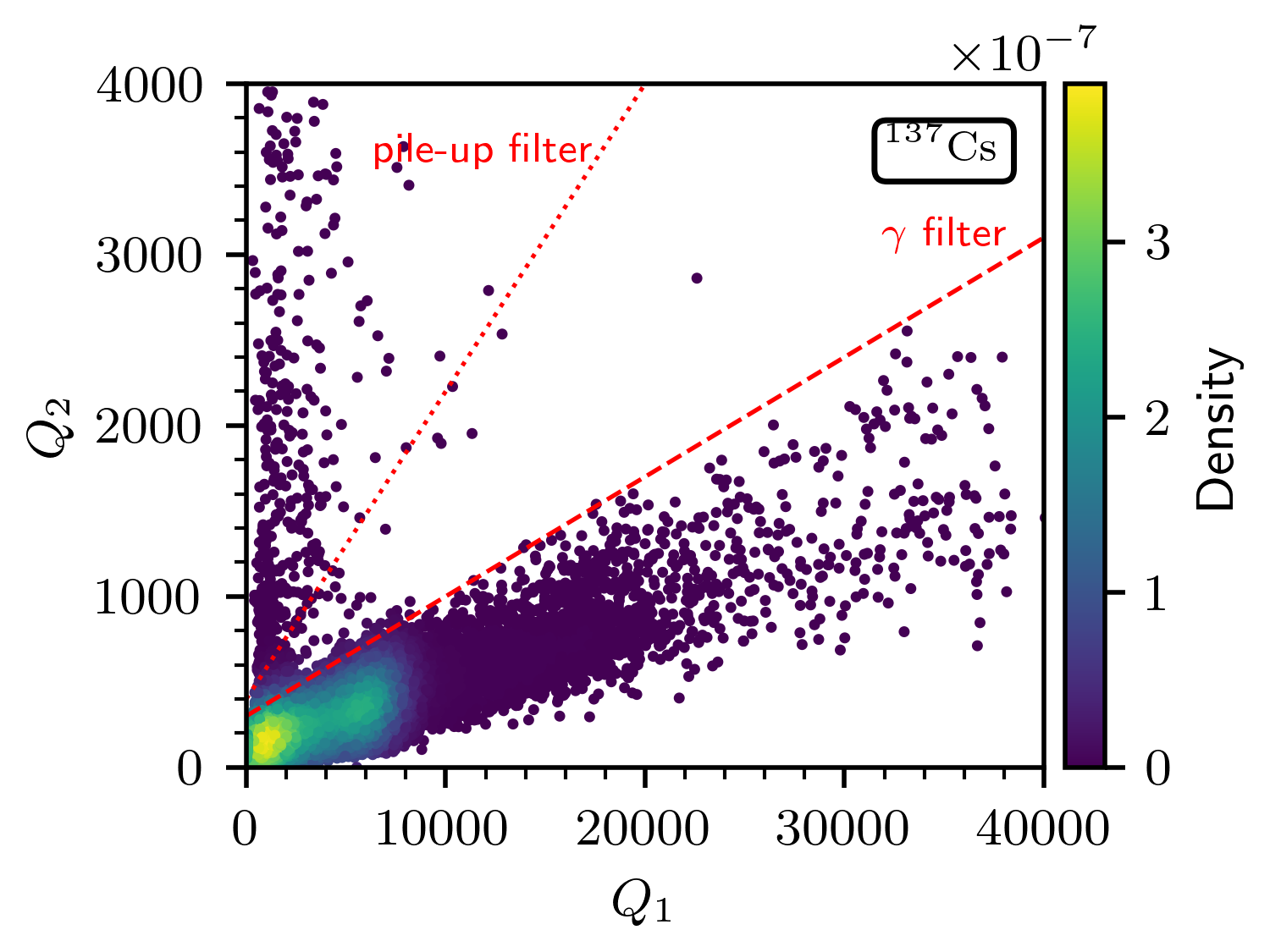} &
    \includegraphics{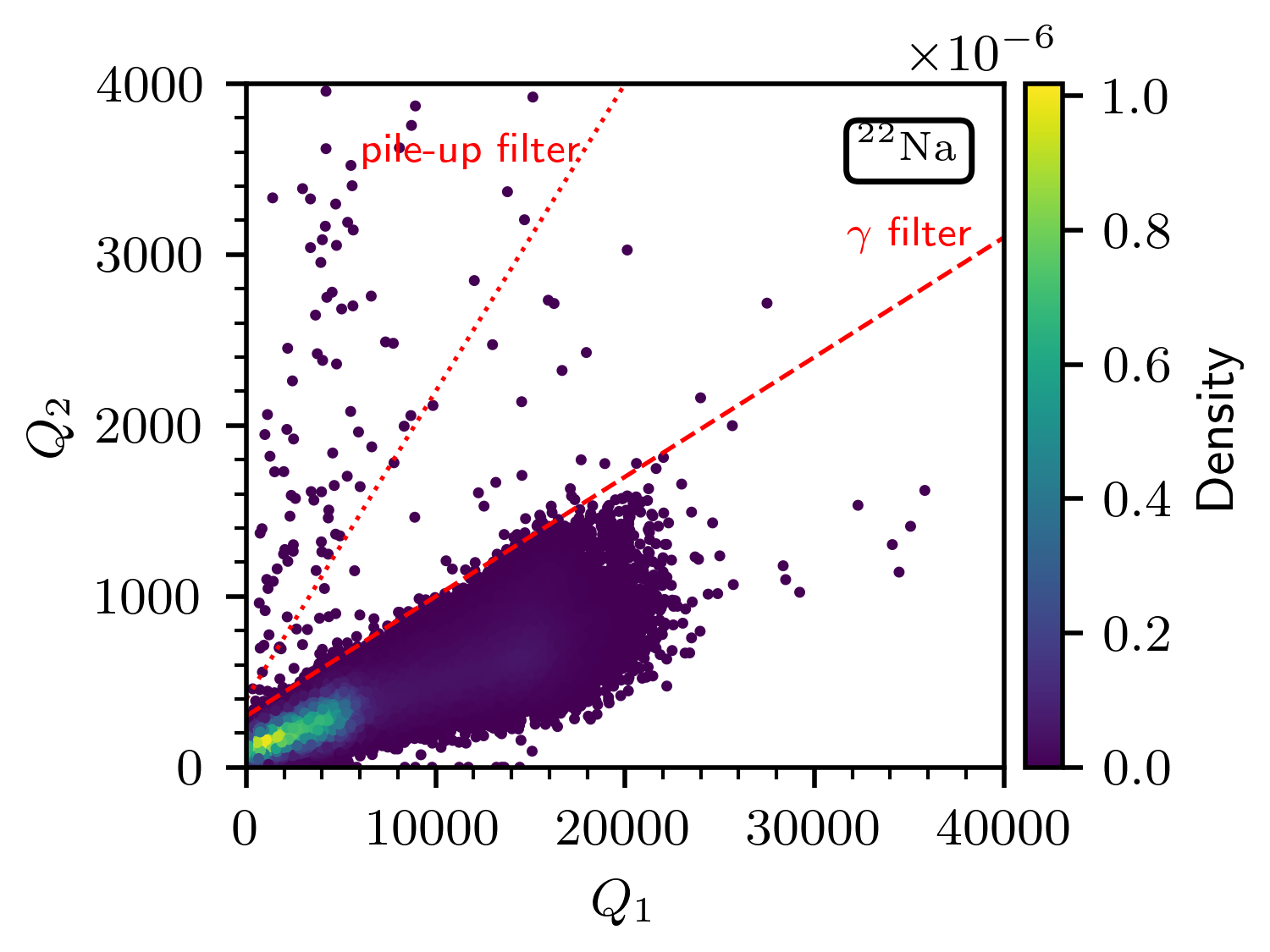}
    \end{tabular}
  \vspace{\floatsep}
    \begin{tabular}{@{}c@{}}
    \centering
    \includegraphics{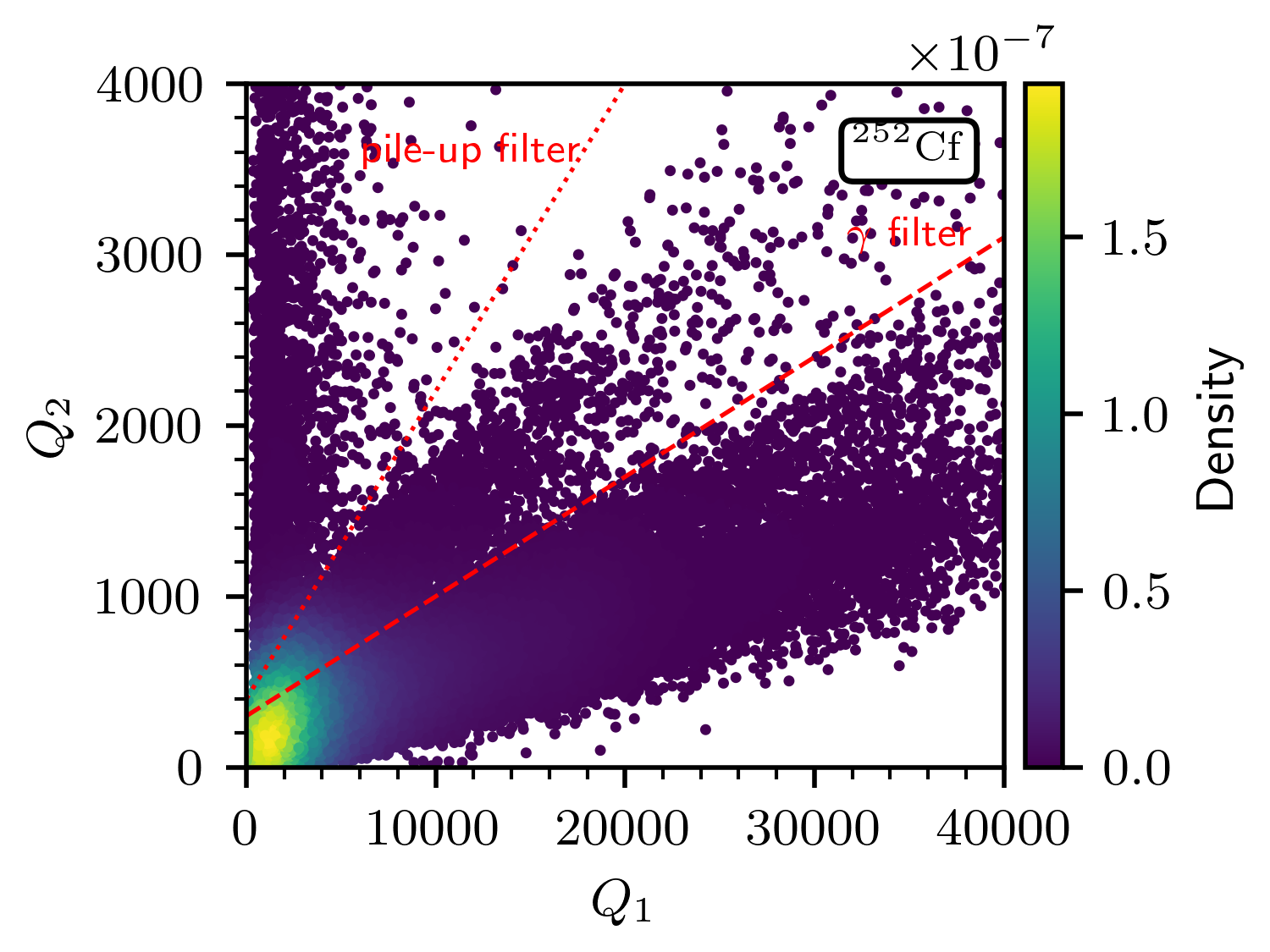}
    \end{tabular}
  \caption{Neutron camera response to $^{137}$Cs, $^{22}$Na and $^{252}$Cf sources. The first two are pure gammas sources and the latter is a mixed neutron-gamma source. Visible in the central region of the $^{252}$Cf response, bounded by the two red lines, and absent in the two top panels, it indicates the response of the scintillator to neutron scatters. The region below the $\gamma$ filter (red dashed) is identified as the response to gammas, and the ``pile-up" filter (red dotted) sets a cutoff to some accidental or environmental coincidences. The high density regions corresponds to interactions where minimal energy is deposited in the scintillator. Taking as a reference figure~\ref{fig:ncameraNormAmp}, the pulse tail merges into the noise and the pulse-shape discrimination becomes impossible. Small variations on the pulses will not be reflected in the Q2 as the integration window is purely noise and the timescale/energy threshold involved obviates any meaningful differences in Q1. A robust energy threshold gets rid of most of these events.
}
  \label{fig:Q1Q2sourceallPMT}
\end{figure}

The Charge Comparison (CC or $Q_1-Q_2$) method distinguishes particles from the integrated charge ratio between two different time windows. The short integral ($Q_1$) starts at the beginning of the pulse up to the optimized boundary time ($t_{\rm{B}}$), scintillator decay time dependent. The long integral ($Q_2$) integrates from $t_{\rm{B}}$ to an optimized end of the pulse tail ($t_{\rm{final}}$). To characterize the n-$\gamma$ PSD of the complete neutron scatter camera in terms of $Q_1$ and $Q_2$, the response of each PMT has been measured using the three radiation sources ($^{22}$Na, $^{137}$Cs and $^{252}$Cf). These responses of all the scintillator volumes are shown in figure~\ref{fig:Q1Q2sourceallPMT} for an illustrative case where $t_{\rm{B}}$=60 ns and $t_{\rm{final}}$=240 ns. The first two panels show the response to pure gammas sources with a clear and dense region at the low $Q_2$ regime. The last panel, shows the response to a $^{252}$Cf source, where a higher gamma presence is expected rather than neutron interactions. There, the visible central lobe contains mostly neutrons and the lower one $\gamma$, hence a boundary between these branches is already an efficient n-$\gamma$ PSD. However, something relevant is the appearance of several events at the low $Q_1$ but high $Q_2$ region. These are ``pile-up" events, accidental coincidences with several particles within the same time window. An extra filter is used to discard these events. The red dashed lines illustrate the linear bounds that one can apply in other to remove the majority of the accidental coincidences and gamma interactions. The lower cut will discriminate most of the gamma events and the higher one the pile-up events, resulting in a selection of mostly neutrons in a mixed environment. After the PSD is completed, primarily the neutron-like events will be tagged as double-scatter candidates. A further step to check the PSD algorithm is to put a gamma shield in front of the $^{252}$Cf source, which will identify only neutrons-like events. We are finishing the design to support the lead bricks and the source to allow its placement in different positions, which include the ground, walls or wardrobes.

\begin{figure}[ph!]
    \centering
    \includegraphics{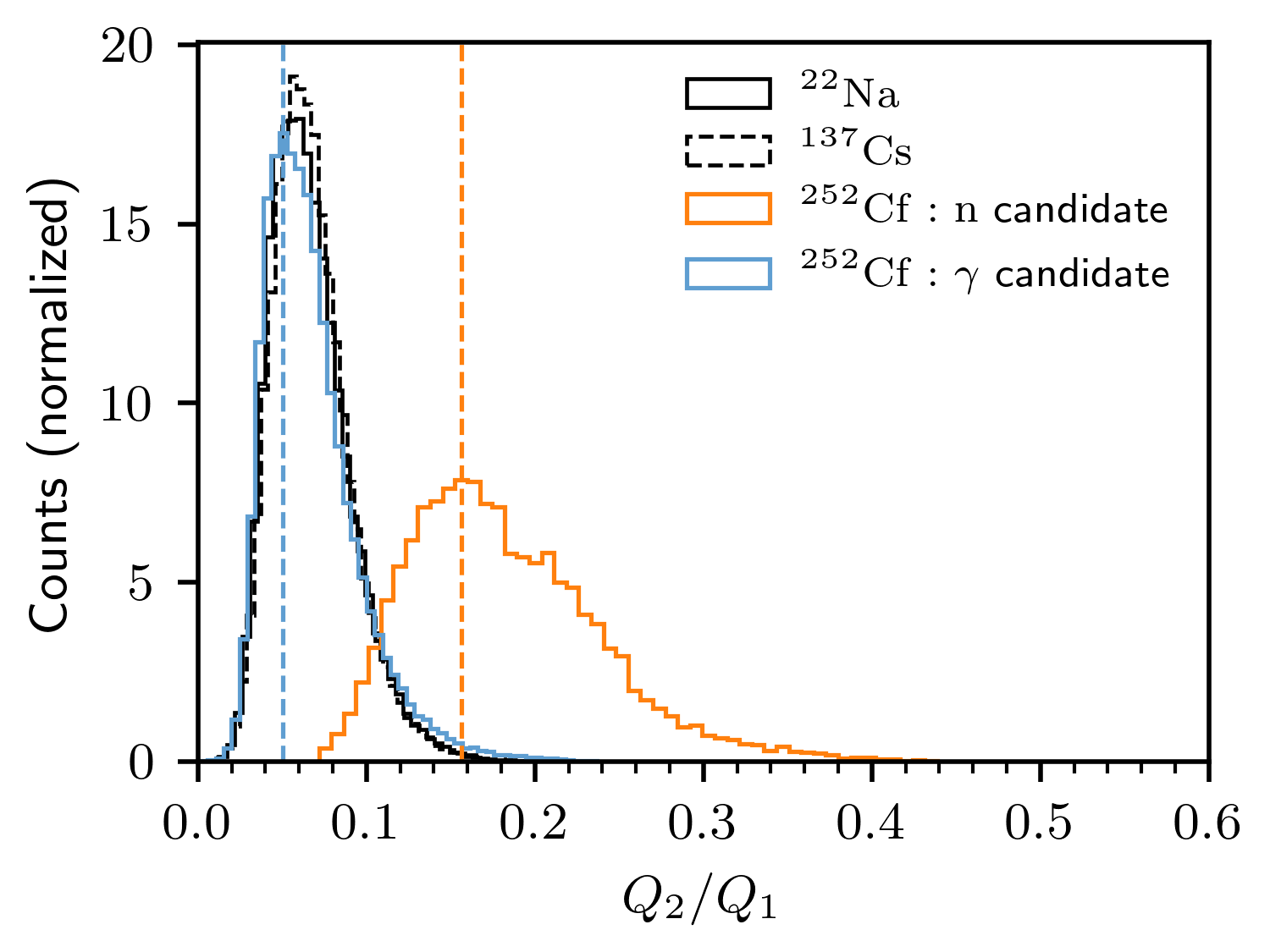}
  \caption{Neutron-gamma discrimination using $Q_1-Q_2$ PSD method. This is the distribution of the discrimination metric $Q_2/Q_1$ for both $^{252}$Cf gamma and neutron candidates together with the equivalent trading for the $^{22}$Na and $^{137}$Cs sources. Being the figure-of-merit the ratio between the separation of the neutron and $\gamma$ peaks and their summed FWHM (equation~\ref{eq:FOM}), the efficiency of the discrimination parameter can be evaluated.}
  \label{fig:Q1Q2_FOM}
\end{figure}

In order to quantify the performance of the n-$\gamma$ discrimination methods, and so optimize them, a figure-of-merit (FOM) is defined as in equation~\ref{eq:FOM}: 

\begin{equation}
    \rm{FOM} = \frac{d}{\rm{FWHM_n} + \rm{FWHM}_{\gamma}},
    \label{eq:FOM}
\end{equation}
where d is the separation between the n and $\gamma$ peaks in the discrimination parameter distribution \cite{luo2014test}. The metric used for such evaluation in this work is $Q_2/Q_1$. Figure~\ref{fig:Q1Q2_FOM} presents the distribution of the metric for the illustrative case in figure~\ref{fig:Q1Q2sourceallPMT}. The events above the ``pile-up" filter were removed. For the $^{252}$Cf source, the neutron event candidates (orange) bounded to the region between the two filters, and the $^{252}$Cf gamma candidates under the $\gamma$ cut (light blue) are shown together with the equivalent trading for the $^{22}$Na and $^{137}$Cs sources. Looking at the peaks and the FWHM in the distribution of the metric $Q_2/Q_1$, from equation~\ref{eq:FOM} one can derive the FOM value and so, the efficiency of the discrimination parameter be evaluated. Usually, a larger FOM value means a better n-$\gamma$ discrimination. 

\begin{figure}[ph!]
    \centering
    \begin{tabular}{@{}cc@{}}
    \includegraphics[width=.5\textwidth]{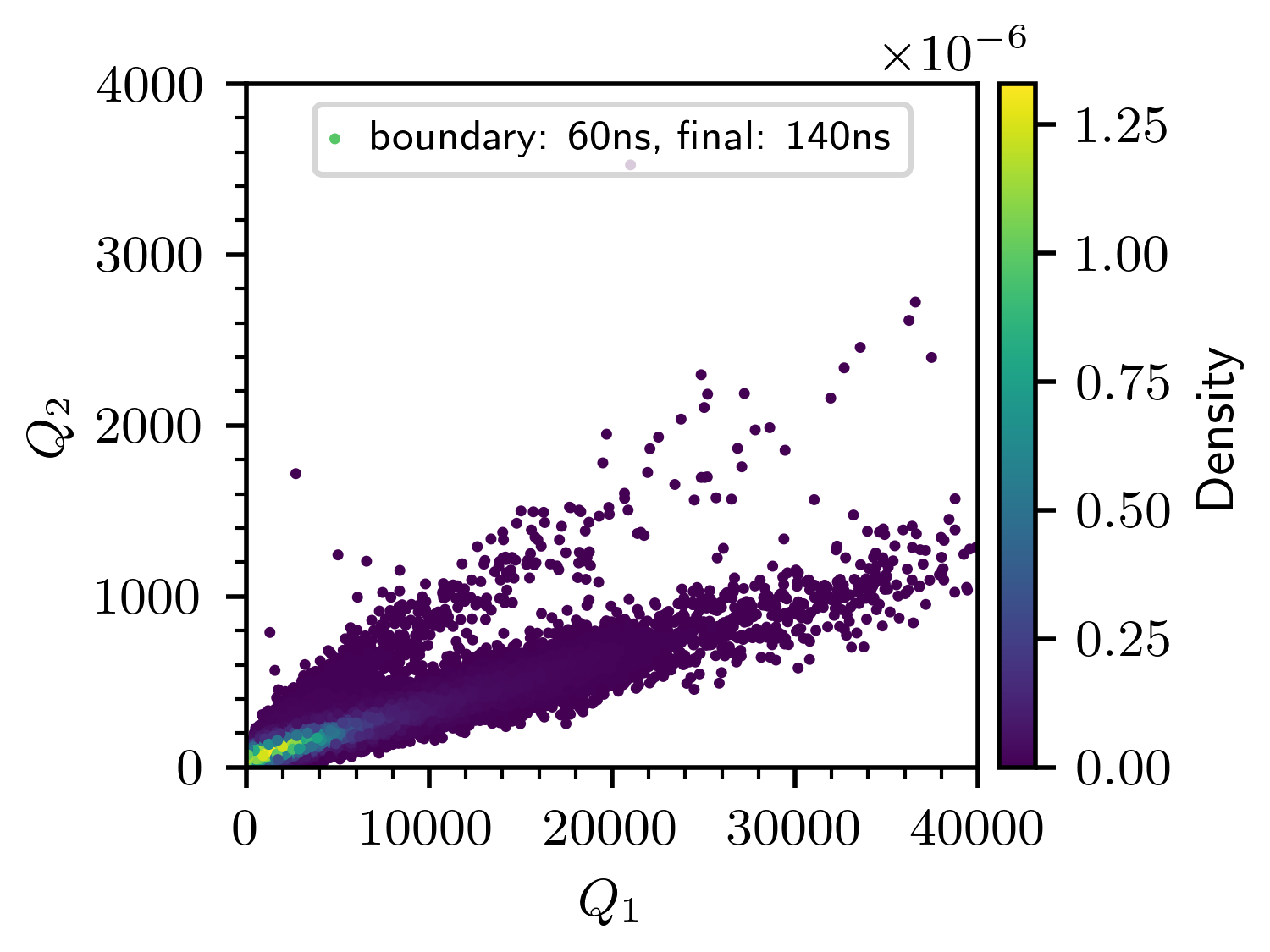} &
    \includegraphics[width=.5\textwidth]{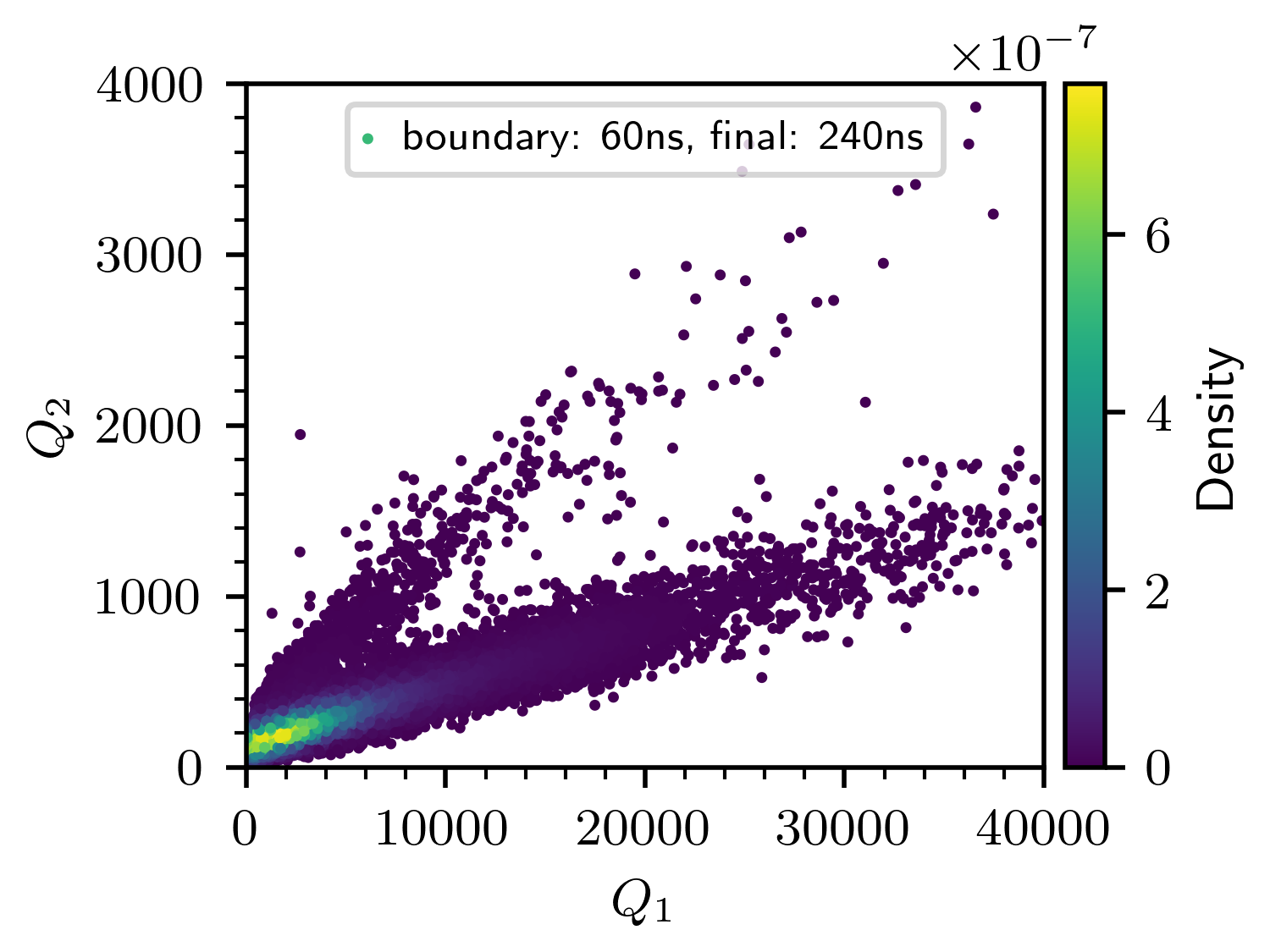}
    \end{tabular}
  \vspace{\floatsep}
    \begin{tabular}{@{}cc@{}}
    \includegraphics[width=.5\textwidth]{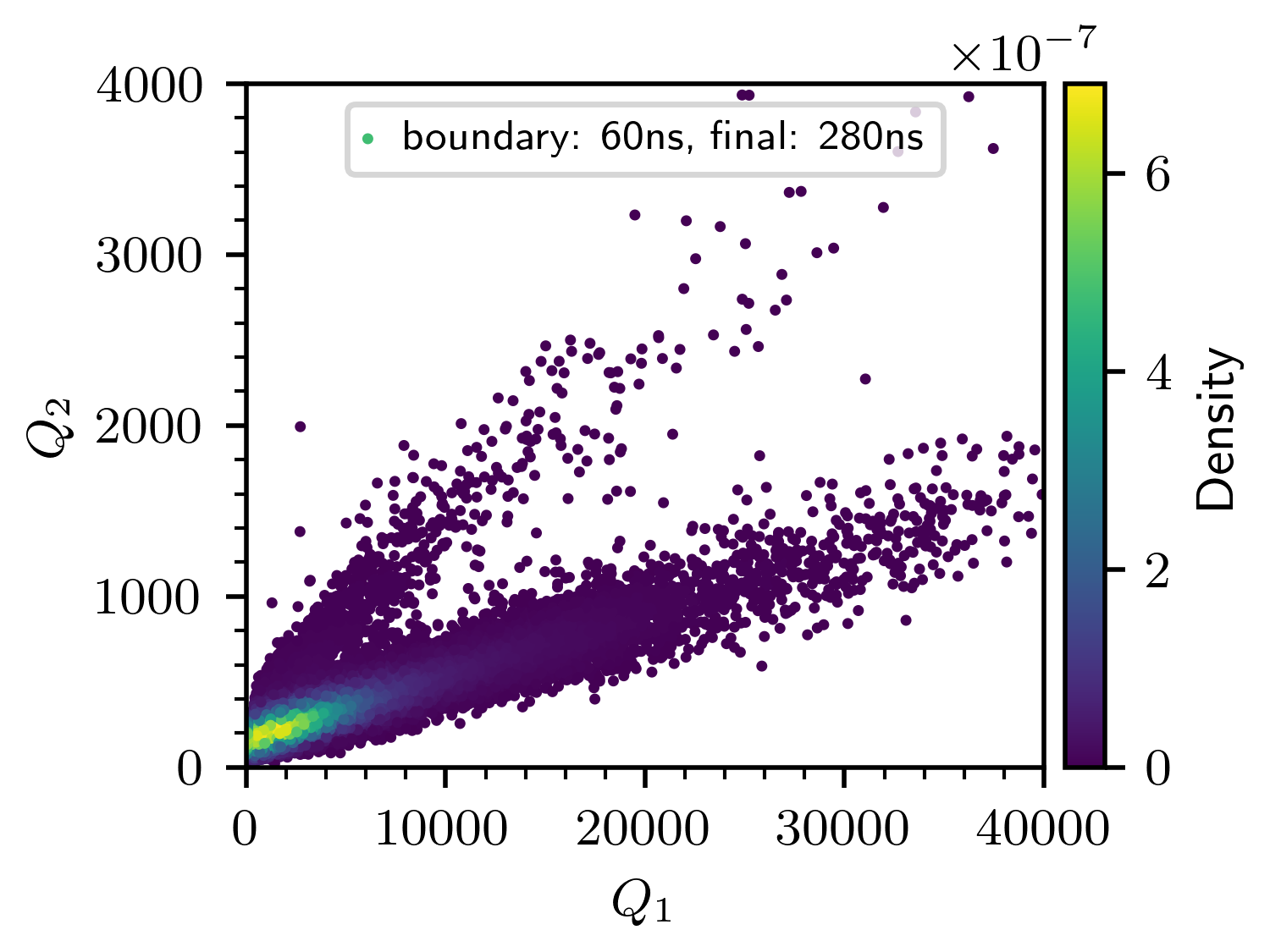} &
    \includegraphics[width=.5\textwidth]{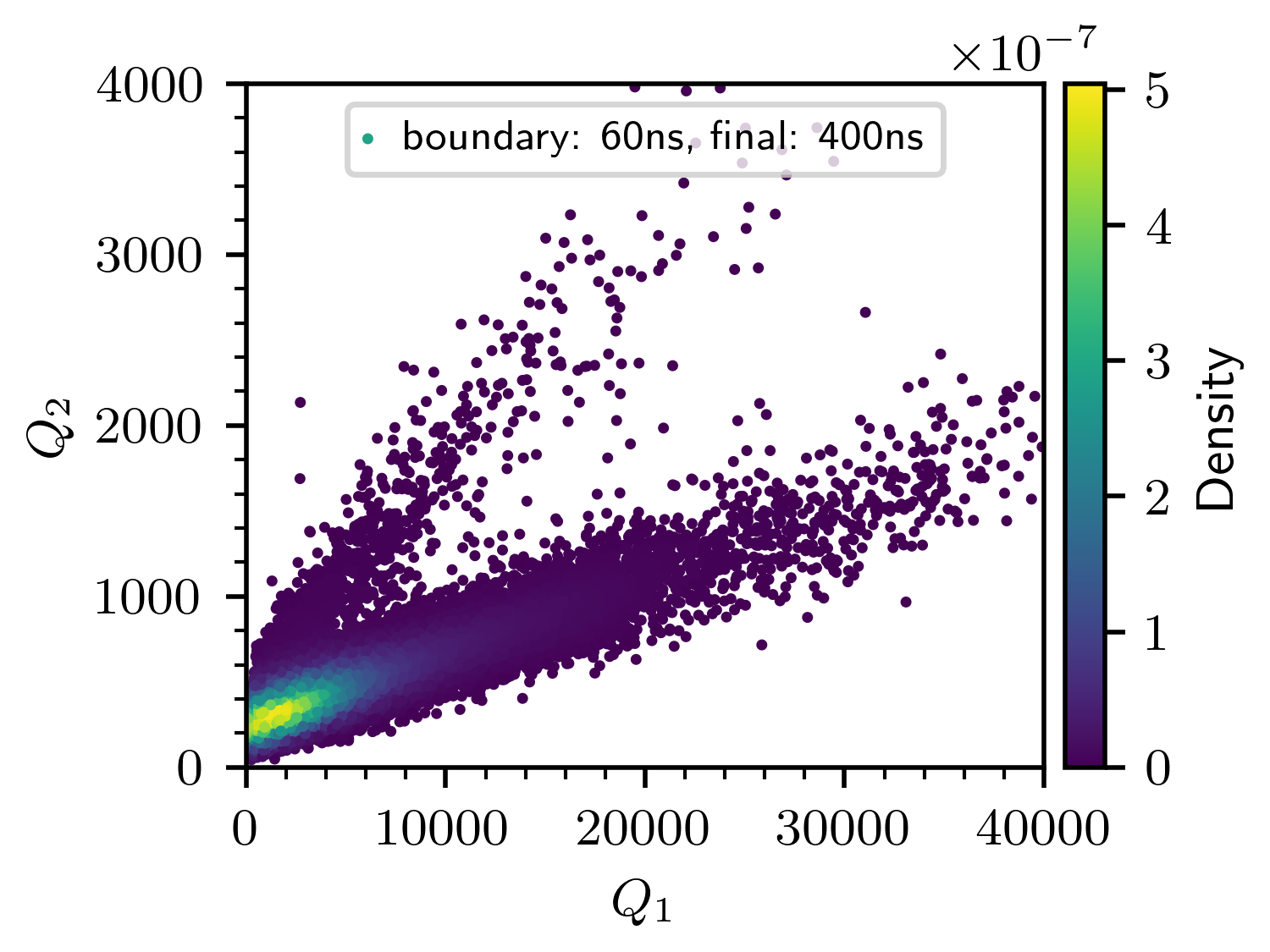}
    \end{tabular}
  \caption{The same neutron camera response to a $^{252}$Cf source but setting different ends on the pulse tail to identify the $Q_1-Q_2$ ratio. Two regions are clearly defined visibly merging towards lower charge values, each one corresponding to neutrons (upper lobe) or gamma rays (lower lobe) interacting with the scintillator. The efficiency of discriminating these two regions will determine the PSD potential. An analysis of the full pulse with minimum noise introduction is desired. Setting $t_{\rm{final}}$ = 140 ns (70 samples) is precipitated from the pulse tails in figure~\ref{fig:ncameraNormAmp}, contrary to $t_{\rm{final}}$ = 400 ns which shows the greatest separation between the regions, but the contribution of noise is larger. So $t_{\rm{final}}$ = 240 ns will ensure the analysis of the full pulse, with minimum noise introduction and a visible distinction of the neutron-gamma regions based on $Q_1-Q_2$.}
    \label{fig:Q1Q2sourceallPMT_combtf}
\end{figure}

To generate this metric and begin discriminating neutron events from gamma events, the first thing is to bound each pulse analyzed, in the same way as the initial bound is based on the onset. This will help reduce excess noise contributions in the slow component of a signal decay, which may blur the separation between the discrimination metric.
For illustration, considering the response of a single scintillator volume from $^{252}$Cf source, figure~\ref{fig:Q1Q2sourceallPMT_combtf} shows the $Q_1-Q_2$ distributions when setting different final bounds $t_{\rm{final}}$ for a fixed $t_{\rm{B}}$=60 ns. Two regions are clearly defined, each one corresponding to neutrons or gamma rays interacting with the scintillator. As before, the efficiency of discriminating these two regions will determine the PSD potential. The case of the longest $t_{\rm{final}}$ (400 ns) shows the greatest separation between the regions. However, the contribution of noise will be larger in this case too. After looking at several waveforms, identifying the tail and the noise introduced, an optimal and conservative final cut in time ($t_{\rm{final}}$) is set to 240 ns. In figure~\ref{fig:ncameraNormAmp}, one can see that the average waveforms end at around 60-70 samples or 120-140 ns, so the $t_{\rm{final}}$ = 240 ns will ensure the analysis of the full pulse, with minimum noise introduction. 

\begin{figure}[ph!]
    \centering
    \begin{tabular}{@{}cc@{}}
    \includegraphics[width=.5\textwidth]{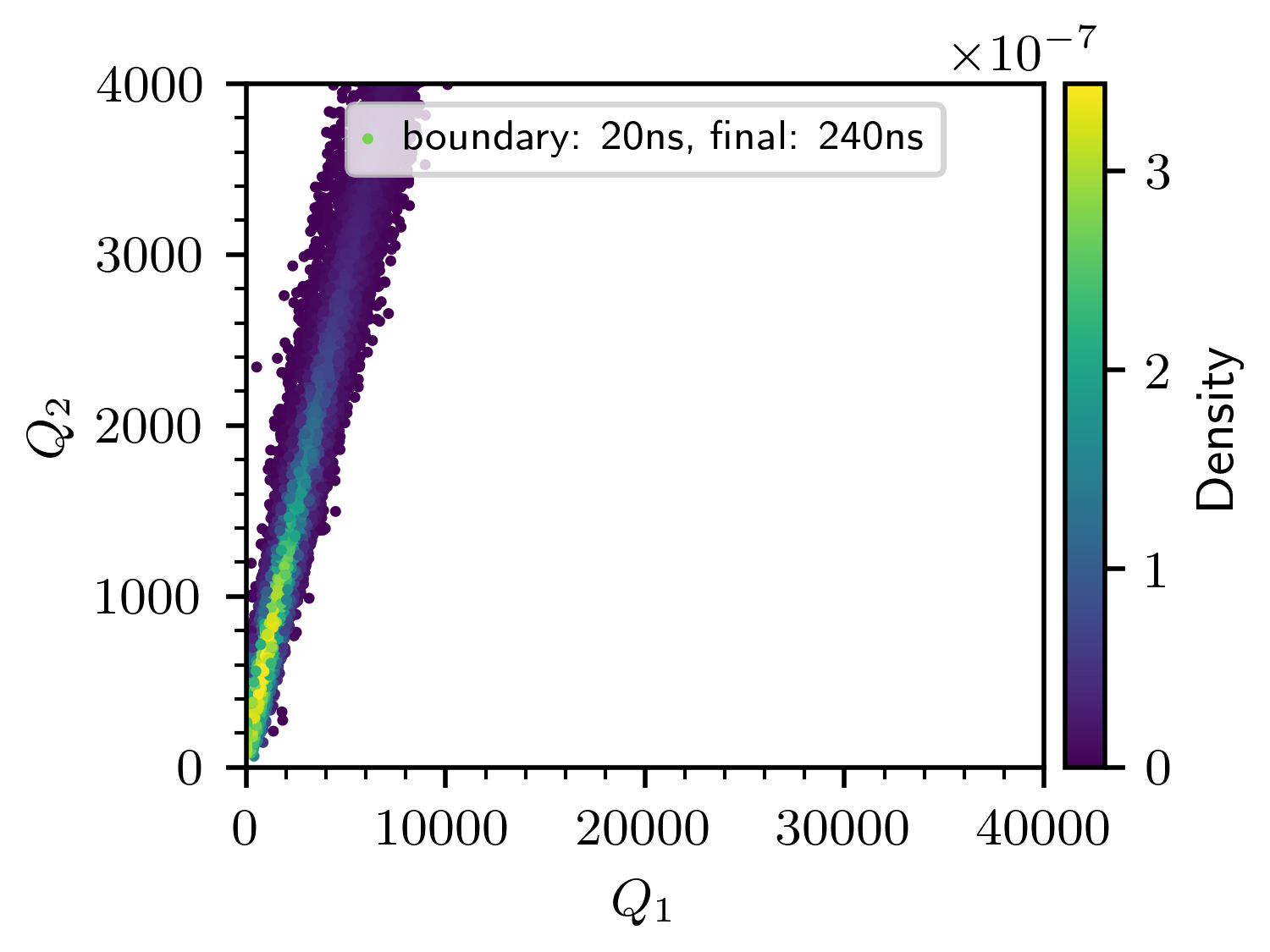} &
    \includegraphics[width=.5\textwidth]{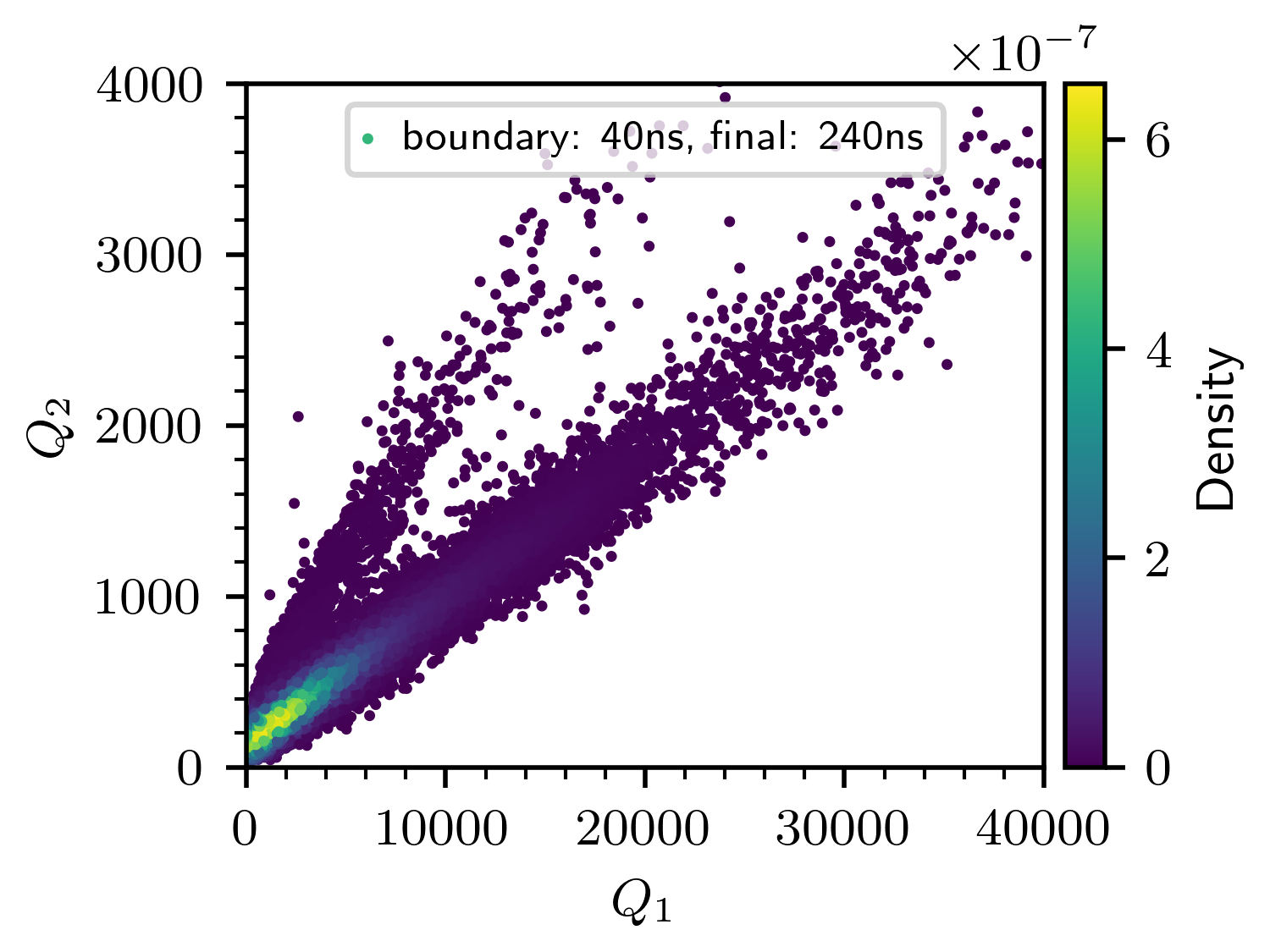}
    \end{tabular}
  \vspace{\floatsep}
    \begin{tabular}{@{}cc@{}}
    \includegraphics[width=.5\textwidth]{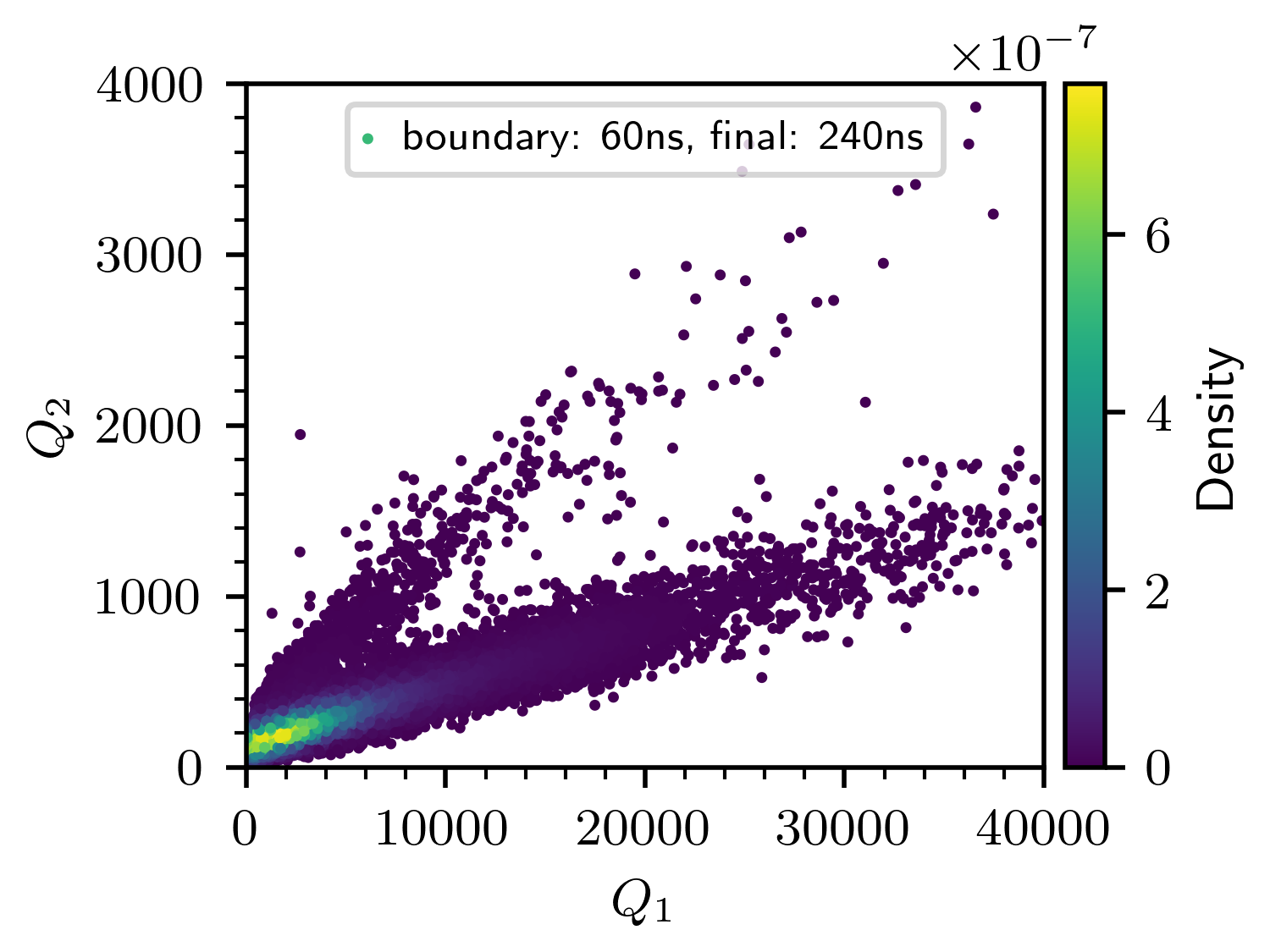} &
    \includegraphics[width=.5\textwidth]{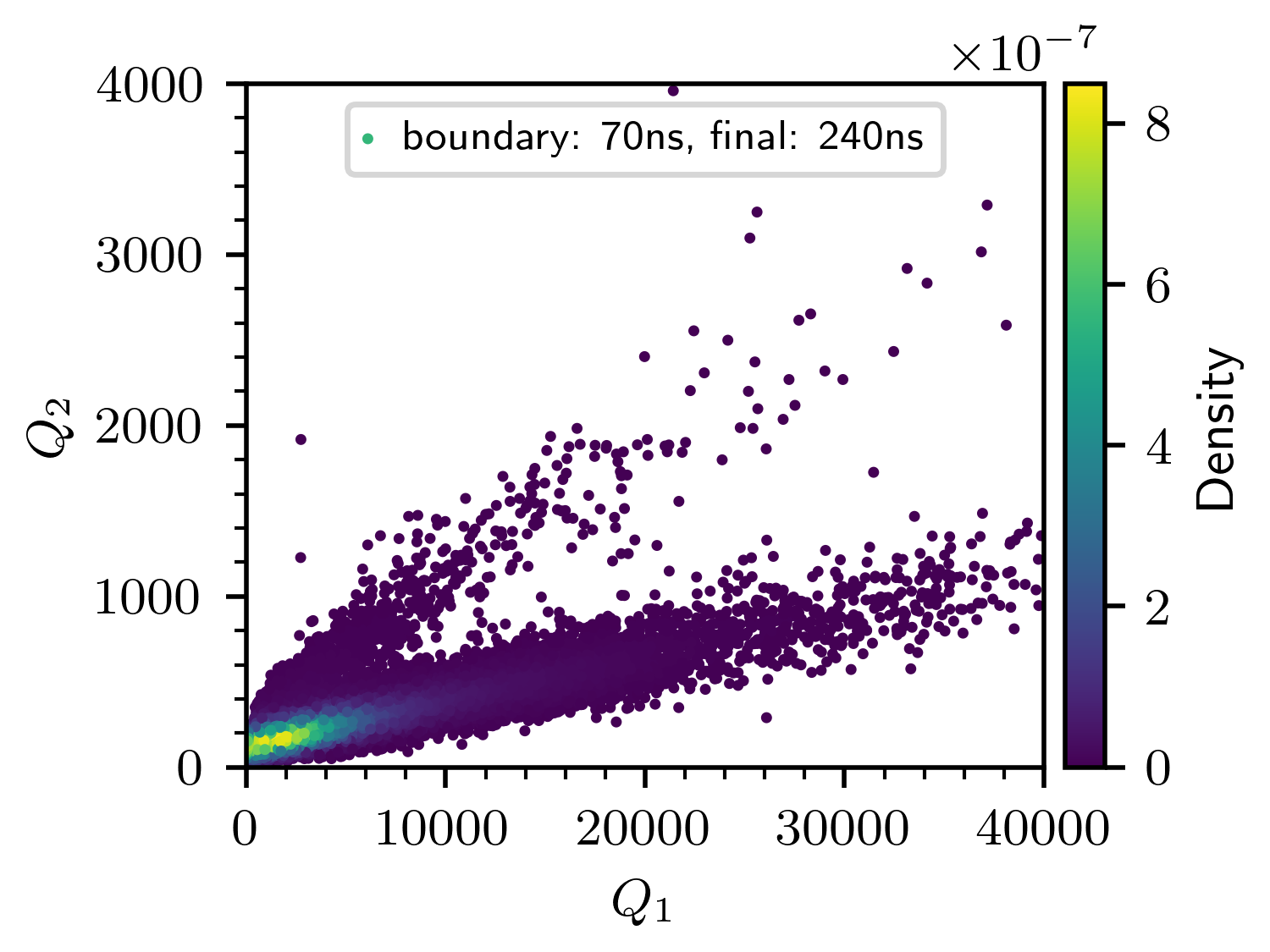}
    \end{tabular}
  \caption{The same neutron camera response to a $^{252}$Cf source but setting different boundary times to identify neutron and $\gamma$ interactions. The upper left plot fails to identify the two expected lobes from the $^{252}$Cf neutron and gammas emissions. The rest show two regions visibly merging towards lower charge values, defining neutron events in the upper lobe and gammas in the lower, as they have faster decay times and thus smaller $Q_2$ when the boundary time $t_{\rm{B}}$ is properly set.}
  \label{fig:Q1Q2sourceallPMT_combtb}
\end{figure}

In the same way, the boundary time must be optimized such that the neutrons' distinctly larger $Q_2$ values are separated from pure ionization signals. If $t_{\rm{B}}$ is set too early in decay time, the response of n and $\gamma$ will overlap, as it is the case of the upper left density plot in figure~\ref{fig:Q1Q2sourceallPMT_combtb}. This figure shows the same dataset as figure~\ref{fig:Q1Q2sourceallPMT_combtf}, having set $t_{\rm{final}}$=240 ns while varying $t_{\rm{B}}$. Except for the upper left panel, there are two regions visibly merging towards lower charge values. As $t_{\rm{B}}$ is increased the distinction of the neutron and gamma lobes is highlighted, up to a point where the slow component of neutrons is also included in $Q_1$ instead of $Q_2$. Therefore, an optimized value for the boundary time should be set. This is a robust process that by evaluating the FOM values in the $Q_2/Q_1$ metric for different $t_{\rm{B}}$ and $t_{\rm{final}}$ combinations, the optimized configurations and filter regions are determined. Overall, we have concluded to set $t_{\rm{B}}$=60 ns and $t_{\rm{final}}$=240 ns as time bounds for an optimized n-$\gamma$ discrimination in the neutron scatter camera, which is the scenario shown the earlier in figures~\ref{fig:Q1Q2sourceallPMT} and \ref{fig:Q1Q2_FOM}. 

\begin{figure}[hbt]
\centering
    \includegraphics{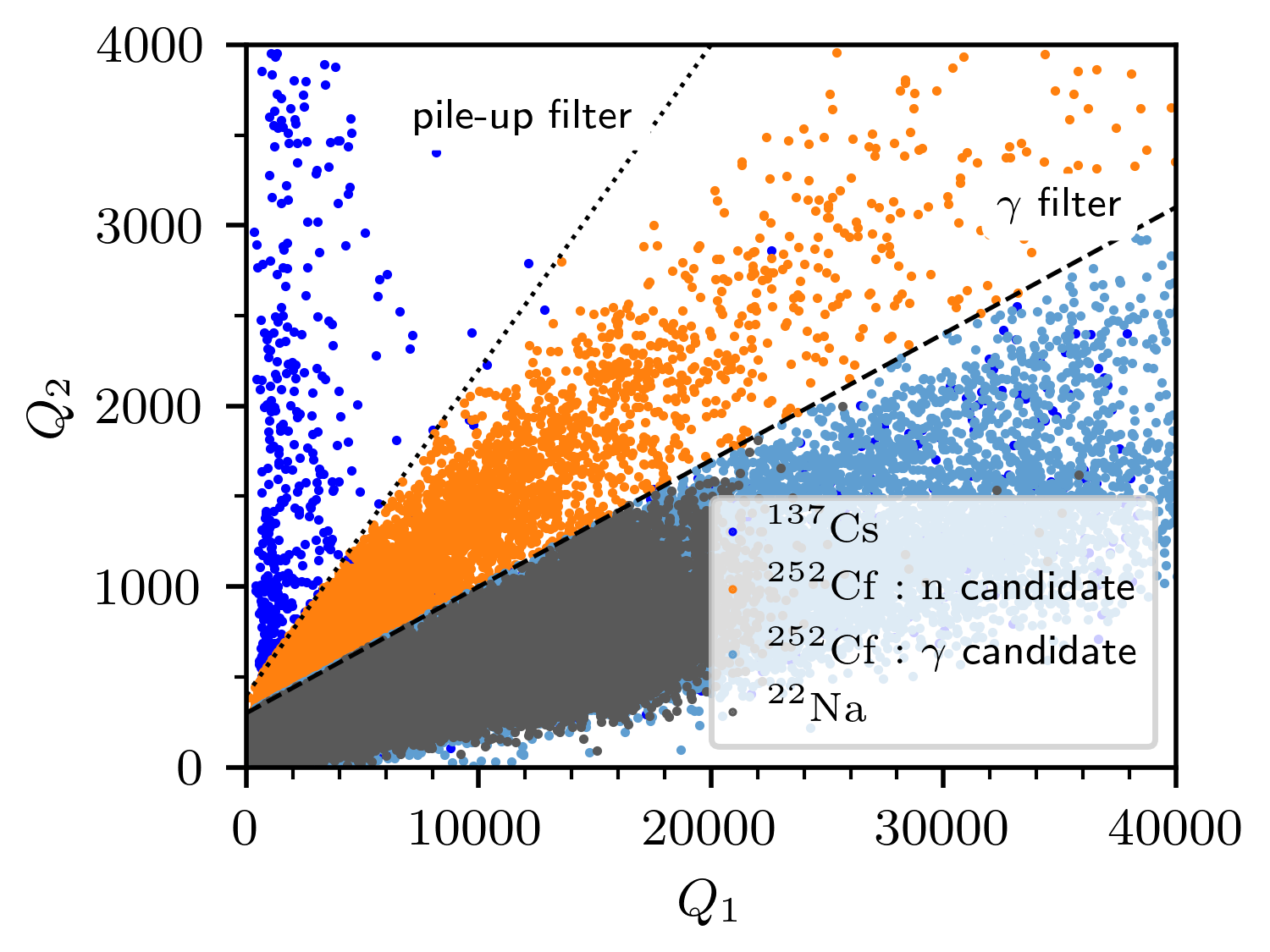}
    \caption{Neutron-gamma discrimination using $Q_1-Q_2$ PSD method. This is the overlap of the response to $^{252}$Cf, $^{22}$Na and $^{137}$Cs from all the detectors. The optimized gamma discrimination criteria is indicated by a dashed black line whereas the dotted black line filters out the rest of the events that are not neutrons. Thus, the remaining $^{252}$Cf neutron-like events are shown in orange and the gamma candidates in light blue. The overlay of the latter and the response of the detector to a $^{137}$Cs source is covered up by the $^{22}$Na source response (light gray).}
    \label{fig:Q1Q2method}
\end{figure}

Likewise, figure~\ref{fig:Q1Q2method} illustrates the impact on event selection due to the filter cuts applied in the dataset from figure~\ref{fig:Q1Q2sourceallPMT}. The response from $^{22}$Na (dark gray dots), $^{137}$Cs (blue dots) and $^{252}$Cf (light blue and orange dots) is overlapped. Once the linear cuts are applied to distinguish possible gamma or neutron events or other events, an orange-dotted region displays the remaining neutron-like events.

Further analysis can be done in the discrimination of neutrons from $\gamma$ and other particles, yet the $Q_1-Q_2$ method for PSD demonstrated a great performance. Furthermore, the imposed detection threshold (see section~\ref{sec:pmtEthres}) will enhance the discrimination as low integrated charge will be cutoff. Consequently, the optimized PSD used throughout the rest of the chapter is illustrated in figure~\ref{fig:Q1Q2method}. 

\section{Simulations to validate the operation of the camera}
\label{sec:ncamerasimG4}

The neutron camera is meant to characterize the spallation neutron background in some of strategic locations, but the first step is to validate the technology and this has been done with an accessible source, precisely, $^{252}$Cf. Therefore, to test if the design for fast neutrons is also effective for lower energy ones, centered at $\sim$1 MeV, several scenarios have been simulated in Geant4~\cite{geant4} and the reconstruction algorithms checked. Energy and position reconstructions were developed, first with a mono-energetic simulation before moving on to a full $^{252}$Cf source evaluation, following the same considerations as in the data analysis introduced in the coming section~\ref{sec:ncameramanual}. 

Ultimately, to evaluate the performance of our neutron camera the comparison with results from the simulations will be the metric. Therefore, the assumption that the neutron scatter is produced with a proton has also been included in the simulation analysis. Though the Geant4 simulation can provide detailed information on the interacting nuclei, the visible energy in the scintillator has no such information deducible and only the electron-equivalent energy is expressed. The quenching factor, which is nuclei-dependent, quantifies the energy dissipated in a nuclear recoil via the ionization channel compared to an equivalent energy electron recoil. Thus, for an interaction energy $E_{\text{nr}}$ the visible energy deposition $E_{\text{ee}}$ reads as $E_{\text{ee}} = \text{QF} (E_{\text{nr}}) \cdot E_{\text{nr}}$. Hence, the quenching factors of hydrogen and carbon are considered independently to bring the simulation to electron recoil-space to match the data. The quenching factor distributions considered are the well-known unimolecular quenching models attributed to Birks~\cite{birks2013theory}.

\subsection{Mono-energetic neutron simulation}

An initial test to validate the Monte Carlo and the reconstruction algorithms was performed by simulating an ideal isotropic and point-like neutron source with an energy of 1 MeV. The source was placed 1 m from the center of the neutron camera orthogonal to one of the faces of the cubic configuration. This is an ideal scenario to check the performance of both Monte Carlo simulation and reconstruction algorithms.
  
\begin{figure}[htb]
  \begin{tabular}{@{}cc@{}}
    \hspace{-0.1in}
    \includegraphics{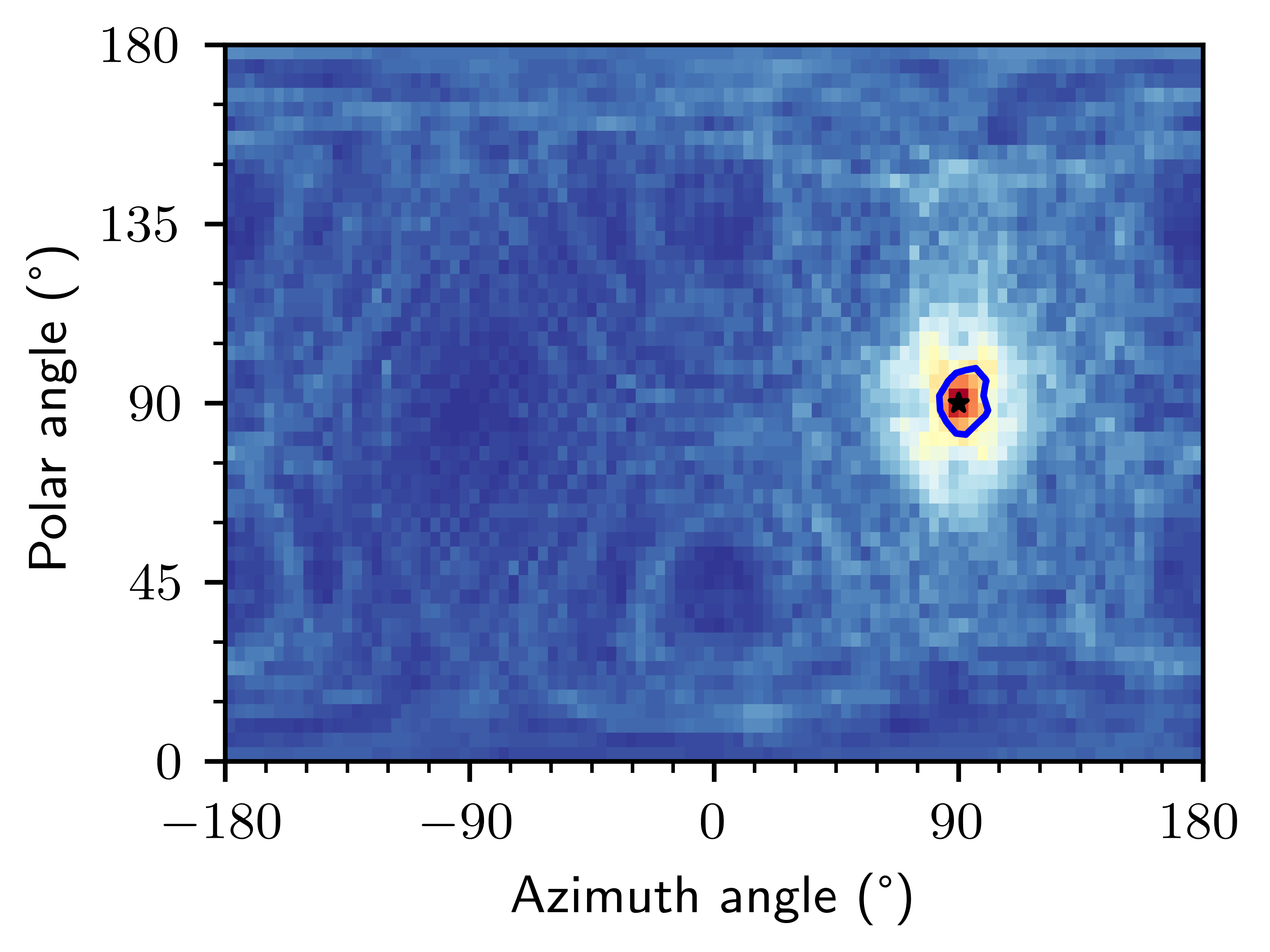} 
    \hspace{0.1in}
    \includegraphics{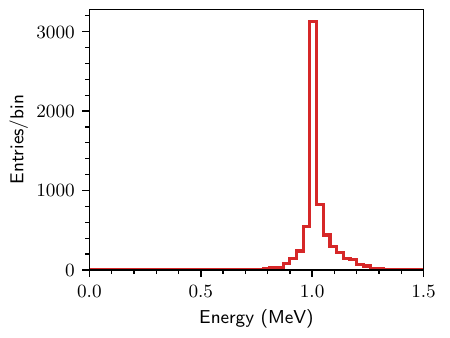}
  \end{tabular}
  \caption{Left: position reconstruction of a 1 MeV isotropic neutron source 1 m from the neutron camera simulated in Geant4. In black the true position of the simulated source and in blue the 2D 1$\sigma$ contour line. As it can be observed the position reconstruction algorithm properly reconstructs the origin of the neutrons. Right: energy reconstruction of the camera for the mono-energetic neutron source placed in the same position. The spread is caused by the uncertainties in the scattered neutron energy calculation.}
    \label{fig:ncameraSIMreco1MeV}
\end{figure}

Based on the theory covered in section \ref{sec:ncameraconcept}, figure~\ref{fig:ncameraSIMreco1MeV} shows the position and energy reconstruction of a 1 MeV isotropic neutron source placed 1 m away from the neutron camera. In the left plot, the overlap of multiple reconstructed cones, the back-projection of each simulated event that hits two scintillators, is shown. Those cones are then projected onto a 2D mapping of the surrounding space. In blue, the contour to 1$\sigma$ is specified and, in black, the real position of the simulated neutron source which indicates excellent coincidence. This overlap between the expected and reconstructed neutron location proves the efficacy of the position reconstruction algorithm, in the most basic context, that will be used to analyze other scenarios. On the right, the energy reconstruction is presented, where there is a clear peak at 1 MeV, but with a spread in energy caused by uncertainties in the scattered neutron energy, exaggerated by assuming a finite size for the scintillators, and by projecting carbon recoils events as if they were hydrogen.

\subsection{$^{252}$Cf source simulation}

The next step was to simulate a realistic $^{252}$Cf source to replicate the datasets that will be later introduced. Though $^{252}$Cf is an intense neutron emitter, it is a hybrid source that decays by spontaneous fission only 3.09$\%$ of the times and 96.91$\%$ by photon emission. For simplicity and because the detectors have n-$\gamma$ discrimination capability, instead of simulating the whole $^{252}$Cf decay chain, only the expected neutron spectrum was assumed. This can be described with the Watt equation as in equation~\ref{eq:Watt} \cite{radev2014neutron}:

\begin{equation}
    N(\rm{E}) \propto e^{-\rm{E}/a} \sinh{(\sqrt{b\rm{E}})},
    \label{eq:Watt}
\end{equation}
where E is the neutron energy in MeV. For $^{252}$Cf, a=1.175 MeV and b=1.034 $\rm{MeV}^{-1}$ with the spectrum peaking at 0.70 MeV and with a mean of 2.13 MeV, reaching energies up to $\sim 10$ MeV.

Moreover, the neutron multiplicity has been included. Dedicated measurements have been performed over decades to determine the density function of the number of neutrons emitted per fission, which has a mean of 3.76 \cite{pronyaev2017new}. Therefore, neutrons were generated following the energy spectrum in equation~\ref{eq:Watt} and with multiplicity following the density function laid out in \cite{pronyaev2017new}.

\begin{figure}[!ht]
\centering
  \hspace{-0.1in}
\includegraphics{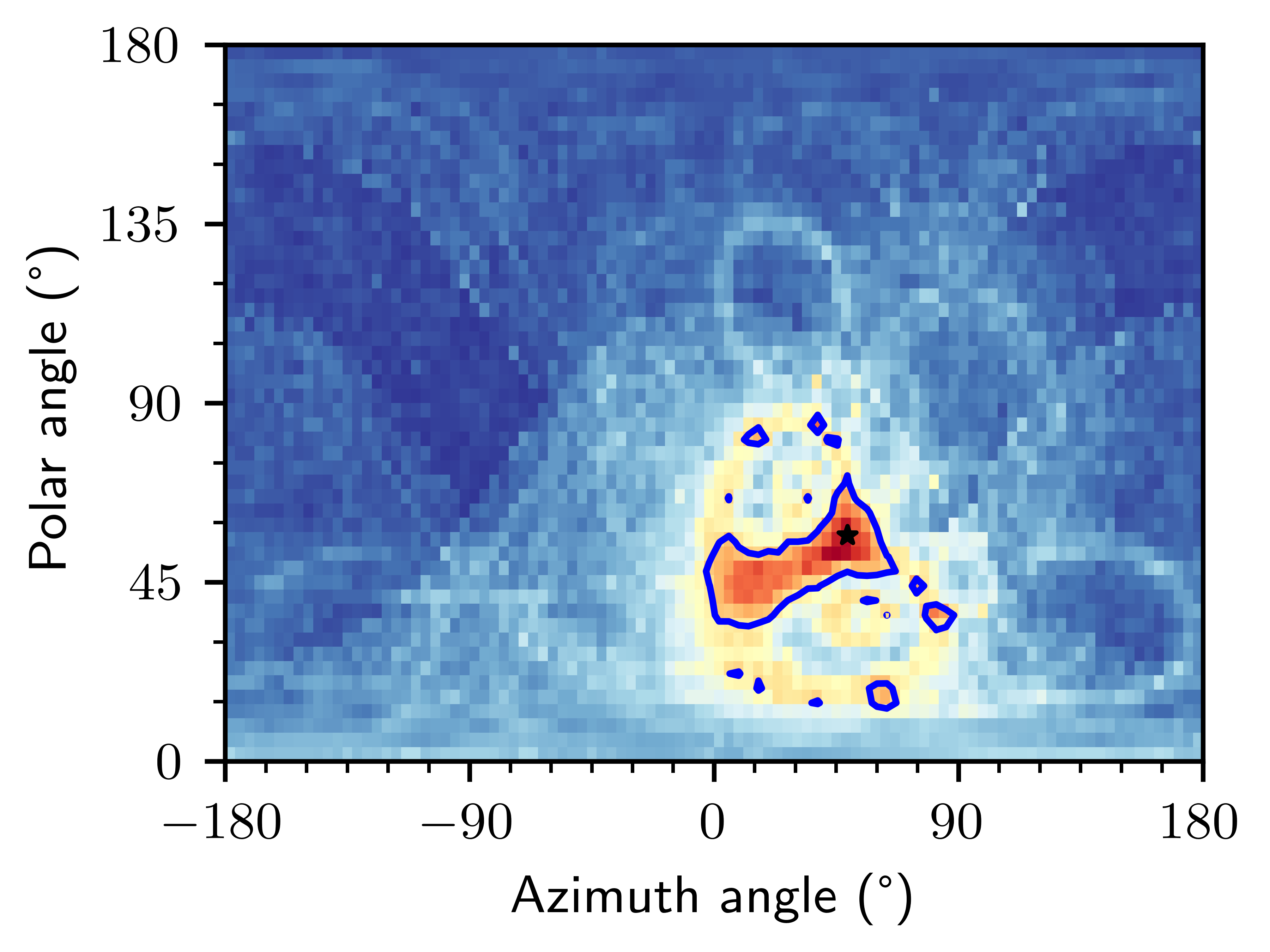} 
  \hspace{0.1in}
\includegraphics{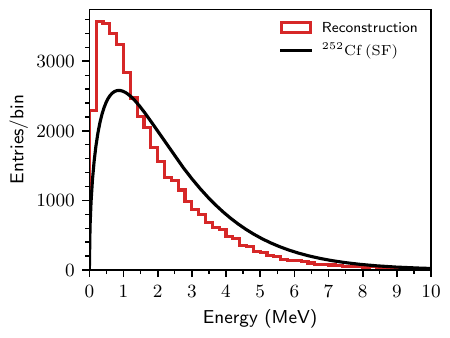}
\caption{Left: position reconstruction of the Geant4 simulated neutrons from $^{252}$Cf placed at one of the corners of the neutron camera frame. In black the true position of the simulated source and in blue the 2D 1$\sigma$ contour line. As it can be observed the position reconstruction algorithm properly reconstructs the origin of the neutrons in this case too. Right: in red, the energy reconstruction of the camera for the $^{252}$Cf source placed in the same position. In black, the expected  $^{252}$Cf Watt spectrum. The reconstructed spectrum shifting towards lower energies is mainly due to the elastic scattering cross-section decreasing as the neutron energy increases, and neutrons scattering off the environment geometry.}
\label{fig:ncameraSIMrecoCf}
\end{figure}

The location and energy reconstruction for a $^{252}$Cf source placed at one of the corners of the neutron camera frame is shown in figure~\ref{fig:ncameraSIMrecoCf}. The reconstructed location of the source in the left panel is in agreement with the expected position (black star) as the blue 2D 1$\sigma$ contour line backs up. The proximity of the source has an indispensable role and taking into account that the corner at which the source was simulated is $\sim$45 cm from the center of the neutron camera, the spread on the angular resolution compared with figure~\ref{fig:ncameraSIMreco1MeV} left was expected. But even if affected by neutron multiplicity the position of the source can be visibly identified by the hot region in the left panel and this will be the metric to evaluate the 4-$\pi$ coverage of the device in section~\ref{sec:ncameralocrecons}. In figure~\ref{fig:ncameraSIMreco1MeV} right, deviations are shown in the energy reconstruction of the $^{252}$Cf Watt spectrum, showing a shift towards lower energies. Even if there might be an underestimation of the energy due to uncertainties, the main contribution to this difference comes from the cross section of MeV-scale neutrons with hydrogen and carbon, as the elastic scattering cross section decreases when the neutron energy increases. Additionally, it is more probable that a low energy neutron double scatters in the set-up and therefore, be considered as an event candidate. Moreover, a neutron can take more complicated paths by scattering off the environment geometry, like the aluminum extrusions, that also lead to under-reconstructing the energy of the incident neutron.

An adequate performance of the location reconstruction algorithm is proved by the 1 MeV neutron and corroborated by the $^{252}$Cf source study. Furthermore, being able to analyze sources in lower energy ranges than spallation neutrons indicates promising results but additional work should be done to characterize the neutron camera in the energy region relevant to a spallation source.

\section{Response of the camera to a $^{252}$Cf source}
\label{sec:ncameramanual}

There are two main stages in the operation of the neutron camera system. The data acquisition trigger system and the double-scattered neutron candidate identification algorithm. 

The eight scintillator volumes act as independent detectors and a coincidence trigger set-up enables the acquisition of double-scatter signals from every combination of these eight detectors. This provides full coverage of the surrounding space, allowing for the identification of a neutron source incoming from any direction.

\subsection{Energy threshold}
\label{sec:pmtEthres}

A common triggering threshold in ADC counts is set for every detector. However, since each baseline is slightly different, one needs to identify the energy threshold for each detector. From the raw waveform one can extract the baseline from the median of the pre-trigger portion of the digitized waveform, the amplitude, and the integrated charge. This way, the amplitude of a pulse can be related to its charge and, therefore, its energy. From projecting the energy downward, the amplitude threshold set can be translated into an equivalent energy threshold. The energy threshold that is set for the data processing is slightly higher than the maximum PMT amplitude threshold that is channel-specific due to the fluctuating baselines. The threshold is defined based on the n-$\gamma$ discrimination criteria using the PSD $Q_1-Q_2$ method described in section~\ref{sec:ncameraQ1Q2}, where the bound to discriminate neutron from gamma responses in the scintillators was imposed by a linear regression with an intercept in $Q_2$ that corresponds to $\sim$400 in charge units, which is translatable into energy. Even if the amplitude threshold is unique for each PMT in the data acquisition, a unified energy threshold is used for the data analysis of 35 $\text{keV}_{\text{ee}}$.

\subsection{General overview of the data processing}

The identification of the coincidences that are neutron-induced requires signal de-noising techniques, peak-prominence ratio filters, energy cutoffs, and time-of-flight limits. From the raw waveforms, a wide range of parameters can be obtained including the start of the pulse, the height, the amplitude and the charge parameters. This set of parameters extracted via LabView\textsuperscript{\textcopyright}, is used to filter, identify and exclude events. 

First, the Pulse Shape Discrimination (PSD) using the $Q_1-Q_2$ method described in section~\ref{sec:ncameraQ1Q2} is applied to each deposition in the detector volumes that triggered in coincidence. This removes many gamma scatters and pile-up events, ensuring that only neutron-like events remain among the double-scatter candidates. The corresponding energy calibration is then used to set a communal minimum energy threshold at a conservative $\text{keV}_{\text{ee}}$ as indicated in section~\ref{sec:pmtEthres}.

Nevertheless, as one can see from the response of the detector, after placing different gamma sources ($^{22}$Na and $^{137}$Cs) as in figure~\ref{fig:Q1Q2sourceallPMT}, some events would not be excluded by the PSD filter and therefore be considered as neutron-like events. This is principally due to the converging distributions in the low energy region. As both gamma and neutron emission spectra from $^{252}$Cf congregate in the low energy region, these energy and discrimination cuts would distort the Watt spectrum reconstruction. These together with background neutron events can results in some accidental coincidences that pass into the dataset. 

A selection on TOF helps discriminate accidental coincidences. For instance, a neutron scatter may initiate the trigger, and within a small time window a background gamma that passed the PSD filter or another neutron, considering the neutron multiplicity of $^{252}$Cf, could interact in a second detector. These false coincidences may happen from no delay between signals, if a scattering gamma, to long timeframes if accidental coincidences. As described in section~\ref{sec:ncameraconcept}, one can relate the time-of-flight and the energy of the scattered particle. Therefore, setting bounds on the energy ranges of interest, the calculated TOF of double scattered interactions, and the deposited energy help reduce accidental coincidences.

One can consider the limit case where just the threshold energy is deposited in the first scatter and derive the minimum TOF expected. A DAQ TOF cut has been applied of $\gtrsim$ 4 ns in order to account for sub-ns gamma scatters and smearing due to PMT response times. From equation~\ref{eq:tofEns}, it translates into maximum scatter energies of $\sim$53 MeV for a distance of $\sim$40 cm between the coincident volumes. A communal upper limit in TOF is also set at 120 ns to match the software limitations of the internal coincidence trigger (see section~\ref{sec:ncameradaq}). The TOF correction for each PMT pair (section~\ref{sec:transittime}) is necessary to establish an accurate separation in the double-scatter events.

\begin{equation}
    \tau(\text{ns}) = \frac{72.3 \, \text{ns}}{\sqrt{E(\text{MeV})}} \cdot d(\text{m})
    \label{eq:tofEns}
\end{equation}

After a great effort to reduce accidental coincidences, the contribution of background neutron events has also been studied, such as cosmic-ray induced neutrons in the atmosphere.
Applying the same event selection method, it is concluded that the background contribution is negligible compared to the signal frequency that an explicit neutron source provides, whether a californium source or the ESS neutron beam in the future. 

One can then reconstruct both the spectrum and the direction of the neutron source from the remaining double scatter events, based on the kinematics and cone back-projection process described in section~\ref{sec:ncameraconcept}. The next sections demonstrate the position and energy reconstructions of a $^{252}$Cf source in different positions around the camera. One should note that the neutron camera is being preliminary validated using radioactive sources of a lower neutron energy range than what it was designed for. The performance of this camera improves as the neutron source is placed further away (point-like reconstruction) and as the initial neutron energies increase, since the design was optimized for spallation neutrons of higher energies with distinct decay characteristics from electron recoil depositions. As a result, this technology validation performed using $^{252}$Cf can be extrapolated to verify the full coverage of the neutron scatter camera, but dedicated measurements with higher neutron energies are planned, alongside full operation at the ESS. 

\section{Spectrum reconstruction}
\label{sec:ncameraenrecons}

Using a $^{252}$Cf source with PSD and the event selection process, the reconstructed incident energy (equation~\ref{eq:En}) of the remaining neutrons is nearly analogous to the spontaneous fission spectrum of $^{252}$Cf. As seen earlier in section~\ref{sec:ncamerasimG4}, there are some discrepancies between this spectrum and the reconstructed energy from the Geant4 simulated neutron camera. As anticipated, several contribution factors overestimate the Watt spectrum at lower energies, such as neutron dissipation in the aluminum frame and the larger neutron-proton elastic scattering cross-section at those energies. However, neutron multiplicity also induce an erroneous energy reconstruction by falsely impersonating the presence of neutrons with higher energies in the event selection process. All in all, to validate the performance of the neutron camera, the reconstructed energy spectrum will be compared to the one derived from an analogous simulation in the energy space of electron-equivalent initial energy depositions and nuclear recoil-equivalent scattered energies. Nevertheless, resolution effects must be taken into account for a proper comparison. The energy smearing depends on the detector's resolution (equation~\ref{eq:Eres}) and the TOF resolution. The latter is the combination of the coincident PMT transit time spread (1.1 ns) and the DAQ timing resolution (1 ns) summed in quadrature. 

\begin{figure}[!ht]
\centering
\includegraphics{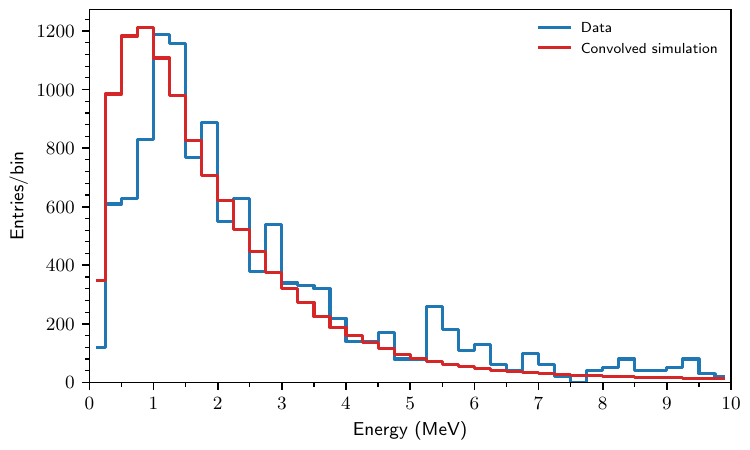}
\caption{Reconstruction of neutron energy spectrum from $^{252}$Cf compared to the simulated spontaneous fission spectrum. The source is placed $\sim$1 away from the center of the neutron camera system. The reconstructed energy is the sum of the initial electron-equivalent energy deposition ($E_{p_{\text{ee}}}$) and nuclear recoil-equivalent scattered energy ($E_{{ns}_\text{nr}}$).}
\label{fig:ncameraErecoCf}
\end{figure}

Figure~\ref{fig:ncameraErecoCf} shows the reconstructed energy of incident neutrons from a $^{252}$Cf source placed $\sim$1 m away. Together with it, the smeared convolved energy spectrum is also shown from the analogous Geant4 simulation with the source at the same distance, demonstrating the system's ability to reproduce the expected spectrum.

In general, the differences between the two spectra are attributable to false-positive double scatter events, statistics, and to the event selection method itself. The event selection accuracy has an impact as some background neutron double scatters or accidental coincident events may make it through. Differences in the lower energy regime are mainly derived through neutron-gamma discrimination criteria. This method is weaker at lower energy particle discrimination visible in figure~\ref{fig:Q1Q2sourceallPMT}, as some neutron events may be discarded in the same way that some gammas may be considered for the analysis. Only the Watt spectrum is simulated and the neutron selection in the measurements will lead to the removal some low-energy neutrons. Nonetheless, it demonstrates an impressive reconstruction of the spontaneous fission neutron emission tail. 

This result represents a turning point in the ability to accurately determine the hardness spectrum of incident neutrons in parity with simulation results. The cameras designed so far mostly focused their functionality on image reconstruction or had sub-par energy spectrum. These results show an impressive ability to identify a neutron flux originating from $^{252}$Cf.

To conclude, the combination of the reconstruction method and design of the camera, which is optimized for higher energies and far-located sources, will not act as a reliable neutron energy spectroscope for low-energy sources. However, the decent approximation to the neutron spectrum of $^{252}$Cf motivates measurements with more applicable neutron sources and thus, validate the technology in an energy regime closer to that of spallation sources. Moreover, it also shows a promising feature to distinguish sources with different neutron emission characteristics, like spontaneous fission with multiplicity as is the case studied in this work with a $^{252}$Cf source, and alpha-neutron ($\alpha$,n) sources such as Am-Be. In particular, there is an ongoing effort to test the neutron scatter camera with an Am-Be source. 

\section{Location reconstruction}
\label{sec:ncameralocrecons}

In order to demonstrate the performance of the neutron scatter camera and in locating neutron sources in a three-dimensional space, the $^{252}$Cf source was placed in several positions and at different distances. The relevant polar coordinates are then used to quantify the field of view from a point near the center of the camera. 

Possible source trajectories are mapped to a circle with a certain thickness on the surface area ($\theta, \phi$) of the enclosing spherical field of view. This band is the uncertainty from the derived cone axial angle (equation~\ref{eq:thetan}), where defines the $\theta$ azimuth angle and $\phi$ the polar angle. The parameter $\sigma_{\theta_n}$, that uncertainty of $\theta_n$, is described via the error of the first energy deposition, the error of the traveled distance between the scatters, and the error in time-of-flight. The annulus of the probability cones projected in the ($\theta, \phi$) plane will have an inner radius of $\theta_n-\sigma_{\theta_n}$ and an outer radius of $\theta_n+\sigma_{\theta_n}$. The deduced source location converged to a point source location as the number of back-projected cones being overlapped increases. In other words, a well-resolved neutron source point of origin will result from a large number of valid double-scatter neutron events.  

The proximity of the neutron scatter camera to the source has an effect on the overall image resolution, as observed from simulations in section~\ref{sec:ncamerasimG4}, in the same manner as on the energy resolution (section~\ref{sec:ncameraenrecons}). The spatial resolution is biased by the finite combination of trajectories making the neutron camera resolve the source environment as the location of the nearby scintillator volumes. This can also exaggerated with sources that are placed further away where one or two detector volumes are directly facing the source, but is in general mitigated. Indeed, this part of the motivation to build a movable device to cover every spot in an area with higher precision. 

\begin{figure}[htb]
\centering
\hspace{-0.1in}
\includegraphics{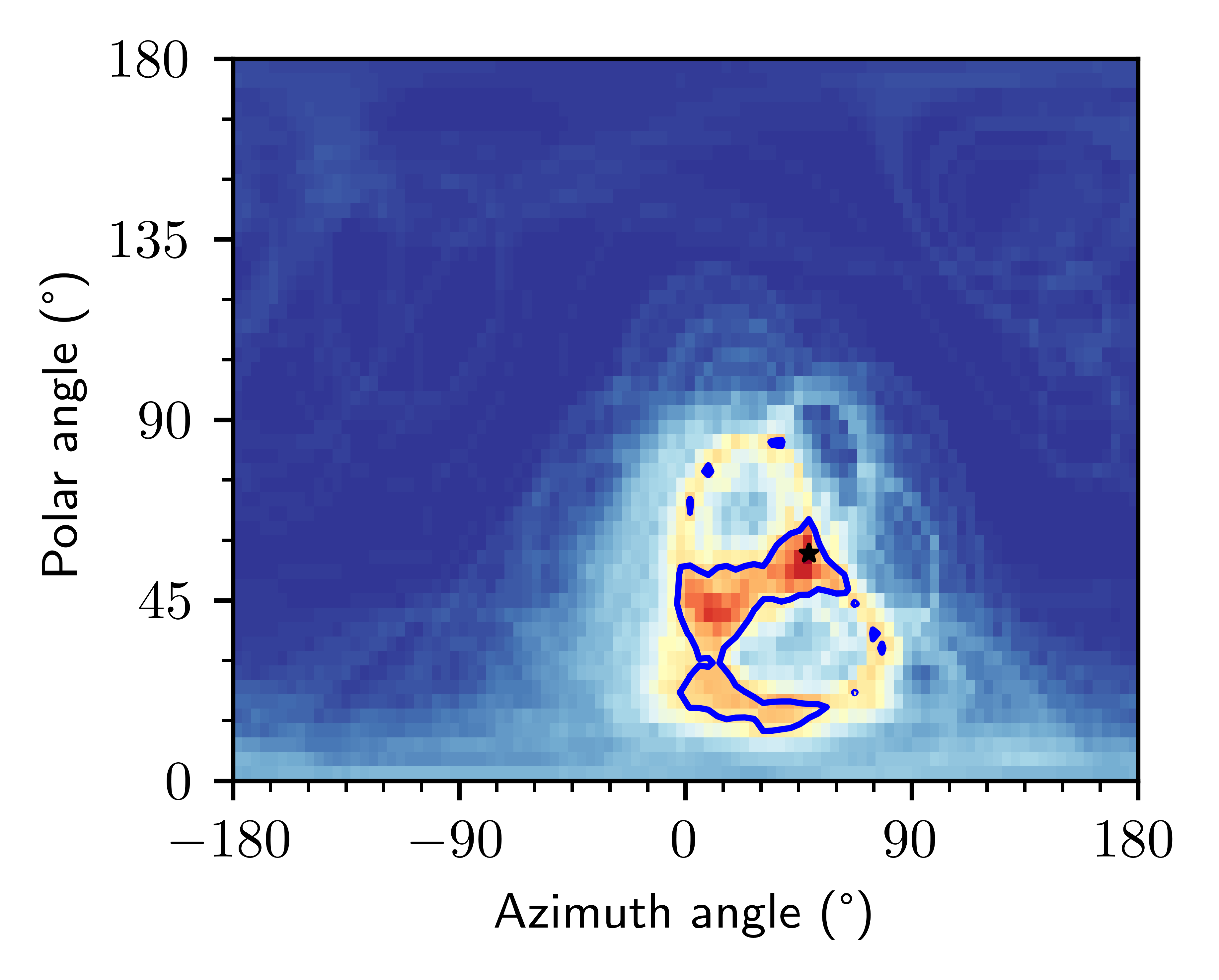}
\hspace{0.1in}
\includegraphics{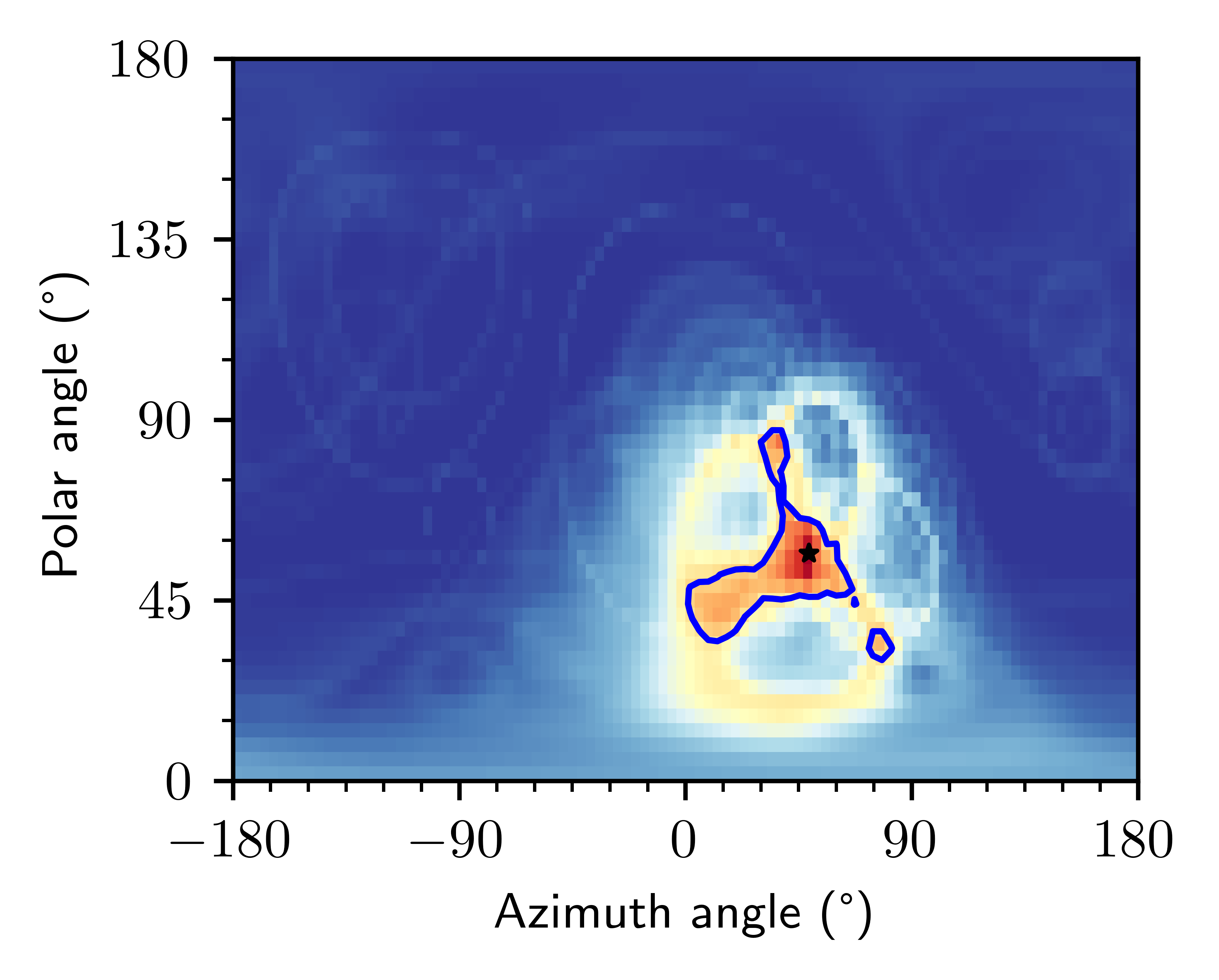}
\caption{Localization of a $^{252}$Cf source placed at a corner of the aluminum frame holding the scintillator volumes. In black the true position of the source and in blue the 2D 1$\sigma$ contour line. Left: Regular neutron image resolved using the back-projection method. Right: Enhanced neutron image after weighting every event and pixel in the polar map with a gaussian centered at each cone's epicenter.}
\label{fig:ncamerarecoCorner13}
\end{figure}

As discussed at the beginning of the chapter, identifying the nuclei in such a scattering process is indistinguishable in the current dataset. Therefore, the ionization-channel portion and trajectory were derived from the hydrogen quenching factor. Taken this into account, figure~\ref{fig:ncamerarecoCorner13} shows the image generated from a $^{252}$Cf source placed $\sim$45 cm away from the center of the camera system. The plot on the left corresponds to the system's ability to resolve the source, arranged in one of the corners of the frame holding the scintillator volumes, via pure back-projection. To do this, the spatial field of view of the detector is pixelated and the area between the inner and outer annulus radius on the pixelated projection encompasses the directionality of the source. This already demonstrates visually the capability of the detector to spot neutron sources. To improve the position reconstruction, each pixel is weighted for a given double scatter event according to a Gaussian distribution centered at the center of the reconstructed annuli, so values closer to that center get higher weights~\cite{steinberger2020imaging}. 
The weights gradually decrease toward the inner and outer edges, following the bell-shaped pattern of the Gaussian distribution. The single event matrix is then convolved with the regular image matrix. The enhancement of the source location is illustrated in the right panel of figure~\ref{fig:ncamerarecoCorner13}. But since the improvement in angular resolution is subtle, the next step, which we are currently working on, is to use the maximum likelihood expectation maximization method as in~\cite{zhang2016image}. This will help mitigate the influence of the energy and intrinsic angular resolutions of the distribution of scintillators in the geometry.

Considering that the camera can identify the source location based on a relatively finite number of back-projected trajectories, measurements placing the source nearby show promising results. The real setting of $^{252}$Cf in ($\theta, \phi$) for each measurement is marked with a black star and the 2D 1$\sigma$ contour lines for the reconstruction are shown in blue. Moreover, this reconstructed directionality is in good agreement with the simulated scenario in figure~\ref{fig:ncameraSIMreco1MeV} left.

\begin{figure}[htb]
\centering
\includegraphics[width=\textwidth]{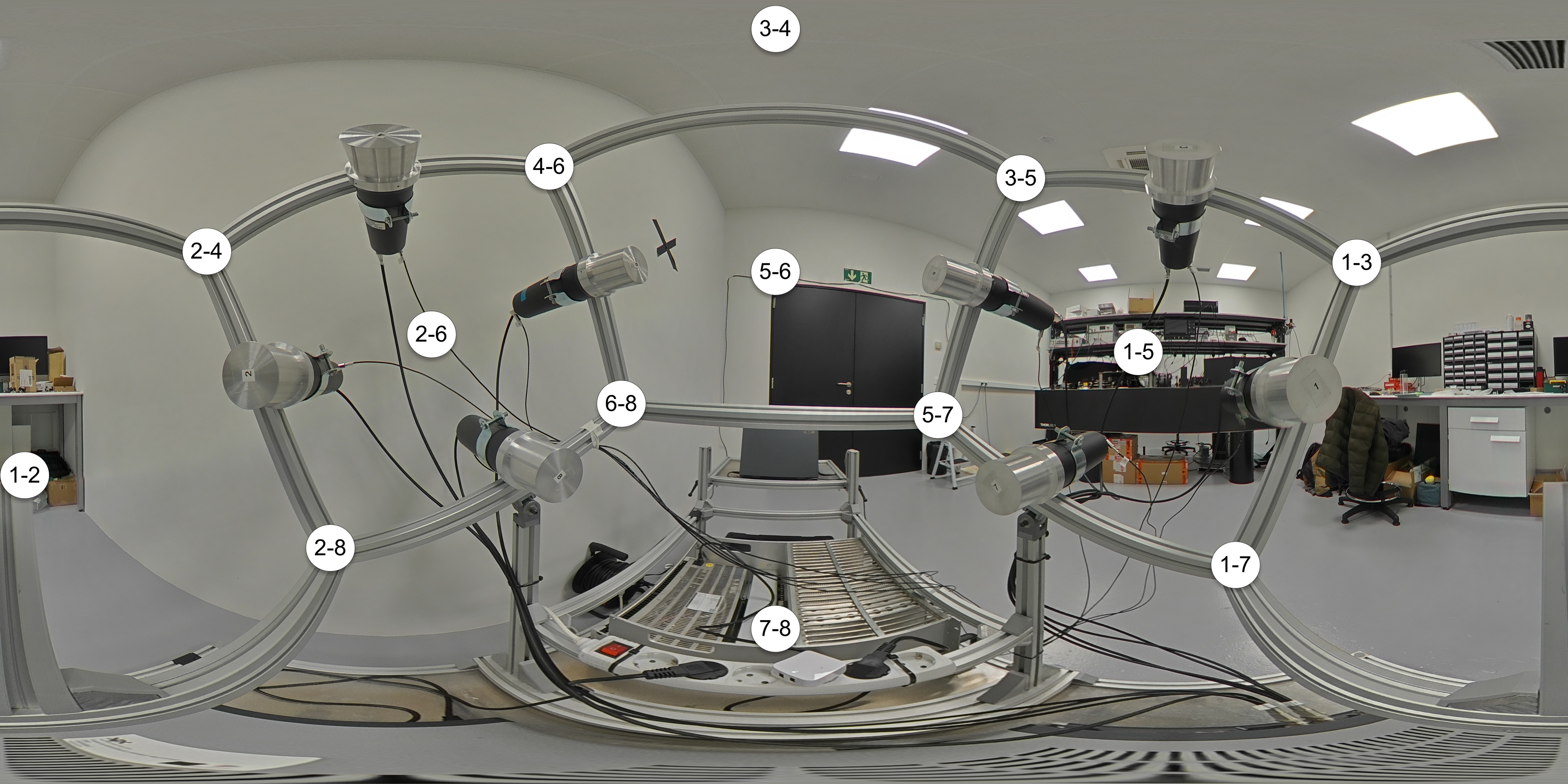}
\caption{Every localization where the $^{252}$Cf source was placed to demonstrate the full coverage ability of the neutron camera. Appendix~\ref{app:appendix} encompasses every neutron image.}
\label{fig:labellingref}
\end{figure}

Overall to validate the technology and demonstrate a full 4$\pi$ coverage, this process has been repeated with the source placed at every corner and across each face of the camera. Figure~\ref{fig:labellingref} illustrates the positions where the $^{252}$Cf was placed and the corresponding reconstructions in polar mapping are shown in Appendix \ref{app:appendix}. These results prove that the neutron camera, even if designed for higher energies and more distant sources, has a novel 4$\pi$ sensitivity. Systematics originate from limiting the pool of possible scattering angles severely, as was the case for source positions at the center of the planes, which will not be relevant to calculating more distant sources.

\section{Resolving the $^{252}$Cf neutron image}
\label{sec:resolve}

As the goal of this neutron scatter camera is to characterize the spallation neutron background in potential zones for neutrino experiments, the set-up should be able to easily spot poorly shielded regions in the locations of interest. This compact system is designed to be a mobile unit that reconstructs the neutron hardness and direction in situ. In order to facilitate a visual mapping of possible hot spots, a 360-degree optical camera has been placed in the structure. 

The addition of the 360$^{\circ}$ camera enables a paired visual of the neutron activity in the possible neutrino experimental rooms. Even if the primary intention is to characterize backgrounds for precision measurements in the sites of interest, the camera will also be used in conjunction to describe and identify unshielded alleys in the whole neutron facility, ensuring radiation-safety conditions for employees. 

\begin{figure}[ht]
    \centering
    \includegraphics[width=0.7\textwidth]{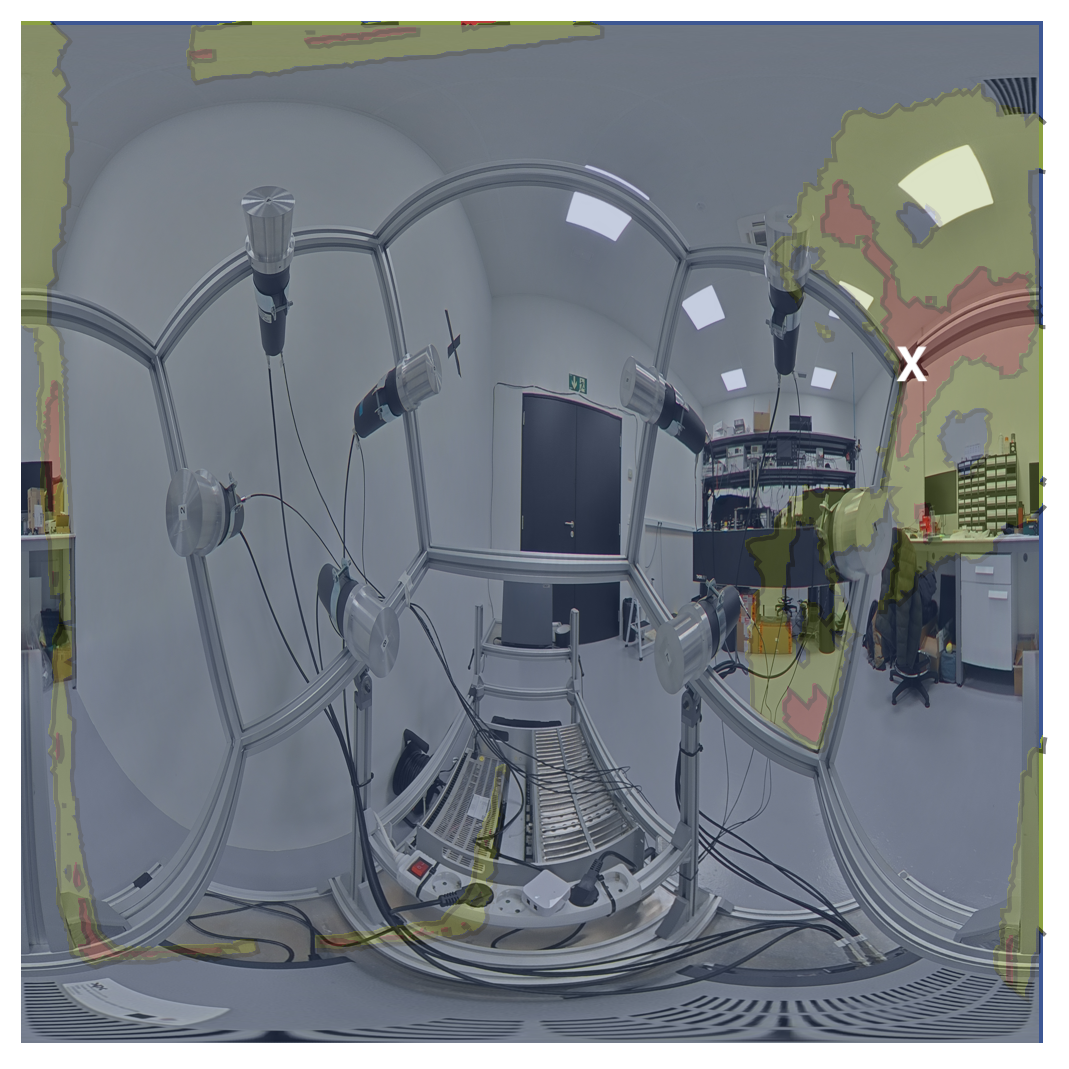} 
    \caption{Overlay of the reconstructed $^{252}$Cf source placed at the ``1-3" corner of the neutron camera following the labeling in photo~\ref{fig:labellingref} and the photo itself taken with a 360$^{\circ}$ camera placed in the set-up. This reconstructed $^{252}$Cf position corresponds precisely to the weighted polar mapping in figure~\ref{fig:ncamerarecoCorner13} right. The color-bar defines intensity contour regions, so the red colored region is the 2D 1$\sigma$ region enclosing the maximum activity and in green the 2$\sigma$ region. The true position of the source is indicated with a with ``X" shows excellent coincidence with the reconstruction.}
\label{fig:ncameraoverlayCf}
\end{figure}

The overlay of the reconstructed neutron image in polar-space as in figure~\ref{fig:ncamerarecoCorner13} and the 360$^{\circ}$ photo shown in figure~\ref{fig:labellingref} is performed in Mathematica\textsuperscript{\textcopyright}~\cite{Mathematica}. Taking established points of reference from the neutron camera structure by taking images with labeled grid systems in the field of view and using the relation between the reconstructed ($\theta, \phi$) coordinates and the pixelated position of the grid points in the optical image, one can overlap any of the reconstructed neutron source locations with an instant photo of the environment (figure~\ref{fig:ncameraoverlayCf}). To avoid the distortion on the 360$^{\circ}$ photo, which will worsen the overlay for some locations, a very fine description of the coordinates has been described. The figure presents the performance of the image overlay for the source placed in the corner labeled ``1-3".

\subsection{Artificial Neural Network implementation} 

\begin{figure}[ht]
    \centering
    \includegraphics[width=0.7\textwidth]{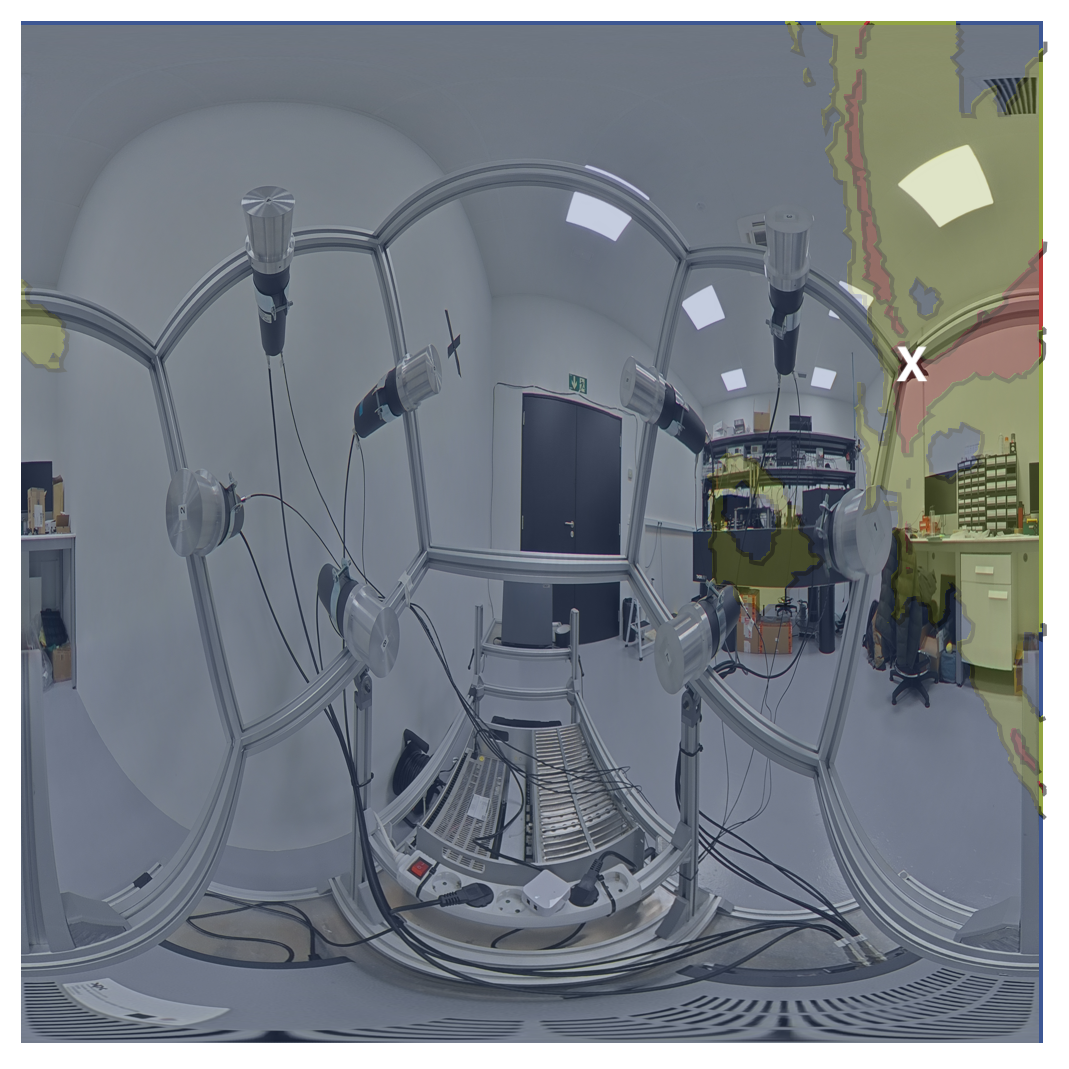} 
    \caption{Overlay of the reconstructed $^{252}$Cf source image (figure~\ref{fig:ncamerarecoCorner13} right) and a 360$^{\circ}$ photo analogous to figure~\ref{fig:ncameraoverlayCf} but this time implementing trained ANN for the coordinate conversion between the optical pixels and the reconstructed ($\theta, \phi$) space. There is an excellent overlap between the reconstructed contour and the source visible in the corner labeled as ``1-3" indicated. The color-bar defines intensity contour regions, so the red colored region is the 2D 1$\sigma$ region enclosing the maximum activity and in green the 2$\sigma$ region. The true position of the source is indicated with a with ``X" shows excellent coincidence with the reconstruction and an enhancement in the interpolation localizing the activity area.}
\label{fig:ncameraoverlayCfANN}
\end{figure}

Artificial Neural Network (ANN) modeling has also been implemented to map a smoother relationship between the pixels in the photo and the azimuth and polar angles ($\theta, \phi$) coordinates from back-projection. To do so, the implemented Neural Net Fitting app from MATLAB\textsuperscript{\textcopyright} workspace~\cite{MATLAB} was used, under the Machine Learning and Deep Learning Toolstrip. Data fitting problems can be solved by a customize trained two-layer feed-forward network. This neural network provides an interpolation of each parameter of the cone back-projection plane, individually, on the spatial distribution of the optical image. The improvement on interpolation using ANN is demonstrated in figure~\ref{fig:ncameraoverlayCfANN}. Figure~\ref{fig:ncameraoverlayCf} shows the overlay of the neutron and the optical image using manually set reference points, whereas figure~\ref{fig:ncameraoverlayCfANN} is the result of the same superposition but including ANN. 

The resulting resolution enhancement to spot neutron sources suggests additional improvements like fine-tuning the overlay calibration and more complicated pixel-weighted techniques. It also engenders an interest towards implementing ANN in neutron-gamma discrimination and pile-up rejection methods described in section~\ref{sec:ncamerapsd}.  

\begin{figure}[ht]
    \centering
    \includegraphics[width=0.7\textwidth]{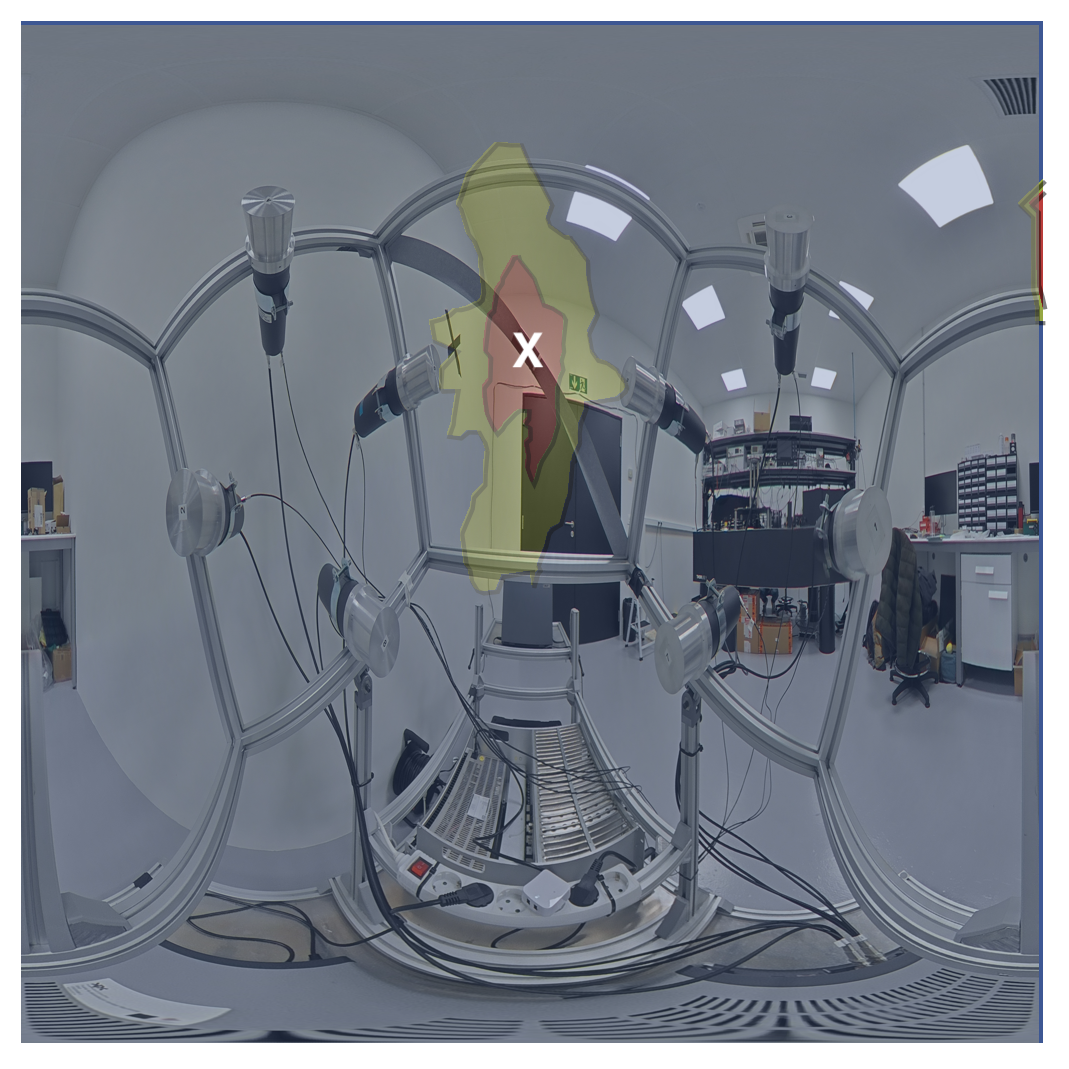}
    \caption{Overlay of the reconstructed $^{252}$Cf source image (Appendix~\ref{fig:appTop} right) and a 360$^{\circ}$ photo after implementing trained ANN for the coordinate conversion between the optical pixels and the reconstructed ($\theta, \phi$) space. The source is visibly identified in the region labeled as ``5-6". The true position of the source is indicated with a with ``X" shows. The reconstructed source location, 2D 1$\sigma$ region in red and 2$\sigma$ region in green, shows a worse angular resolution biased by the 2 detectors that are placed next to $^{252}$Cf.}
\label{fig:ncameraoverlayCfANN_top}
\end{figure}

Another notable demonstration of the remarkable performance of the source identification algorithm developed is shown in figure~\ref{fig:ncameraoverlayCfANN_top}. This presents the image overlay of the $^{252}$Cf source placed in the region labeled as ``5-6" and its corresponding polar map in Appendix~\ref{fig:appTop} right. 

To conclude the validation of the full coverage neutron scatter camera built, the $^{252}$Cf source also placed farther away outside the actual frame of the camera. Seeking for a point-like reconstruction several measurements have been performed. To be concise, just one of the measurements performed will be detailed in this section and two others illustrated in appendix~\ref{app:appendixANN}.

\begin{figure}[ht]
    \centering
    \includegraphics[width=0.7\textwidth]{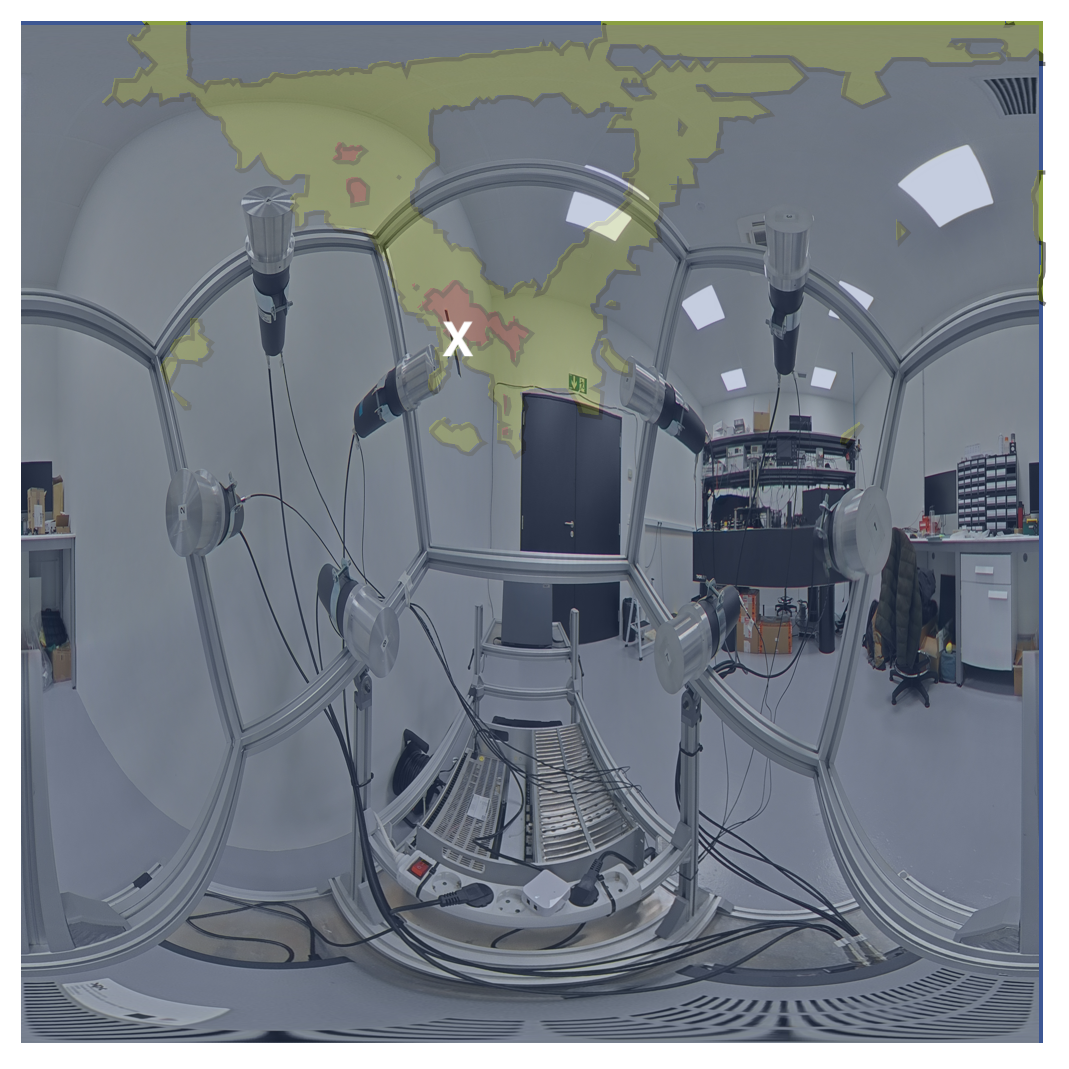}
    \caption{Overlay of the reconstructed source image and a 360$^{\circ}$ photo after implementing trained ANN for the coordinate conversion between the optical pixels and the reconstructed ($\theta, \phi$) space. The $^{252}$Cf source was placed in the ``X" mark on the wall. The reconstructed hot-spot 2D 1$\sigma$ region (red) comes in agreement with the true position.}
\label{fig:ncamerarecoCf2}
\end{figure}

The only information known is that the $^{252}$Cf source was located somewhere in the room and was at least slightly off-set from the neutron scatter camera. Following the event selection process as before, a neutron image is resolved and afterward overlapped with the 360$^{\circ}$ photo. The resulting reconstruction is shown in figure~\ref{fig:ncamerarecoCf2} in an attempt to identify the source in the room. Later, the position was marked with a cross to more directly compare the stress-test of the assembly.

\section{Conclusions and future directions}

To characterize the neutron background in the possible neutrino experimental halls at a spallation source, the full coverage neutron scatter camera has been constructed. Briefly summarized, this chapter details the neutron scatter camera's relevant simulations, design, assembly, and performance. We have built a compact and inexpensive neutron camera capable of identifying the position of a $^{252}$Cf source far afield and reconstructing the energy spectrum of this source. Although sources placed near a specific scintillator volume severely bias the possible trajectories of the incident neutron and limit directionality vectors, this device aims to localize sources placed further away and has been prove to be very efficient. Algorithms based on maximum likelihood expectation maximization method as in~\cite{zhang2016image} and currently being implemented to improve the angular resolution. Furthermore, an optical 360$^{\circ}$ camera has also been integrated into the system to provide a more user-friendly interface for positional results. A meticulous algorithm to overlap the image with the reconstructed contour plots, including an implementation of ANNs, results in a sophisticated and innovative representation of the high activity regions.

Taking all this into account, this compact and full coverage device is positioned as one of the pioneering neutron scatter cameras as it is capable of identifying neutron sources from any direction simultaneously with a good angular resolution. There is an ongoing effort to improve pixel-weighted techniques. It also reconstructs the energy spectrum in agreement with simulated predictions within some tolerance mostly exaggerated by n-$\gamma$ discrimination at low energies. One should note that the first validation of the camera has been done with a source of lower energy than what it was designed for, namely spallation neutrons. This keeps the door open to continue characterizing the device with higher energy sources before going to spallation sources, and an excellent candidate is an Am-Be beam. Apart from bringing the device to such facility, conducting measurements at nuclear reactors or medical radiology installations to check the shielding is also being discussed. 

An eventual quality-of-life improvement will be the implementation of a real-time reconstruction algorithm and a visual overlay.
\renewcommand{\arraystretch}{1.5}

\begin{savequote}[65mm]
``Azkar eta ondo egiten du usoak hegan"
\qauthor{-- Old proverb in Basque language.}
\end{savequote}

\chapter{The Gaseous Detector for Neutrino Physics at the ESS (GanESS)}
\label{sec:ganess}

\section{Introduction}
\label{intro}

The first \cenuns\ observation was received enthusiastically as it opened the path to many phenomenological proposals based on the process, which would strongly benefit from improved measurement (see chapter~\ref{sec:cenuns} for details). Still, current measurements are heavily limited by signal statistics and further progress is needed to fully explore all the possibilities derived from \cenuns\ measurements.

The development of new spallation neutron sources, like the European Spallation Source (ESS), and the upgrade of existing ones offers an opportunity in the next years to fully exploit the physics of the \cenuns\ process with new technologies that could guarantee measurements not limited by statistics (see chapter~\ref{sec:cevnsess} for details).

The Gaseous Detector for Neutrino Physics at the ESS (\ganess) experiment will develop a high-pressure noble gas time projection chamber with electroluminescence amplification (HPNG EL-TPC) to observe the \cenuns\ process from spallation neutrinos (pion decay-at-rest process). See section~\ref{sec:detcon} for details on the detection concept. The \ganess\ detector aims to observe for the first time the \cenuns\ process from spallation neutrinos in xenon and operate the detector also with different noble gases to allow full exploitation of measurements not limited by statistics, having the same systematics but different nuclei. The technique offers several attractive traits. First, it can be operated with different noble gases without any major intervention to the system and thus, yields to \cenuns\ measurements in different targets with the same detector, xenon and argon being the main candidates for its operation. This will be greatly beneficial to constrain the parameter space for different physics scenarios \cite{baxter2020coherent}. 
Second, optical amplification of the ionization signal, via electroluminescence (EL, see section~\ref{sec:el}), enables a potential detection threshold as low as the energy required to form an electron-ion pair (for reference, 22.1 eV in gaseous Xe \cite{RevModPhys.52.121}). Third and finally, while noble liquid-gas dual-phase detectors are affected by charge trapping and delayed release in the inter-phase between liquid and gas \cite{RED-100:2019rpf, LUX:2020vbj, Baur:2022sel}, that is not the case for single-phase detectors. In addition, gas-phase detectors are easier to operate since they do not require a cryogenic system to keep the noble gas in a liquid state. It should be noted that operation in the gaseous phase implies a considerably lower interaction rate when compared to other \cenuns\ detectors. However, this circumstance is mitigated thanks to the large neutrino flux at the expected at spallation sources and the operation at high pressure.

\begin{figure}[!ht]
\centering
  \includegraphics[width=0.6\textwidth]{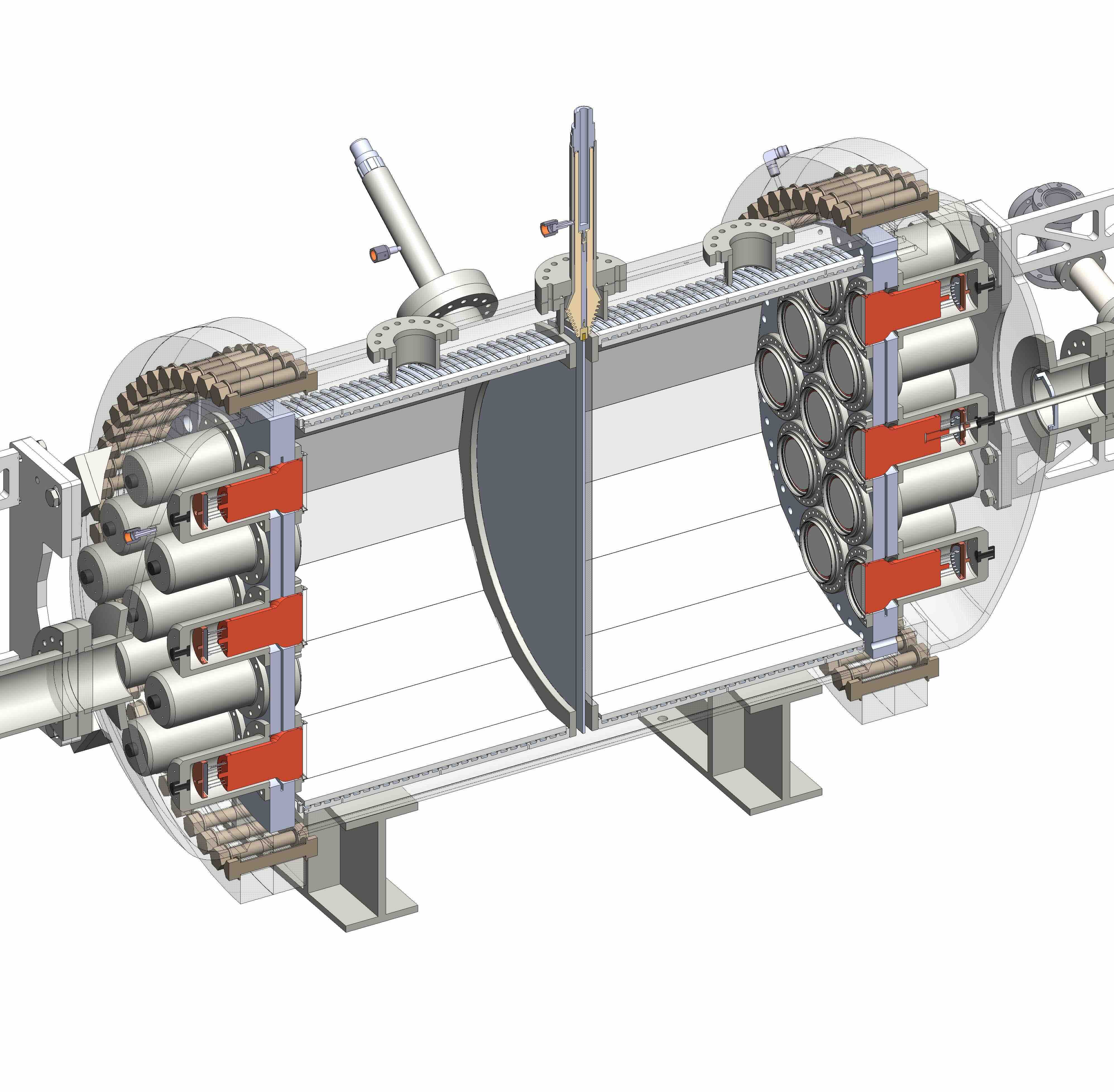} 
\caption{Conceptual design of the \ganess\ detector. It consists of two symmetric TPCs with a central solid cathode with 60 cm diameter and 30 cm drift length. Two planes of PMTs will detect with high efficiency the light produced in the EL regions right in front of them.}
\label{fig:ganess}
\end{figure}

The current design of the \ganess\ detector is shown in figure~\ref{fig:ganess}. It consists on two symmetric TPCs of 30 cm drift distance with a shared cathode placed in the center of the vessel. The detector will have two EL regions at the extremes of the detector of 60 cm diameter where the S2 light will be produced. This light will then be read by a plane of 19 3-inch PMTs placed closely to the anode mesh. These PMTs (Hamamatsu R11410-20) need to be protected from pressure as they do not stand more than 2 bar before degradation. The solution implemented here is different than the one in the NEXT experiment, where a large copper plate was used to isolate the whole PMT volume and hold it at vacuum. 

\begin{figure*}[!ht]
\centering
  \includegraphics[width=0.4\textwidth]{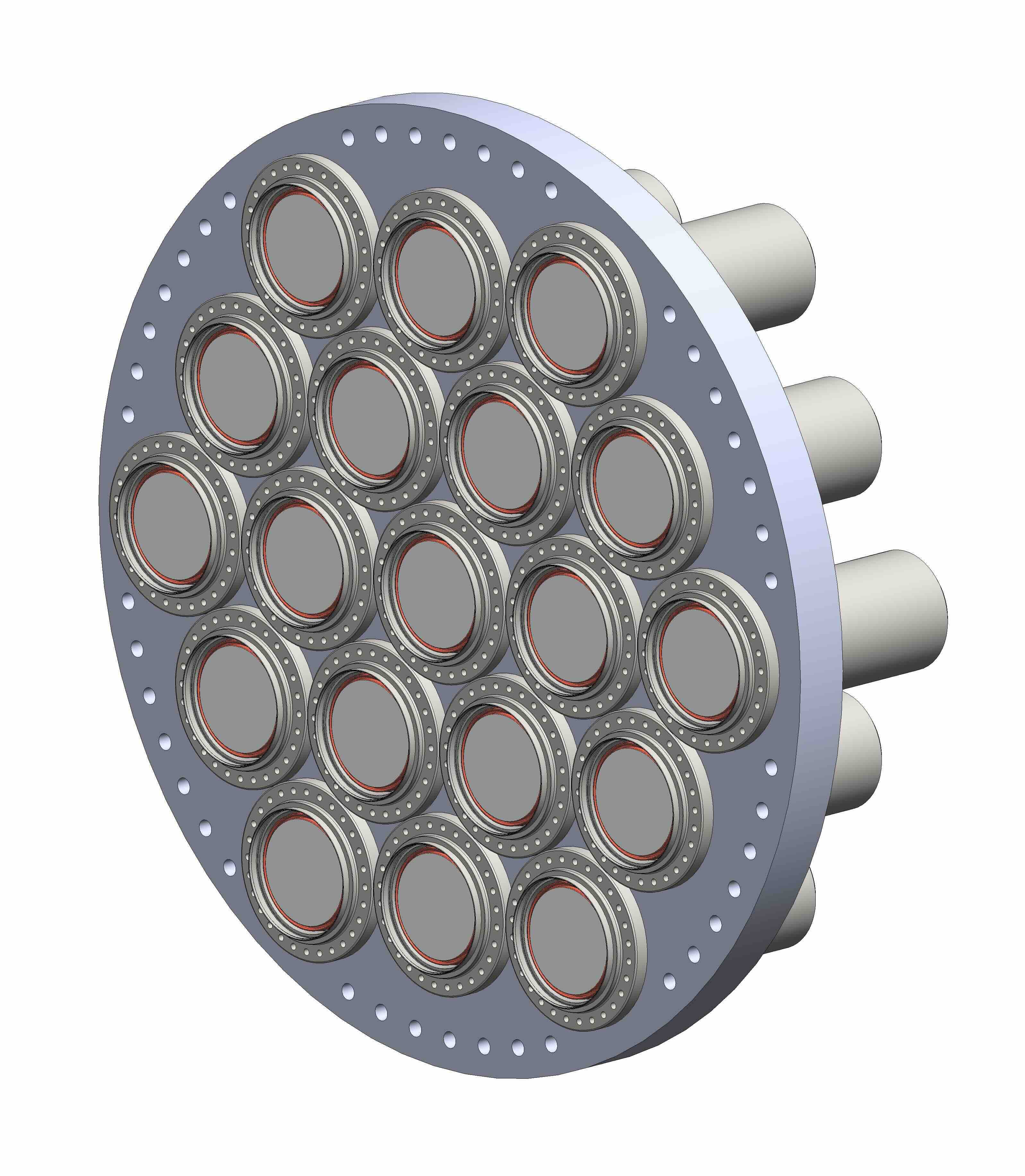}
  \hspace{2.0cm}
  \includegraphics[width=0.4\textwidth]{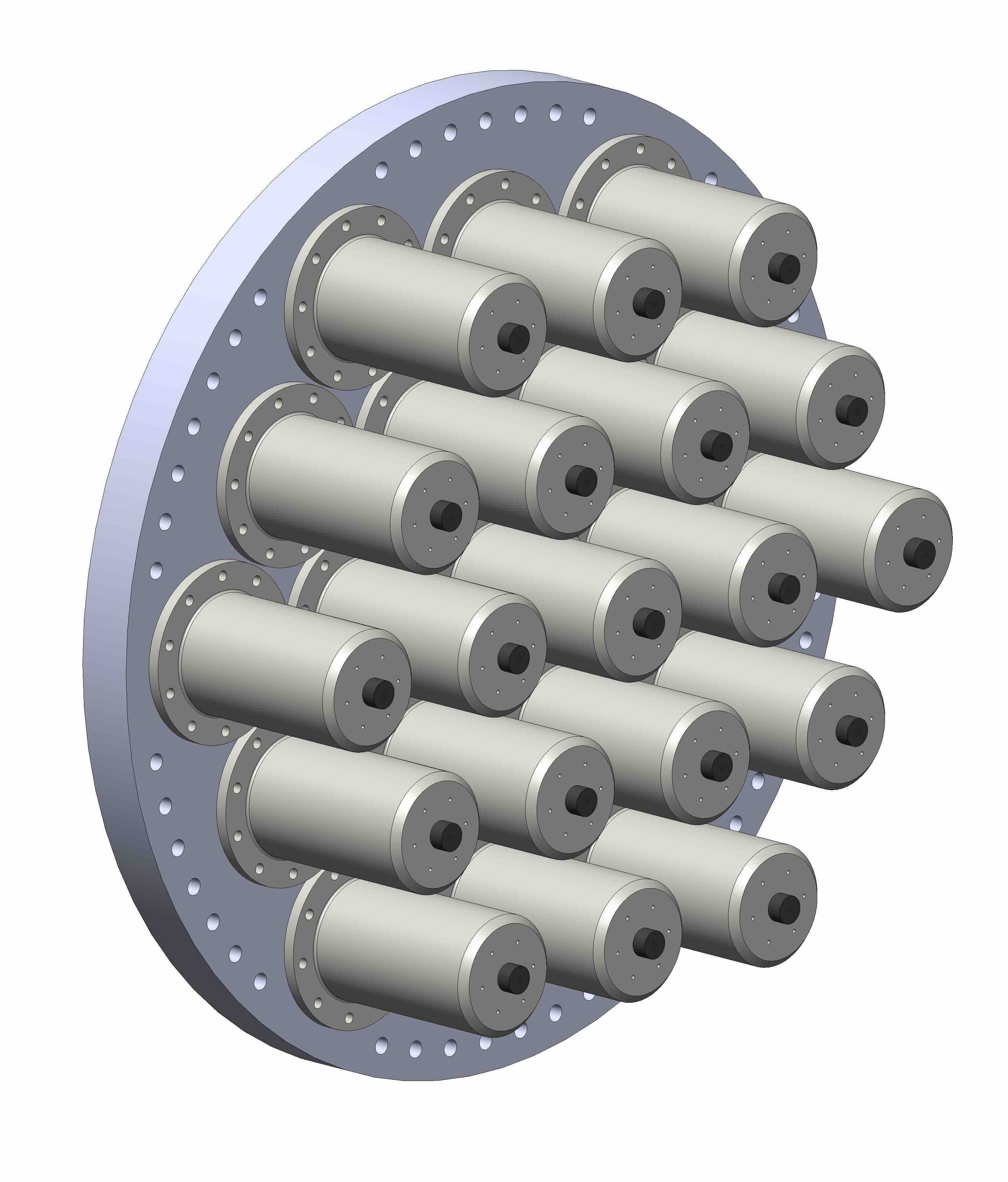}
\caption{Conceptual design of one of the two identical PMT plates in the \ganess\ detector. 19 3-inch PMTs are placed closely to the anode mesh attached to a holder plate of stainless steel and each volume is individually sealed with a sapphire window (left panel) and a steel cap (right panel) to allow vacuum in their volumes.}
\label{fig:ganesspmts}
\end{figure*}

In the case of GanESS, each PMT will have its own volume sealed with a sapphire window on the front and with a steel cap on the back attached to a holder plate of stainless steel that will connect the different volumes of the PMTs to allow to get vacuum in their volumes (figure \ref{fig:ganesspmts}). The PMT signals will be extracted using a home made epoxy feedthrough in each one of the PMTs caps and then using a commercial feedthrough at the vessel exit. The PMTs will be coated with PEDOT, a resistive polymer used in already NEXT-White, to prevent charge accumulation and electric field leakage into the PMT, and then with TPB in order to shift the light from the noble gas being operated to blue, where sapphire is very transparent and the PMTs more resistive. The active volume will also be covered by PTFE panels coated with TPB to enhance light collection. 

This geometry will hold $\sim$20 kg of Xe when operating at 20 bar. With this mass, when the ESS operates at its full power, $\sim$7,000 \cenuns\ Xe events per year will be detected at a distance of 20 m from the ESS target with a threshold of 1 $\rm{keV}_{\rm{nr}}$ assuming a 20\% quenching factor \cite{baxter2020coherent}. The number is reduced to $\sim$700 events per year when operating with Ar at the same pressure.

\section{NuESS detector team} 
\label{sec:NuESS}

The GanESS experiment is part of a larger collaboration effort aiming to observe the \cenuns\ signal at spallation sources and reactors, the NuESS collaboration that involves groups from DIPC, IFIC and University of Chicago. The NuESS is developing different instrumentation techniques that will allow the measurement of the \cenuns\ with high statistics and so allowing precision physics analysis. We can divide the technologies of the NuESS collaboration in three groups: Scintillators, Ge-ppc detectors and noble gases TPCs.

The scintillator technology will be based on CsI detectors, similars to the ones used for the first observation of the \cenuns\ process at the SNS, with enhanced light collection by the use of wavelength-shifting molecules in front of the light sensor to allow for a much reduced energy threshold. In addition, the production of CsI crystals has improved in the last years and we can find larger crystals, allowing for larger mass detectors with the same number of readout channels. In particular, the CoSI proposal aims to build a 44 kg detector divided in 7 crystals.

On the other hand, the Ge-PPC detectors have an ultra low energy threshold (at the level of 120 $\rm{eV}_{\rm{ee}}$) that, together with the fact that the Germanium is a relatively light nuclei, is the most promising technology for observation of the \cenuns\ process from reactor anti-neutrino, with some claims already published \cite{colaresi2021first, colaresi2022measurement, conus2025}.
Currently the groups at DIPC and U.Chicago are operating a 3 kg detector in one of the tendon galleries at the Vandellós reactor. The detector, placed at an average distance from the reactor of 20.1 m but in a very well shielded environment with negligible backgrounds. This detector has been operating for more than 6 months at the moment of writing this work and the final results should be available as soon as the reactor stops and a period of data with reactor OFF can be taken, to allow for proper background subtraction. The reactor OFF period is expected along summer 2025.

Last, a big part of the effort of the NuESS collaboration is based on the development of noble gases single phase TPCs with electroluminescence amplification. We are currently exploring two complementary possibilities. First, a more traditional one, based on high pressure gas, the GanESS experiment that is based on the technology developed by the NEXT collaboration on high pressure xenon and it is expanding it to higher pressures, lower energies and different gases such as argon. This technology is largely explained in this thesis.

On the other hand, a more novel approach is the use of noble TPCs in liquid but producing electroluminescence in the liquid instead of having liquid-gas interphase, the COLINA experiment. This prevents charge accumulation in such interfere, a problem that could limit this kind of detectors when operated on surface. On the other hand, in order to produce electroluminescence photons in liquid one needs to reach very high fields. The solution proposed here is the use of ultra-thin wires in such a way that the 1/r effect on the gradient of the voltage will produce such large fields with relatively moderate voltages. This way of producing EL light has already been demonstrated \cite{aprile2014measurements, martinez2024first}. However, the use of this thin wires or micro strips limits the surface that can be instrumented or used for amplification, limiting in principle the detector’s total mass in the active volume. In order to bypass this limitation, the COLINA experiment has proposed the use of conical-shaped TPC that will combine a larger volume and at the same time a small region for the amplification of the electrons. This approach, similar to the spherical TPCs, has the benefit of a more uniform field across the drifting region with a slower decrease of this intensity, improving in principle the characteristics of the electrons during drift such as diffusion and electron life-time.

\section{Detecting particles using amplification} 
\label{sec:detectors}

In general, a gaseous detector consists of pure noble gases or mixtures contained in what is usually known as active volume. Independent of the particle's nature or its energy, excited and ionized molecules and/or atoms of the gas are produced. Nonetheless, the detector design depends on the interactions undergone by the incident particle in the gas medium. The simplest way to detect radiation is directly collecting the charge generated by an incident particle without any further signal amplification. However, as it is the case of CE$\nu$NS recoil signals, amplification is required. In gaseous detectors, this can happen through charge amplification mechanisms (\textit{Townsend avalanches}) and light amplification (\textit{electroluminescence}). This section only covers detectors using electroluminescence amplification to give an overview of the GanESS detector technology: a high-pressure noble gas time projection chamber with electroluminescence amplification. 

\subsection{Gas Scintillation Proportional Counter}

The Gas Scintillation Proportional Counter (GSPC) introduced by C. A. N. Conde and A. J. P. L. Policarpo in 1964 uses electroluminescence amplification \cite{conde1967gas}.

\begin{figure}[!ht]
\centering
  \includegraphics[width=1\textwidth]{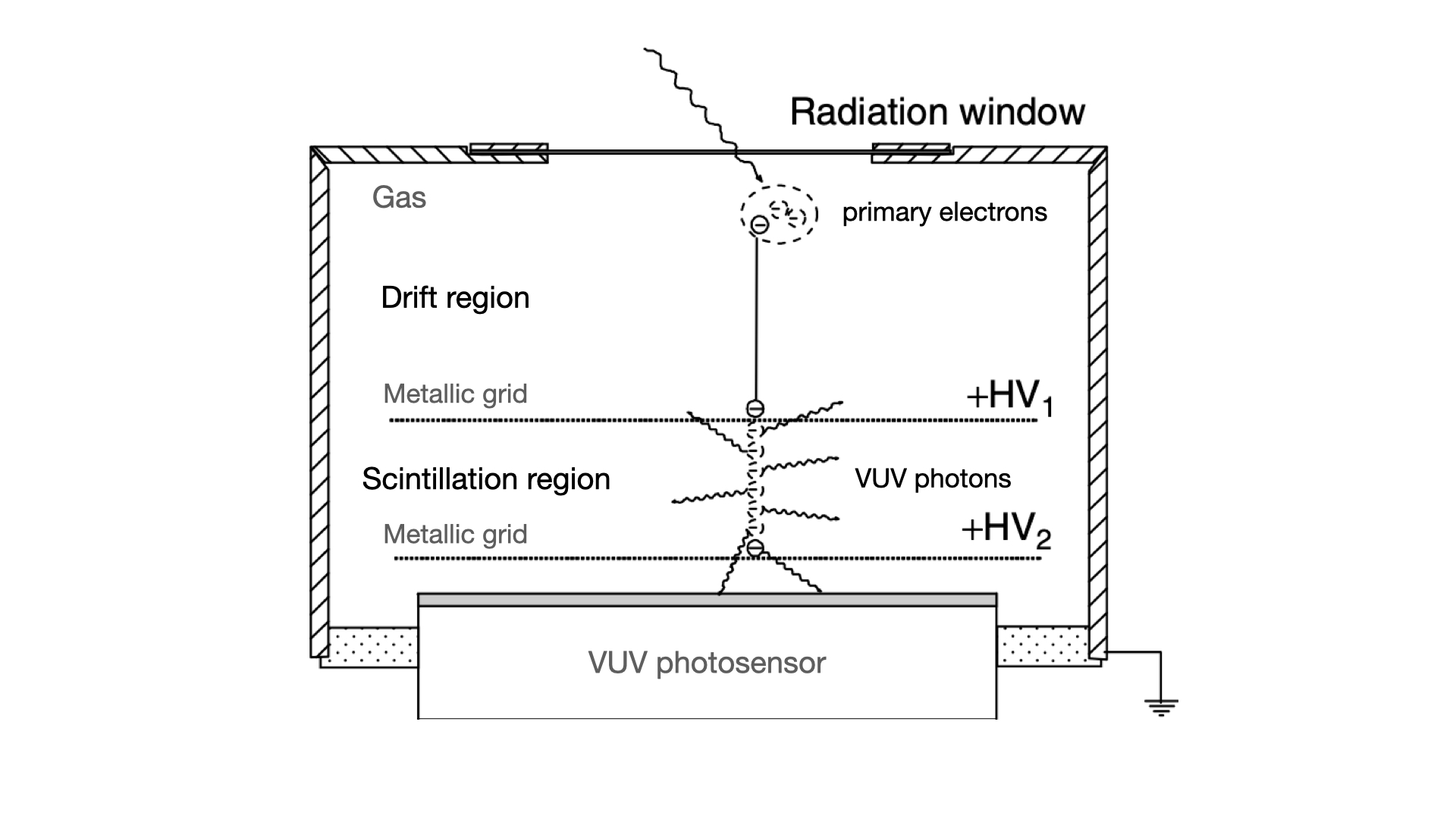}
\caption{Schematic representation of the amplification principle GPSCs, adapted from~\cite{dosSantos2001_Eres}.}
\label{fig:GPSC}
\end{figure}

One of the main applications of this device, X-ray spectrometry, will be described.
Considering the most used type of GSPC, figure~\ref{fig:GPSC} summarizes the operation principle of a GSPC with a nearly uniform electric field created by parallel meshes. A X-ray enters through the chamber window and it is absorbed producing ionization electrons in the drift region, a region of weak electric field ($>$ 0.8 kV/cm/bar). The primary charge is drifted to the scintillation or electroluminescence region (EL region), which is a region of moderately high electric field ($\sim$3 – 4 kV/cm/bar range). In the scintillation region, primary electrons are accelerated, exciting but not ionizing the gas atoms or molecules, which consequently emit VUV light. The so-called secondary scintillation is then detected by a photosensor, usually a photomultiplier tube. 

Since the number of primary electrons is proportional to the energy of the incoming X-ray, and each primary electron produces on average a constant number of secondary scintillation photons, the signal amplitude is nearly proportionally related to the X-ray energy. Hence, the device is called gas proportional scintillation counter.

When the electric field strengths and the spatial widths of the EL region are appropriately selected, the number of secondary scintillation photons produced by each primary electron can reach up to several thousand photons per electron. Therefore, the intensity of the secondary scintillation light is an amplified signal compared to the primary scintillation. However, because secondary scintillation occurs as the electrons drift, it has a much longer delay compared to that of primary scintillation, and its rise time is significantly slower — lasting a few microseconds, whereas primary scintillation occurs in just a few nanoseconds.

The predecessor of GPSC is the Proportional Chambers (PC) introduced in the late 1940s~\cite{simpson1947precision}. The detection concept is similar but uses charge amplification. In a PC the primary electrons are drifted towards a high electric field region – the multiplication region –, usually in the vicinity of the anode wire. The electric field, radially dependent, is generated from a positively supplied anode thin wire (typically 25 $\mu$m diameter). In this region, the avalanche multiplication begins as electrons engage in ionizing collisions and ultimately, a charge signal is produced from all secondary electrons. The primary electrons, on average, undergo the same multiplication gain if space charge effects are neglected. Therefore, the charge signal at the end of the avalanche is proportional to the number of primary electrons and thus, proportionally related to the energy of the incoming particle too. Since the gain comes from a scintillation process with minimal fluctuations in GPSCs, and PCs should account for large multiplication gain fluctuations, the energy resolution of GPSCs is better. For GPSCs, the typical value for energy resolution from 5.9 keV X-rays is 8$\%$ FWHM \cite{dosSantos2001_Eres}.

\subsection{Scintillation Drift Chamber}

The Scintillation Drift Chamber (SDC) was invented in 1978 \cite{charpak1975scintillating}. An SDC is a time projection chamber (TPC) with EL amplification instead of electron avalanche multiplication in the gas that leads to charge gain. 

The TPC was invented by David R. Nygren in 1974 \cite{Nygren:1976fe} (see \cite{Nygren:2018sjx} for a detailed description of the origin of the idea), which allowed to collect the information produced by ionizing particles after meter-long drifts. This breakthrough invention offered thus, the closest representation of the long-sought idea of three-dimensional reconstruction of a particle trajectory and the determination of the interaction point in the volume. A time projection chamber consists of a gas or liquid filled detection volume where high-voltage endplates are used to establish an electric field that allow the drift and collection of the electrons enabling a position-sensitive system. 

Three years after its invention, C. Rubbia proposed the use of a similar concept as a neutrino detector but replacing the gas mixture with liquid argon \cite{Rubbia:1977zz}. Afterwards, a large SDC filled with xenon, composed of two parallel meshes creating a uniform electric field and 19 PMTs \cite{bolozdynya1997high} demonstrated excellent energy resolution for high energy X-rays operated at high pressure (9 bar). 

Since then, although the TPC was first used in accelerator detectors, it has had an enormous impact on rare event searches, as it offers the possibility to scale up the detector and topological signature, a major feature for neutrinoless double beta decay ($0\nu\beta\beta$) experiments. This idea combined with the electroluminescence (EL) amplification and an optimal energy resolution are behind the \ganess\ detection concept (see section~\ref{sec:detcon}).

\section{Detection concept}
\label{sec:detcon}

\begin{figure}[hbt!]
    \centering
    \includegraphics[width=\textwidth]{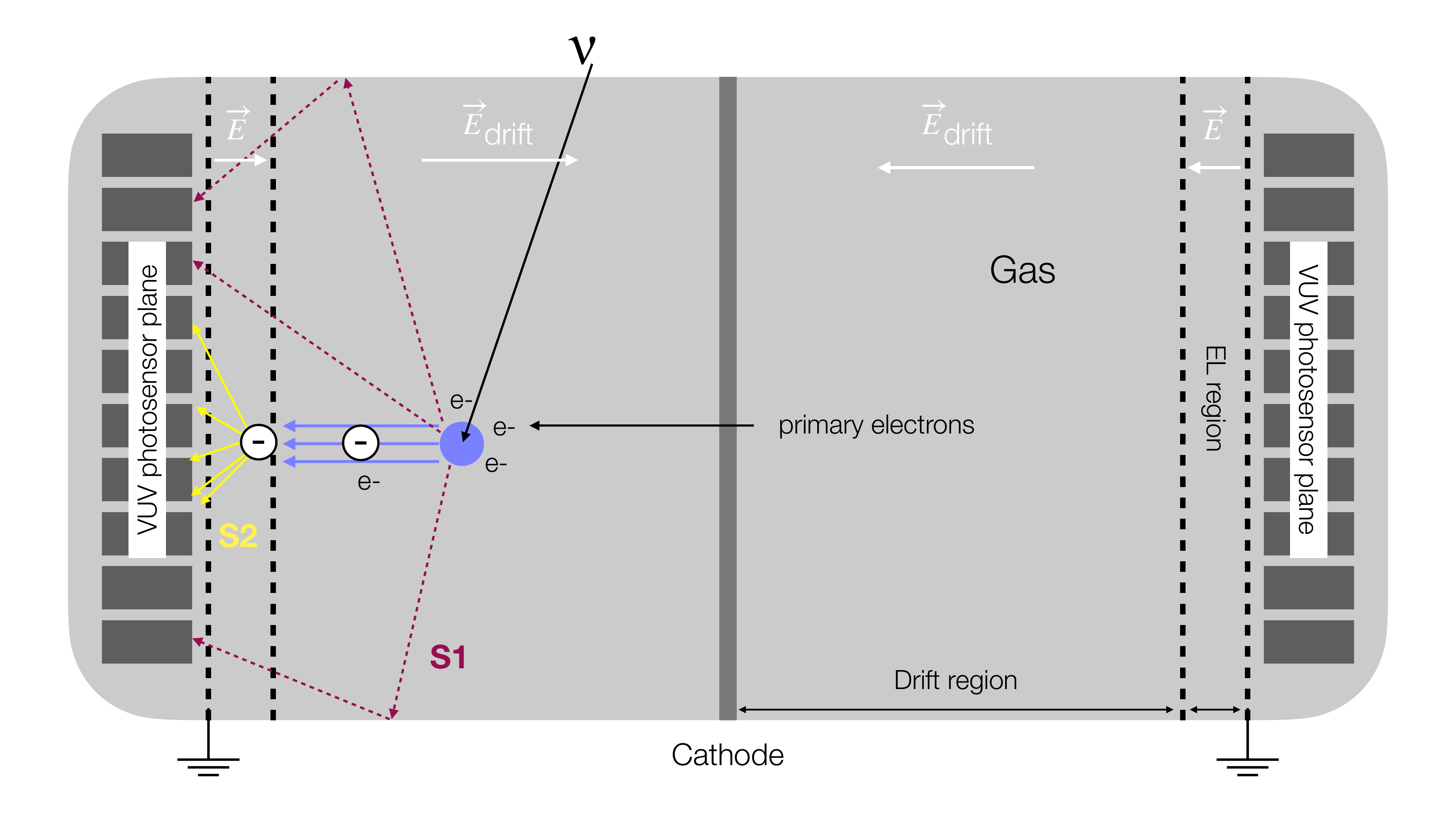}
    \caption{Schematic illustration of the S1 and S2 signal production and collection in a symmetric HPNG EL-TPC like GanESS.}
    \label{fig:ganessconcept}
\end{figure}

The GanESS detection process illustrated in figure~\ref{fig:ganessconcept} works as follows.

When a massive neutral particle interacts with the gaseous medium producing a nuclear recoil, it excites and ionizes the target atoms. Via elastic scattering the energy of the incoming particle is transferred into the recoiling nucleus. Even if some of the recoiling energy is also related to atomic motion (heat), unlike other techniques, this energy will not be detected by a TPC. 

After the incident particle excites the atoms, $\rm{R}^{*}$, the excitation energy is released as prompt in the vacuum ultraviolet (VUV) scintillation light. The process is well described in equation~\ref{eq:xe_excitation}. 

\begin{equation}
     \rm{R}^{*} \xrightarrow[]{\text{+R}} R_2^{*} \xrightarrow[]{} 2 \rm{R} + h\nu.
    \label{eq:xe_excitation}
\end{equation}

\begin{figure}[hbt!]
    \centering
    \includegraphics[width=.5\textwidth]{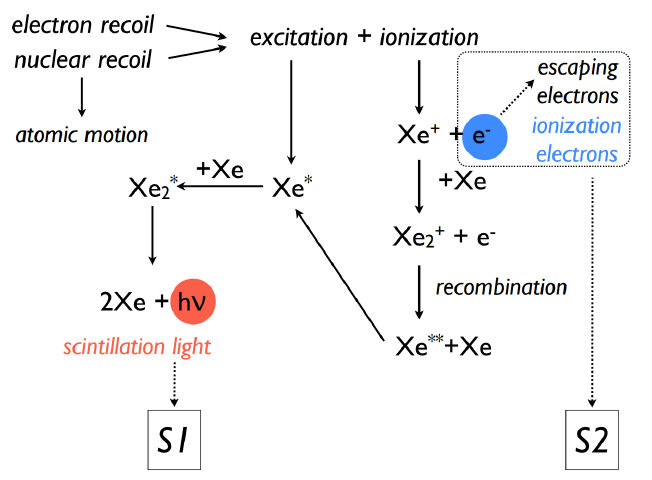}
    \caption{Schematic illustration of the S1 and S2 signal production and collection, extracted from \cite{Manzur_2010}.}
    \label{fig:S1S2}
\end{figure}

During this process, a free ionized electron may recombine with an ionized atom from the medium, forming an excited atom which then returns to its ground state. It is known as the recombination of electron-ion pairs (one secondary electron and one ion $\rm{R}^{+}$) and it also contributes to the prompt scintillation (S1) signal. For noble gases at atmospheric pressure or higher, collision between the ions and the atoms in the ground state quickly ($<$5 ns) form molecular ions, $\rm{R}_2^{+}$~\cite{suzuki1982time}. Followed by the \textit{dissociative recombination}, highly excited atoms are formed, which colliding with ground atoms result in excited atoms, as those produced through direct electron impact that conclude with the emission of VUV photons~\ref{eq:xe_excitation}. The process is described in~\ref{eq:xe_ion}, which is visually represented in figure~\ref{fig:S1S2} for Xe. This recombination process will only take place if the ionization electrons remain at the interaction site. The ionization process leaves behind tracks of positive ions and free electrons, which are kept from recombining by an applied electric field ($\mathcal{O}(1)$kV/cm), thereby reducing the recombination contribution to the S1 signal. 

\begin{equation}
    \rm{R}^{+} + e^{-} \xrightarrow{+2\rm{R}} \rm{R}_2^{+} + \rm{R} \xrightarrow{+e^{-}} 2\rm{R} + \rm{R}^{**} \xrightarrow{} 2\rm{R} + \rm{R}^{*} + \rm{heat} \xrightarrow{(\ref{eq:xe_excitation})} 4\rm{R} + h\nu + heat.
    \label{eq:xe_ion}
\end{equation}

Meanwhile, the primary electrons are drifted (drift region) towards an amplification interface with a much stronger electric field ($\mathcal{O}(10)$kV/cm) before reaching the positively biased TPC anode. The ionization electrons are extracted into this electroluminescence (EL) region with sufficient energy to excite but not ionize the gas atoms (details in section~\ref{sec:el}). Thus, in this region, additional VUV photons are emitted isotropically through electroluminescence. As a result, both scintillation and ionization produce optical signals, which are detected by a group of photosensors, usually PMTs, positioned behind the anode (after the EL region). The detection of the initial scintillation light (S1 signal) marks the start of the event, while the electroluminescence light (S2 signal) is a delayed signal that provides an energy measurement, proportional to the charge (electrons) released in the interaction. The high voltage plates are highly transparent meshes to ensure that the scintillation light reaches the photomultipliers. The measurement of the proportional light provides an amplified S2 signal and therefore a lower energy threshold than with a direct charge measurement from ionization. 

For low-energy events topological information is not significant since these events are always point-like. Therefore, increasing the detector's operating pressure does not lead to any loss of information. This makes high pressure a compelling option for increasing the number of target nuclei providing effective signal amplification while keeping the detector size unchanged.

\section{Electroluminescence}
\label{sec:el}

The electroluminescence process in noble gases is associated with their atomic energy structure. As electrons drift through the gas, they collide with its atoms, transferring energy through motion, excitation, or ionization. In the latter two, the excited atoms or ions undergo various physical processes, ultimately leading to the emission of photons in two different ranges: VUV and Visible-IR (Vis-IR).

\subsection{Electroluminescence modes}

Drifted electrons are accelerated and their kinetic energy is increased as they interact with gas atoms. When the electric field is low and the energy is below the excitation potential, these electrons primarily undergo elastic collisions, transferring a very small fraction of the electron kinetic energy to the atom. Elastic collisions only allow energy transfers that do not produce any light emission. 

When the electric field is moderate, the energy transfer from the drifting electrons to the atoms may excite the atoms without producing ionization. If excitation occurs, the atom can enter one of the several possible excited states and later return to the ground state emitting a VUV photon (description in~\ref{sec:vuv}). After an excitation collision, the electron can generate other excitations and thus, the VUV emission approximately increases linearly with the increase in electric field, since the possible electron energies for excitations are directly proportional to the applied electrical potential.

Photons in the Visible or Infra-Red (IR) range can also be emitted from excited atoms to even more energetic levels. These will conclusively decay with the emission of another photon in the VUV range. Therefore, these excited atoms lead to the emission of two photons. In the same way as VUV yield, the Vis-IR yield grows linearly but with a more gradual slope. The atomic emission scintillation wavelengths in the near-infrared (NIR) are between 580-725 nm, 695-845 nm, and 755-895 nm in neon, argon and krypton respectively~\cite{lindblom1988atomic}, and for xenon ranges between 800 nm and 1600 nm \cite{bressi2001infrared}. Even if the threshold for Vis-IR EL production is reported to be in the operation field ranges of our detector (see measurements in~\cite{buzulutskov2011infrared,belogurov2000infrared}), the spectral sensitivity range of conventional light readout devices is below these wavelengths and thus, only VUV EL yield is expected. Regarding references to the light emission in the electroluminescence process during this work, only the VUV yield will be considered. 

When the electric field is increased above the ionization threshold, secondary particles are produced too. Undergoing the same acceleration as primary electrons, excitation of the gas atoms may occur, which produces VUV or Vis-IR EL photons together with new ionizations. Ultimately, ions may recombine with free electrons and produce VUV photons as described in the previous section via~\ref{eq:xe_ion}. In contrast to the drift region, the recombination is not suppressed by the applied electric field. Although in principle, the electric field is accelerating the electrons and thus, there are not thermalized electrons, this reaction can happen with primary electrons from the drift region. The expected contribution in light yield from the avalanche multiplication is exponential, as the number of secondary electrons increases exponentially with the field.

\subsection{VUV emission}
\label{sec:vuv}

Focusing on the excitation process, as pressure increases above a few tens of Torr, the average time between collisions of excited atoms and surrounding gas atoms decreases, being lower than their natural radiative decay time \cite{oliveira2011monte}. Consequently, these atoms predominantly form excimers, electronically excited molecular states that later decay via scintillation which is characterized by a continuum spectra. Indeed, molecular emissions progressively replace the atomic emissions \cite{suzuki1979mechanism}. Therefore, the main channel of de-population of excited atoms is the excimer formation~\cite{leichner1976two}. Excimers, $\rm{R}_2^{**}$, are formed through three-body collisions between one excited atom, $\rm{R}^{*}$, and two atoms in the ground state, $\rm{R}$:

\begin{equation}
    \rm{R}^{*} + 2\rm{R} \xrightarrow{} \rm{R}_2^{**} + \rm{R}.
    \label{eq:R_el}
\end{equation}

The excimers responsible for the VUV electroluminescence are mainly $^1\Sigma_u^+$ and $^3\Sigma_u^+$ \cite{mulliken1970potential,koehler1974vacuum}. The formed excimers through process~\ref{eq:R_el} are in high vibrational states and can either decay into the molecular ground state, $^1\Sigma_g^+$, emitting a VUV photon,

\begin{equation}
    \rm{R}_2^{**} \xrightarrow{} 2\rm{R}  + \rm{h}\nu_1,
    \label{eq:R_unrelaxedphoton}
\end{equation}

or can lose vibrational energy through a two-body collision between these vibrational unrelaxed excimers ($\rm{R}_2^{**}$) and ground state atoms, described in~\ref{eq:R_relaxed}:

\begin{equation}
    \rm{R}_2^{**} + \rm{R} \xrightarrow{} \rm{R}_2^{*} + \rm{R}.
    \label{eq:R_relaxed}
\end{equation}

The latter leads to relaxed or low vibrational excimers ($\rm{R}_2^{*}$) that also emit a VUV photon:

\begin{equation}
    \rm{R}_2^{*} \xrightarrow{} 2\rm{R}  + \rm{h}\nu_2.
    \label{eq:R_relaxedphoton}
\end{equation}

The energy of the VUV photon is slightly higher from the $\rm{R}_2^{**}$ radiative decay rather than originated from $\rm{R}_2^{*}$. Among the radiative transitions, the singlet-singlet transitions are more probable than triplet-singlet transitions. Therefore, the former transition has short radiative decay rates of 4.2 ns in argon and 5.5 ns in xenon; and the latter has lifetimes of 3.2 $\mu$s in argon and 96 ns in xenon \cite{suzuki1979mechanism,keto1974production}.

Following the Franck-Condon principle for excimer's electronic transitions \cite{banwell2017fundamentals}, a continuum emission spectrum is attributed to the transitions from $^1\Sigma_u^+$ and $^3\Sigma_u^+$ states to the repulsive ground state $^1\Sigma_g^+$. Two peaks are experimentally observed for low gas pressures in these continua, usually called ``first continuum" at higher frequencies, and ``second continuum" at lower frequencies. However, at high pressures (above a few hundred Torr), due to the increase in the number of atom collisions, the process \ref{eq:R_relaxed} is preferred to \ref{eq:R_unrelaxedphoton} \cite{suzuki1979mechanism}. In this way, only the ``second continuum" is observed in the proportional electroluminescence spectrum. 
The second continuum is approximately Gaussian and peaks at a wavelength of 128 nm, 147 nm, and around 172 nm for gaseous argon, krypton and xenon, respectively \cite{suzuki1979mechanism, takahashi1983emission}.

Being an amplification mechanism with no gain, EL is more stable than avalanche amplification, making it ideally suited for high pressure operations.
\renewcommand{\arraystretch}{1.5}

\begin{savequote}[65mm]
``Nola soinu, hala dantza" 
\qauthor{-- Old proverb in Basque language.}
\end{savequote}

\chapter{The Gaseous Prototype (GaP)}
\label{sec:gap}
The HPNG EL-TPC technique, widely used in neutrinoless double beta decay searches as part of the NEXT experiment \cite{NEXT:2018rgj, NEXT:2015wlq}, has mainly been developed for moderate pressures up to 15 bar, which is lower than the \ganess\ goals. Additionally, similar detectors have been optimized for signals in the few MeV range, higher than the energy region of interest for \cenuns, which lies in the sub-keV to the few-keV range. The performance of noble gas TPCs in this lower energy regime is yet to be explored and requires thorough characterization to fully evaluate the technique's potential for low-energy searches.

\begin{figure}[!tbh]
\centering
\resizebox{0.75\linewidth}{!}{%
  \includegraphics{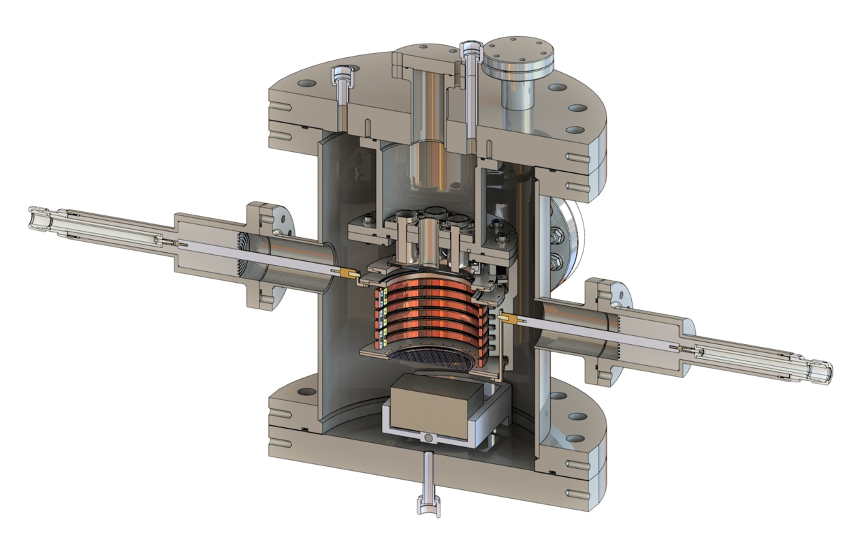}
}
\caption{Render image of the inner parts of the GaP detector. The field cage and high-voltage feedthroughs are visible with the EL region on the upper side of the detector. Just in front of the EL, there is a plane of seven 1" cylindrical PMTs. While the detector is designed to allow for pressure isolation of the PMTs, in this initial run they were in contact with the gas and thus, the operation has been limited to pressures below 10 bar. The drift volume is limited by the cathode and gate, which are separated by 8.7 cm, and the amplification region determined by the gate and the anode has a nominal separation of 1.02 cm.}
\label{fig:gap1}
\end{figure}

To assess many of these aspects and validate the \ganess\ design for high pressure operations the Gaseous Prototype (GaP) was built in Donostia International Physics Center (DIPC) in 2023. The detector is a small time projection chamber illustrated in figure~\ref{fig:gap1}. In this chapter, the GaP detector is described in detail. 

See~\cite{larizgoitia2025gaseous} where we described in detail for the first time the Gaseous Prototype (GaP) detector at DIPC and its first results with gaseous argon at various pressures up to 10 bar.

\section{Gas system}
\label{gap:gas}

\begin{figure*}
\centering
  \includegraphics[height=7.5cm]{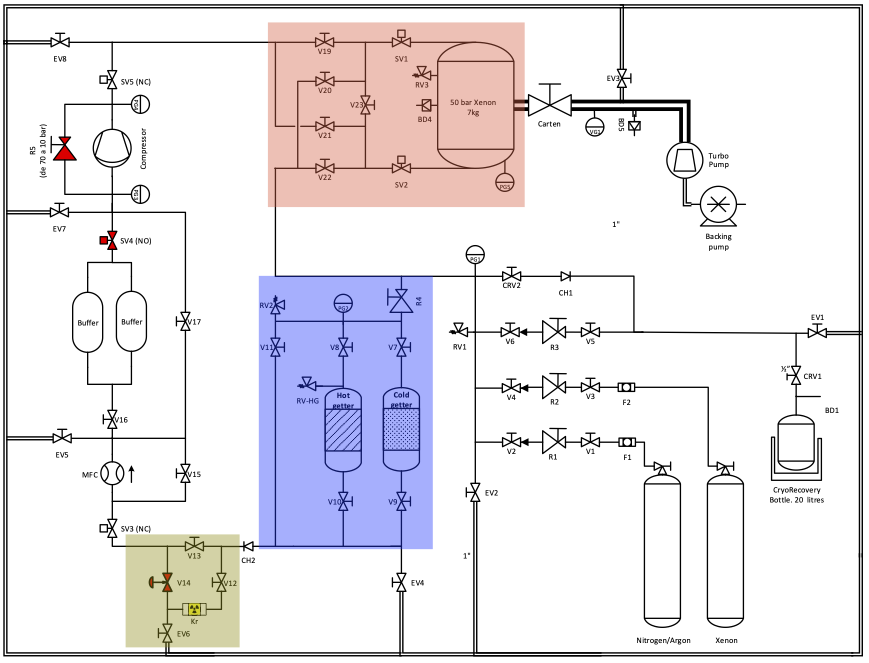}
  \hspace{0.7cm}
  \includegraphics[height=7.5cm]{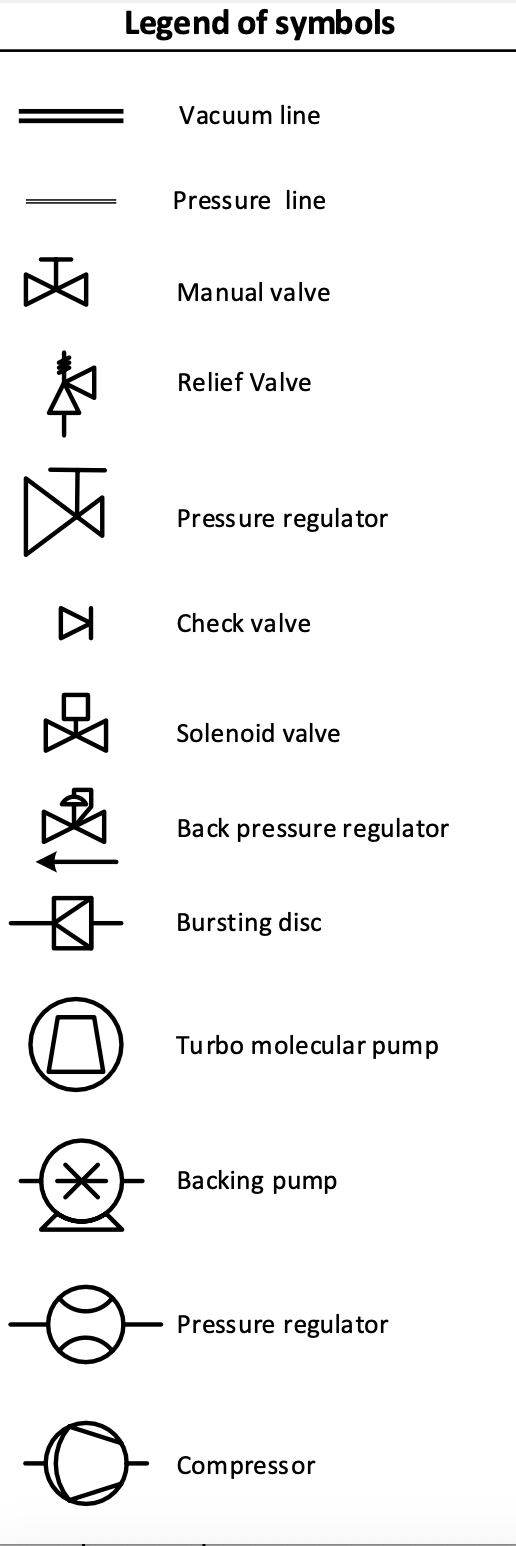}
\caption{GaP gas system scheme. The red area covers the GaP vessel and the valve system that allows to change the gas circulating direction in the detector. The blue area covers the getters section of the gas system. As this section cannot operate at pressures above 10 bar it is equipped with a regulator at the entrance of it and with a non-return valve at the exit allowing for a separate pressure section. The green area represents a part of the gas system where a gaseous Rb/Kr source can be installed.}
\label{fig:gas_system}
\end{figure*}

Operation with clean gas is mandatory in noble gas TPC as impurities reduce the electron lifetime and quench VUV light. To maximize gas purity, the pressure vessel, which houses the TPC, is connected to a recirculation loop, detailed in figure~\ref{fig:gas_system}, which includes gas purifiers. During operation, gas is in continuous recirculation through this loop. Recirculation is forced by one of two different pumps, each used to operate in a different pressure regime. For low pressure, up to 10 bar, a smaller single diaphragm pump is used. At higher pressures, recirculation is produced by a 2-stage compressor by SERA company, which operates to a maximum of 50 bar. As the purifiers cannot operate above 10 bar, the gas system is designed with a low-pressure region where the gas circulates below this pressure and is purified. The gas is then pumped up by the SERA compressor and re-introduced into the detector. In order to prevent large decreases in pressure at the entrance of the compressor, two buffer volumes are installed at its entrance increasing the volume of the low-pressure region.
The system counts with a series of vacuum lines and connections that allow for an efficient evacuation of the whole volume prior to its filling with a noble gas using a combination of a scroll and turbo pumps. After air evacuation, gas is introduced into the system via 3 different inlets, each connected to a different bottle, facilitating gas changes without major interventions. An outlet for cryo-recovery is also available and it will be used when operating with pure xenon.

An ambient getter (Sigma Technologies MC1500-902 FV model) and a hot getter (Sigma Technologies PS4 MT15 R2 model) are placed in parallel in the recirculation loop, allowing operation with either one or both of them at the same time. They are placed before the inlet of the recirculation pump as they are not able to operate at the maximum operational pressures of GaP. Before them, a pressure regulator reduces the pressure to a maximum of 10 bar. This regulator, combined with a non-return valve at the exit of the getters region, allows to protect this part of the gas system and ensures the safe operation of the getters when operating at higher pressure.
The gas system also includes a section that allows to introduce a \kr\ source, used for detector characterization \cite{NEXT:2018sqd}. A general vacuum line is connected to the gas system at various points. This configuration, in combination with the different valves distributed throughout the system, allows individual pumps of different sections if required. Pressure and vacuum are monitored at different points of the system using a set of gauges read by a custom slow control module developed on LabView\textsuperscript{\textcopyright}.

\section{Pressure vessel} 

\begin{figure}
\centering 
\includegraphics[width=0.75\linewidth]{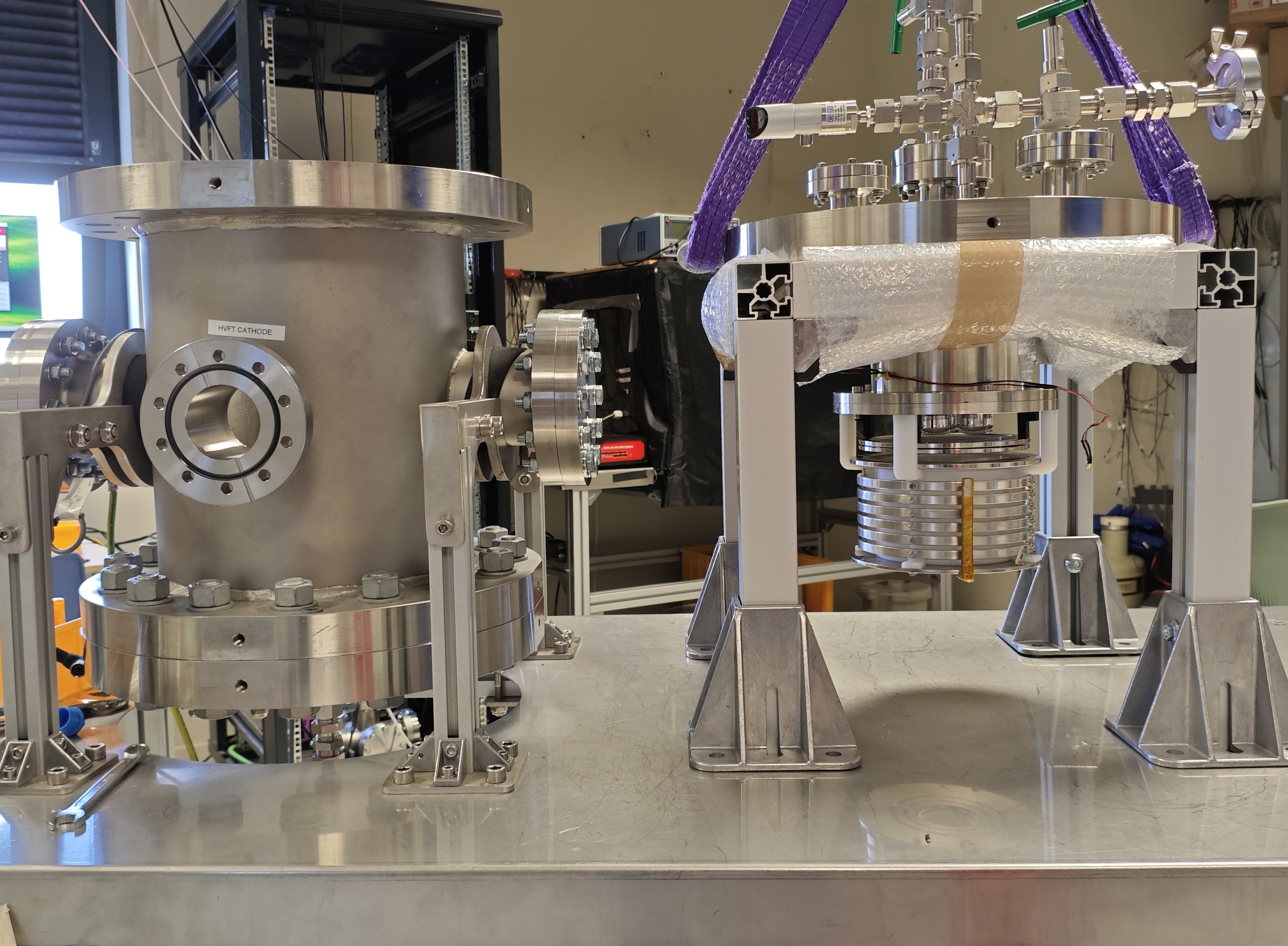} \hspace{1cm}
\caption{GaP pressure vessel and inner detector. The lateral ports are used for the HVFTs. The top endcap, attached to the field cage in the image, the central port is used to power the PMTs and extract their signal while the lateral ones are used to pulse an internal LED for calibration. The field cage is shown for comparison with the vessel size.}
\label{fig:vessel} 
\end{figure}

The TPC is housed inside a 6-mm-thick cylindrical stainless steel pressure vessel of 13.8 cm internal radius and 38.65 cm length, capable of operating up to 50 bar, figure~\ref{fig:vessel}. An additional cylindrical inner chamber is fixed to the top cap of the vessel. This volume has seven holes in the bottom to accommodate an equal number of photomultiplier tubes (PMTs) for light detection. Although not currently installed, a quartz window can be coupled to isolate the sensors, rated for up to 20 bar, from higher pressures. In addition, the TPC hangs from this inner cylinder.

The vessel is connected to the recirculation system through two gas lines, one located at the bottom and another at the top. The recirculation direction can be changed by manipulating a system of valves. The top connection is done through a cross, which is also connected to a pressure gauge, to monitor the pressure vessel, and a KF port connected to a vacuum pump, to clean the vessel prior to filling the detector. An additional gas line is connected to the volume of the inner cylinder and it is used to monitor the vacuum inside its volume.

The vessel has 4 lateral ports distributed uniformly, i.e. each port is at $\pm$ 90 degrees from its neighbors. Two of the ports, located each in front of the other, are used for the high-voltage feedthroughs, while the other two are left available for future use. Three additional ports are in the top cap. One of them is centered, connecting to the inner cylinder, and it is used as a feedthrough for the photosensor supply and signal. The other two are placed at a larger radius, communicating directly to the main volume. 

\section{Time projection chamber}
\label{gap:tpc}

\begin{figure}[!tbp]
\centering
\includegraphics[height=5.5cm]{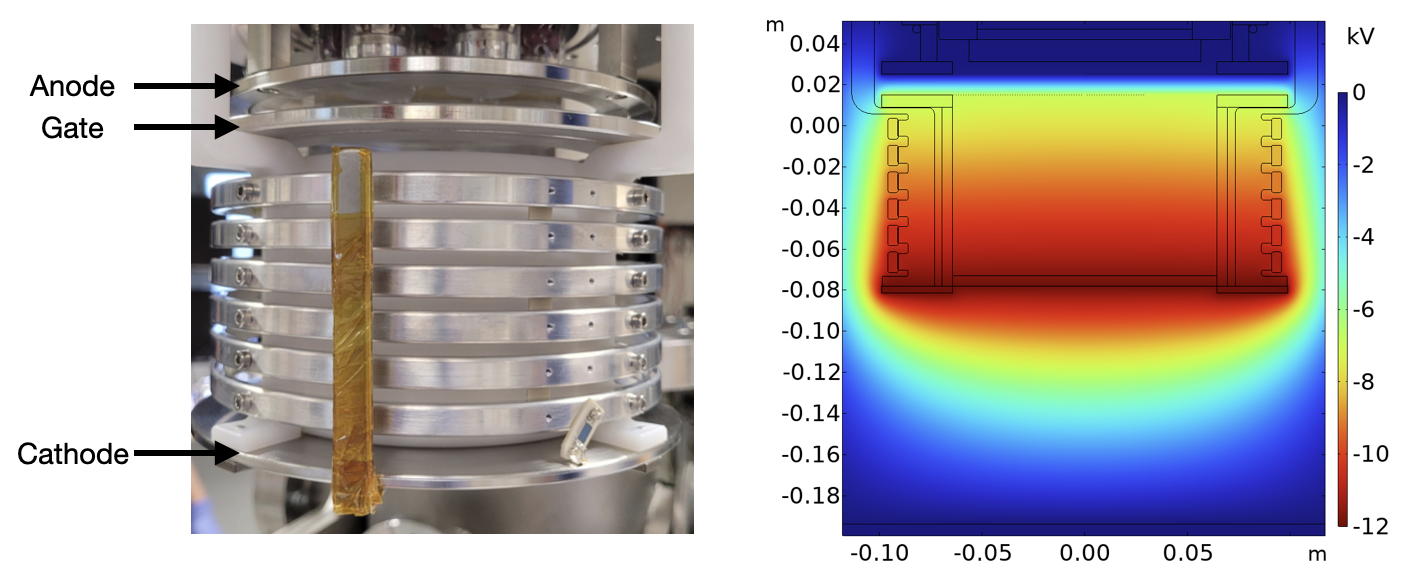}
\caption{Left: Full view of the GaP TPC. The cathode, gate and anode are visible while 6 rings connected by resistors guarantee the creation of a uniform electric field in the central part of the detector. The metallic piece wrapped with kapton allows to connect the High-Voltage feedthrough to the cathode.
Right: Comsol simulation of the electric field inside the GaP vessel. The simulation shows the homogeneity of the electric field in the active region of the detector generated by the field cage shown in this figure.}
\label{fig:gap_TPC}
\end{figure}

\begin{figure*}
\centering
  \includegraphics[height=7.5cm]{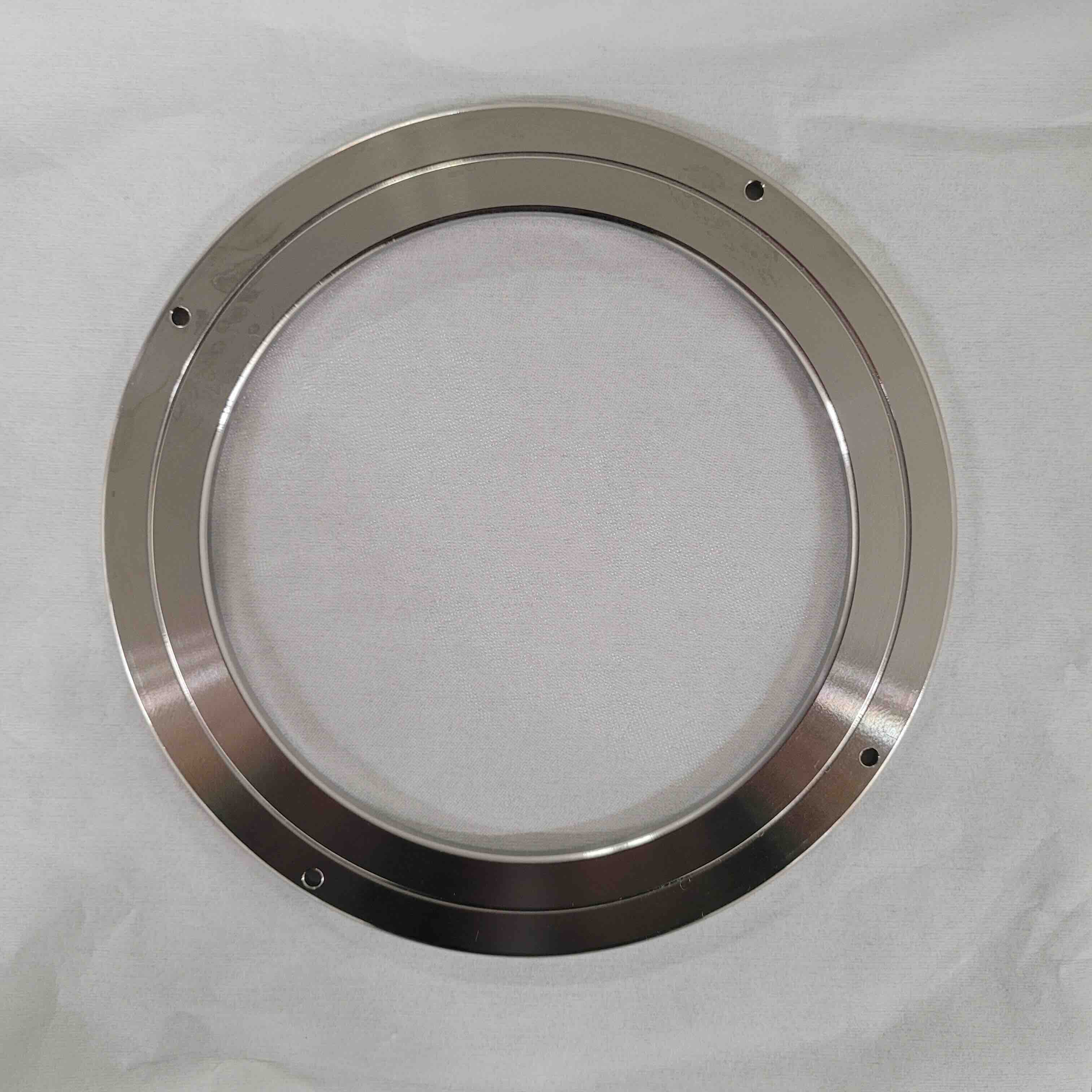}
  \hspace{0.7cm}
  \includegraphics[height=7.5cm]{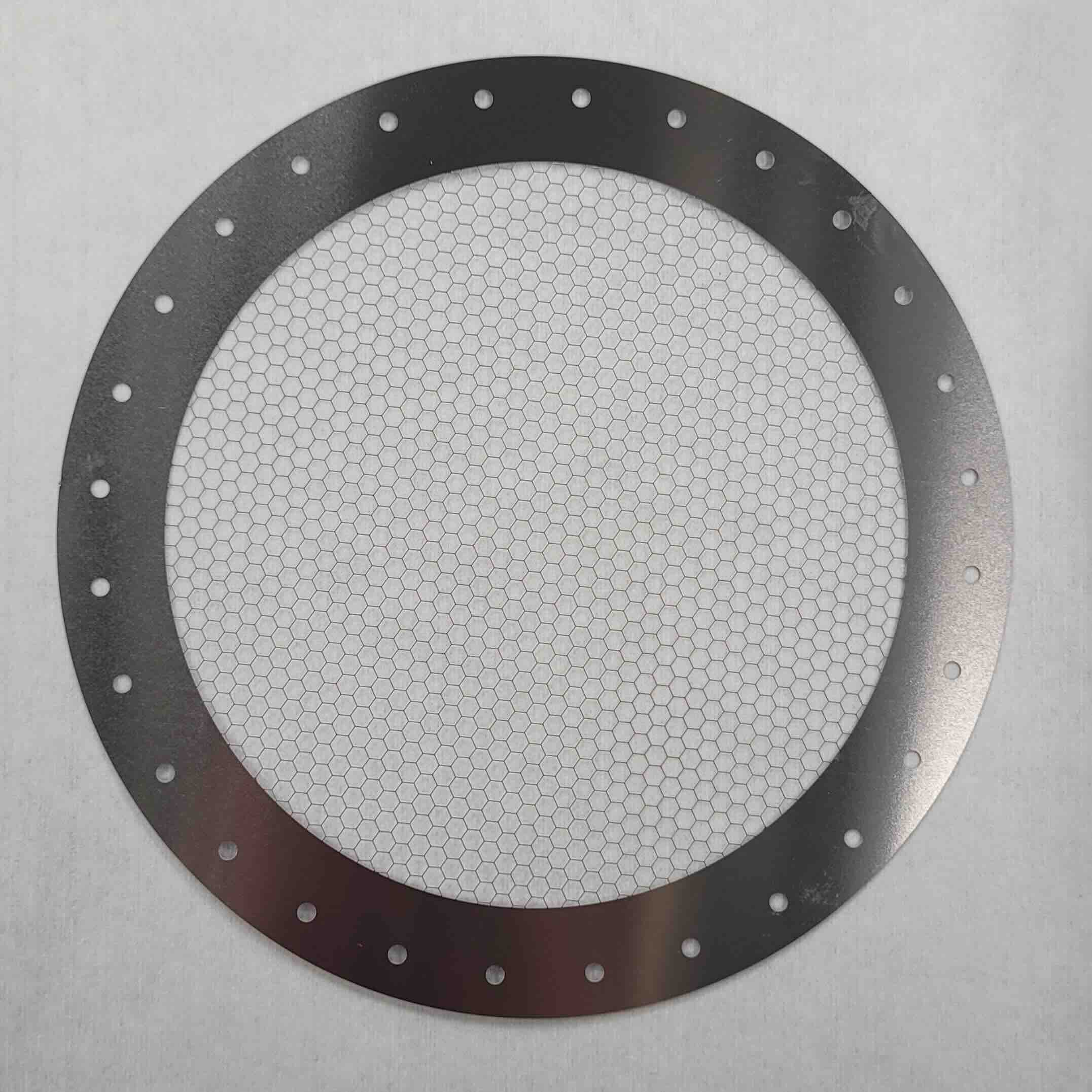}
\caption{Left: Thin wire mesh of 50 $\mu$m diameter wires with a 500 $\mu$m spacing (0.81 transparency) which is later cryo-fitted to two concentrical rings acting as the gate electrode between drift and the electroluminescence regions. Right: 75 $\mu$m thick photoetched hexagonal grid used in the anode. Its hexagons have a 1.467 mm side and their contour is 150 $\mu$m wide resulting in a 0.89 transparency.}
\label{fig:mesh}
\end{figure*}

The time projection chamber, shown in figure~\ref{fig:gap_TPC} left, is composed by three stainless steel electrodes: a cathode, a gate and a grounded anode. The cathode is a solid plate with mechanized apertures to facilitate the recirculation of the gas within the chamber volume. It also has a central hole used to place radioactive sources, which are held in place by an additional plate. The gate is a thin wire mesh (see figure~\ref{fig:mesh} left) of 50 $\mu$m diameter wires with a 500 $\mu$m spacing (0.81 transparency) that has been cryo-fitted to two concentrical rings acting as holders. The anode is a 75 $\mu$m thick photoetched hexagonal grid (see figure~\ref{fig:mesh} right) attached to a holder ring with kapton tape in the edges. Its hexagons have a 1.467 mm side and their contour is 150 $\mu$m wide resulting in a 0.89 transparency. 

\begin{figure}
\centering 
\includegraphics[width=0.5\linewidth]{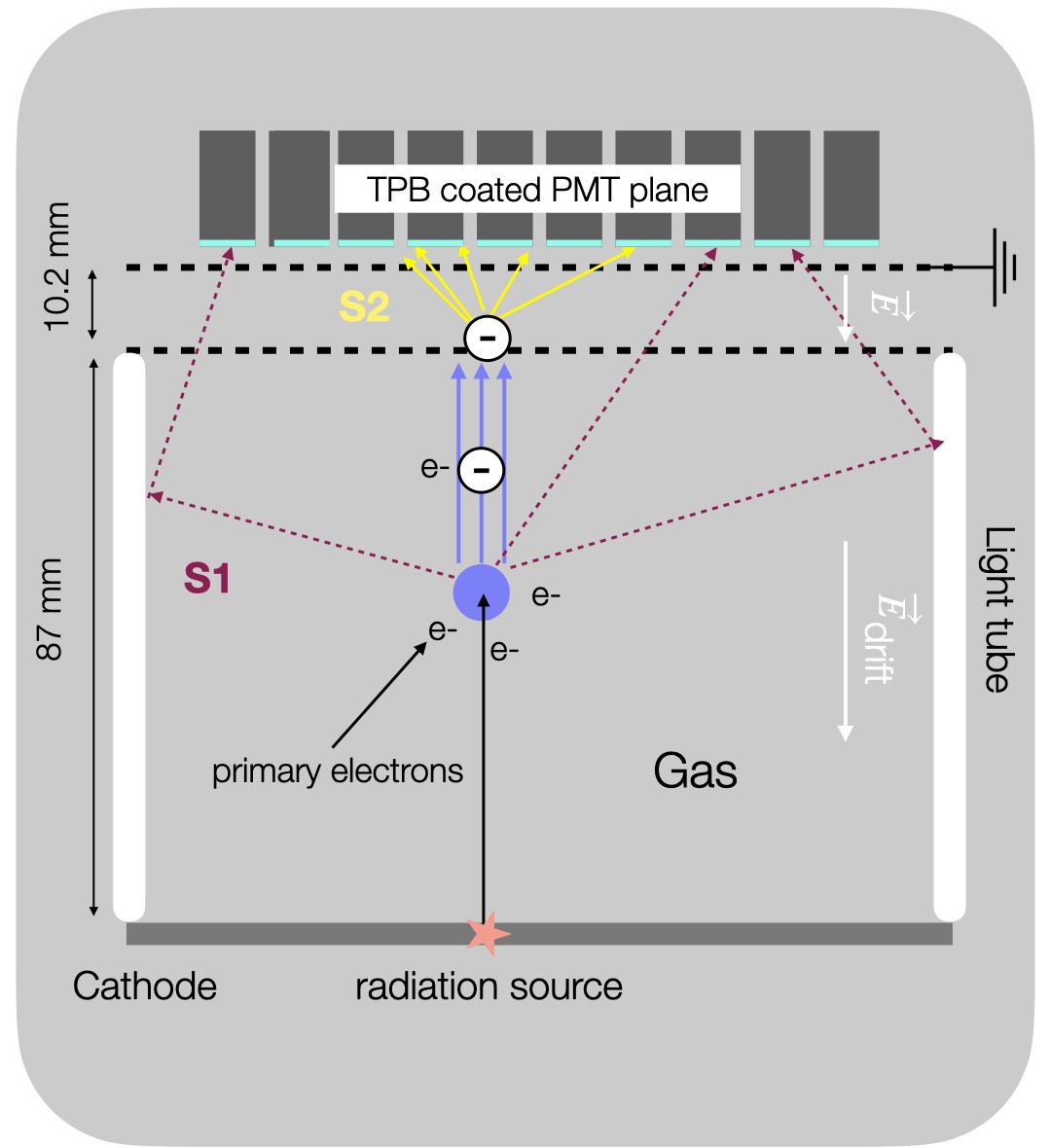}
\caption{Schematic illustration of the S1 and S2 signal production and collection in a vertical HPNG EL-TPC like GaP. The light tube is a single piece of PTFE coated with TPB.}
\label{fig:detconGaP} 
\end{figure}

The three electrodes divide the chamber into two areas: the drift and the electroluminescence regions. The drift volume is limited by the cathode and gate, which are separated by 8.7 cm. A moderate voltage difference between them results in an electric field that guides the electrons toward the electroluminescence (EL) region, avoiding electron recombination following an ionization event. To ensure uniformity, the cathode and gate are electrically connected via a resistor chain of 100 MOhms resistances. Starting from the cathode and ending in the gate, each resistance is connected in series to six field-shaping aluminum rings of 130 mm inner diameter, 5 mm thickness and 10 mm width. The electroluminescence region, limited by the gate and the anode, spans over a nominal separation of 1.02 cm. There, a much higher electric field is applied, which accelerates the electrons enough to excite the gas and produce secondary scintillation via electroluminescence. Following the detection concept described in section~\ref{sec:detcon}, figure~\ref{fig:detconGaP} illustrates the process in the GaP vertical HPNG EL-TPC. Both the drift and electroluminescence field have been simulated with COMSOL Multiphysics\textsuperscript{\textcopyright} finite element software and are shown in figure~\ref{fig:gap_TPC} right.

Two Fug HCP35-35000 High-Voltage power supply modules are used to apply voltage to the cathode and gate. These modules are controlled and monitored with a custom slow control system, built in LabView\textsuperscript{\textcopyright}, which allows increasing the voltage gradually in small steps to avoid discharges in the ramp-up. To protect the system, the software turns off the voltages automatically in the event of a discharge.

\begin{figure}
\centering 
\includegraphics[width=0.5\linewidth]{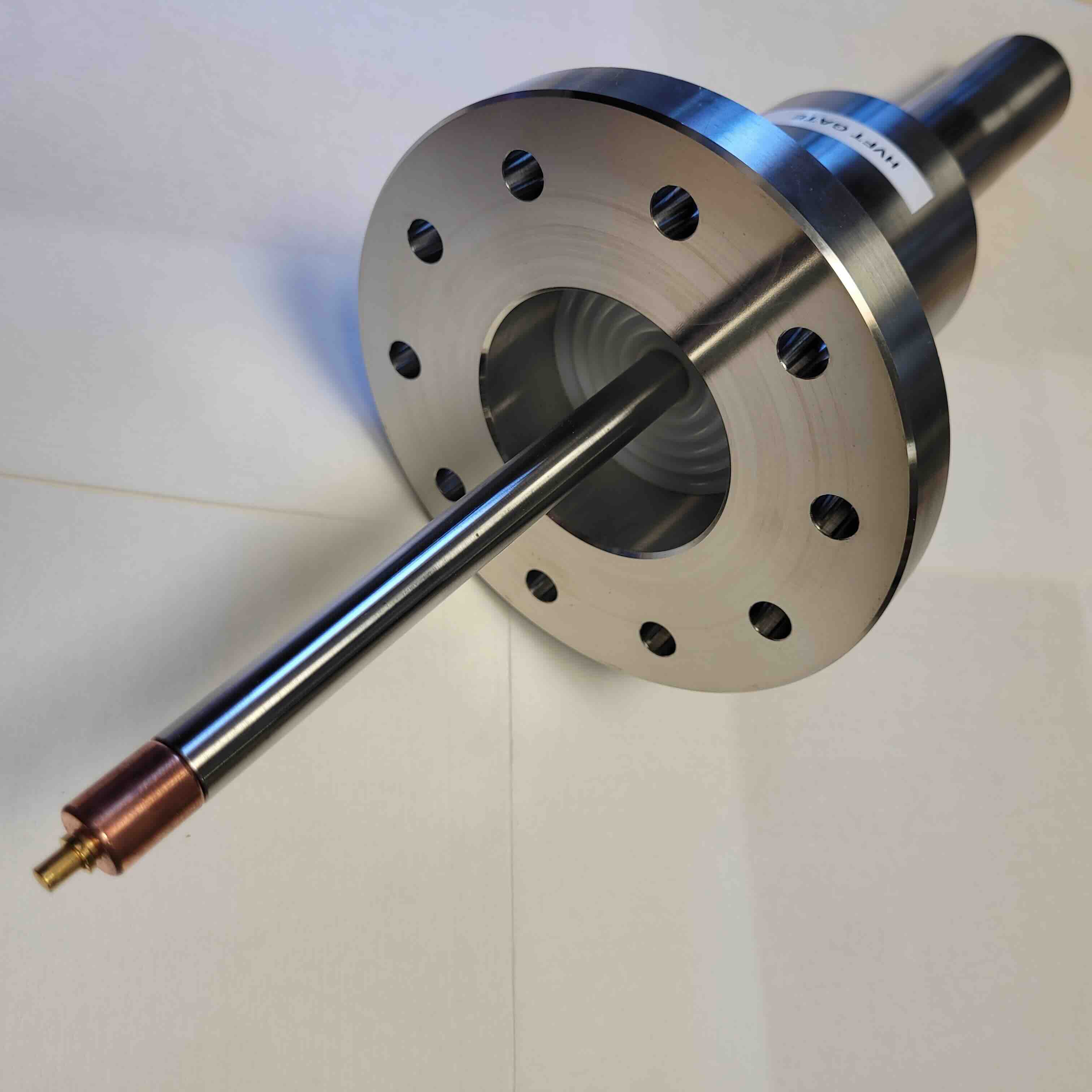}
\caption{The GaP high voltage feedthrough based on the previously used HVFT for NEXT-White's gate.}
\label{fig:HVFT} 
\end{figure}

Two identical high voltage feedthroughs (HVFTs) are used to connect the power supplies to the cathode and gate (see figure~\ref{fig:HVFT}). Given the short drift length, the difference in voltage between the cathode and gate is marginal compared to the electroluminescence requirements; therefore, the voltage requirements are defined by the operation of the electroluminescent region. As the goal of GaP is to operate up to 50 bar, they must be able to hold up voltages of at least $\sim$42 kV, slightly above the 0.83 kV/cm electroluminescence threshold of Xe \cite{Monteiro:2007vz}. The design of the feedthroughs is based on the one used for NEXT-White's gate \cite{NEXT:2018rgj}, which was able to hold up to 22 kV at 15 bar. The design relies on the gas itself to provide insulation and employs a metal rod with a spring contact to the electrodes. The connection to the gate is done directly with the feedthrough. However, that is not the case for the cathode, as its position does not coincide with the port position. A metallic l-shaped piece, screwed to the cathode, makes the connection between the cathode and the spring contact. Outside the contact regions, the piece is wrapped in kapton to avoid electrical discharges with the field rings. The anode is electrically connected by 4 metallic pieces, which also act as holders, to the rest of the pressure vessel, which is grounded. The rest of the TPC is mounted vertically on 4 high-density polyethylene rods which are used to attach the TPC to the pressure vessel without making electrical contact.

\section{Light collection and Photosensors}
\label{gap:sensors}

A hollow cylindrical PTFE tube is inserted inside the drift volume to maximize light collection (light tube). The top side, facing the EL region, is open while the bottom one is solid with the same apertures as the cathode, where it sits. The thickness of the cylinder is 5 mm and its inner radius is 60 mm. 

\begin{figure*}
\centering
  \includegraphics[height=7.5cm]{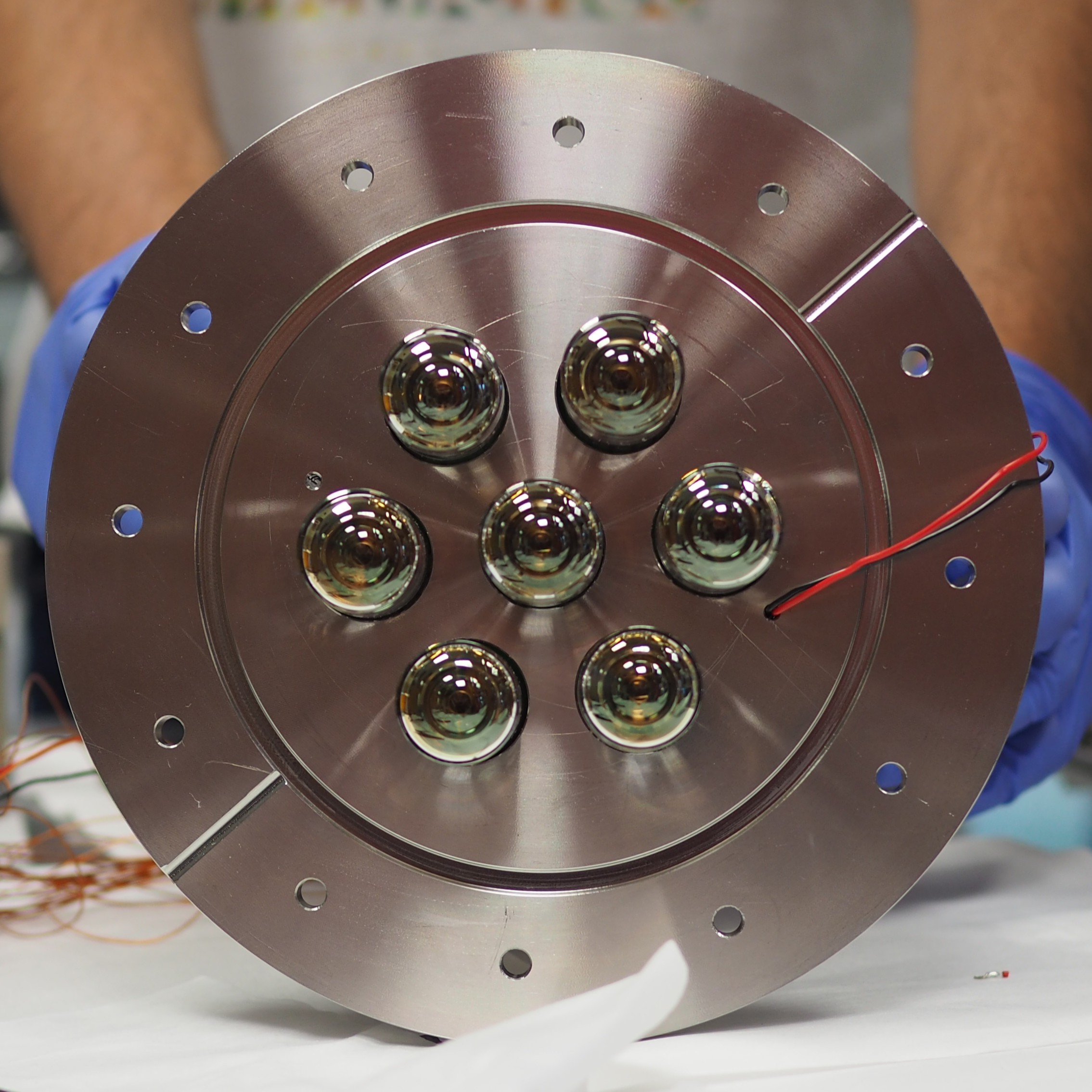}
  \hspace{0.7cm}
  \includegraphics[height=7.5cm]{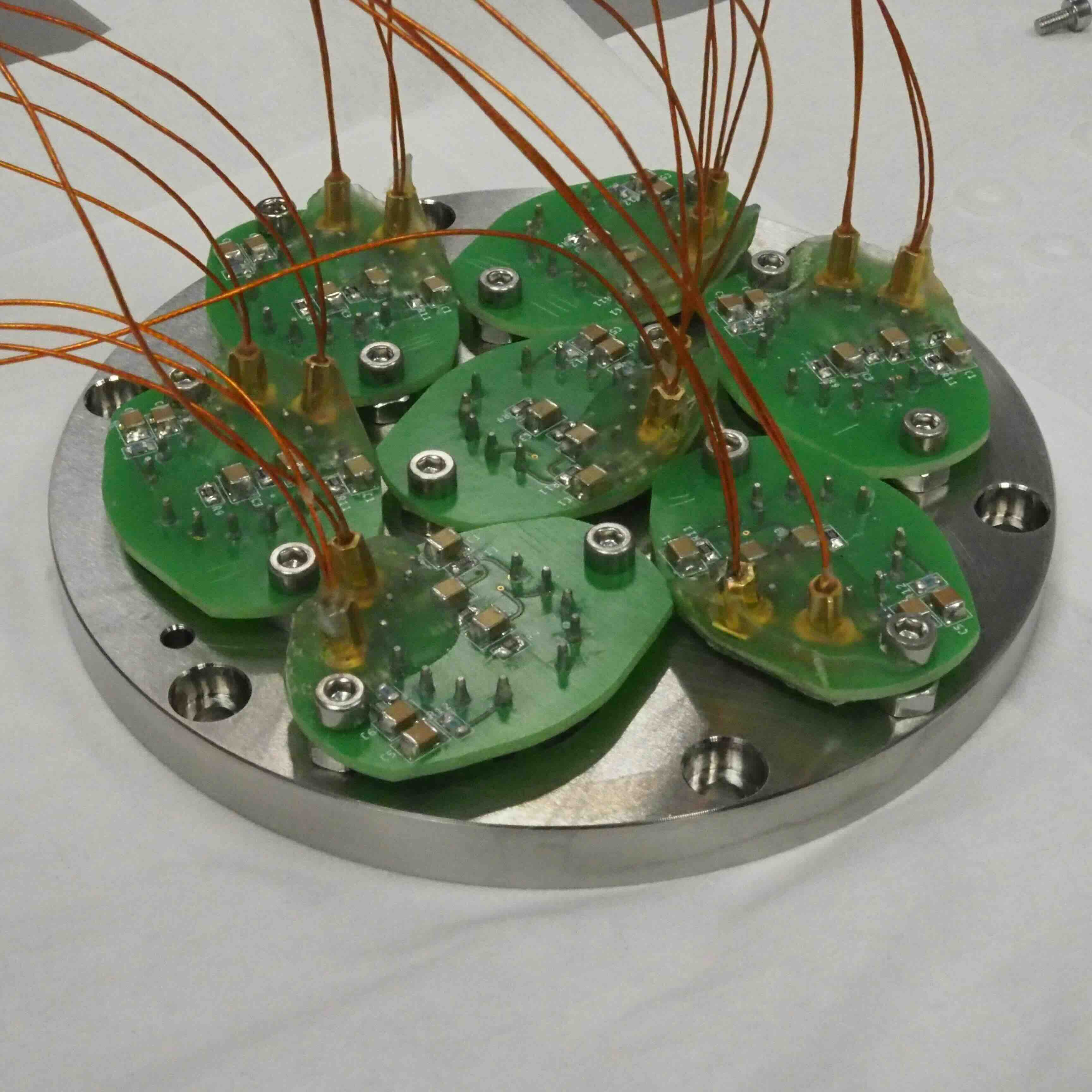}
\caption{Left: GaP PMTs distribution. Right: The PMT bases used in the GaP detector.}
\label{fig:gap_pmts}
\end{figure*}

Light is detected by seven 1-inch head-on Hamamatsu R7378A photomultiplier tubes located behind the anode grid. The sensors are distributed hexagonally, with one of them placed in the center of the hexagon and the rest at 36.37 mm from the center, see figure~\ref{fig:gap_pmts} left. The distance between the sensors and the grid can be adjusted thanks to the space available inside the inner cylinder, where the PMT bases are fixed. For the current configuration, where their maximum rated pressure (20 bar) was not exceeded, the PMTs were partially placed inside the inner cylinder, with their face at 2 mm away from the anode grid with the goal of maximizing the solid angle covered by the sensors. 

The PMT bases design follows the manufacturer's recommendations, with several resistors, capacitors and pin receptacles mounted on FR-4 boards, figure~\ref{fig:gap_pmts} right. The bases are fully covered in epoxy to avoid dielectric breakdown at low pressures. Four kapton wires, two for biasing and two to extract the signal are connected to each base. These are connected to a 36-pin feedthrough attached to the central top port. A 375 nm LED, intended for sensor calibration, is linked to a lateral feedthrough already tested and available to allow operations up to 50 bar. The PMTs are negatively biased with a CAEN A7236SN power supply and operated at 1250 V, for an average gain of 10$^6$. 

\subsection{Coating}

\begin{figure}
\centering
\resizebox{1\linewidth}{!}{%
  \includegraphics{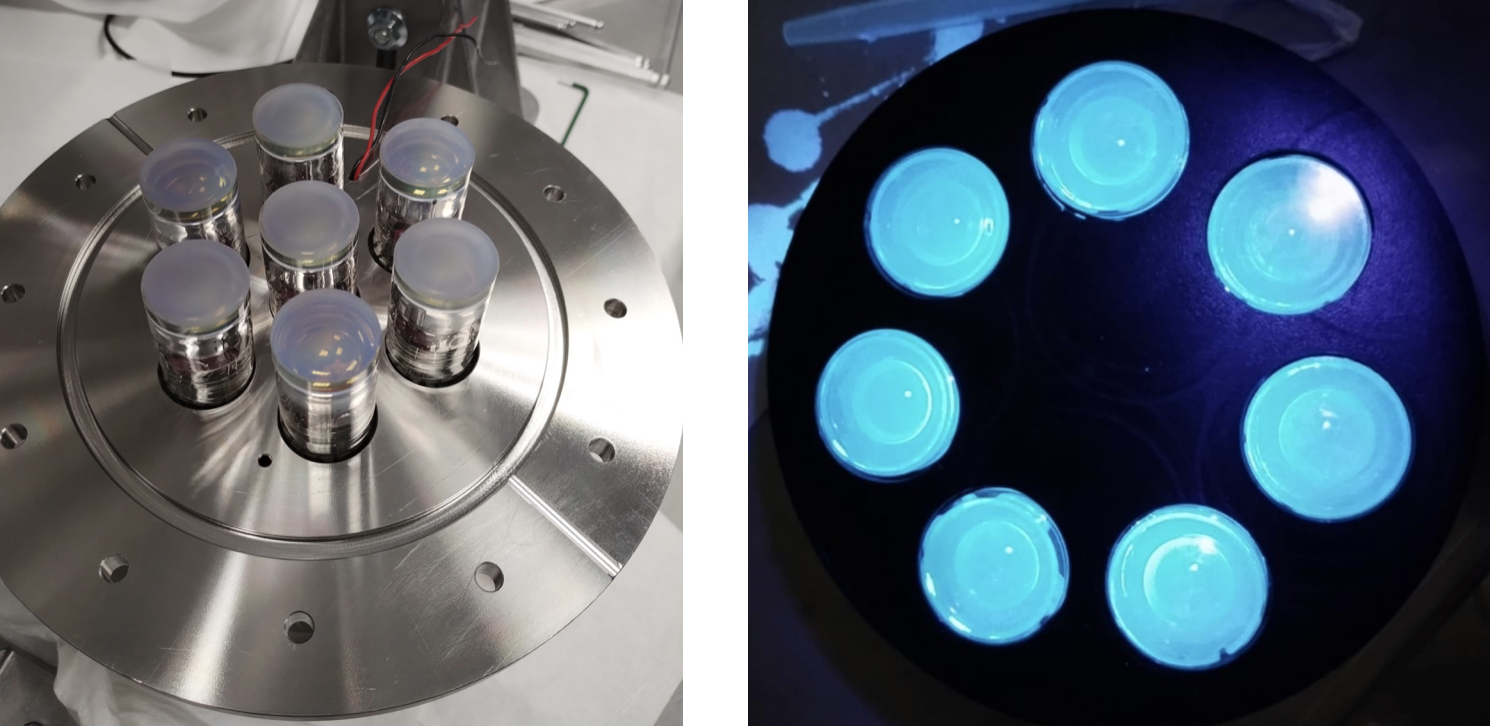}
  }
\caption{Left: GaP PMTs distribution with TPB deposition. Right: PMTs illuminated with UV light after TPB deposition.}
\label{fig:gap_pmtstpb} 
\end{figure}

\begin{figure}
\centering 
\includegraphics[width=0.5\linewidth]{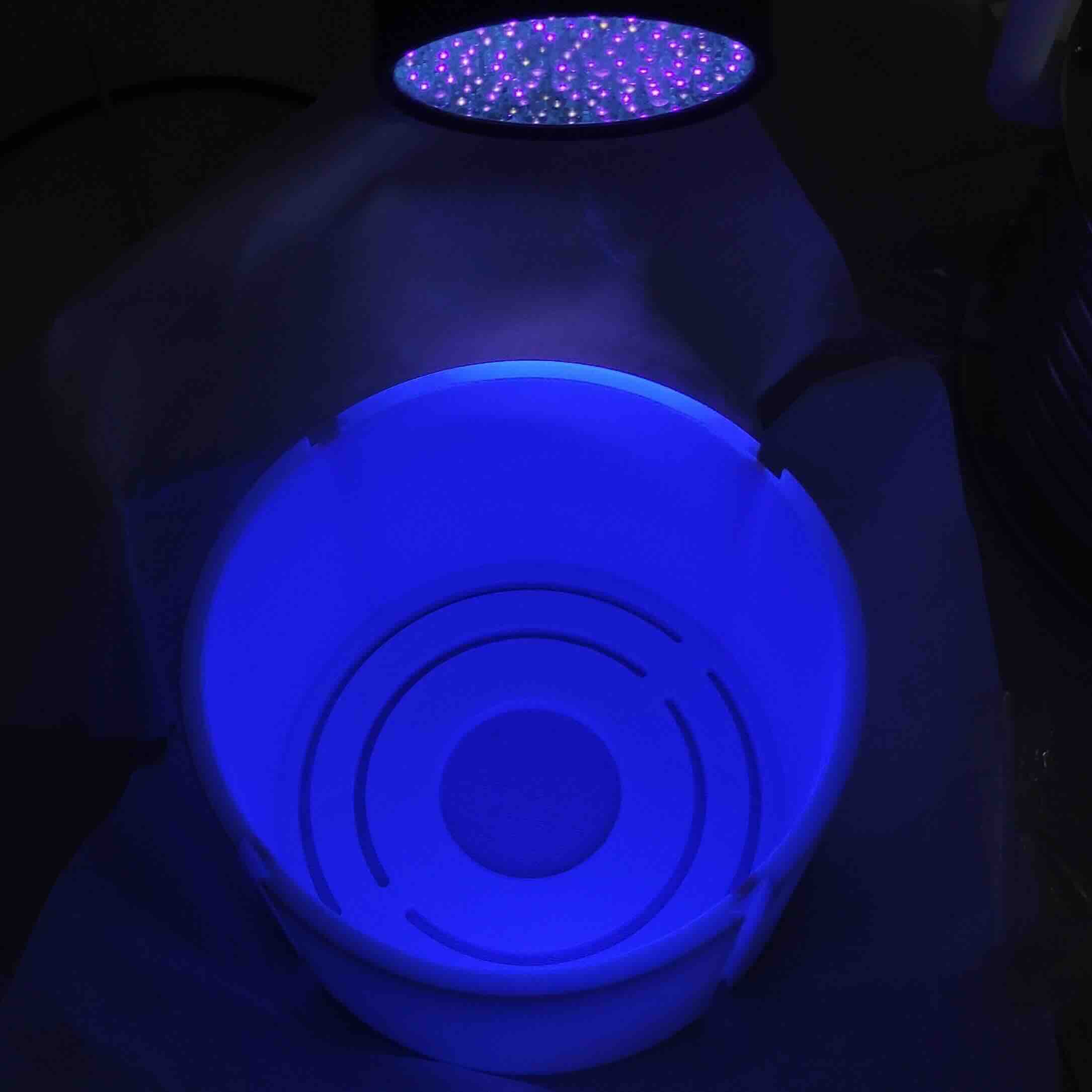}
\caption{GaP's PTFE light tube illuminated with UV light after TPB deposition.}
\label{fig:lighttube} 
\end{figure}

Since the quantum efficiency of the sensors at the Xe scintillation wavelength is rather low, $\sim$10\% and non-sensitive to Ar light, it is necessary to use a wavelength shifter coating. Because of its emission properties, Tetraphenyl Butadiene (TPB) has been used. TPB is a wavelength shifter commonly used to shift the VUV light from noble gas scintillation towards blue ($\sim$425 nm) in order to maximize reflectivity and light collection efficiency \cite{MCKINSEY1997351}. The emission spectrum of TPB peak results to be in the region of highest quantum efficiency of the Hamamatsu R7378A photomultiplier tubes used to operate the detector. For this reason, we evaporated TPB directly into the PMT window, see figure~\ref{fig:gap_pmtstpb}. On the other hand, the light emitted towards the cathode will be shifted on the PTFE reflector walls, also coated with TPB as in figure~\ref{fig:lighttube}. Although the system lacks precise control of the coating thickness, a few micrometer thickness coating is expected, sufficient for efficient shifting \cite{Benson:2017vbw}. 

\subsection{Photosensor calibration}
\label{gap:sensorscalib}

To ensure a reliable and stable measurement of the detector's energy resolution, the photosensors need to be calibrated so that the number of photo-electrons detected by each sensor can be precisely determined from the digitized response. This calibration must also be continuously monitored during operation, as the calibration constants can be affected by external factors such as temperature and pressure, as well as by photodetector degradation over time or occasional sparks in the TPC. Thus, a sensor calibration was performed periodically with the internal LED mounted to a lateral feedthrough available to operate up to 50 bar.

An external trigger signal, in coincidence with the LED emission, was used to acquire the data with a pretrigger of 2 $\mu$s. The LED's intensity is carefully adjusted to minimize the probability of detecting more than one photoelectron and thus, the LED voltage was first tuned so $\sim$10\% of the triggers produced an observable signal in the PMT. As the distance from the LED to each sensor is different, additional data was taken around such voltages to make sure to have suitable datasets available for single photoelectron (SPE or 1PE) calibration in each PMT, i.e. with minimum contamination from multiphoton detection.

\begin{figure}
\centering 
\includegraphics{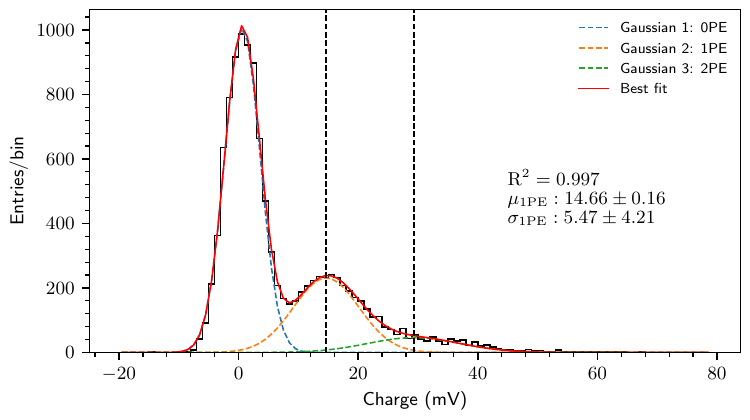}
\caption{Typical single photoelectron spectrum of one of the PMTs used in GaP. The three Gaussian best fit curve together with the contribution of each Gaussian shown separately. The first vertical dashed line corresponds to the 1PE peak and the second to the 2PE peak. The estimated single photoelectron conversion factor is also shown.}
\label{fig:pmt_calib} 
\end{figure}

The procedure to extract the SPE spectrum is: 1) the baseline of each PMT is individually calculated as the average of the waveform outside the integration window; 2) the baseline is subtracted from the waveform; 3) the waveform is integrated within a window of 200 ns, starting 80 ns prior of the trigger position; 4) the signal integral is histogrammed and yields the SPE spectrum shown in figure~\ref{fig:pmt_calib}; and 5) the spectrum is fitted to a three Gaussian distribution, accounting for the system noise, the single photon and the two-photon peaks. The difference in the peak centroid between consecutive peaks corresponds to the sensor gain and is fixed as a free parameter. The centroid of the system noise should be around 0 as no light is collected, also known as the pedestal. The centroid of the single-photon peak, provides the conversion factor between photoelectrons and charge measured (figure~\ref{fig:pmt_calib}). The position of the two-photon peak relative to the single-photon peak serves as a check on the measurement's robustness. The sigma of the two-photon peak is set as $\sigma_{2\rm{PE}}=\sqrt{2}\cdot\sigma_{1\rm{PE}}$, being $\sigma_{1\rm{PE}}$ the sigma of the SPE peak.

The fit was done with \textit{The Model} class in the lmfit Python library, which provides tools for non-linear least-squares minimization and curve fitting. The quality of the fit is given by the floating point $R^2$ statistic, defined for data $y$ and best-fit model $f$ as:

\begin{equation}
    R^2 = 1 - \frac{\sum_i (y_i-f_i)^2}{\sum_i (y_i-\bar{y})^2}.
\end{equation}

The PMT gain is calculated by estimating the charge of the SPE pulse based on the characteristics of the DAQ system.
This involves estimating the pulse area (in volt-seconds), which is the product of the integrated charge and its duration. The DAQ used is terminated with a resistor of $R = 50~\Omega$ and it has a $\Delta t$ sampling time of 8~ns. So, the next step is to convert this voltage-time product into charge (in coulombs) using the relation Q = V$\cdot$ $\Delta t$ / R. Finally, the PMT gain is the number of electrons per pulse, calculated by dividing the pulse charge by the elementary charge $e$= $1.602 \times 10^{-19}$~C.

This calculation provides a practical way to convert an observed signal into a physical charge measurement and thus determine the PMT gain. Accordingly, table~\ref{tab:pmtgain} shows the gain of the 7 PMTs at the time of the data taking for the results shown in chapter~\ref{sec:gapresultschapter}, which is in agreement with the specifications of Hamamatsu R7378A PMT. Similarly, at the operating voltage of 1250 V the anode dark current (DC) is 20 nA, which for the average gain in table~\ref{tab:pmtgain} corresponds to $\sim$11 kHz. This translates into $\sim$0.78 DC at a 10 us window for all 7 PMTs.

\begin{table}[h!]
\centering
\begin{tabular}{|c | c |} 
 \hline
 PMT number & Gain \\ [0.5ex] 
 \hline\hline
 1 & $7.45\cdot 10^{6}$  \\ 
 \hline
 2 &  $5.51\cdot 10^{6}$   \\
 \hline
 3 &  $1.57\cdot 10^{7}$   \\
 \hline
 4 &  $8.59\cdot 10^{6}$   \\  
 \hline
 5 &  $1.24\cdot 10^{7}$   \\ 
 \hline
 6 &  $1.72\cdot 10^{7}$   \\  
 \hline
 7 &  $1.20\cdot 10^{7}$   \\ 
 \hline
\end{tabular}
\caption{The gain for each PMT inside the high pressure gaseous TPC prototype at the time when the measurements for the results shown in chapter~\ref{sec:gapresultschapter} were performed. PMT number 1 corresponds to the central sensor in GaP with the others surrounding it, following the distribution in figure~\ref{fig:gap_pmts} left.}
\label{tab:pmtgain}
\end{table}

\subsection{Sensor gain}
\label{sec:sensor_gain}

\begin{figure}
\centering
\includegraphics{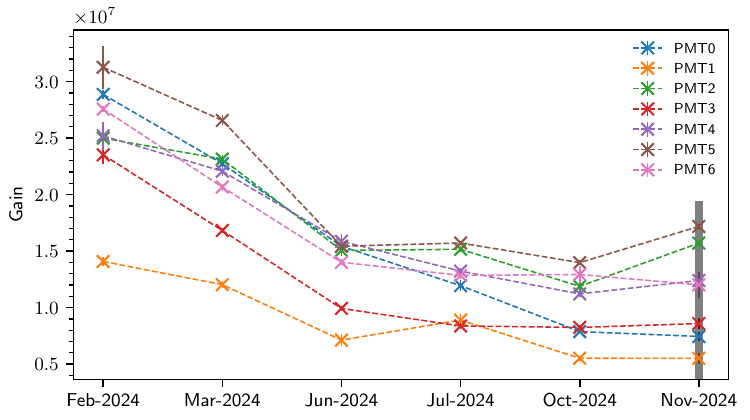}
\caption{Variation of the PMT gain during the several months operation of GaP. Over the span of a few months, a clear degradation of the PMTs is observed while a stability region is reached around June 2024. The grey bar in November 2024 marks the achieved gain values at the time the results presented in this work were obtained.}
\label{fig:pmt_gain} 
\end{figure}

Calibration was performed intermittently throughout radioactive source data collection. This allowed continuous stability monitoring and quick identification of any changes in conditions. A degradation of the sensor gain was observed at the beginning while a stabilization is visible in figure~\ref{fig:pmt_gain}. This circumstance demands performing sensor calibration before and after each major data-taking campaign and considering the variation in gain as a systematic uncertainty. The source of such degradation is not currently understood and will demand further investigation. The possibility of Ar leaking into the PMT volume was considered. Should that be the case, an increase of afterpulsing (AP) population should be observed. The presence of afterpulsing at $\sim$250 ns was clearly observed with an average AP rate of $\sim$6.3$\%$. However, this does not appear to be the case following studies in a dedicated setup in which a fresh PMT unit was exposed to 10 bar of Ar for two months. Here the gain was characterized weekly and, while a decrease in yield was still apparent, no increase in afterpulsing was observed.

\section{Digital trigger system}
Signals are directly read with a CAEN DT2740 digitizer. It is a 64-channel tabletop unit with a 16-bit 125 MS/s ADC, i.e. 8 ns time binning, and an input range of 2 V$_{pp}$. While the system allows for custom programming of the FPGA, the WaveDump2 \cite{wavedump} software was used to acquire data. It allows for simultaneous acquisition of all active channels following an edge trigger signal, independently configurable for each channel, and maximum software pretrigger of 16 \mus\ (2000 samples) is allowed, while the buffer size can be expanded over several hundreds of \mus. 
\renewcommand{\arraystretch}{1.5}

\begin{savequote}[65mm]
``Ez duk larrosik arantza gaberik" 
\qauthor{-- Old proverb in Basque language.}
\end{savequote}

\chapter{Operation and first results of the Gaseous Prototype (GaP)}
\label{sec:gapresultschapter}

Since being mounted, GaP has been operated with Ar at pressures ranging roughly from 1.5 to slightly above 8.5 bar. Different calibration sources have been used to characterize the detector response. Moreover, data was taken with and without the light tube. 

While specific details will be given in the following, generally speaking, data was taken in short periods, up to 15 minutes obtaining approximately $5\cdot10^4$ events per run, for each given configuration to avoid time-dependent variations. Each configuration had its own set of operational parameters which included a fixed pressure, cathode and gate voltages. Gas was continuously recirculated during data-taking and initiated daily at least one hour prior to the start of data-taking in order to remove impurities outgassed during the night when the recirculation loop was stopped to prevent possible unsupervised damage to the system. The estimated time for the gas to complete a whole loop through the system is about 15 min, with one hour of circulating before the initial data taking the gas proven to be clean enough and in stable conditions. With the current data what we have explored is the stability of the total charge observed with time, with runs under the same conditions after several hours of operation with the hot getter. The result is that no difference on the total detected S2 is observed. We have slow controls monitoring the pressure, HV and supply of the PMTs and the detector has been operating for over a year now stable at high pressures. 

\section{Data taking with \fe\ internal source}
\label{subsec:Fe}

\begin{figure}
\centering
\includegraphics [trim={0.2cm 0 0 0},clip]
{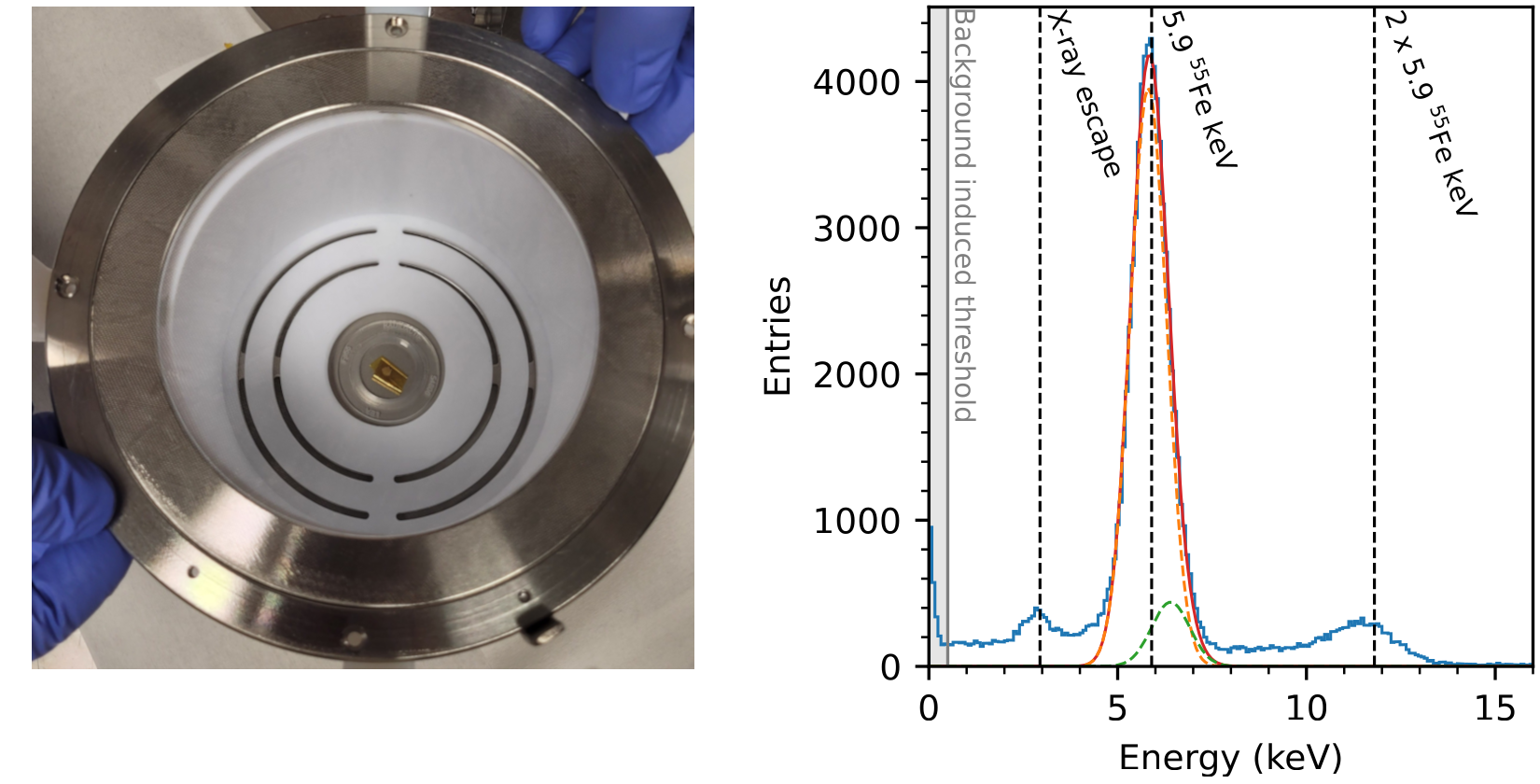}
\caption{Left: Image of the position of the \fe\ source in the detector. The source is attached to the cathode with a small kapton tape. The position corresponds to the radial center of the active volume. Right: Energy spectrum obtained from the \fe\ source. A clear peak, corresponding to an energy of 5.9 keV, the dominating line, is seen. The red line corresponds to the result of the two Gaussian fit described in the text, while the dashed lines correspond to the 5.9 (orange) and 6.5 (green) keV populations. A duplicate of the peak is seen at twice the energy, corresponding to events which had accidental coincidences due to the high activity of the source. At 2.94 keV a peak corresponding to the escape of Ar $K_{alpha}$ 
X-rays are also seen. In gray, a low-energy background population appears at high electric fields and limits the detection threshold to $\sim$0.5 keV. The detection threshold will be discussed in \ref{subsec:det_threshold}.}
\label{fig:fe_spectrum} 
\end{figure}

A \fe\ source was placed in the cathode's central hole. \fe\ decays into $^{55}$Mn via electron capture. The decay is followed up by the emission of Mn characteristic X-rays, with energies 5.9 and 6.5 keV corresponding to the $K_{\alpha}$ and $K_{\beta}$ lines. 
The source, provided by ORANO, is a 3 mm diameter radionuclide disk placed between 2 thin polyester foils of 75 \mum\ thickness, mounted on a 38 mm diameter plexiglass ring (figure \ref{fig:fe_spectrum}). The original activity of the source was 20 kBq when bought (June 2023) and is estimated to be 15.85 kBq at the beginning of data-taking campaign (May 2024). 

Data was acquired by triggering on the central PMT as we expect it to be the sensor with a larger signal, being closer to the radial position of the source. The trigger threshold was set to a fixed value slightly above the electronic noise. Given the low energy of the gammas, the trigger was only sensitive to secondary scintillation. 

The mean free path of the gammas will range between 0.24 and 1.1 cm, depending on the operational pressure. This means that their interaction point will be extremely concentrated around the source. 
However, the distance between the interaction point and amplification region precludes the observation of primary scintillation. The drift velocity for Ar at these pressure ranges was calculated using Magboltz \cite{Biagi:1999nwa}, the calculation indicates a drift velocity between 1 and 3 mm/\mus. At these conditions, the primary scintillation will reach the photosensors several tens of \mus\ before the start of the secondary scintillation. As the maximum pre-trigger of the system is 16 \mus, the primary scintillation cannot be observed when triggering on S2. With this in mind and to maximize the acquisition rate, only limited by the data throughput, the acquisition window was set to 40 \mus\ with a pretrigger of 10 \mus.

\begin{figure}
\centering
\includegraphics{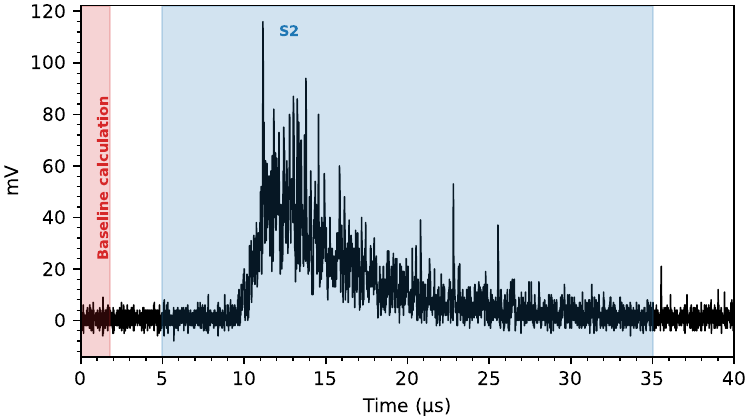}
\caption{A typical waveform signal produced by the secondary scintillation of \fe\ induced interactions. Data is triggered at 10 \mus, close to the start of the pulse. The red region covers the 225 samples used to calculate the baseline. The integration window, shaded in blue, expands from 5 to 35 \mus. This interval contains more than 98\% of the integrated charge of the full waveform while the losses are related to delayed emissions, like the small peak at 35.5 \mus. The waveform signal corresponds only to the central PMT inside the GaP detector. The sum waveform of all the PMTs is very similar since this central PMT is the one that sees most of the light, as it is the photosensor placed straight on top of the \fe\ source.}
\label{fig:fe_wvf} 
\end{figure}

A simple data processing scheme is applied to create analysis datasets. For each event, the baseline of each channel is calculated, as the average of the first 225 samples (first 1.8 \mus ), and is later subtracted from the signal. The signal of each channel is later converted to photoelectrons based on calibration factors periodically obtained with a LED as explained before. It should be noted that a constant decrease in such conversion factors was observed over several months. As mentioned earlier in section~\ref{sec:sensor_gain}, this variation is considered a systematic uncertainty in the analysis. 

Following conversion into photoelectrons, the waveforms of the 7 PMTs are added together. The integral of such sum in the interval [5, 35] \mus\ is taken as the recorded energy of the event. The considered interval is enough to cover the signal, as shown in figure~\ref{fig:fe_wvf}. The resulting spectrum obtained is shown in figure~\ref{fig:fe_spectrum} left. A fit to two Gaussians, corresponding to the peaks at 5.9 and 6.5 keV, is applied. The centroid of the higher energy peak is constrained to be a factor 6.5/5.9 larger than the lower energy peak. The centroid value of the lower energy peak is considered as the average number of detected photons, $N_{det}$, for 5.9 keV depositions, which can be used, as discussed later, to evaluate the charge yield of the detector.

The energy resolution, obtained from the fit to the energy spectra at different pressures and electroluminescence fields, is shown in figure~\ref{fig:reso}. It shows a minimum value at a reduced electric field of 1.3-1.5 kV/cm/bar with a considerable worsening for low pressures mostly related to the lower number of photons produced in the amplification process. The resolution obtained is worse than anticipated by the number of photoelectrons detected per electron, this will be discussed in section \ref{subsec:eres}.

\section{Results}
\label{sec:results}

\subsection{Summary of the data runs}

Table~\ref{tab:el_field_1.5bar} indicates the specific configurations set for measurements done at 1.5 bar. The drift voltage is 3 kV to avoid voltage induction in the gate, which corresponds to $\sim$0.345 kV/cm. However, for the rest of the measurements with pressures ranging from around 2.5 bar to slightly above 8.5 bar an electric field of $\sim$0.575 kV/cm was set. As an illustration, table~\ref{tab:el_field_4.5bar} also shows the operational parameters for the 4.5 bar dataset. Accordingly, table~\ref{tab:el_drift_fields} gives an overview of the different high voltage configurations required to operate the detector at different pressure but similar light amplification conditions. 

\begin{table}
\centering
\begin{tabular}{|c|c|c|c|}
\hline
\thead{\textbf{Pressure (bar)}} & 
\thead{\textbf{Reduced EL} \\ \textbf{field (kV/cm/bar)} } & 
\thead{\textbf{Voltage} \\ \textbf{ in cathode (kV)} } & \thead{\textbf{Voltage} \\ \textbf{ in gate (kV)} } \\
\hline
1.51 & 1.039 & -4.60 & -1.60 \\
1.51 & 1.169 & -4.80 & -1.80 \\
1.51 & 1.299 & -5.00 & -2.00 \\
1.51 & 1.428 & -5.20 & -2.20 \\
1.51 & 1.558 & -5.40 & -2.40 \\
1.51 & 1.688 & -5.60 & -2.60 \\
1.51 & 1.818 & -5.80 & -2.80 \\
1.51 & 1.948 & -6.00 & -3.00 \\
1.51 & 2.078 & -6.20 & -3.20 \\
1.51 & 2.208 & -6.40 & -3.40 \\
1.51 & 2.337 & -6.60 & -3.60 \\
1.51 & 2.467 & -6.80 & -3.80 \\
1.51 & 2.597 & -7.00 & -4.00 \\
\hline
\end{tabular}
\caption{Voltage readings at various reduced EL field values where electroluminescence yield measurements are performed for pressure 1.51 bar with an EL length of 1.02 cm. The drift voltage is fixed to 3 kV to ensure that there is no voltage induced from the cathode to the gate. As GaP is designed with 8.7 cm of drift length, the drift electric field is set to $\sim$0.345 kV/cm.}
\label{tab:el_field_1.5bar}
\end{table}

\begin{table}
\centering
\begin{tabular}{|c|c|c|c|}
\hline
\thead{\textbf{Pressure (bar)}} & 
\thead{\textbf{Reduced EL} \\ \textbf{field (kV/cm/bar)} } & 
\thead{\textbf{Voltage} \\ \textbf{ in cathode (kV)} } & \thead{\textbf{Voltage} \\ \textbf{ in gate (kV)} } \\
\hline
4.5 & 0.861 & -8.95  & -3.95 \\
4.5 & 0.937 & -9.30  & -4.30 \\
4.5 & 1.013 & -9.65  & -4.65 \\
4.5 & 1.089 & -10.00 & -5.00 \\
4.5 & 1.166 & -10.35 & -5.35 \\
4.5 & 1.242 & -10.70 & -5.70 \\
4.5 & 1.318 & -11.05 & -6.05 \\
4.5 & 1.394 & -11.40 & -6.40 \\
4.5 & 1.471 & -11.75 & -6.75 \\
4.5 & 1.547 & -12.10 & -7.10 \\
4.5 & 1.623 & -12.45 & -7.45 \\
4.5 & 1.699 & -12.80 & -7.80 \\
4.5 & 1.776 & -13.15 & -8.15 \\
4.5 & 1.852 & -13.50 & -8.50 \\
\hline
\end{tabular}
\caption{Voltage readings at various reduced EL field values where electroluminescence yield measurements are performed for pressure 4.5 bar with an EL length of 1.02 cm. The drift voltage is fixed to 5 kV this time so for a 8.7 cm drift length the drift electric field is set to $\sim$0.575 kV/cm. For the rest of the measurements with pressures ranging from around 2.5 bar to slightly above 8.5 bar the drift voltage is fixed to the same value.}
\label{tab:el_field_4.5bar}
\end{table}

\begin{table}
\centering
\begin{tabular}{|c|c|c|c|c|}
\hline
\thead{\textbf{Pressure (bar)}} & 
\thead{\textbf{Reduced EL} \\ \textbf{field (kV/cm/bar)} } & 
\thead{\textbf{Reduced E\textsubscript{drift}} \\ \textbf{field (kV/cm/bar)} } & 
\thead{\textbf{Voltage} \\ \textbf{ in cathode (kV)} } & \thead{\textbf{Voltage} \\ \textbf{ in gate (kV)} } \\
\hline
1.51 & 1.168 & 0.228 & -4.80 & -1.80 \\
2.61 & 1.126 & 0.220 & -8.00 & -3.00 \\
3.55 & 1.160 & 0.162 & -9.20 & -4.20 \\
4.50 & 1.165 & 0.128 & -10.35 & -5.35 \\
5.50 & 1.132 & 0.104 & -11.35 & -6.35 \\
6.58 & 1.140 & 0.087 & -12.65 & -7.65 \\
7.63 & 1.156 & 0.075 & -14.00 & -9.00 \\
8.62 & 1.160 & 0.067 & -15.20 & -10.20 \\
\hline
\end{tabular}
\caption{The operating voltages of the detector as a function of pressure for a similar reduced electroluminescence field values together with the corresponding reduced drift field. Note the high voltages required for only a reduced EL field of $\sim$ 1.15 kV/cm/bar at higher pressures.}
\label{tab:el_drift_fields}
\end{table}

\subsection{Charge yield and light collection efficiency}
\label{subsec:charge_yield}

The charge yield \gtwo\ of the system is defined as the number of photoelectrons detected per ionization electron. It is one of the key aspects of the detector as it affects the energy resolution and the detection threshold being the threshold inversely proportional to \gtwo. We define \gtwo\ as:

\begin{equation}
    g_2 = \frac{N^{S2}_{det}}{N_{e^-}} = \frac{N^{S2}_{det}\cdot W_i}{E}
    \label{eq:g2}
\end{equation}

being $N^{S2}_{det}$ the number of detected photoelectrons from the secondary scintillation light production for depositions of a given energy $E$, $N_{e^-}$ the number of produced electrons by such energy deposition and $W_i$ the average energy to produce an ion-electron pair, also known as the $w$-value, which is taken as 26.27$\pm$0.14 eV \cite{wi_argon}. On the other hand, in our system, $N^{S2}_{det}$ is given by the product of the secondary scintillation light collection efficiency ($LCE_{S2}$, defined as the probability of detecting a photon produced via electroluminescence), the electroluminescence yield ($Y_{EL}$, number of photons produced per ionization electron) and the number of electrons arriving at the amplification region: 

\begin{equation}
    N^{S_2}_{det} = LCE_{S2}\cdot Y_{EL}\cdot N_{e^-}
    \label{eq:LCE}
\end{equation}

It follows that, ultimately, \gtwo\ is proportional to $LCE_{S2}$ and $Y_{EL}$. The light collection efficiency is a purely geometrical factor and should not, to first order, change with neither the gas pressure nor the electroluminescence voltage. On the other hand, the electroluminescence yield has been commonly described as a linear process that depends on the electric field applied in the amplification region. 

\begin{figure}
\centering
\includegraphics{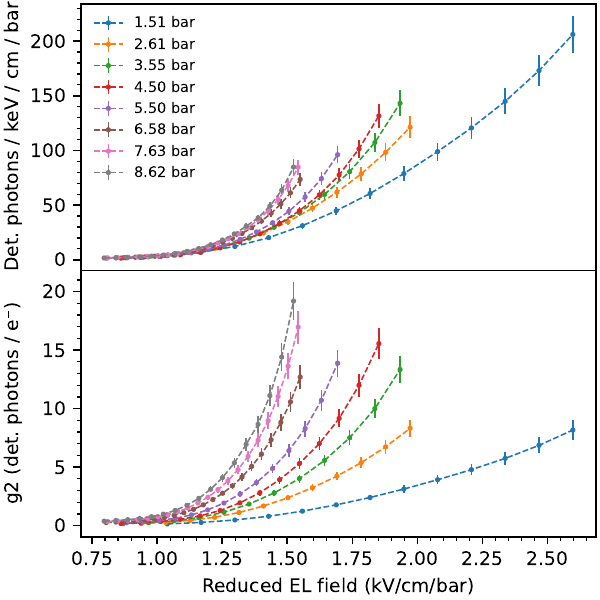}
\caption{Detected photons per keV (top) and \gtwo\ (bottom) as a function of the electroluminescence field for different operating pressures. An exponential trend is observed at all pressures. The difference in the number of det. photons per bar indicates that the reduced yield depends not only on the reduced field but also on the absolute pressure, indicating an extra light production mechanism in the EL region.}
\label{fig:charge_yield} 
\end{figure}

However, we observe an exponential increase of \gtwo\, illustrated in figure~\ref{fig:charge_yield}, as a function of the electroluminescence field. As explained above, this behavior is counter-intuitive as \gtwo\ should follow the same trend as the electroluminescence yield, which should behave linearly up to at least $\sim$3 kV/cm/bar. 

An exponential behavior could be related to operation at sufficiently high electric fields to produce charge amplification, at least in some parts of the amplification structure like wires or the borders of the electroformed mesh. In an attempt to understand the trend, the electric field has been simulated with COMSOL\textsuperscript{\textcopyright} \cite{comsol2024} and used as an input for Garfield++ \cite{GarfieldPP} to evaluate the light yield and possible regions where ionization is more likely to happen even at lower nominal fields. The details of this study are discussed in chapter \ref{sec:garfield}. 

Another possible explanation for the evidence of extra light amplification in the Ar detector other than the electroluminescence process could be the presence of photo-ionizing impurities in the gas system. These candidates should have an ionization potential smaller than the one of argon VUV scintillation photons (128 nm). We are currently working on identifying such impurities, if any. 
An element currently under consideration is the TPB. The ionization potential of the TPB used to coat the PMTs for maximizing the quantum efficiency is very likely to be lower than the argon VUV photon energy \cite{segreto2015evidence}. This could cause that each absorbed photon excites more than one TPB molecule at a time explaining the observation of the non-linear behavior. The removal of wavelength shifters has been planned by modifying the detector set-up and operating GaP with Silicon Photomultipliers (SiPMs). The SiPMs that will be used are the 2x2 quad modules VUV4 MMPCs from Hamamatsu (model S13371-6050CQ-02) \cite{VUV4MPPC}. 
 
The LCE for the GaP detector has not been estimated from the data in this work. As the non-linear behavior of the electroluminescence yield does not agree with literature, the determination of the LCE would not have been realistic. However, considering the GaP geometry, a LCE of 2.7$\%$ was estimated when the VUV light of argon (128 nm) was simulated at the center of the EL region. As mentioned, this should be a purely geometrical factor and depending on the origin of the photoemission the LCE will vary. Overall, the average LCE was estimated to be 1.85$\%$ considering also different emission positions in between PMTs and closer to the walls.

\subsection{Evaluation of the detection threshold}
\label{subsec:det_threshold}

Notwithstanding the exponential behavior, the observed \gtwo\ is sufficiently large to achieve a detection threshold extremely competitive. Concretely, the maximum charge yield is achieved at 8.62 bar and an electric field in the electroluminescent region of 1.52 kV/cm/bar. A value of 711.6$\pm$58.6 photons per keV is observed with the uncertainty being fully dominated by the PMT gain decrease commented on section \ref{sec:sensor_gain}. This translates into a \gtwo\ value of 18.7$\pm$1.7 detected photons per electron, which should be more than sufficient to detect single electrons. 

However, the spectrum exhibits the presence of a low-energy population which results in an effectively higher threshold. For example, this population becomes dominant below $\sim$0.5 keV when operating at the maximum charge yield. The reconstructed energy for this population decreases with the electroluminescent yield, which points towards instrumental effects related to instabilities at high voltage (i.e. glows and/or field emission from the gate grid). 

Aside from this limitation, the optimal way to demonstrate the detection threshold would be by using calibration sources with the lowest energy possible. Unfortunately, the lowest energy source available was \fe. Still, a reasonable assessment of the threshold can be given by evaluating the calibration peak at different yields and assuming that the number of photons detected at minimum yield is an effective threshold. Concretely, the rationale is that if the \fe\ peak can be observed at different electroluminescent yields, then the ratio between the yields can be used to estimate the minimum detection threshold at the higher yield. For example, given two datasets taken with a factor 10 difference in yield, the fact that the 5.9 keV can be observed in the lower yield dataset means that 0.59 keV events should be detected when operating at the higher yield. 

\begin{figure}
\centering
  \includegraphics{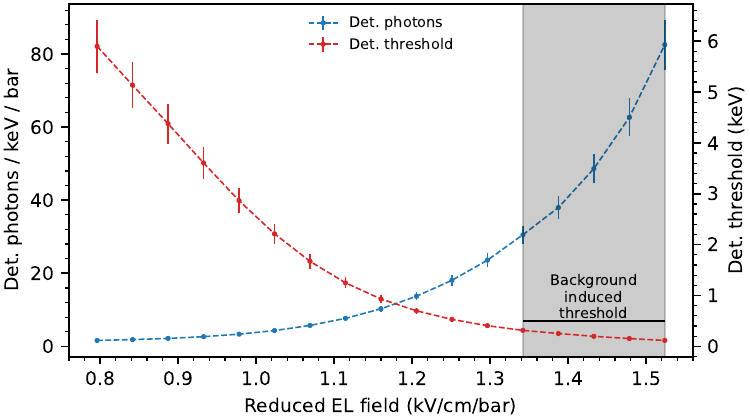} 
\caption{In blue, detected photons per keV and bar extracted from \fe\ peak at 8.62 bar. In red, the detection threshold assuming the minimum detectable signal corresponds to the number of detected photons per keV (14.00 $\pm$ 1.15 phot./keV) in the lower yield conditions. The shadowed gray area marks the point where the instrumental background dominates the threshold.
}
\label{fig:det_threshold} 
\end{figure}

The estimates for the extrapolated detection threshold are illustrated in figure~\ref{fig:det_threshold}, accompanied by the number of photons detected per keV and bar, for the data acquired at 8.62 bar. The lowest threshold corresponds to 0.117 $\pm$ 0.010 keV, around 4.5 ionization electrons, for an electroluminescence field of 1.52 kV/cm/bar. However, the low background population dominates the threshold down to a field of 1.34 kV/cm/bar. This population becomes subdominant at a field of 1.30 kV/cm/bar, where the detection threshold is estimated to be 0.42 $\pm$ 0.04 keV, which corresponds to approximately 16 ionization electrons. Figure~\ref{fig:fe_spectrum} shows an example of the threshold level in a spectrum. 

\subsection{Energy resolution}
\label{subsec:eres}

 The expected energy resolution of a gaseous detector with electroluminescence amplification was defined in \cite{dosSantos2001_Eres} and fundamentally depends on the intrinsic resolution of the gas (Fano factor) and the fluctuations in the amplification and photon detection process. Mathematically this results in equation \ref{eq:reso}:

\begin{equation}
    R_e = 2.35 \sqrt{\frac{F}{N_{e^-}} + \frac{\sigma_{EL}^2}{N_{e^-}\cdot Y_{EL}^2} + \frac{1 + (\sigma_{q}/q)^2}{N_{det}} + \frac{\sigma_b^2}{N_{det}^2}}
    \label{eq:reso}
\end{equation}

In equation.~\ref{eq:reso} each summand accounts for different fluctuations. Thus, the first describes fluctuations in the number of ionization electrons produced, with $F$ being the Fano factor, 0.23 $\pm$ 0.05 in gaseous Ar \cite{Ar_Fano}. The second term describes fluctuations in the electroluminescence process with $Y_{EL}$ being the absolute yield per electron and $\sigma_{EL}$ its variance. Its contribution is generally much smaller than the Fano factor and can be considered negligible as shown in \cite{OLIVEIRA2011217}. The third element defines fluctuations related to the sensor charge resolution. It is determined by the number of photons detected, $N_{det}$, and the sensor relative variance $(\sigma_{q}/q)^2$ being $q$ the average charge produced by a single photon and $\sigma_{q}$ its standard deviation. The fourth term is introduced to account for variations in the waveform baseline during the integration window, with $\sigma_b^2$ being the standard deviation anticipated based on the variation observed in each event's waveform, computed on a run-by-run basis. 

\begin{figure}[tbp]
\centering
  \hspace{-0.1in}
  \includegraphics{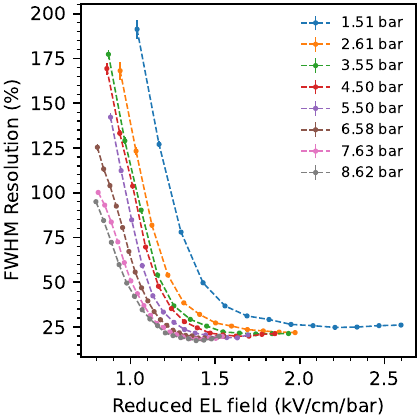}
  \hspace{0.1in}
  \includegraphics{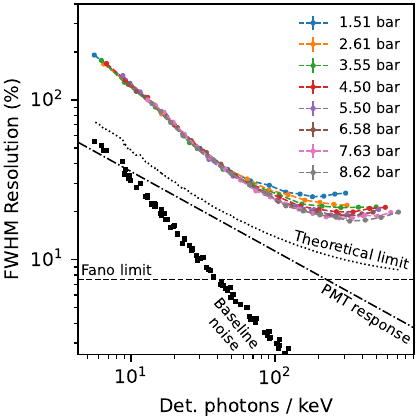}
\caption{Left: Dependence of the energy resolution with the reduced field in the EL region for different pressures. One can see that the resolution saturates for values of the reduced field in the region [1.25, 1.75] kV/cm/bar where the light produced is enough to minimize the statistical fluctuations. Right: Energy resolution as a function of the number of detected photons per keV for different pressures and its comparison with the theoretical limit calculated using \ref{eq:reso}. It can be observed that there is a clear difference between the observed and the expected resolution.}
\label{fig:reso} 
\end{figure}

In figure \ref{fig:reso} we compare the theoretical expected value for the resolution at the \fe\ peak as a function of the photons detected per keV. 
We can observe two effects. First, the resolution at low reduced electric fields evolves with a higher slope than the theoretical value, our main hypothesis for this is the fact that very few photons per electron are produced at these configurations and electrons can travel a few millimeters inside the EL gap before producing any excitation, as it will be observed in figure \ref{fig:z_excitation}. In this situation the position of the few photons produced by the different electrons will be very different and, as the detector plane is close to the EL region, the collection efficiencies for different points of the EL are quite different. The combination of these two factors will account for an extra fluctuation parameter that disappears once the distance in between excitations inside the EL is reduced and therefore the light production is more homogeneous.
Second, at higher reduced fields, we observe a saturation of the energy resolution at a value that is about 80\% worse than the theoretical value.
We associate this difference with the non-linearity of the amplification region at large fields.
Both effects are planned to be investigated in the future with a combination of simulations and detector modifications.

\section{Summary of the results}

The presented results mark the first demonstration of the low detection threshold that can be achieved with the HPNG EL-TPC technique, confirming its enormous promise for \cenuns\ detection. Concretely, a threshold as low as 0.117$\pm$0.010 $\rm{keV}_{\rm{ee}}$, has been estimated when operating at 8.62 bar, the maximum pressure considered. However, instrumental backgrounds, likely related to the the amplification region itself and the high voltage operation, limit such threshold to 0.42$\pm$0.04 $\rm{keV}_{\rm{ee}}$. In fact, this could be the reason why we observe a non-linear amplification in the EL region. It should be noted that a more careful peak selection may filter out the background population and allow to recover the lower detection threshold. In any case, assuming a quenching factor in gaseous Ar similar to values reported for gaseous Xe \cite{renner_qf_xe}, the threshold would correspond to a nuclear recoil of 2.32 $\rm{keV}_{\rm{nr}}$ and, if the instrumental backgrounds can be suppressed, 0.65 $\rm{keV}_{\rm{nr}}$. These estimates will be further evaluated and corroborated in the future by using lower energy sources and measuring the quenching factor of gaseous Ar. Moreover, the studies will soon be extended to higher pressures as well as using Xe and ArXe mixtures. See section~\ref{gap:future} for future modifications and measurements to fully characterize the GaP detector. 

\section{Future plans}
\label{gap:future}
\subsection{Low energy threshold characterization}

\subsubsection{X-ray emission from an \fe\ source with an aluminum target}

Previously introduced in section~\ref{subsec:Fe}, \fe\ source emits characteristic Mn X-rays $K_{\alpha}$ ($\sim$5.9 keV) and $K_{\beta}$ ($\sim$6.5 keV) lines via electron capture. These are considered significantly high energies to characterize a detector that aims to operate in the region of sub to few keV. For this reason, an extension to the \fe\ source X-ray emissions is planned adding a thin (few $\mu$m) aluminum target in front of the source providing X-ray fluorescence (XRF). When the Mn $K_{\alpha}$ and $K_{\beta}$ X-rays from \fe\ interact with the aluminum target, several processes occur.

Attenuation occurs due to scattering and photoelectric absorption. From the X-ray attenuation coefficients for aluminum, the presence of an aluminum foil with such thickness experiences partial absorption, reducing the transmission values of Mn X-rays. Some of the transmitted Mn X-rays can also ionize Al K-shell electrons (binding energy $\sim$1.56 keV), leading to fluorescence emissions: Al $K_{\alpha}$ ($\sim$1.49 keV) and $K_{\beta}$ ($\sim$1.56 keV) lines. $K_{\alpha}$ is the dominant fluorescence line. However, due to aluminum’s low atomic number, Auger electron emission dominates fluorescence. Therefore, it is more probable that, instead of emitting X-rays, the energy from the electron transition is transferred to another electron, which is then ejected. Additional interactions include Compton scattering, which produces a lower-energy X-ray continuum, and Rayleigh scattering, where Mn X-rays elastically scatter without energy loss. 

In this \fe\ and aluminum setup, the detector will primarily observe Mn $K_{\alpha}$ and $K_{\beta}$ lines from \fe, alongside Al $K_{\alpha}$ and $K_{\beta}$ fluorescence induced by Mn X-rays. Additional contributions from interactions and reabsorption processes may modify the overall spectrum. All in all, this setup will provide a detector calibration system in a lower energy regime.

\subsubsection{Determination of the energy threshold from individual free electrons}

Attaching a needle to the cathode inside the TPC, where field emission would be more likely, provides a well-defined point-like source of ionization that can also be used for detector calibration. When a high voltage is applied to the cathode and thus, to the needle, it can induce controlled field emission or corona discharge, generating localized ionization electrons. These electrons drift toward the anode under the influence of the electric field in the GaP detector. This will mimic the signal of particle interactions with the advantage that follows a Poisson distribution, so with these controlled free individual charges, one can characterize the detector's sensibility and electroluminescence gain to observe individual electrons. In the same way, the drift velocity, electric field uniformity, electron lifetime, and gain variations can also be monitored. This is an effective tool that provides a stable and repeatable reference signal without requiring radiation sources.

\subsection{Quenching factor measurements in noble gases}

The use of high pressure gas TPCs has not been considered for \cenuns\ observation before, and as a result, the response in gas to nuclear recoils and its comparison with electromagnetic interactions (quenching factor) is not well understood at such low energies. The response of detector materials to these nuclear recoils is difficult to predict using theoretical models and it varies from target to target. Since the nuclear recoils from \cenuns\ interactions have low energies (few-keV and sub-keV), the experimental characterization of the quenching factor (QF) is complicated. As such, it is also a motivation for the low energy threshold characterization methods described previously. 

This is a fundamental parameter for a proper interpretation of the data, so the determination of the QF is another the main goal of the GaP prototype. It is essential for accurate energy reconstruction, background discrimination, and sensitivity optimization. Reliable QF data enhances cross-section measurements and searches for new physics.

Photo-nuclear radioactive sources have been suggested as an effective technique for generating monochromatic neutrons, which can induce a well-defined spectrum of low-energy nuclear recoils \cite{2013Collar}. The successful application of $^{88}$Y/Be and $^{124}$Sb/Be photo-nuclear sources has been demonstrated \cite{PhysRevD.94.122003,PhysRevD.94.082007,collar2021germanium}, with the most recent use during the calibration of the XENON-nT detector \cite{aprile2024low} before its first observation of solar neutrinos via \cenuns\ \cite{aprile2024first}.

A challenge in using these sources is their extremely high gamma-neutron ratio (on the order of $10^5$). This requires the use of substantial gamma shielding, which can slightly affect the otherwise highly predictable neutron component. Moreover, it leads to the need to conduct differential measurements: subtracting the only-gamma response to neutron-gamma response \cite{2013Collar}.

A recent proposal \cite{biekert2023portable} suggests combining photo-nuclear $^{124}$Sb/Be with an iron filter \cite{barbeau2007design} to create a compact ``neutron Howitzer" capable of generating a high-purity 24 keV neutron beam. This beam would be ideal for inducing sub-keV nuclear recoils in future \cenuns\ detector materials within the NuESS team (section~\ref{sec:NuESS}) while overcoming some of the challenges associated with current methods. The only other known technique for producing nuclear recoils in this energy range involves specialized beams at nuclear reactors \cite{soum2023study}. 

When properly designed, this Howitzer offers a convenient, table-top setup with no radioprotection concerns. Inexpensive $^{124}$Sb sources can be made by irradiating high-purity Sb samples at experimental reactors. Combined with suitable ``backing detectors", such as those based on $^6$Li \cite{barbeau2007design,biekert2022backing} to tag scattered neutrons in coincidence with recoils in the detector, this "facility" can provide valuable calibration data for future \cenuns\ targets (Xe, Ar, Ge, CsI) in the challenging sub-keV recoil energy range without bringing the detector to a thermalized neutron beam at a nuclear reactor. 

In the near future, a ``neutron Howitzer" beam is planned to be built at DIPC in order to perform accurate quenching factor measurements in sub-keV recoil energy range. Namely, the GaP detector will be used to characterize the noble gas target materials.

\subsection{Technological modifications}

Recalling the main objective of the GanESS project, to fully exploit the \cenuns\ process at the European Spallation Source, the goal of the GaP detector is to develop the most suitable high pressure gaseous TPC for this experiment and characterize its technology. For this reason, GaP is planned to undergo several modifications. 

\subsubsection{Silicon Photomultipliers (SiPMs)}

The initial design described in chapter~\ref{sec:gap} is based on the current technology used in the NEXT experiment. Therefore, among other similarities, the energy plane is composed of TPB coated PMTs. One of the changes already under study is the replacement of these Photomultiplier Tubes with Silicon Photomultipliers (SiPMs). 

There are several reasons why the performance of the detector would benefit from using SiPMs as photosensors. While PMTs are glass enclosures that may be prone to damage under high pressure making them very fragile, SiPMs are compact solid-state devices more resistant to mechanical stress. They can operate at high pressures without any protection window to isolate them in a low pressure region. Another key advantage is the finer spatial resolution that can be obtained by placing these sensors in an array with minimal dead space in the design. Especially at shorter wavelengths, such as the vacuum ultraviolet emission crucial in a gaseous TPC, SiPMs are more sensitive and thus, the absence of wavelength shifters can be evaluated. The SiPMs that will be used are the 2x2 quad modules VUV4 MMPCs from Hamamatsu (model S13371-6050CQ-02) \cite{VUV4MPPC}. These SiPM modules have sensitivity for 178 nm, referring to LXe VUV emission. Even if the company reports several improvements compared to the 3$^{\text{rd}}$ generation VUV-MPPC, sensitivity goes down to 120 nm with $\sim10 \%$ photon detection efficiency. Therefore, independent studies will be performed in the house to determine if wavelength shifters are needed, especially since VUV emission in Ar peaks at 128 nm.

The main drawback however is the temperature sensitivity. SiPMs are highly sensitive to temperature fluctuations, and their performance can degrade if the temperature is not carefully controlled. 

SiPMs are more susceptible to dark counts (random signals caused by thermal noise) than PMTs. This can become a significant issue, particularly in high-pressure TPCs where low-light signals are often present, as dark counts can introduce noise and reduce signal-to-noise ratios. The current design of both GaP and GanESS intends to operate at room temperature so the possibility to locally cool down the SiPMs will need to be evaluated. Dedicated studies are currently undergoing to evaluate the SiPMs' gain and determine the optimal operating temperature, considering instrumental limitations. Preliminary results indicate that cooling down the SiPMs to -$10^\circ$C reduces the dark count rate sufficiently for an adequate signal-to-noise ratio, expecting a detection threshold between 1$e^{-}$ and 100 eV. Assuming a group of 4 near SiPM modules (each one with 4 SiPM channels) separated by 5 mm, 6 photoelectrons are expected per module within 10 $\mu$s for the desired 0.1 $\rm{keV}_{\rm{ee}}$ energy threshold in our \cenuns\ detectors. The intrinsic dark current rate (DCR) of each SiPMs module is 4 Mcps. Additionally, the DCR of the module has a dependence on temperature and so, the noise of a SiPM module will be derived from $DCR (T) \cdot \Delta t$ for a given time window. Then, the corresponding DCR per SiPM module at -10$^{\circ}$C is $\sim$0.165 Mcps. So for 10 $\mu$s, the signal-to-noise ratio for 4 near SiPM modules is the following:

\begin{equation}
    \frac{\text{signal}}{\sqrt{\text{noise}}} = \frac{4\cdot 6 \text{PE}}{\sqrt{4 \cdot 1.65 \text{DC}}} = 9.34.
\end{equation}

The analogous study to be sensitive of a 1$e^-$ threshold indicates a signal-to-noise ratio of 2.57. So while a detection threshold of 1$e^-$ may be ambitious at these operating characteristics, sensitivity to 0.1 $\rm{keV}_{\rm{ee}}$ energy threshold looks promising from these preliminary studies.

Overall, SiPMs have excellent single-photon sensitivity, a well-defined gain, and benefits from their photon detection efficiency at VUV emission. Once the SiPMs' temperature is determined, the cooling system will be assessed. Two options are currently being considered. The first one is a cold finger, which allows to cool down the SiPMs in an isolated and controlled way. The second is to cool the gas directly. The latter would involve minimal mechanical changes but is only feasible if the temperatures required to reduce the dark current are not too extreme.

\subsubsection{GALA: Large amplification with Gas Electron Multipliers (GEMs)}

It is important to ensure the highest possible signal gain and therefore the signal is amplified via electroluminescence in GanESS and GaP, which take as reference the NEXT experiment. Given the low rate and/or high background nature of this experiment, the best performance requires the design of a larger-volume TPC that will introduce new challenges. 

The conventional design would consist of at least a few-m$^2$ meshes or wires in the EL region in GanESS and the ton-scale NEXT detector would require even larger mesh surfaces. Apart from increasing the chance of encountering practical problems with the mesh handling, alignment and stretching, the number of PMTs for VUV scintillation readout needed increases, leading to higher background levels and gain loss. One of the modifications currently ongoing is the replacement of PMTs with SiPMs as previously mentioned. In parallel, seeking detector stability and improved energy resolution, an alternative amplification method is also being evaluated. Even if this would require a major modification in the structure, it is a high optical gain technique worth exploring in the GaP detector to evaluate its performance and scalability to larger gaseous TPCs. 

\begin{figure}[tbp]
\centering
  \includegraphics[height=6.5cm]{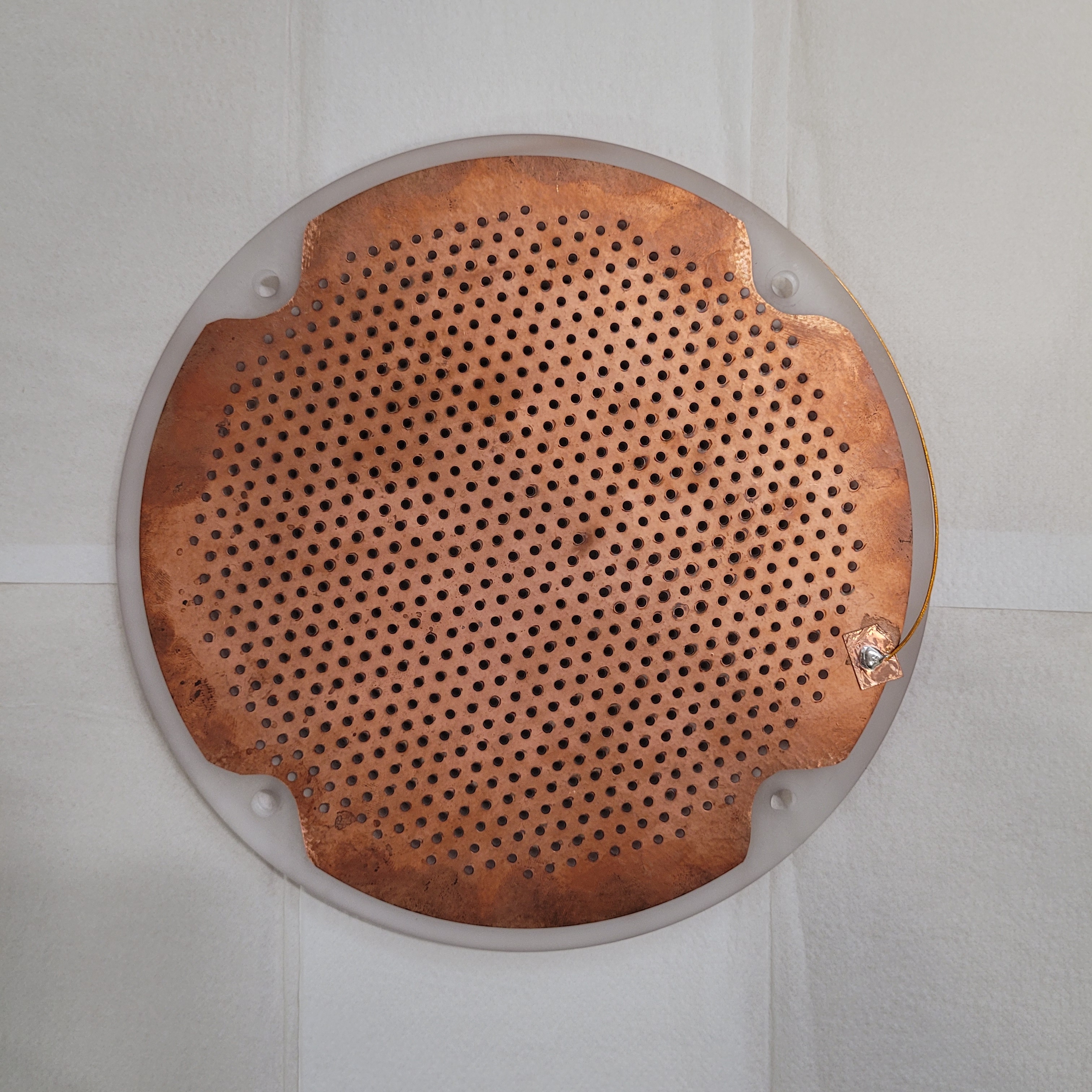}
  \hspace{0.7cm}
  \includegraphics[height=6.5cm]{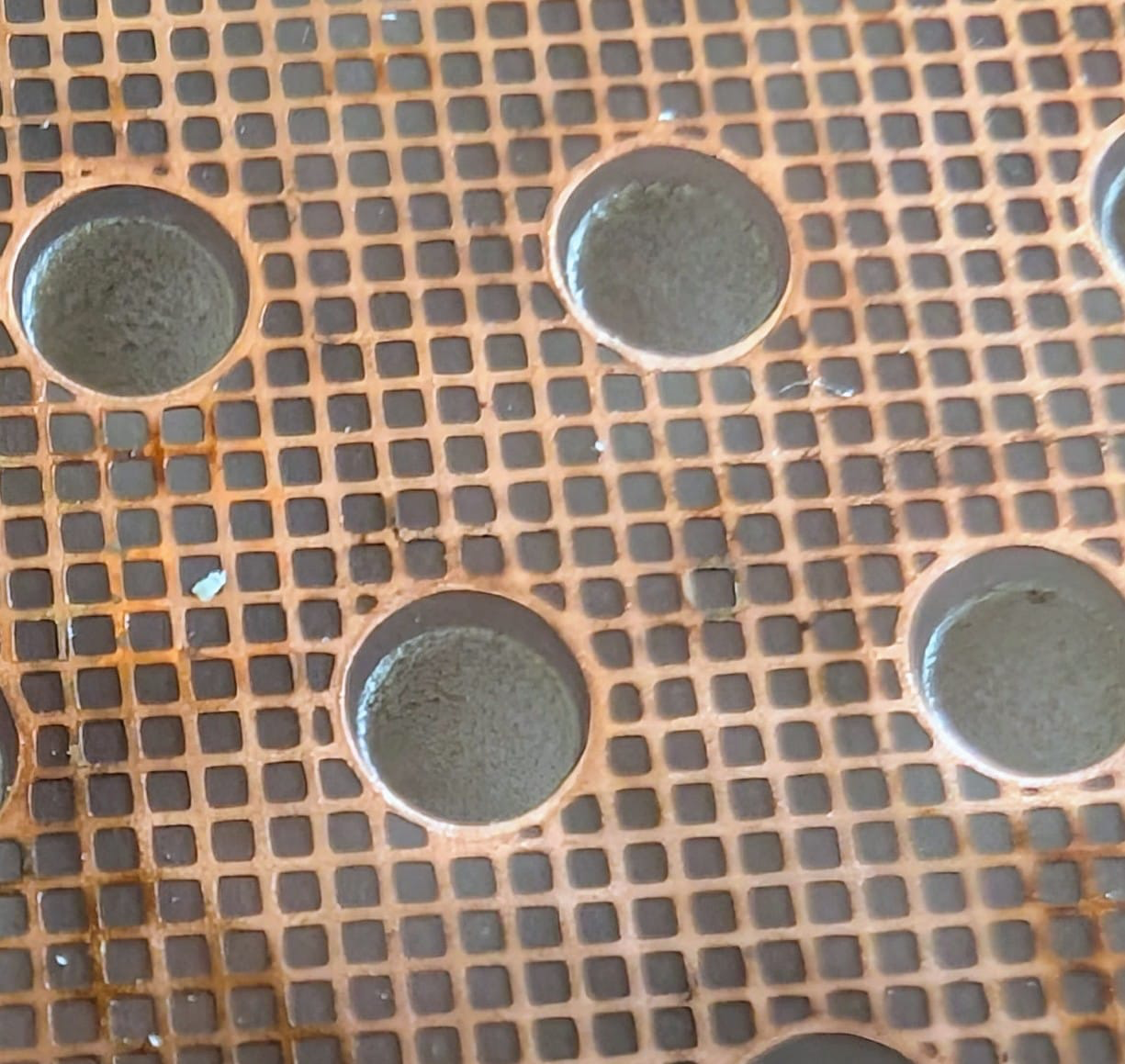}
\caption{Left: Image of the 2 mm-hole/5 mm-pitch FAT-GEM that will be installed in the GaP detector. The base is an acrylic (Poly(methyl methacrylate), PMMA) plate with two thermally bonded copper electrodes on the sides. On one of the sides  a hatched pattern is made by photo-lithography, to ensure a degree of translucence. Right: Closer view to the hatched pattern and the FAT-GEM holes.}
\label{fig:gala} 
\end{figure}

Introduced in \cite{gonzalez2020new} and its operation demonstrated in pure xenon for several pressures, the FAT-GEM (Field-Assisted Transparent Gaseous Electroluminescence Multiplier) is an alternative technique to the meshes used at present for scintillation amplification, see photograph in figure~\ref{fig:gala}. As its name indicates, it is a modified version of the original GEM device introduced in 1996~\cite{sauli1997gem}. Since GEMs can operate at relatively low voltages, incorporating them to the TPC system would allow us to reach stably higher amplification regions. This potential integration to the GaP detector for signal amplification has adopted the name GALA (GAseous Large Amplification), and for its photon readout system SiPMs will be implemented. 

In the same way as the EL region in the current GaP structure, GEMs exhibit uniform amplification characteristics across their surface, which is essential for maintaining consistent a performance of the detector. To characterize the light yield and compare it to the present performance of GaP and thus, study the feasibility of implementing this new amplification technique, Garfield++ simulations are being carried out. This toolkit will evaluate different alignments of GEMs and quantify the light yield produced in the amplification process together with the light collection under different detector geometries.

\renewcommand{\arraystretch}{1.5}

\begin{savequote}[65mm]
``Zenbat buru, hainbat aburu" 
\qauthor{-- Old proverb in Basque language.}
\end{savequote}

\chapter{Electroluminescence simulations}
\label{sec:garfield}

As indicated in the previous chapter, the number of photons produced per primary electron (electroluminescence yield) is described as approximately linear to the electric field applied in the amplification region as long as the ionization probability in the EL region is low. Nevertheless, results from GaP operated with argon at several pressures follow a non-linear increase in light yield, shown in figure~\ref{fig:charge_yield} (section~\ref{sec:results} from the previous chapter). 

Microscopic simulations of electron interactions with the gas were conducted to assess the possibility of charge amplification in the EL region near the wires caused by a rapid increase in the electric field due to its geometrical shape. To do so, a combination of Garfield++ libraries~\cite{GarfieldPP} and the electric field in the EL region was used. The latter was calculated using the finite element software COMSOL Multiphysics~\cite{comsol2024}.

We built a versatile simulation that provides a parameterization of the electroluminescence and ionization processes involved in gaseous detectors.

\section{COMSOL simulation}

\begin{figure}[tbp]
\centering
  \hspace{-0.1in}
\includegraphics[width=.45\textwidth]{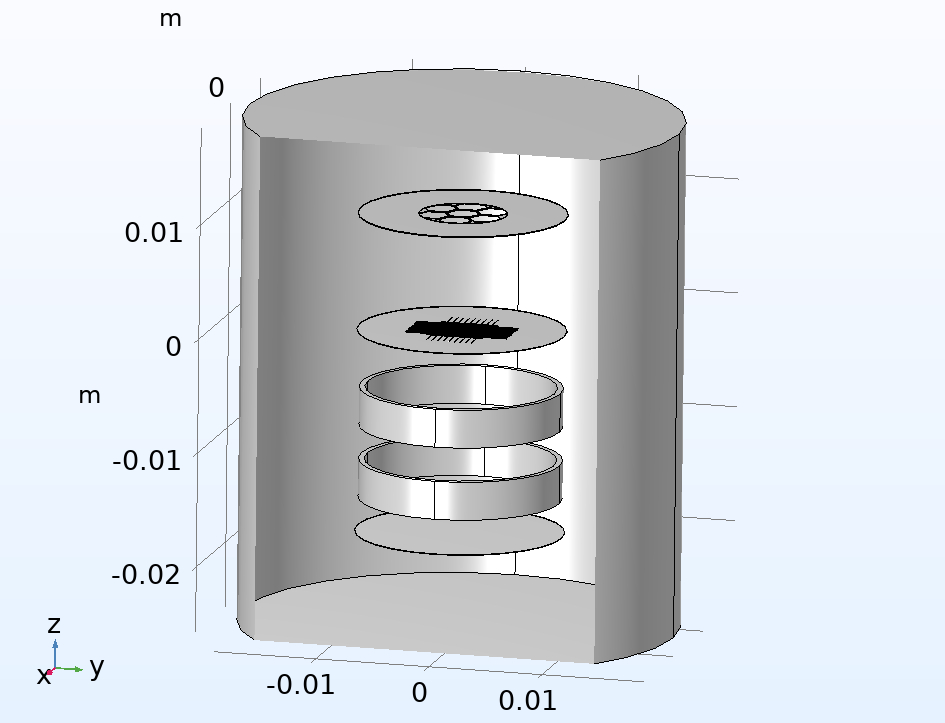}
  \hspace{0.1in}
\includegraphics[width=.45\textwidth]{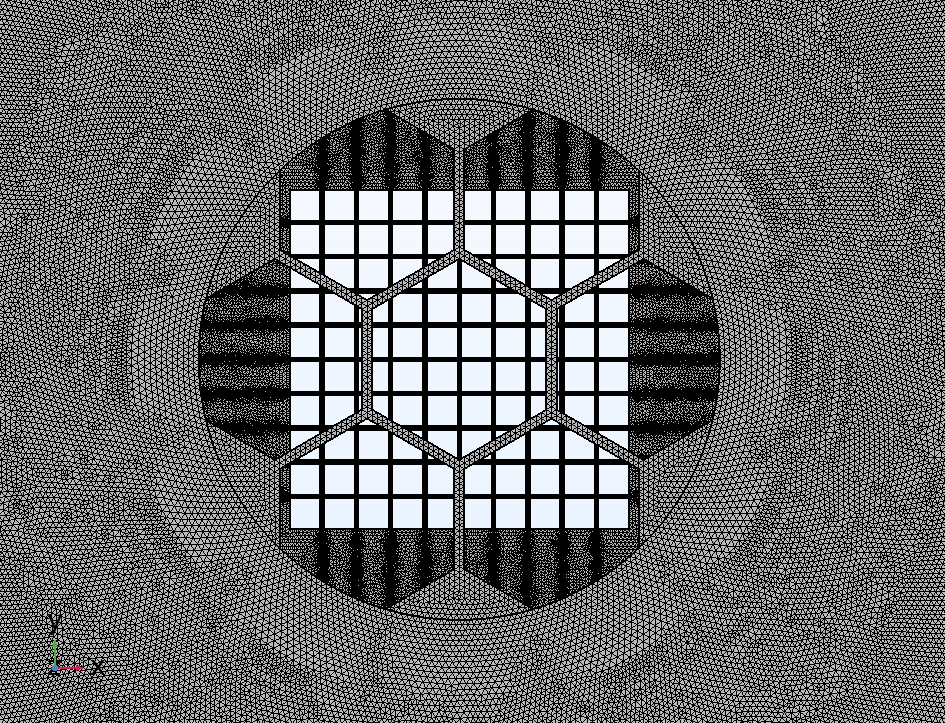}
\caption{Left: Implemented geometry in COMSOL of a small TPC to simulate a simplified version of the GaP detector. Right: Top perspective of overleap of the two reduced meshes used to simulate the EL region.}
\label{fig:comsol} 
\end{figure}

To balance the geometrical accuracy with the GaP detector and a computationally feasible COMSOL simulation, since interest is focused on the EL region, a reduced version of the detector was built. As the goal was to evaluate the performance of the high-voltage mesh-electrodes inducing a controlled electric field that would set off an amplification signal, it is reasonable to reduce the diameter of the detector system and the drift length too, as long as the EL distance is maintained and the voltages for the three simulated stainless steel electrodes are accordingly applied to reproduce the electric fields generated in the GaP detector. Therefore, the characteristics of the meshes used in the data taking have been used, a hexagon mesh in the anode and a woven mesh in the gate separated by the corresponding EL gap distance as described in section \ref{gap:tpc}. Figure~\ref{fig:comsol} left represents the geometry simulated in COMSOL. The right panel shows the top perspective of the two reduced EL region meshes overlapped. The Garfield++ simulation is localized in a central circle of 1.8 mm radius.

\begin{figure}[tbp]
\centering
  \hspace{-0.1in}
\includegraphics[width=.45\textwidth]{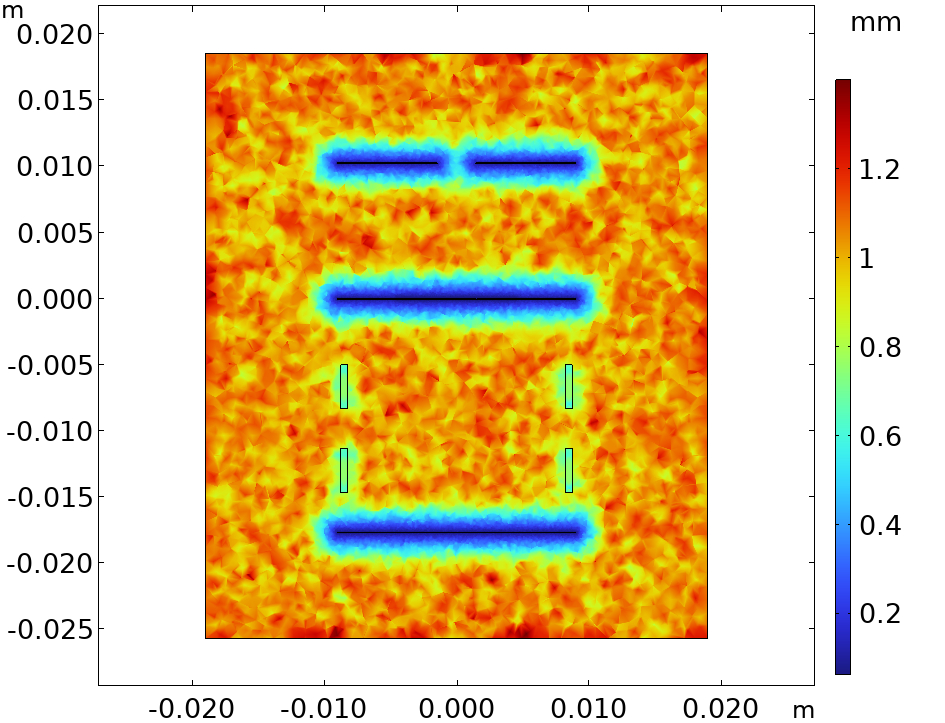} 
  \hspace{0.1in}
\includegraphics[width=.45\textwidth]{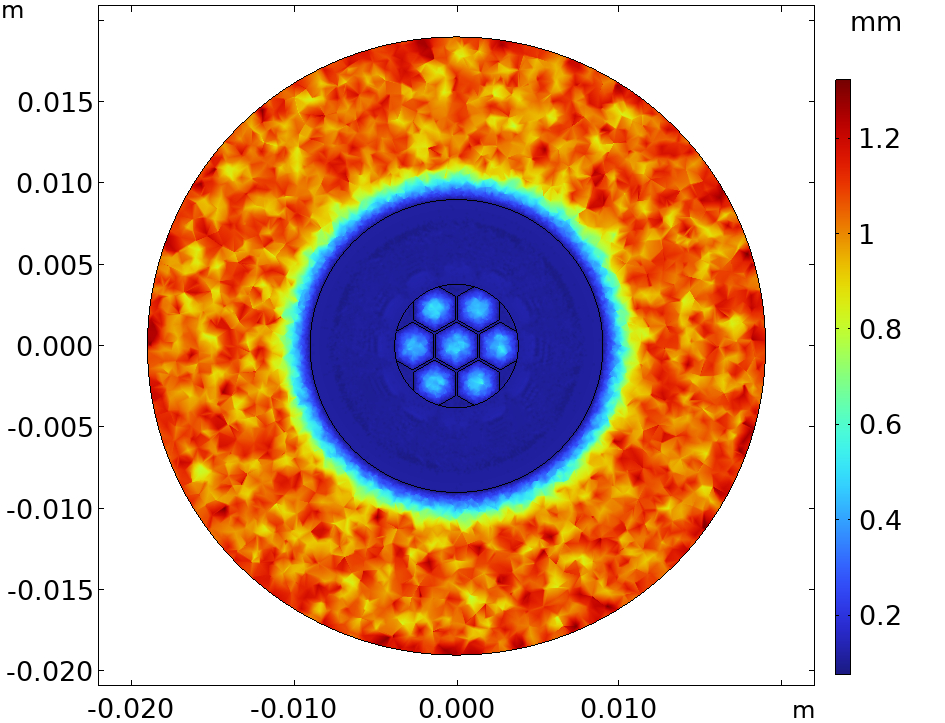}
\caption{COMSOL mesh distribution along the geometry defined in figure~\ref{fig:comsol}. Left: Frontal perspective of a plane cut at the center of the system. Right: Top view of the element size distribution at a cut plane in the anode's position.}
\label{fig:comsolmesh} 
\end{figure}

COMSOL generates a virtual mesh to evaluate the electrostatics along the 3D volume. This mesh had as the smallest size a fraction of the thickness of the hexagons to guarantee proper electric field calculation in these areas. Figure~\ref{fig:comsolmesh} shows the mesh element size distribution used for the corresponding figure~\ref{fig:comsol} simulation.

\begin{figure}[tbp]
\centering
  \hspace{-0.1in}
 \includegraphics[width=0.5\linewidth]{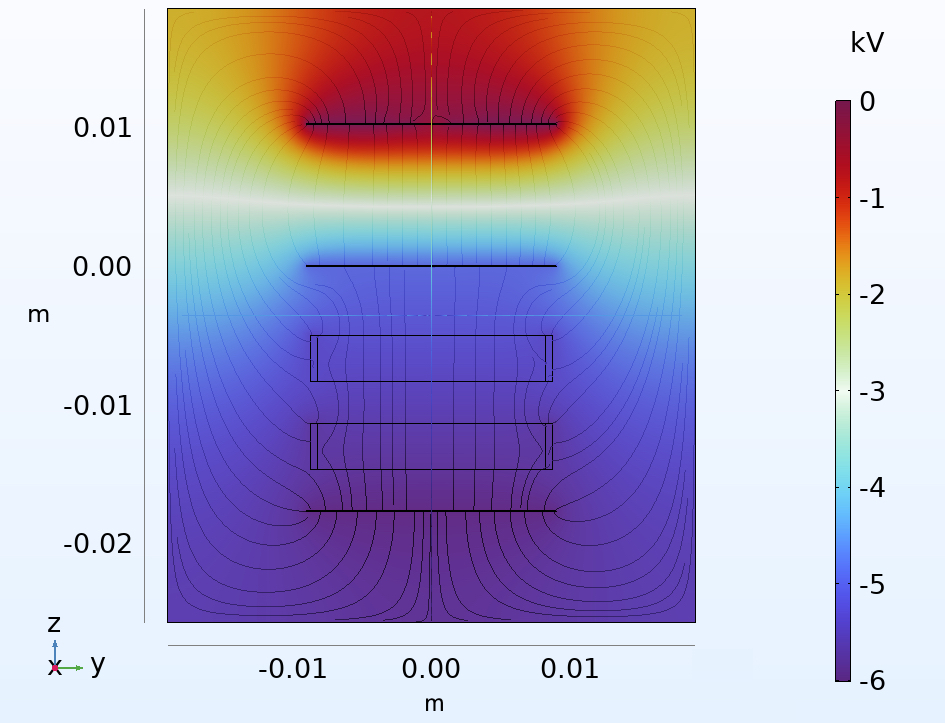} 
  \hspace{0.1in}
 \includegraphics[width=0.4\linewidth]{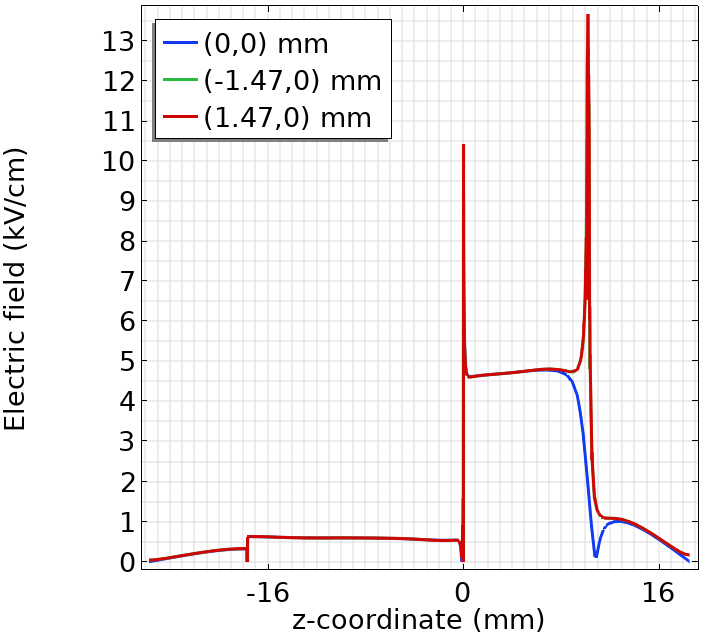}
\caption{COMSOL simulation output with the gate electrode at -5 kV. Left: Frontal cut at the center of the system picturing the electric potential and the corresponding electric field lines. Right: The electric field along the vertical axis for three lines radially separated, two across the anode wires and the other along the central hexagonal gap.}
\label{fig:comsolVE} 
\end{figure}

The COMSOL simulation has been run for different voltage supplies to mimic the reduced electric field in the EL from data taken for different pressures. Figure~\ref{fig:comsolVE} illustrates the scenario of the simplified GaP TPC supplied with -5 kV at the entrance of the EL region (gate). On the left, a frontal cut at the center of the system is applied picturing the electric potential and the corresponding electric field lines. To check the performance of the calculation, the right panel shows the electric field along the vertical axis for three lines radially slightly separated. They were chosen so that one could see the possible ambiguous effects in electric field lines across the wires in the anode mesh. Therefore, two of them are displayed crossing the wires of the mesh and the other one goes through the central hexagonal gap. 

As expected from the operation of a TPC, the electric field is mostly constant along both the drift and amplification region, the latter at an increased electric field. The deviation of the electric field in the ramping-up region before the EL and in the close neighborhood of the anode wires, may produce some fluctuations in electroluminescence gain within a small region. However, as it is shown later, these fluctuations in the yield are negligible and the light yield obtained from such simulations behaves mostly linearly as anticipated. Moreover, one should note that at the edges of the TPC electric field lines curve because of border effects, so as mentioned before, the Garfield++ simulations were performed localized in a reduced central region to ensure parallel electric field lines. These lines were then exported in a Garfield++ suitable format for several configurations of voltages to compare with the data at similar operation conditions in our detector. 

\section{Garfield++ simulation}

The simulation of the detailed physics processes in the gaseous detector was developed using Monte Carlo Garfield++ toolkit. The microscopic transport properties of electrons in a gas medium are calculated using an interface to the Magboltz program~\cite{BiagiWEB}. Based on the gas properties from Magboltz, Garfield simulates each collision as the electrons are transported along a virtual electric field and magnetic field. In our case, only the electric field is considered and it is imported from the computed field maps using COMSOL (see previous section). The present simulation aims to determine the electroluminescence signal properties in this gaseous TPC to compare it with the results in section~\ref{sec:results} from the previous chapter. Therefore, each collision with the gas atoms is classified as elastic, excitation or ionization. The available information on the excitation collisions (location, time and excitation level) enables the electron tracking and the characterization of electroluminescence.

\subsection{Magboltz}

Magboltz calculates the electron transport parameters in gas mixtures contained in electric or magnetic fields~\cite{biagi1999monte}. It includes an extensive database of electron interaction cross sections with a large number of detection gases. The electron drifting parameters are stored in a \textit{gas table} that can later be imported into Garfield++. This feature makes the Garfield++ simulation more efficient. In such a manner, Magboltz was run to generate the argon noble gas table for every pressure of the data taking.

\subsection{Model description}

With an initial electron energy according to the calculated Magboltz energy distribution, the microscopic electron transport simulation was generated by sampling primary electrons 2 mm before entering the EL region randomly distributed within a 1.8 mm radius. The influence of the electric field imported from COMSOL drifts the electrons to the gate mesh, where the EL region starts. 

To evaluate the process, the emission of one VUV photon from a second continuum excimer decay is considered per excited atom. Additionally, if charge multiplication occurs, it is also assumed that the ionization process emits one VUV photon~\cite{saito2002absolute}.

\section{Results and discussion}

The present section aims to give an overview of Garfield++ simulations performed that lead to an exhaustive description of the EL process in GaP evaluating all collisions between electrons and argon gas atoms that produced light, together with a discussion on the awaited functioning of a noble gas TPC. 

The simulations were run for different pressures and reduced electric fields to compare with the data at similar operation conditions to those in the detector. The number of excitations produced per primary electron (corresponding to the light yield) and the number of ionizations have been checked to explore possible gain effects.

\begin{figure}[tbp]
\centering
  \hspace{-0.1in}
\includegraphics{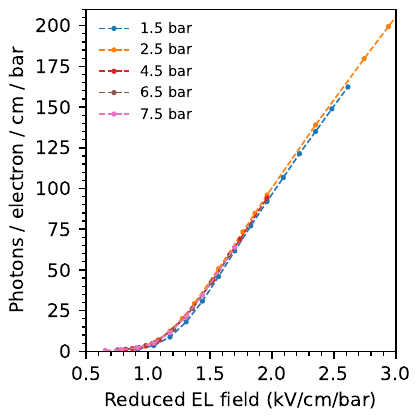}
  \hspace{0.1in}
\includegraphics{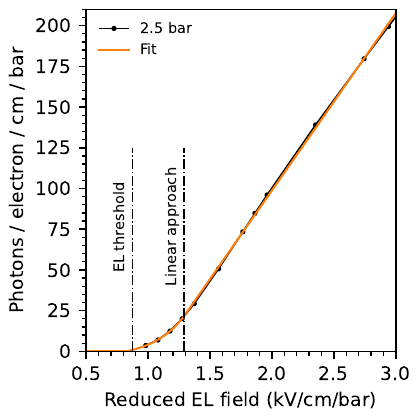}
\caption{Left: Light yield in the EL as a function of the reduced electric field for different pressures according to Garfield++ simulation. Right: The determined EL threshold is 0.87 $\pm$ 0.29 kV/cm/bar showing a fast rise region up to  $\sim$1.29 kV/cm/bar, where we can observe the expected linear behavior at least until 3 kV/cm/bar - theoretical charge multiplication threshold.}
\label{fig:garfield_scanEL} 
\end{figure}

The reduced electroluminescence yield (Y/p) is defined as the number of photons emitted per primary electron and unit of EL drift path divided by the density of the gas in pressure units. In this way, the reduced EL electric field (E/p) is defined in units of electric field E divided by operating pressure (kV/cm/bar). Electroluminescence occurs when the electric field provides sufficient energy to electrons for excitation, causing a sudden rise in the cross section and leading to a rapid increase in light yield. See section~\ref{sec:el} in chapter~\ref{sec:ganess} for details on the process. From literature, the behavior of Y/p with E/p is expected to be linear at a low electric field, above the minimum energy required to produce one excitation but before the contribution of other processes such as ionization is significant. This agrees with the results obtained from our simulation shown in figure~\ref{fig:garfield_scanEL}. Changes in the linear behavior are visible at high values of the electric field when fluctuations in secondary charge particles start to dominate over the electroluminescence process. Therefore, even at the characteristic charge multiplication threshold of the gas, $\sim$3 kV/cm/bar for argon \cite{monteiro2008secondary,OLIVEIRA2011217}, the EL yield preserves the linear behavior, as the ionization probability is low.

From these results in figure~\ref{fig:garfield_scanEL}, one can observe that the predicted number of photons produced per primary electron is linear even at rather large reduced electric fields. This consolidates the overall agreement with \cite{OLIVEIRA2011217} that the main channel of light production at moderately reduced electric fields is through excimer decay, refusing the hypothesis of extra ionizations contributing to the total light in the regions near the wires. 

\begin{figure}
\centering
  \hspace{-0.3in}
  \includegraphics{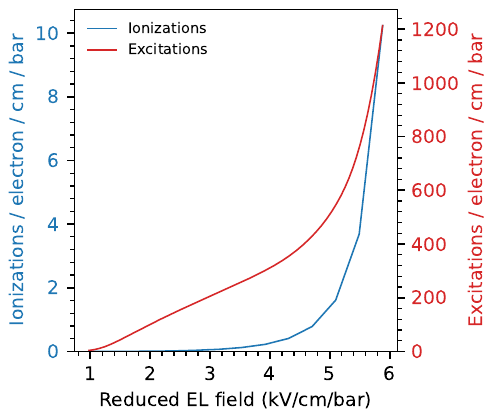}
\includegraphics{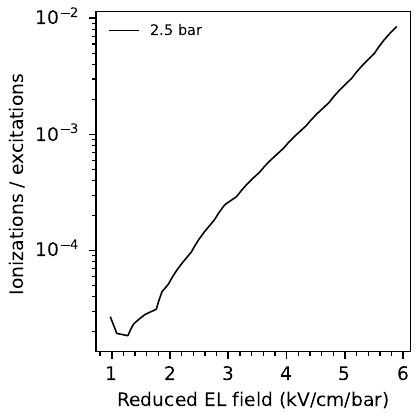}
\caption{Left: Number of predicted ionizations (blue) and excitations (orange) as a function of the reduced electric field. Up to the charge multiplication threshold ($\sim$3 kV/cm/bar) only excitations occur. When ionizations are produced the reduced light yield increases exponentially as it is given in terms of primary electrons. Right: Fraction of the number of predicted ionizations over the number of excitations for different values of the reduced electric field. We can observe that this fraction, even while it increases rapidly with the field remains negligible (less than 10$^{-2}$ until very high fields, near 5 kV/cm/bar. Visible fluctuations in the low EL field range were to be expected as it is below the charge multiplication threshold. 
}
\label{fig:ioni_exc_fraction} 
\end{figure}

Figure \ref{fig:ioni_exc_fraction} narrates the evolution of the photon production as a function of the reduced electric field, showing the fraction of ionizations potentially contributing to light production. Indeed, figure \ref{fig:ioni_exc_fraction} left shows the contribution to the light yield from excitations and ionizations separately, assuming that each ionization produces one VUV photon in addition to successor excitation processes that contribute to the signal multiplication. One can observe that ionizations are relevant at very high reduced electric field and accordingly excitations increase exponentially as expected. On the right, the fraction of ionizations over excitations is explicitly shown. This remains very small for moderate electric fields, even at higher EL fields than the charge multiplication threshold ($\sim$3 kV/cm/bar). These fractions emphasize again that the predicted ionization contribution to the total light production is negligible in our detector electric field operation range. 

Figure~\ref{fig:garfield_scanEL} right carefully describes the simulated behavior of the secondary scintillation light for argon at low E/p: a fast variation on light emission which approximates to a linear increase with E/p. We have determined that the electroluminescence threshold is 0.87 $\pm$ 0.29 kV/cm/bar for argon by extrapolating the electroluminescence yield from fast rise region shown. This is in good agreement with \cite{monteiro2008secondary,feio1982thresholds,dias1986unidimensional}. Furthermore, as described in~\cite{feio1982thresholds} the extrapolation to no yield from the linear part of the curve sets another threshold often used too. In such case, the instrumentation threshold obtained corresponds to 1.04 $\pm$ 0.03 kV/cm/bar which also agrees with the Garfield study derived in~\cite{OLIVEIRA2011217}.

\subsection{Electron topology}

\begin{figure}[tbp]
\centering
  \hspace{-0.1in}
\includegraphics{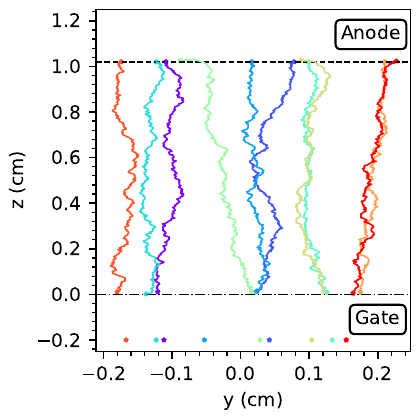}
  \hspace{0.1in}
\includegraphics{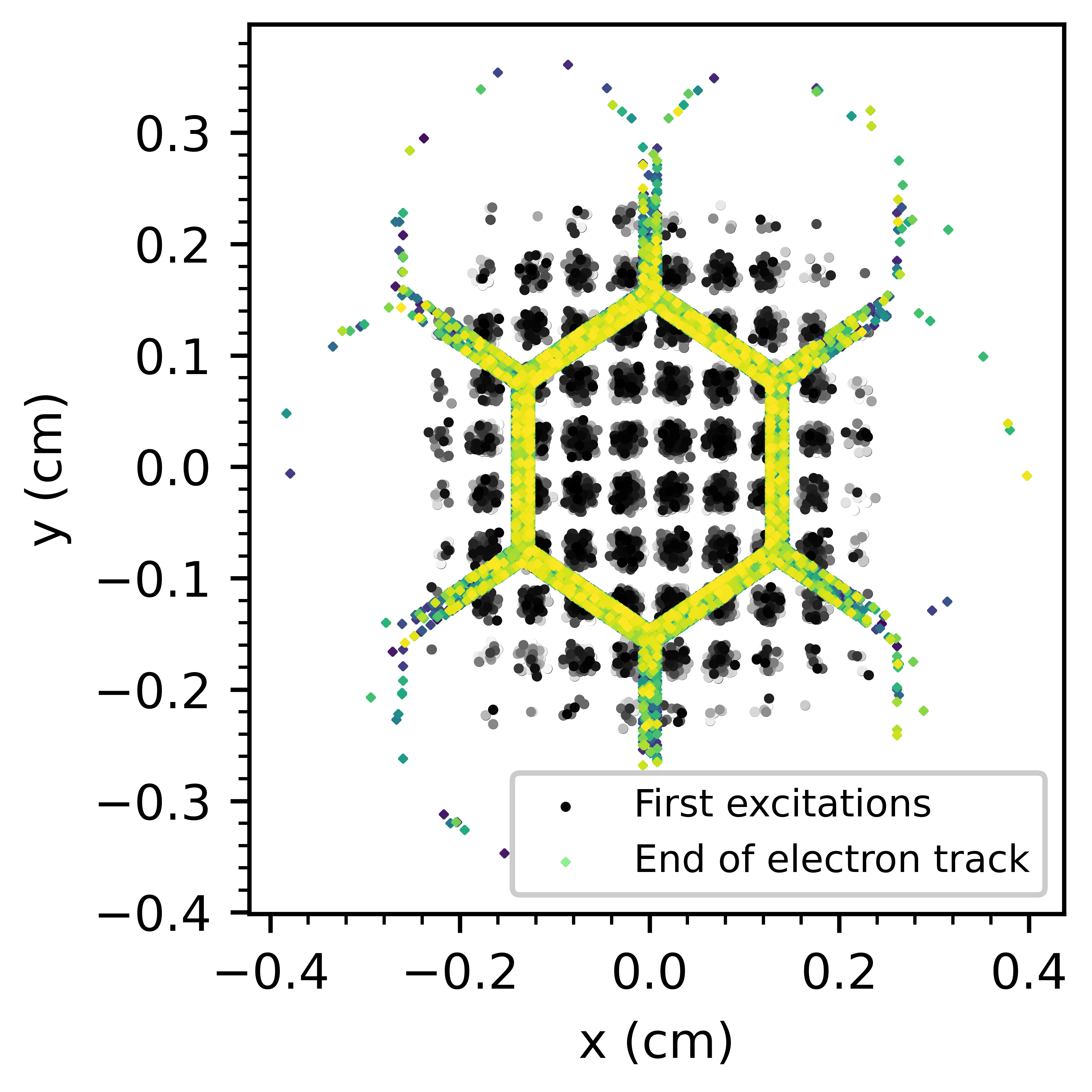}
\caption{Left: Illustration of the electron trajectories through the EL region simulated by Garfield++ for different starting points of the electrons. The Gate and Anode are represented by the bottom and top discontinuous lines, respectively. The plot represents every interaction of the electron in the gas where it produces an excitation. The simulation is run for 1.7 kV/cm/bar at an absolute pressure of 7.5 bar.  
Right: Top view of the radial location of the electrons when the first excitation occurred (gray scale) and its final position implying the end of its transport (green-yellow scale). The latter suggests the shape of the hexagonal photoetched mesh in the anode such that the electrons are attracted to the grounded electrode as expected.}
\label{fig:garfieldtracks} 
\end{figure}

An example of different electrons propagating across the EL is shown in figure~\ref{fig:garfieldtracks} left, together with a scatter plot on the right indicating the radial location of the electrons when the first excitation occurred (gray scale) and its final position implying the end of its transport (green-yellow scale). As expected, for electric fields above the EL threshold, the electrons initiate excitation processes as they go through the high voltage gate plate (start of the EL region) and continue exciting the gas atoms until the anode - connected to ground - is reached. Shown in figure~\ref{fig:comsol} and commented in the chapter~\ref{sec:gap}, a woven mesh is used in the gate and a hexagonal photoetched mesh in the anode. It is visible in figure~\ref{fig:garfieldtracks} right that the electrons pierce the wire mesh holes and excite a gas atom. In the same way, it can be observed that the electron transport ends at the hexagonal anode frame (green-yellow scale). 

\begin{figure}[tbp]
\centering
  \hspace{-0.1in}
\includegraphics{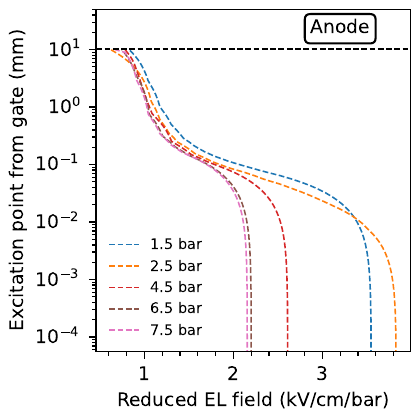}
  \hspace{0.1in}
\includegraphics{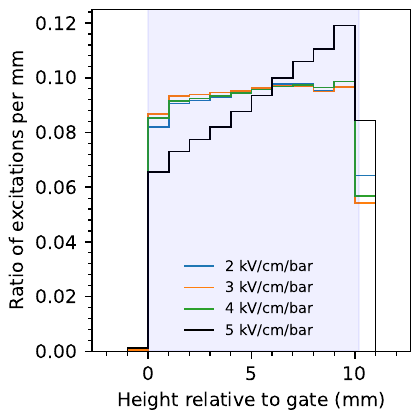}
\caption{Left: Mean value of distance from the gate for interactions producing an excitation. The image shows that above $\sim$ 1 kV/cm/bar the first excitation occurs in the first millimeter of the EL. Above this value, the simulation predicts interactions every fraction of a millimeter. 
Right: Fraction of excitations produced per millimeter for different reduced electric fields. The amount of excitations is almost constant until very high fields, where ionization effects appear and the amount of light produced is larger near the anode.}
\label{fig:z_excitation} 
\end{figure}

It should be mentioned that figure~\ref{fig:garfieldtracks} shows the scenario of an electroluminescence dominant region (1.7 kV/cm/bar) for 7.5 bar argon. In contrast, for low reduced EL fields (e.g. 0.9 kV/cm/bar) the first excitations occur anywhere in the EL region but with a preference in the first half. Additionally, even if most of the electrons reach the anode, some are stopped earlier. In this direction, we evaluated the vertical position of the first excitation produced to check for possible optical differences in the photon detection. Figure \ref{fig:z_excitation} shows that the mean point of the first excitation is similar for all configurations and the number of excitations produced is also very similar along the EL region (figure \ref{fig:z_excitation} right), discarding also any possible optical effect at larger electric fields.

\section{Outlook of the electroluminescence yield studies}

\begin{figure}
\centering
\includegraphics{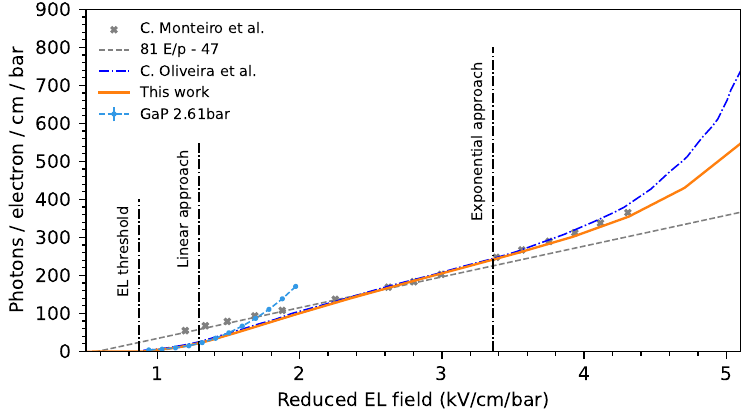}
\caption{Number of photons per primary electron per EL length per pressure unit as a function of the reduced electric field. This work's simulation is shown in orange, the analogous analysis in \cite{OLIVEIRA2011217} in dark blue, together with the measurement and the derived light yield function in~\cite{monteiro2008secondary} (gray). The measurements in our GaP detector are also shown in light blue.}
\label{fig:garfield_coimbra} 
\end{figure}

As mentioned before, even when some ionization is produced, $\sim$3 kV/cm/bar for argon, the linear behavior of Y/p remains. However, according to our simulations, above an E/p value of 3.36 kV/cm/bar, the extra electrons generated by ionization lead to a steeper increase in yield, and therefore, the linear behavior becomes exponential. Y/p is calculated per primary electron so when the additional electrons (avalanche multiplication) involved are relevant, the light tends to an exponential behavior. As illustrated in figure \ref{fig:garfield_coimbra}, there is an apparent discrepancy in the exponential approach between this work and the EL simulations in \cite{OLIVEIRA2011217} and thus, the E/p value reported for the curve variation is a bit higher than in this work. Nevertheless, the discussion on the measurement from \cite{monteiro2008secondary,dias1986unidimensional} agrees with the present study on the E/p value where additional scintillation from ionizations is exhibited. 

Once ionization is present, the number of electrons will increase with the EL drift distance. Therefore, the reduced light output also depends on this distance. Such influence could explain the visible differences between the two simulations along the charge multiplication region. In other words, detector geometry differences may be inducing small fluctuations near the wires, where the electric field is very large, and consequently inducing such variations in the light yield. But overall, at reduced electric fields where our detector will operate, before the charge multiplication threshold, both EL Garfield++ studies have a strong pairing. 

To conclude the overview of the electroluminescence yield studies up to date, the broadly used curve from~\cite{monteiro2008secondary} to describe Y/p as a function of E/p is also shown. This work emphasizes the discrepancies seen between that measurement for argon and the results obtained from the microscopic electron transport studies. Even if the linear tendency followed by an exponential increase is visible, below $\sim$2.25 kV/cm/bar the prediction underestimates the measurement. The effect of the operating temperature was discarded and possible impurities in the active target or incorrect calibrations brought up. To date, no further explanations have been reported. 

In this way, figure \ref{fig:garfield_coimbra} also includes the measurements obtained from the GaP detector (light blue). See the section \ref{sec:results} from the previous chapter for more details. As already mentioned in chapter~\ref{sec:gapresultschapter} and the main reason to develop the EL simulation toolkit detailed in this chapter, it shows a completely different behavior compared to that described before: the linear approach is absent. One could think that the charge amplification is reached earlier than expected caused by an extra contribution inducing an electric field other than the high voltage supplies. This brings up the possibility of having an induced field from the photomultipliers, as they are very close to the anode. An extra anode plate was added to the assembly next to the PMTs. The corresponding data and the modified COMSOL simulations completed show that it does not justify the difference. 

Several other changes in the detector have been carried out, but the response is unchanged. The non-linear behavior of the detector at E/p values lower than those observed in other detectors and in this electroluminescence simulation remains unsolved. Future work includes implementing modifications to our setup to test the possibility of photoelectric effect in the gate mesh which is not included in Garfield++ simulations and the evaluation of possible photo-ionizing impurities. 

Overall, a deeper study on argon gas detectors is needed from the community to ensure a correct comprehension of the electroluminescence yield, as it is the main detection channel of many rare event searches, as dark-matter dual-phase TPCs or $0\nu\beta\beta$ TPC experiments. 
\renewcommand{\arraystretch}{1.5}

\begin{savequote}[65mm]
``Aditzaile onari, hitz gutxi"
\qauthor{-- Old proverb in Basque language.}
\end{savequote}

\chapter{Outlook and conclusions}
\label{sec:conclusion}

Coherent elastic neutrino-nucleus scattering (\cenuns) is the dominant interaction channel for neutrinos with energies below a few tens of MeV. However, despite its relative prominence at these energies, it remains a weak-scale process and is inherently challenging to detect. One major difficulty arises from the small energy transfer to the target nucleus — typically only a few keV. This challenge is further compounded by the inefficiency of converting such low-energy nuclear recoils into detectable signals. 

The discovery of \cenuns\ less than a decade ago has opened exciting new avenues for exploring physics beyond the Standard Model, spanning both particle and nuclear physics. Studying \cenuns\ will expand our incipient knowledge of neutrino properties in several different ways, and each of these approaches could lead to breakthroughs. Therefore, this work sets the stage for understanding the motivation behind pursuing new detection strategies, emphasizing the advantage of exploring the process with different - reactor, spallation and solar - neutrino sources and using a wide variety of nuclei targets. The synergy between Xe and CsI of detector mediums is a centerpiece for the next generation \cenuns\ precision experiments.

In the attempt to describe the neutron background at ESS installation, this thesis presents a comprehensive investigation into the feasibility and optimization of observing \cenuns. The optimal conditions of signal-to-background ratio should be found at this facility in an available experimental site $\sim$24 m from the tungsten target by adding a minimal shielding of $\sim$20 cm of HDPE. It evaluates the ESS as a promising site for \cenuns\ experiments, highlighting the advantages of spallation sources in this context. Nonetheless, due to construction delays, it concludes that other installations such as JPARC-MFL are also a great opportunity to pursue \cenuns\ measurements.

To support this assessment, the compact, low-cost, full-coverage neutron scatter camera developed to establish the experimental feasibility is introduced. Even if it was designed for measuring spallation neutrons in the future, this system has proven to determine also the direction and energy spectrum of neutrons from sources like $^{252}$Cf, and includes a 360$^{\circ}$ optical imaging component for identifying unshielded areas. Despite limitations when sources are too close, the camera performs well for distant sources. Advanced algorithms, which encompass detailed attention to neutron-gamma discrimination and the addition of neural networks, have been implemented for source localization and rapid visualization. The initial validation is done with an available $^{252}$Cf source, but future tests with sources like Am-Be beams and reactor environments are planned before going to a spallation source. Real-time reconstruction capabilities are also being pursued.

This work also sets the focus on the novel high-pressure noble gaseous time projection chamber with electroluminescence amplification named GanESS, specifically designed for \cenuns\ detection that allows the use of multiple targets such as argon, xenon, or krypton. It strategically undertakes all the previous knowledge acquired by the NEXT experiment. A key advantage is that, for such low-energy events, topological information is not important since the interactions are always point-like. This means that increasing the detector’s operating pressure does not result in any loss of information, making it an attractive strategy to boost the number of target nuclei without changing the detector's physical size. New challenges such as the detector stability and performance at high pressure, or the characterization of the quenching factor in the low-energy regime will be approached with the GaP detector prototype as a proof of concept for the GanESS detector. Here, the construction and an insight into the detector’s first experimental performance is provided. An energy threshold as low as 0.42$\pm$0.04 $\rm{keV}_{\rm{ee}}$ has been estimated when operating at 8.62 bar, confirming its enormous promise for \cenuns\ detection. 

Making particular emphasis on electroluminescence amplification, the counter-intuitive behavior observed, below the ionization threshold, could be related to charge amplification caused by the structure of the electroformed mesh. Motivated by the need to understand the observed phenomena, a Garfield++ toolkit to detail the physics processes in the gaseous detector has been developed. The care to detail during the description of the electric field in GaP is of crucial importance, and therefore, the finite element software COMSOL Multiphysics was used. The electroluminescence threshold is estimated to be 0.87$\pm$0.29 kV/cm/bar for argon and the ionization threshold $\sim$3.36 kV/cm/bar. These findings further support previous Garfield simulations and experimental measurements. However, the non-linear behavior of the GaP detector at low E/p values remains unsolved. Yet, enhanced efforts are still underway. 

Taken together, the studies presented in this thesis establish a solid foundation for \cenuns\ measurements at spallation sources and contribute meaningfully to the development of novel detector technologies with broader applications in neutrino physics and low-background experiments, seeking new physics searches. The two major achievements of this work are the following: the original, portable and pioneer 4$\pi$ coverage neutron scatter camera with an integrated optical camera, and the promising results obtained from the GaP prototype both indicating a low energy threshold and demonstrating stability at high pressure operations.

\begin{appendices}
\chapter{Neutron scatter camera: 4$\pi$ coverage demonstration}
\label{app:appendix}

This appendix shows the extension of the results described in section~\ref{sec:ncameralocrecons} proving the performance of the  position reconstruction algorithm developed. 

\begin{figure}[htb]
\centering
\hspace{-0.1in}
\includegraphics{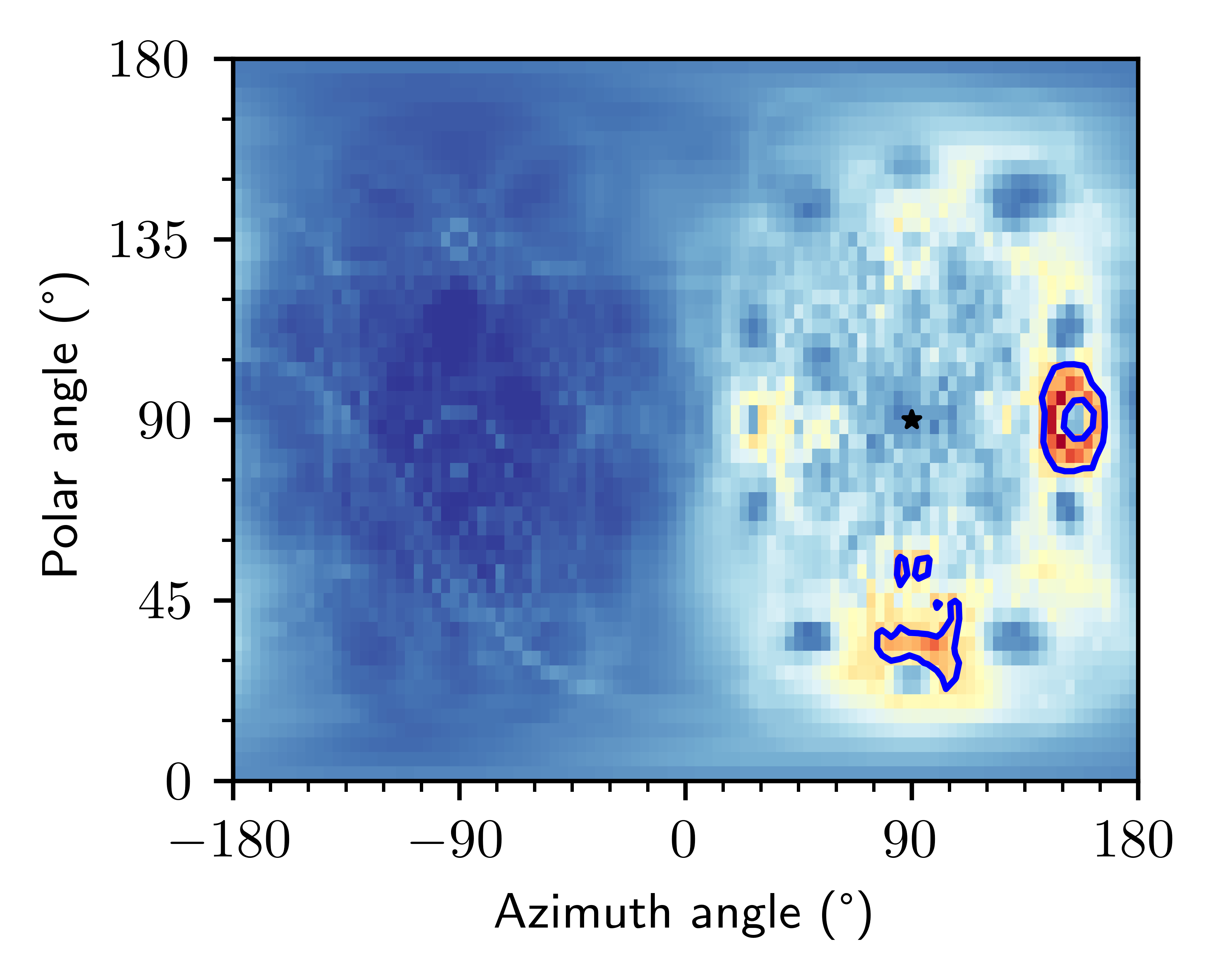}
\hspace{0.1in}
\includegraphics{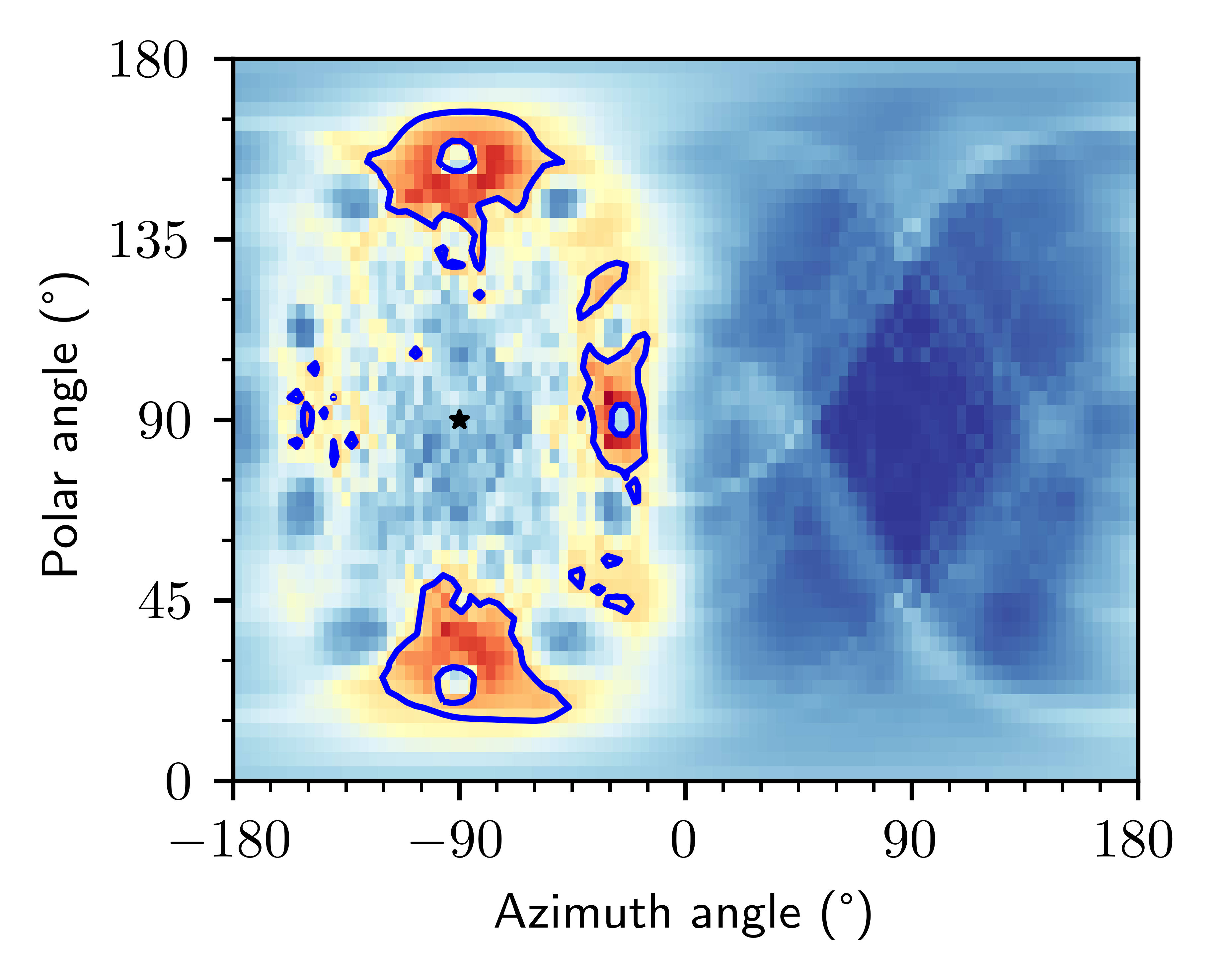}
\caption{Localization of a $^{252}$Cf source placed at the opposing faces of the aluminum frame holding the scintillator volumes, which correspond to centers ``1-5" and ``2-6" respectively. In black the true position of the source and in blue the 2D 1$\sigma$ contour line. Several globes appear due to the fixed position of the detector volumes, resulting in identifying the biased position of the scintillators of the first energy deposition, instead of the source that was placed between 4 scintillator volumes.}
\label{fig:appOddEven}
\end{figure}

The polar results of the position reconstructions of the $^{252}$Cf source placed at the corners of the aluminum frame that holds the scintillator volumes as labeled in figure~\ref{fig:labellingref} is presented here. One could have anticipated that the system's ability to resolve the source located between 4 scintillator volumes is very limited resulting in identifying the coordinates of the surrounding detectors instead of the source in the center. So, except for figure~\ref{fig:appOddEven}, where the reconstruction is biased by the intrinsic position of the detector volumes surrounding the source, in general, the camera performs a well-bounded position reconstruction of the true source position, recalling that the device was designed for neutrons in a higher energy range and coming from further away. The proximity of the source has strong impact on the position reconstruction, resulting in a more point-like localization those $^{252}$Cf sources placed further from the neutron camera device, as observed in section~\ref{sec:resolve}. Nonetheless, the polar maps shown here demonstrated that the device we built is capable of localizing the $^{252}$Cf source positioned across each face of the every the neutron camera, validating the full 4$\pi$ coverage feature.

\begin{figure}[htb]
\centering
\hspace{-0.1in}
\includegraphics{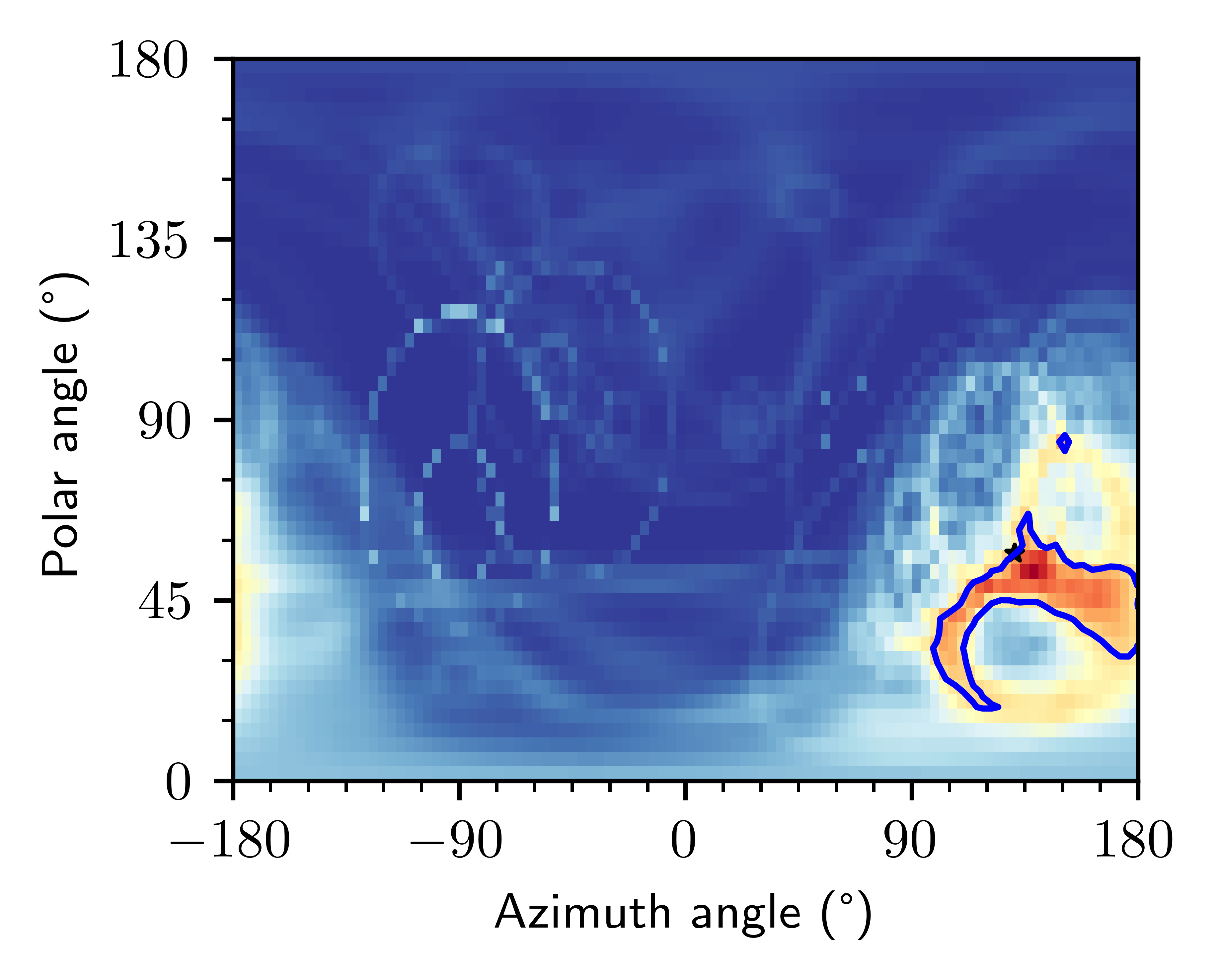} 
\hspace{0.1in}
\includegraphics{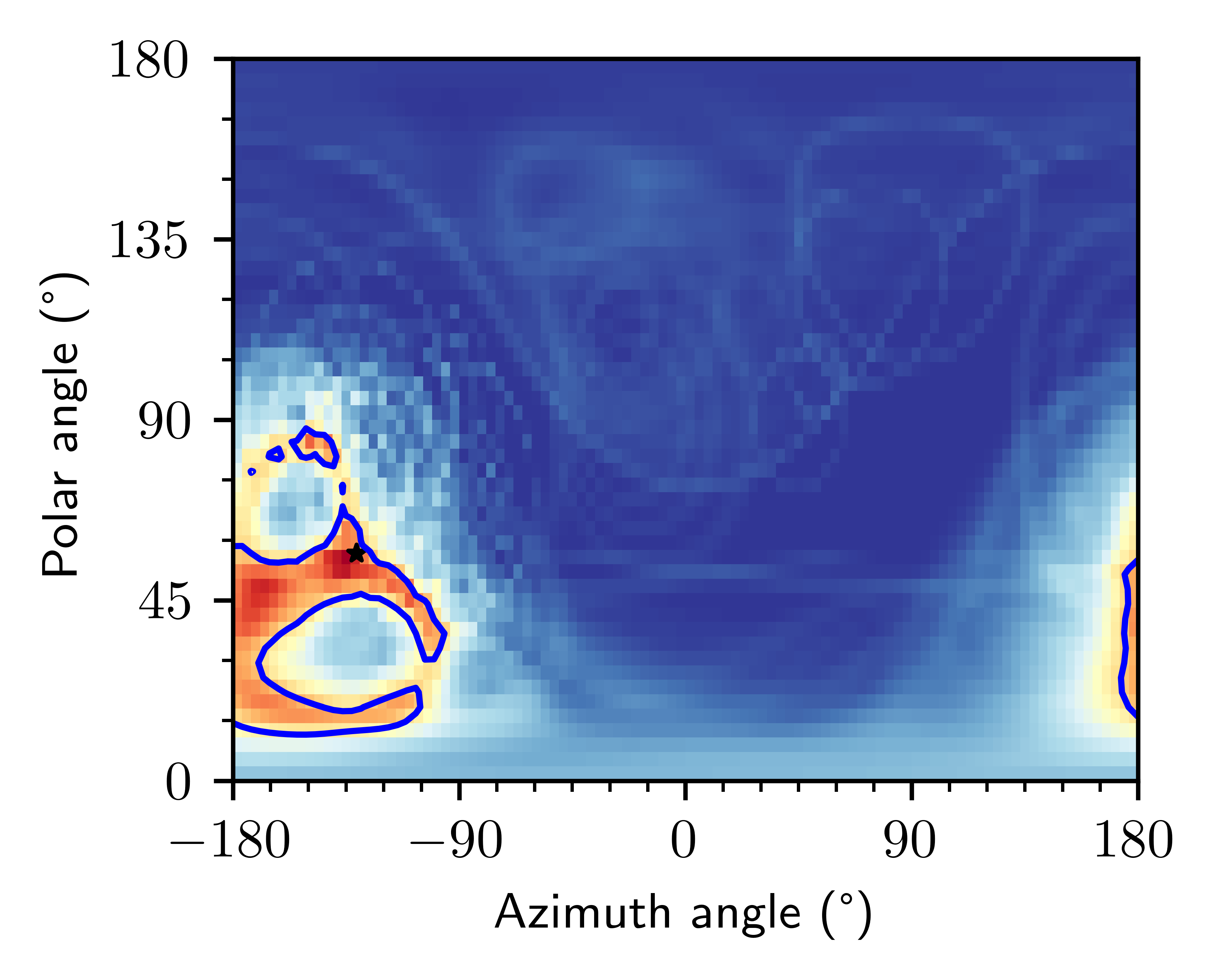}
\caption{Localization of a $^{252}$Cf source placed at the opposing corners of the aluminum frame holding the scintillator volumes, which correspond to corners ``1-7" and ``2-8" respectively. In black the true position of the source and in blue the 2D 1$\sigma$ contour line.}
\end{figure}

\begin{figure}[htb]
\centering
\hspace{-0.1in}
\includegraphics{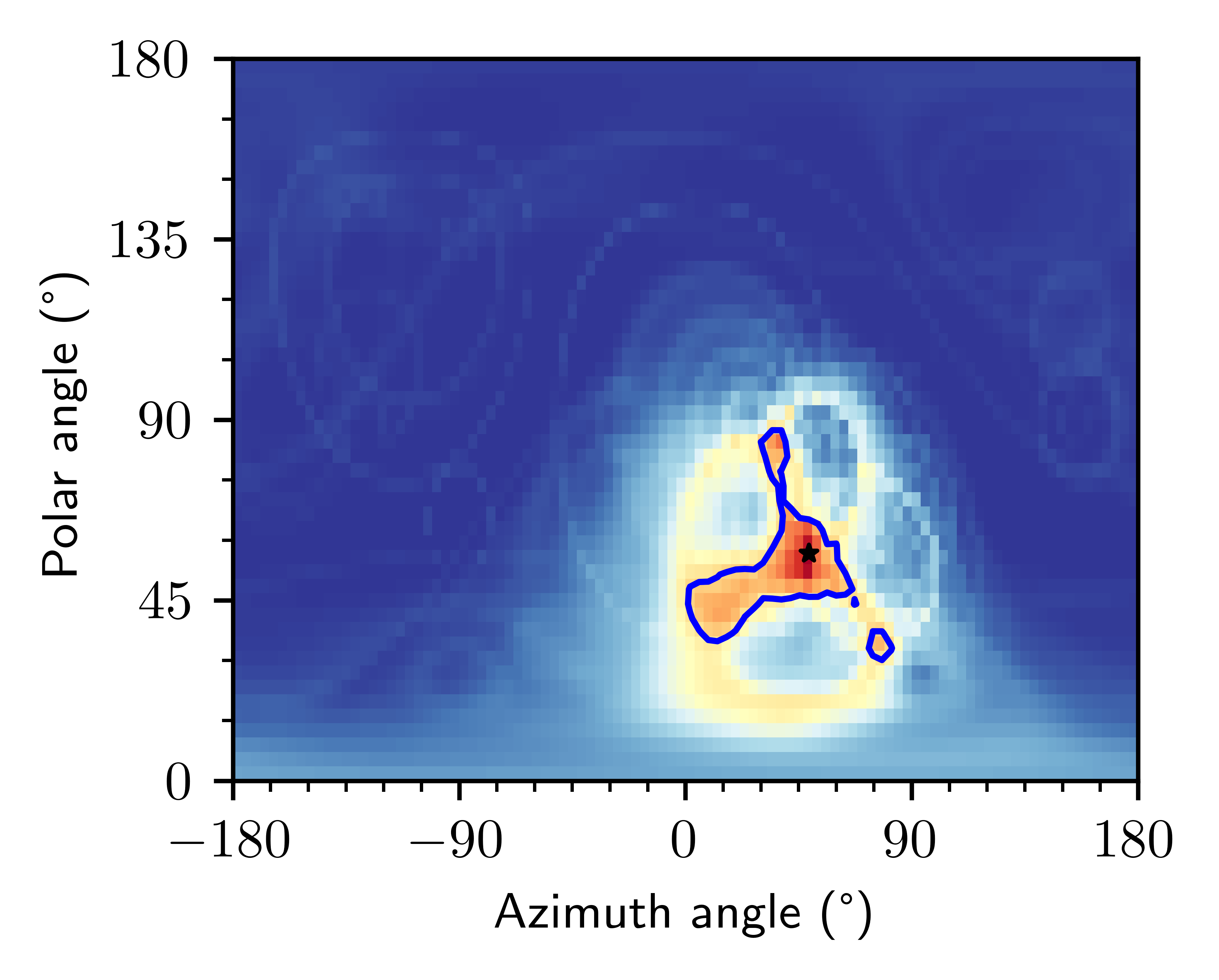}
\hspace{0.1in}
\includegraphics{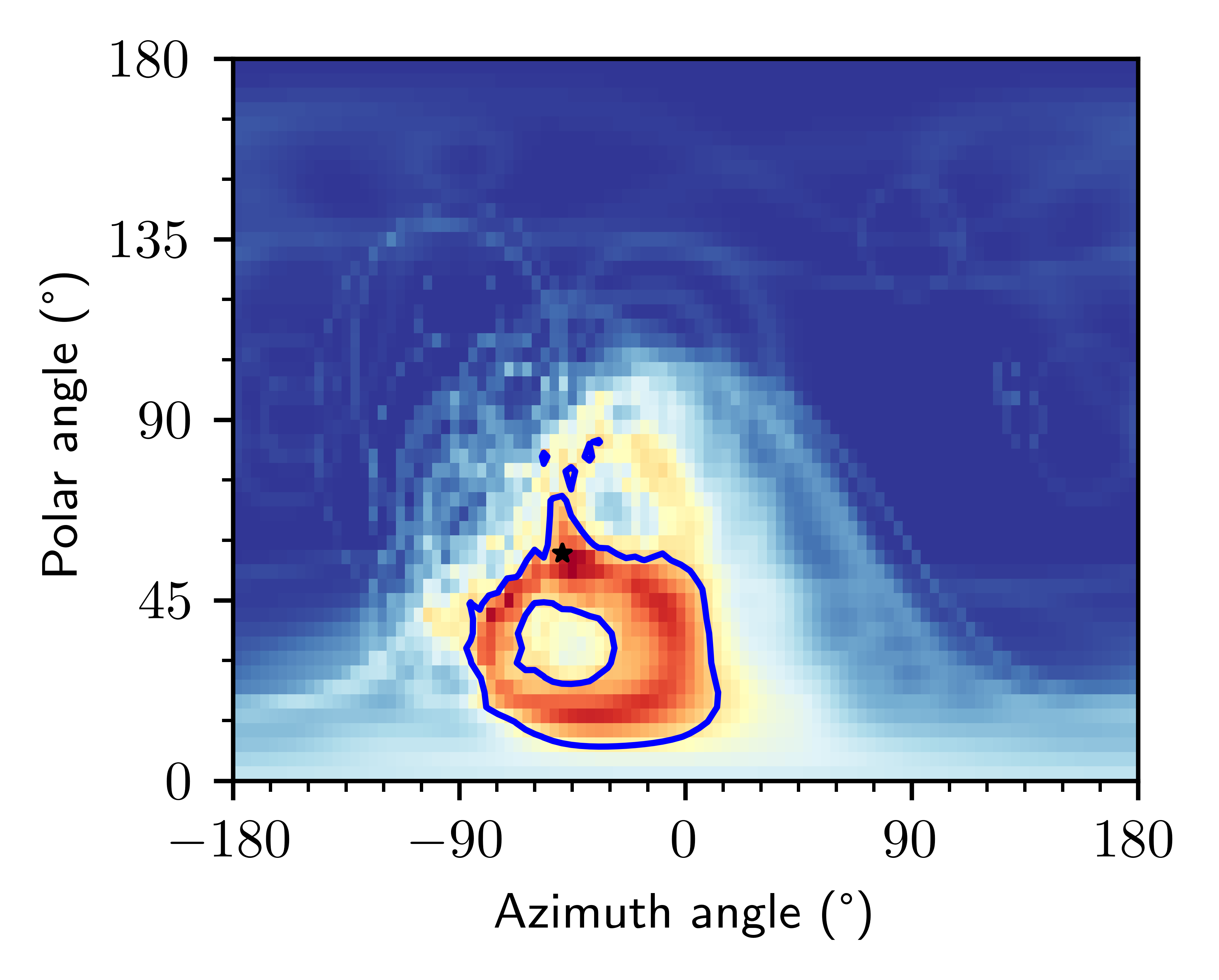}
\caption{Localization of a $^{252}$Cf source placed at the opposing corners of the aluminum frame holding the scintillator volumes, which correspond to corners ``1-3" and ``2-4" respectively. In black the true position of the source and in blue the 2D 1$\sigma$ contour line.}
\end{figure}

\begin{figure}[htb]
\centering
\hspace{-0.1in}
\includegraphics{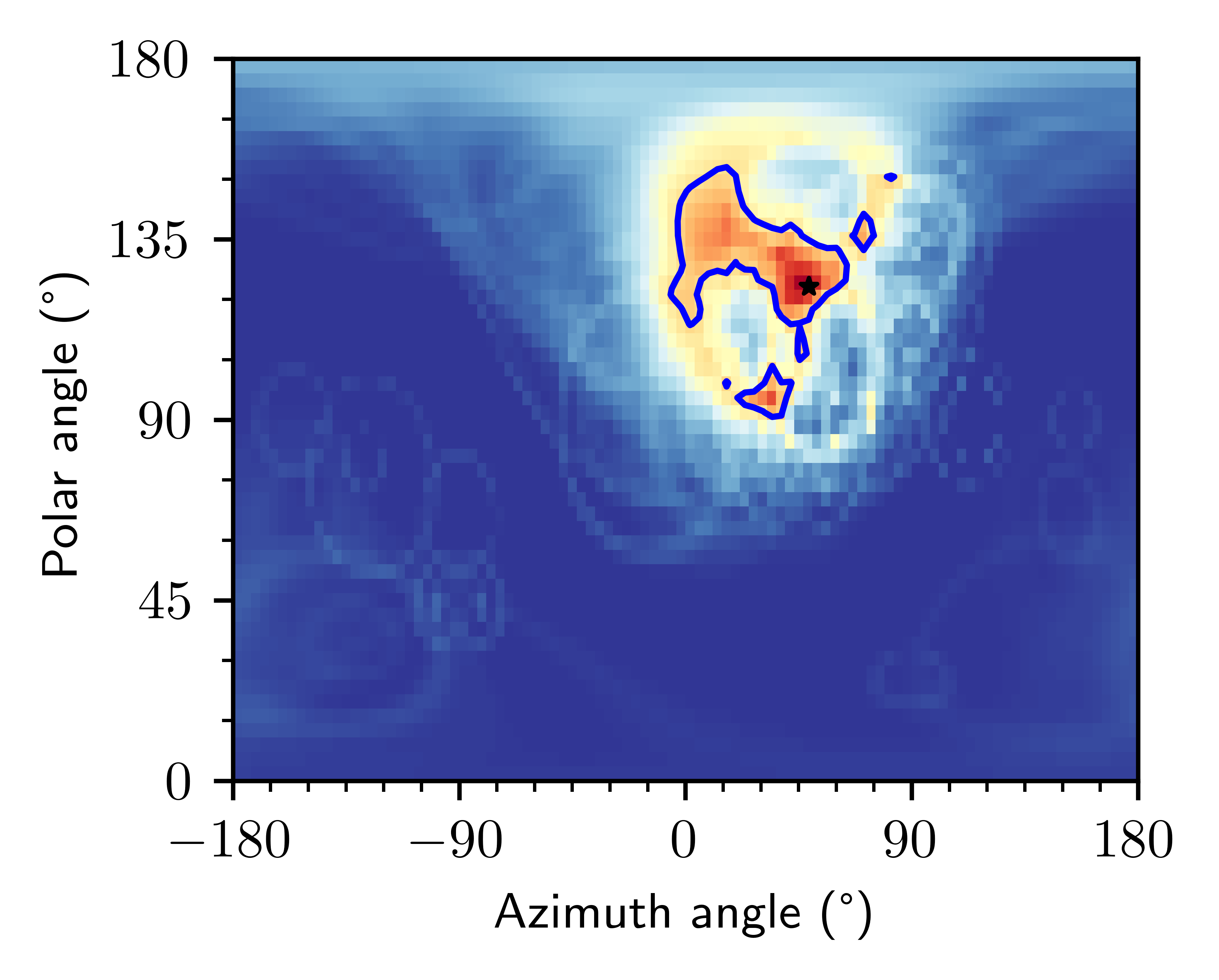}
\hspace{0.1in}
\includegraphics{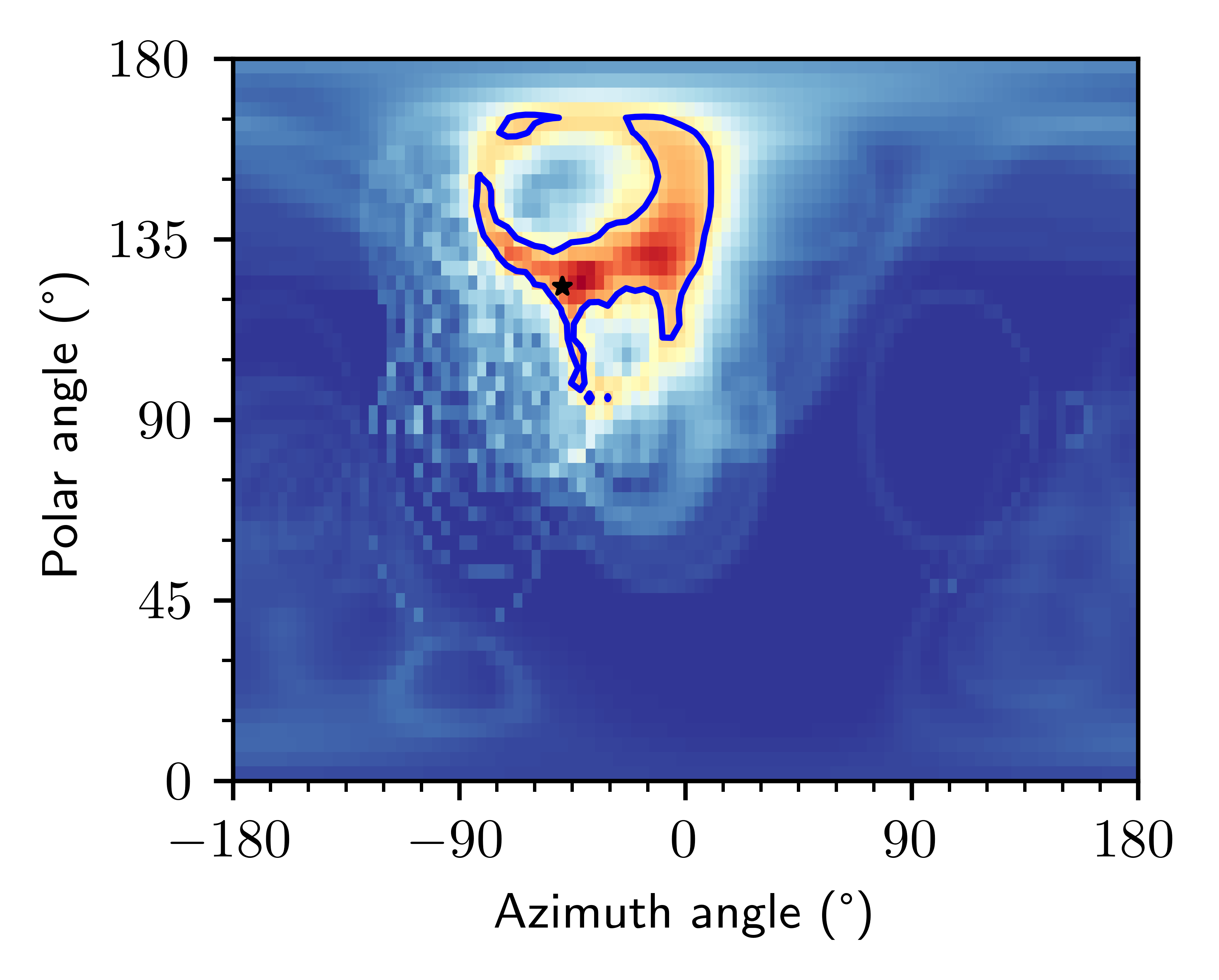}
\caption{Localization of a $^{252}$Cf source placed at the opposing corners of the aluminum frame holding the scintillator volumes, which correspond to corners ``3-5" and ``4-6" respectively. In black the true position of the source and in blue the 2D 1$\sigma$ contour line.}
\end{figure}

\begin{figure}[htb]
\centering
\hspace{-0.1in}
\includegraphics{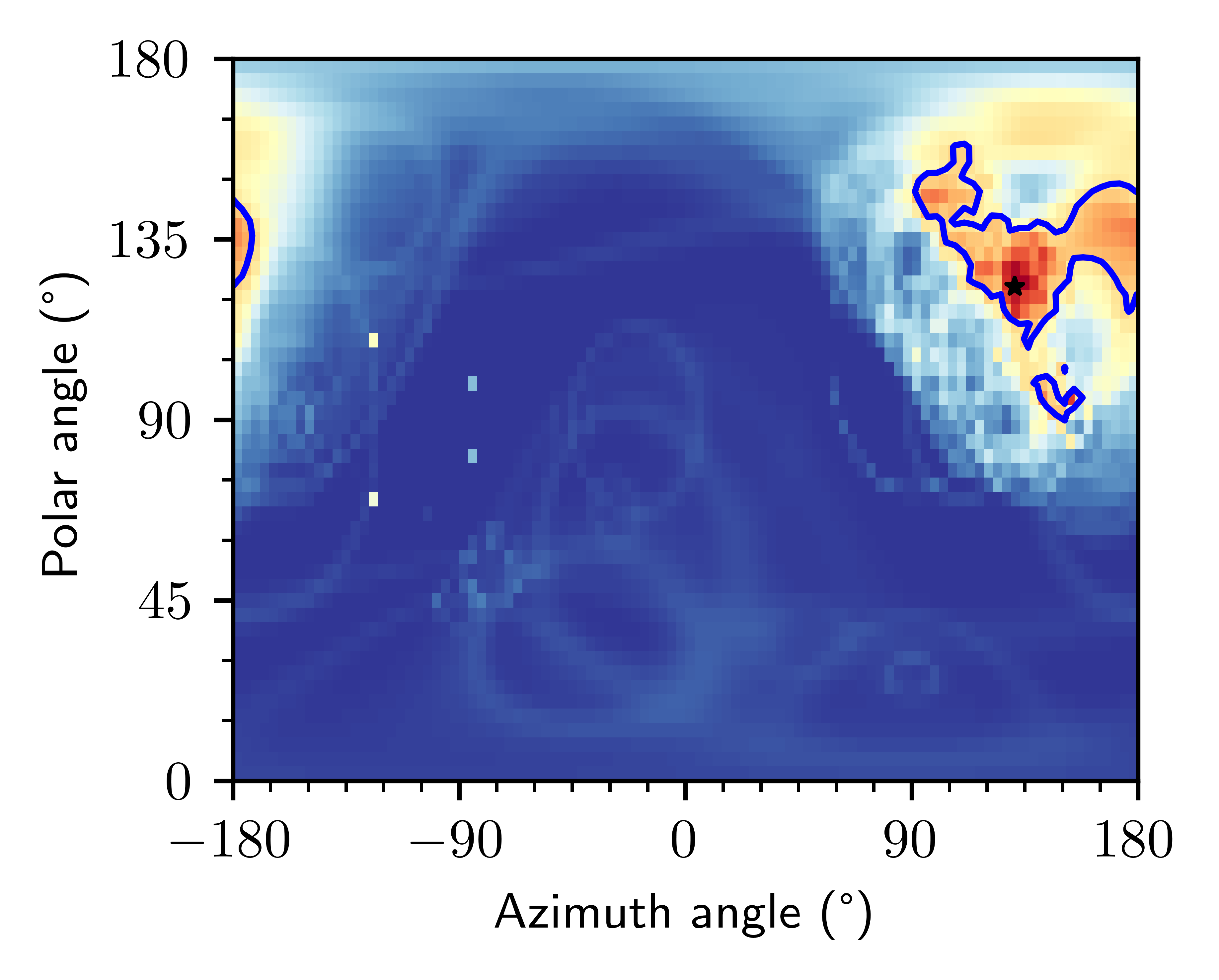}
\hspace{0.1in}
\includegraphics{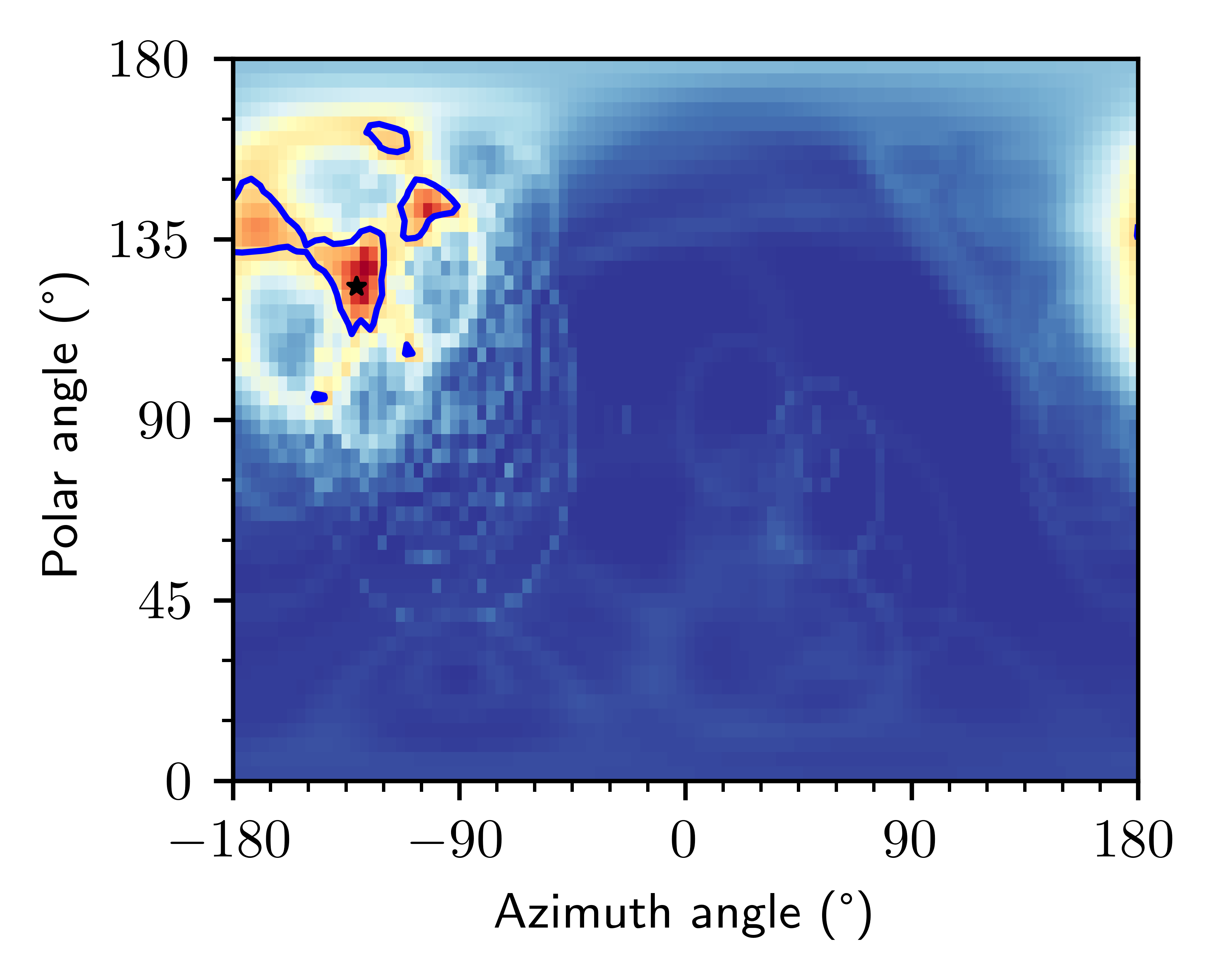}
\caption{Localization of a $^{252}$Cf source placed at the opposing corners of the aluminum frame holding the scintillator volumes, which correspond to corners ``5-7" and ``6-8" respectively. In black the true position of the source and in blue the 2D 1$\sigma$ contour line.}
\end{figure}

\begin{figure}[htb]
\centering
\hspace{-0.1in}
\includegraphics{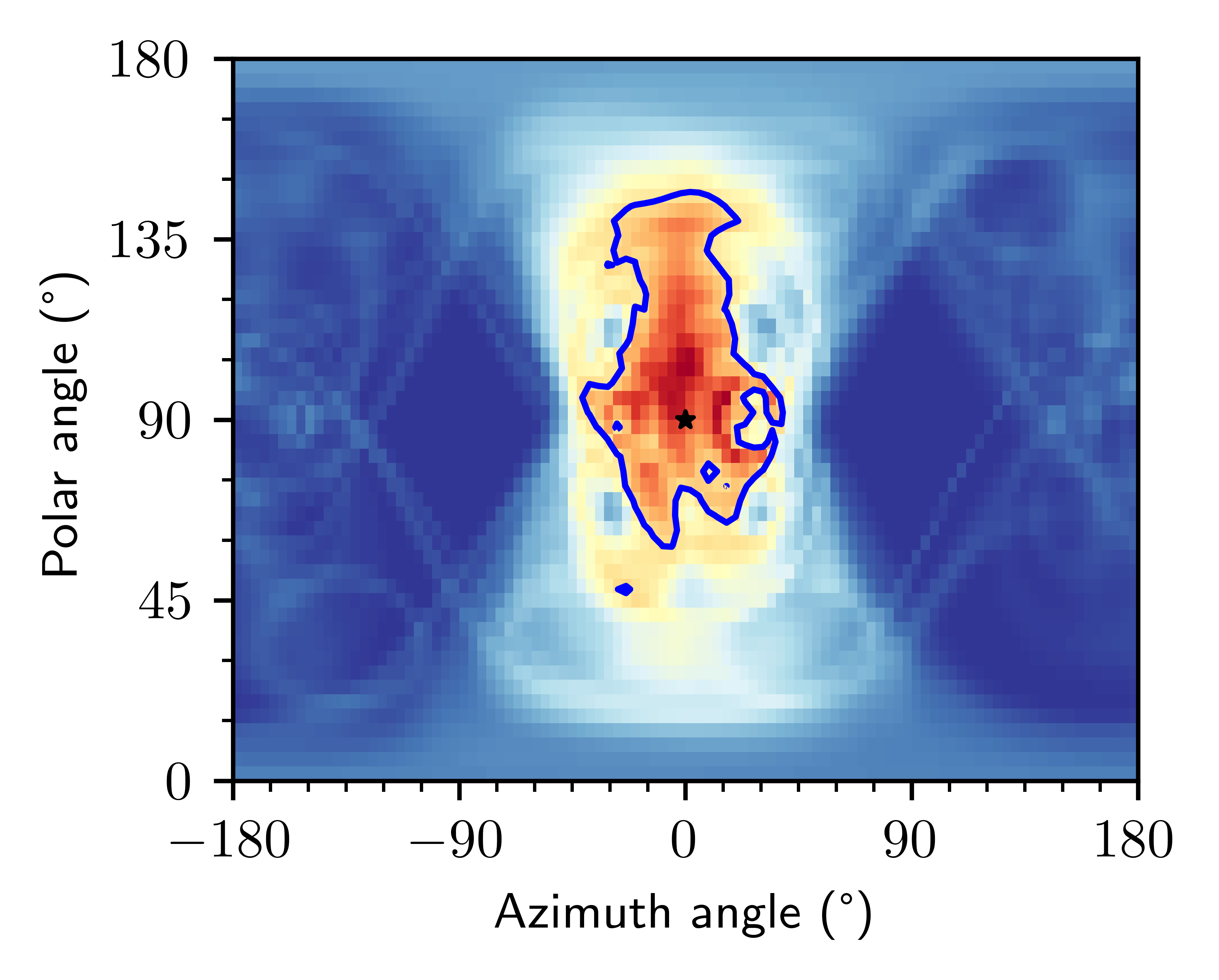}
\hspace{0.1in}
\includegraphics{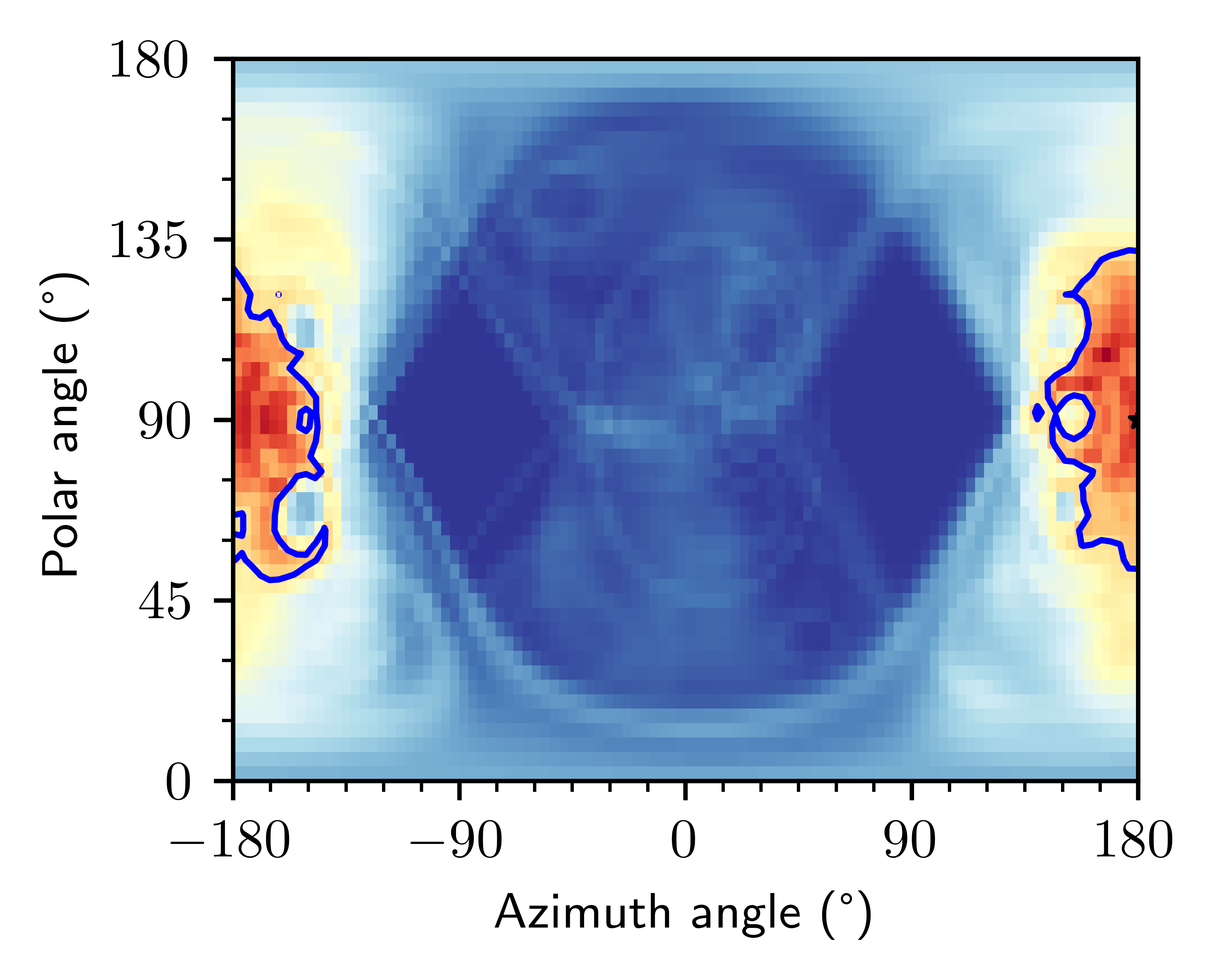}
\caption{Localization of a $^{252}$Cf source placed at the opposing faces of the aluminum frame holding the scintillator volumes, which correspond to centers ``3-4" and ``7-8" respectively. In black the true position of the source and in blue the 2D 1$\sigma$ contour line.}
\end{figure}

\begin{figure}[htb]
\centering
\hspace{-0.1in}
\includegraphics{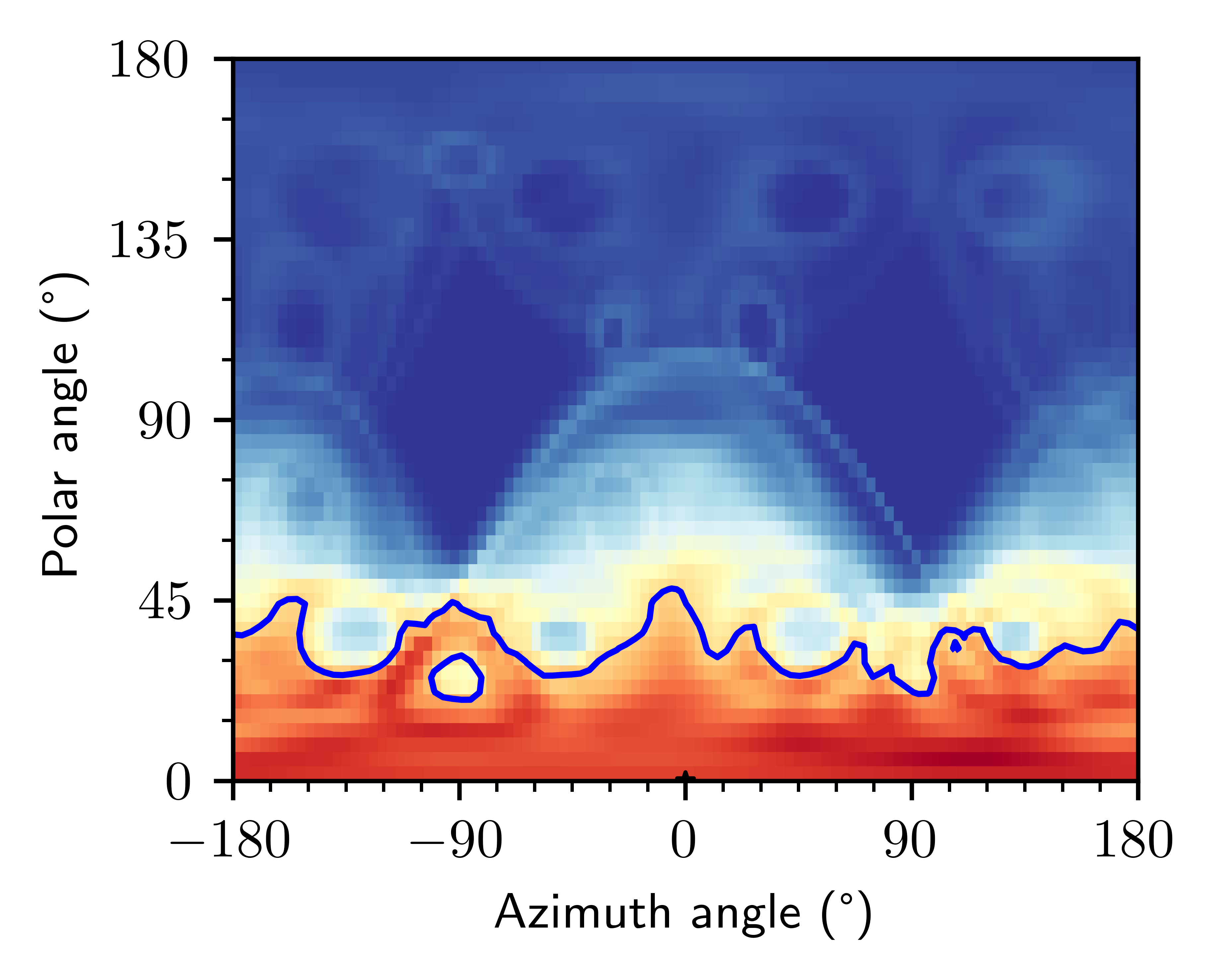}
\hspace{0.1in}
\includegraphics{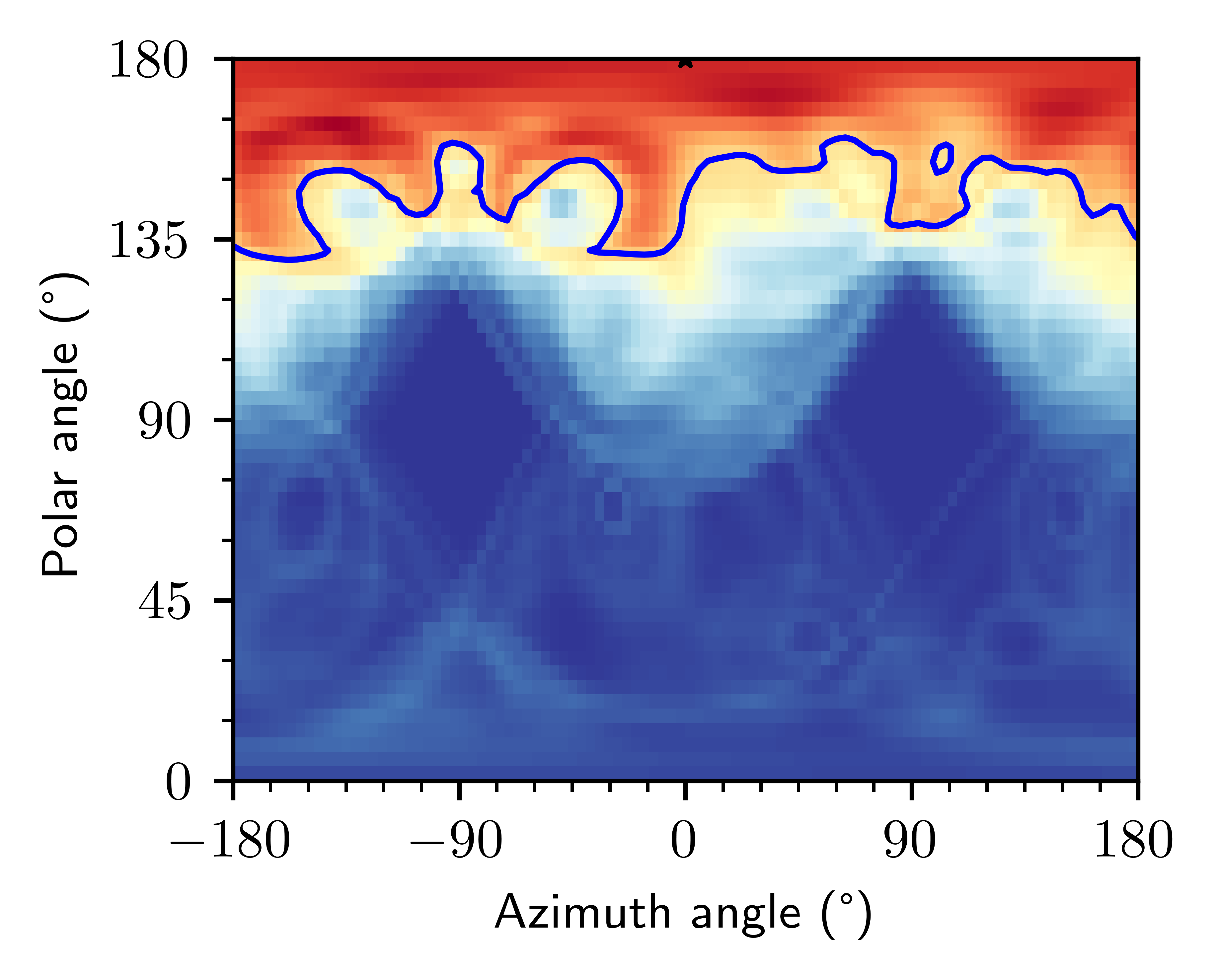}
\caption{Localization of a $^{252}$Cf source placed at the opposing faces of the aluminum frame holding the scintillator volumes, which correspond to centers ``1-2" and ``5-6" respectively. In black the true position of the source and in blue the 2D 1$\sigma$ contour line.}
\label{fig:appTop}
\end{figure}

\chapter{Spotting the hidden $^{252}$Cf source}
\label{app:appendixANN}

This appendix is intended to point out two major upgrades that we are currently working on in order to improve the operation of neutron scatter camera introduced in chapter~\ref{sec:ncamera}. Concretely, these changes aim to tackle the distortions that the 360$^{\circ}$ image generate when showing the reconstructed source localization overlapped to the photo, and the weakening of the back-projection process from the indirect paths that neutron scatters with the environmental geometry cause. 

In addition to figure~\ref{fig:ncamerarecoCf2}, the reconstruction and the performance of the neutron scatter camera is shown here an attempt at identifying a $^{252}$Cf source that was placed somewhere on the overall system and thus, closer to the detector volumes than when the source was place on the wall. Therefore, one would expect a worse reconstruction than the further position.  

\begin{figure}[tbh!]
    \centering
    \includegraphics[width=0.7\textwidth]{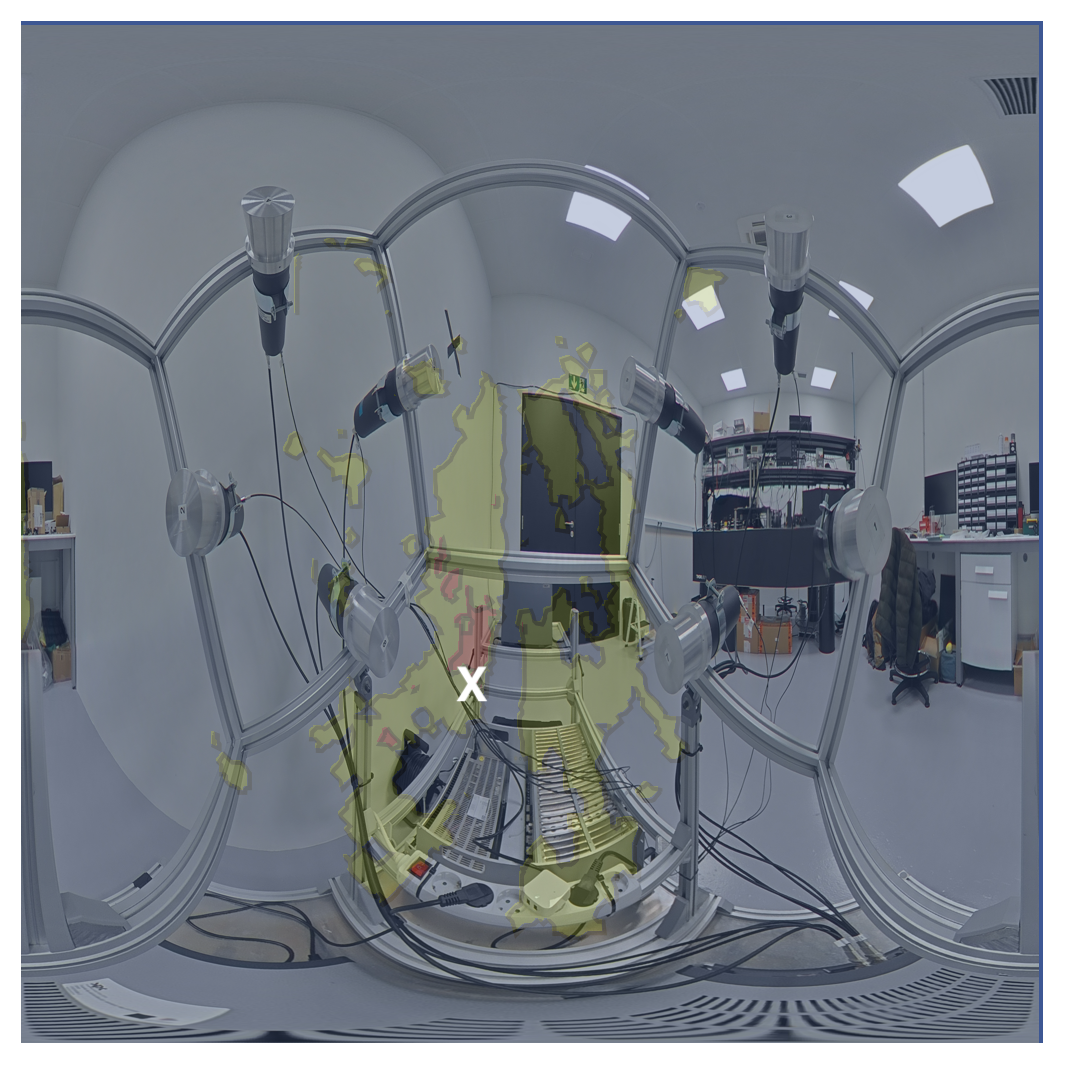} 
    \caption{Overlay of the reconstructed source image and a 360$^{\circ}$ photo after implementing trained ANN for the coordinate conversion between the optical pixels and the reconstructed ($\theta, \phi$) space. The $^{252}$Cf source was placed in the ``X" mark next to the laptop-table. The reconstructed source location enclosed by 2D 1$\sigma$ region in red and 2$\sigma$ region in green comes in agreement with the true position. So even if the reconstruction is well-defined, the overlap with the photo is displaced due to the distortion of photo itself in the region where the source was placed, and the limitations in the interpolation.}
\label{fig:ncamerarecoCf1}
\end{figure}

Figure~\ref{fig:ncamerarecoCf1} shows how our device intercepts a source that was located next to the laptop table. Even if the contour map is well bounded to a very small region (red region), there is a small mismatch between the reconstructed contour and the real position in the image. This is because the source was placed in a region where the 360$^{\circ}$ image is highly distorted, which introduces weaknesses in the interpolation of the reference points to perform the overlap of the optical image and the reconstructed contour map. The result show an good reconstruction of the source, however displaced in the photo due to the interpolation limitations. So, this reinforces the performance of the camera and the reconstruction algorithms, but points out the need to define more in detail overlay calibration.  

\begin{figure}[tbh!]
    \centering
    \includegraphics[width=0.7\textwidth]{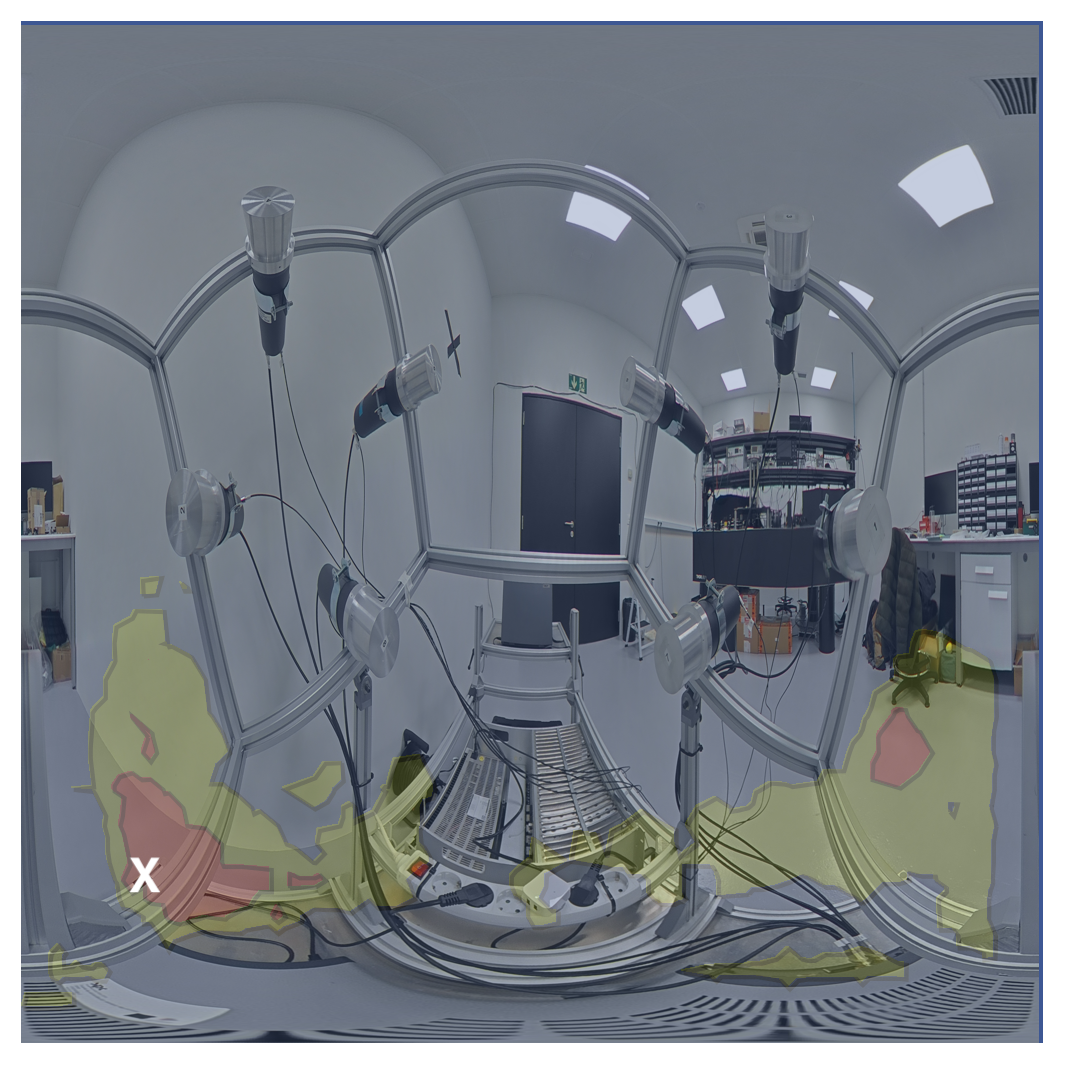} 
    \caption{Overlay of the reconstructed source image and a 360$^{\circ}$ photo after implementing trained ANN for the coordinate conversion between the optical pixels and the reconstructed ($\theta, \phi$) space. The $^{252}$Cf source was placed in the ``X" mark on the base of the neutron camera chart. The reconstructed source location enclosed by 2D 1$\sigma$ region in red and 2$\sigma$ region in green does not show a good reconstruction of camera at these conditions. The low angular resolution is mainly caused by neutron scatters with the environmental geometry, such as the room-floor or the aluminum chart base, prior to the first detectable scatters in a scintillator.}
\label{fig:ncamerarecoCf3}
\end{figure}

Similarly, figure~\ref{fig:ncamerarecoCf3} demonstrates the performance of pinpointing a source placed on one of the corners of the neutron camera chart base. Even if the measurement obtained shows lower angular resolution, this result brings up several shortcomings and stresses the need to address them as part of the upcoming improvements. The distortion of the image has an effect, and as mentioned when discussing figure~\ref{fig:ncamerarecoCf1}, a finer interpolation grid on the photo and an improved ANN will help solving deviations or spreads in the contour regions. Nevertheless, the primary reason for such a wide reconstructed region comes from the placement of the source itself. Apart from being closer to the detector volumes, and therefore, being biased by the finite detector position, the source was placed on the chart base. The base, which is an aluminum plate, together with the room-floor act as an interacting medium for the $^{252}$Cf source neutrons before the first detectable scatter in one of the scintillators. This blurs the neutron path, and as a consequence, the reconstruction of the neutron trajectories lacks of precision. The combination of neutron scatters with the environmental geometry and the direct ones results in the reconstruction shown in figure~\ref{fig:ncamerarecoCf3}, which deviates from the expected point-like localization of the $^{252}$Cf source. More realistic simulations, including room foundations for instance, could help understanding the impact this has on the reconstruction and address it. In addition, we are working on improved pixel-weighted techniques in order to pinpoint the source with better angular resolution.

\end{appendices}


\iffull
\fi


\iffull
\fi

\makebibliography


\end{document}